\documentclass[
	 floatfix,
        amsfonts,
        aps,
        prd,
        12pt,
        eqsecnum,
        bibtotoc,
        tightenlines,
        secnumarabic,
        preprint,
        nofootinbib,
        showpacs]{revtex4}

\usepackage{graphicx}
\usepackage{epsfig,psfrag}
\usepackage[american,german]{babel}
\usepackage{subfigure}
\usepackage{color}

\usepackage{amsmath,amssymb,amstext,bbm}
\usepackage{bm}                         

\newcommand{\F}{\mathcal{F}}
\newcommand{\tr}{\mathrm{tr}}
\newcommand{\N}{\mathcal{N}}
\newcommand{\A}{\mathcal{A}}
\newcommand{\B}{\mathcal{B}}
\newcommand{\wn}{\mathfrak{w}}
\newcommand{\mn}{\mathfrak{m}}
\newcommand{\qn}{\mathfrak{q}}
\newcommand{\R}{\mathfrak{R}}

\newcommand{\dd}{\mathrm{d}}

\newcommand{\del}{\partial}
\newcommand{\dt}{\tilde{d}}
\newcommand{\ft}{\tilde{f}}
\newcommand{\vrho}{\varrho}

\renewcommand{\O}{\mathcal{O}}
\renewcommand{\S}{\mathcal{S}}
\renewcommand{\L}{\mathcal{L}}

\def\RE {I\kern-6pt R    }
\def\Z  {Z\kern-13pt Z   }

\def\bi {\begin{itemize} }

\def\ei {\end{itemize}   }
\def\gtwid{\mathrel{\raise.3ex\hbox{$>$\kern-.75em\lower1ex\hbox{$\sim$}}}}
\def\ltwid{\mathrel{\raise.3ex\hbox{$<$\kern-.75em\lower1ex\hbox{$\sim$}}}}
\input epsf

\begin{document}

\selectlanguage{american}
\preprint{MPP-2008-94}
\title{Holographic quark gluon plasma with flavor}
\author{Matthias Kaminski}
\email{kaminski@mppmu.mpg.de}
\affiliation{
         Max-Planck-Institut f\"ur Physik (Werner-Heisenberg-Institut),
         F\"ohringer Ring 6,
         80805 M\"unchen, Germany}

\pacs{11.25.Tq, 11.25.Wx, 12.38.Mh, 11.10.Wx}

\begin{abstract}
In this work I explore theoretical and phenomenological implications 
of chemical potentials and charge densities inside a strongly coupled 
thermal plasma, using the gauge/gravity correspondence.
Strong coupling effects discovered in this model theory are interpreted geometrically and may 
be taken as qualitative predictions for heavy ion collisions at RHIC and LHC.
In particular I examine the thermodynamics, spectral functions, 
transport coefficients and the phase diagram of the strongly coupled plasma. 
For example stable mesons, which are the analogs of the QCD Rho-mesons, 
are found to survive beyond the deconfinement transition.
This paper is based on partly unpublished work performed in the context of my PhD thesis.
New results and ideas extending significantly beyond those published until now are stressed.
\end{abstract}

\maketitle

{\it 
 {\bf How to read this:}
This paper is based on the author's work partly published 
in~\cite{Erdmenger:2007ap,Erdmenger:2007ja,Erdmenger:2008yj,Dusling:2008}.
New results extending significantly beyond those published until now are reported in 
sections~\ref{sec:anaHydroIso},~\ref{sec:thermoB&I},~\ref{sec:hiNf},~\ref{sec:peakTurning},~\ref{sec:isospinDiffusion},
and~\ref{sec:charmDiffusion}. Completely new ideas are developed in the three outlook 
sections~\ref{sec:molecular},~\ref{sec:mesonSpectraB&I} and~\ref{sec:diffusionMatrix}.
}\bigskip\\

\tableofcontents

%

\section{Introduction} \label{sec:introduction}
The {\it standard model} of particle physics is a theory of the four known fundamental forces of nature 
which has been tested and confirmed to incredibly high precision~\cite{PDBook}.
Unfortunately the standard model treats gravity and the remaining three forces on different footings, 
since gravity is merely incorporated as a classical background.
{\it String theory} is a mathematically well-defined and aesthetic theory successfully unifying gravity with all other 
forces appearing in string theory~\cite[for example]{Polchinski:1998rq,Polchinski:1998rr}, which unfortunately lacks 
any experimental verification until now. 
In this respect string theory and the standard model of particle physics can be seen as complementary approaches which had been 
separated by a gap whose size even was hard to estimate. The advent of {\it AdS/CFT} or more generally the
{\it gauge/gravity correspondence}~\cite{Maldacena:1997re}~(explained in chapter~\ref{sec:adsCft}) and its
intense exploration during the past ten years now provides us with the tools to build a bridge over this gulch, 
a bridge to connect the experimentally verified gauge theory called the standard model with the consistently unifying
novel concepts of string theory. AdS/CFT amends both string theory and the standard model. In particular
the {\it duality}-character of the gauge/gravity correspondence can be used to extend our conceptual understanding 
to thermal gauge theories at strong coupling~\cite{Son:2007vk} such as those found to govern the thermal plasma 
generated at the Relativistic Heavy Ion Collider~(RHIC) at Brookhaven National Laboratory~\cite{Gyulassy:2004zy}.

{\bf The standard model and its limitations}
In order to set the stage for our calculations and to fit them into the `terra incognita' on the currently accepted
map of particle physics, we start out by reviewing the {\it standard model} and its limitations.
At the time the {\it standard model} of particle physics~\cite[for an introduction]{Peskin:1995ev} is a widely accepted model 
for the microscopic description of fundamental particles and their interactions. It claims that in nature two sorts
of particles exist: matter particles~(these are fermions, i.e. they carry spin quantum number~$1/2$) and
exchange particles~(these are vector bosons, i.e. they carry spin quantum number~$1$). The matter particles interact
with each other by swapping the exchange particles. This means that the exchange particles mediate the attractive
and repulsive forces between the matter particles. The matter particle content of the standard model is given by
table~\ref{tab:smMatter}. As seen from this table the matter particles are organized into three families of so called 
{\it leptons} and {\it quarks} which differ by their mass and quantum numbers. In this thesis the behavior of these 
quarks~\footnote{To be more precise we have to
take in account that the theory we will be using in this work as a computable model for strong coupling 
behavior is the supersymmetric~$\N=4$ Super-Yang-Mills theory coupled to a~$\N=2$ fundamental 
hypermultiplet. This hypermultiplet contains both fermions and scalars due to supersymmetry and we will refer
to both of them as quarks.} will be studied in a regime where a perturbative expansion of the standard model is not possible. 
In particular in chapter~\ref{sec:thermalSpecFunc} we will study how quarks are bound into quark-antiquark states 
(called {\it mesons}) inside a plasma at finite temperature. Furthermore we will examine 
the transport properties of quarks and mesons inside a plasma in chapter~\ref{sec:transport}.
\begin{table}
\begin{tabular}{|c|c|c|c|c|}
\hline
Fermions & Family & Electric charge & Color charge & Weak isospin \\
 &1\; 2 \; 3 & & & left-handed \hfill right-handed \\
\hline
\hline
Leptons & $\nu_e\; \nu_\mu\; \nu_\tau$ & 0 & / & $1/2$\hfill  /  \\
  & $e\; \mu\; \tau$ & -1 & / & $1/2$ \hfill 0 \\
\hline
Quarks & $ u\; c\; t $ & $+2/3$ & r, b, g & $1/2$ \hfill 0 \\
  & $ d\; s\; b$ & -1/3 & r, b,  g & 1/2 \hfill 0 \\
\hline
\end{tabular}
\caption{ \label{tab:smMatter}
The matter particles of the standard model carrying spin~$\frac{1}{2}$ grouped into families by their masses~\cite{PDBook}.
}
\end{table}

The exchange particles given in table~\ref{tab:smExchange} are responsible for the mediation of the three fundamental 
forces: the electromagnetic force, the weak force and the strong force. 
\begin{table}
\begin{tabular}{|c|c|c|c|c|}
\hline
Interaction & couples to & Exchange particle & Mass (GeV) & $J^P$ \\
\hline
\hline
strong & color charge                    & 8 gluons & $0$ & $1^{-1}$ \\
electromagnetic & electric charge & photon    & $0$ & $1^{-1}$ \\
weak & weak charge                     & $W^{\pm},\,Z^0$ & $\sim 10^2$ & $1$ \\
\hline
\end{tabular}
\caption{ \label{tab:smExchange}
The exchange particles of the standard model carrying spin~$1$, the interaction or force they mediate and 
the charge to which they couple~\cite{PDBook}. 
}
\end{table}

Technically the standard model is a {\it quantum field theory} and as such incorporates the ideas of quantum mechanics,
field theory and special relativity. Starting from the classical theory of electrodynamics it is clear, that if we want to 
apply it to the small scale of fundamental particles, we need to consider effects appearing at small scales which are
successfully described by quantum mechanics. From this necessity
{\it quantum electrodynamics}~(QED) emerged as the unification of field theory and quantum mechanics describing
the {\it electromagnetic force}. Next it was 
discovered that the force which is responsible for the beta-decay of neutrons in atomic nuclei, called the {\it weak force}
can be described by a quantum field theory as well. The standard model unifies these two quantum field theories
to the electro-weak quantum field theory. The third force, the {\it strong} one is described by {\it quantum chromodynamics}~(QCD)
which the standard model fails to unify with the electro-weak theory. Both electro-weak theory and QCD are based 
on the concept of {\it gauge theories}. This means that the quantum field theory is gauged by making its symmetry
transformations local~(i.e. dependent on the position in space-time). 
By gauging a theory new interactions among matter particles and gauge bosons arise (e.g. the electromagnetic,
weak and strong interaction in the standard model).
This kind of gauge theories is the one which is studied in the AdS/CFT correspondence -as described in chapter~\ref{sec:adsCft}- 
which may also be called gauge/gravity correspondence.

Up to now we have introduced the standard model as an interacting quantum field theory but in this setup none of the 
particles has a nonzero mass, yet.
Thus one important further ingredient to the standard model which is not yet experimentally confirmed is the Higgs boson.
This particle is a spin~0 field which is supposed to generate the masses for the standard model particles via the
Higgs mechanism~\cite{Higgs:1964pj}. 

The standard model leaves many questions open of which we mention only three: 
The weak force is $10^{32}$~times larger than gravity. Where does this hierarchy in coupling strengths come from? 
Due to its modeling character the standard model has (at least)~18 parameters~(masses and coupling constants) 
which need to be put in by hand. What are the physical mechanisms fixing the values of these parameters?
How can gravity be incorporated into the gauge theory framework?

Some of these problems are theoretically solved by extensions of the standard model: The {\it minimal supersymmetric
standard model}~(MSSM)~\cite[for a status report]{Djouadi:1998di} explains the force 
hierarchy~(and also yields dark matter candidates). Some further 
phenomenologically studied extensions contain extra-dimensions~\cite[for a review]{Hooper:2007qk}, 
the non-commutative standard model with
non-commuting space-time coordinates~\cite{Calmet:2001na}~(recent progress may be found 
in~\cite{Najafabadi:2008sa,Alboteanu:2007bp,Chamseddine:2007ia,Buric:2007qx}) 
and the addition of an unparticle sector governed by conformal 
symmetry~\cite{Georgi:2007ek} which thus is closely related to the conformal theories we will review in section~\ref{sec:CFT}.
But the most developed and consistent theory known to incorporate gravity in the same conceptual way as all other forces 
is string theory~(note, that {\it loop quantum gravity}~\cite[for a recent review]{Ashtekar:2007tv} has the same goal). 

Finally, the standard model is computed as a perturbative expansion in the gauge coupling coefficients. Therefore this 
description relies on the coupling coefficients to be small. Due to the fact that the coupling 
constants are running~\cite[for pedagogical treatment]{Peskin:1995ev}~(i.e. they change as the energy
at which the particle collision is performed)
there are regimes where the standard model perturbation series is not applicable. The most prominent 
example of physics in such regimes is the quark gluon plasma generated in heavy ion collisions at the RHIC 
collider~\cite[for example]{Back:2004je,Fischer:2006xq}. Also the ALICE detector at the Large Hadron Collider~(LHC)
currently under construction will soon produce data from those strong coupling regimes. Exactly these regimes 
of gauge theories are now accessible~(with certain restrictions) by virtue of the AdS/CFT correspondence as described in
section~\ref{sec:stateAdsCft} and methodically introduced in chapter~\ref{sec:holoMethods}.

{\bf String theory}
{\it String theory} can solve some of the problems mentioned above mainly because of its 
fundamental and mathematically structured character. In string theory the fundamental objects
are not point-like particles but {\it strings}, i.e. one dimensional objects,
characterized by only one single parameter: the string tension~$\alpha'$. These strings have to be
embedded into ten-dimensional space-time. Furthermore, they have to satisfy certain boundary 
condition just like a classical guitar string. Closed strings are loops which can propagate through 
space-time, whereas the end points of open strings are confined to hyperplanes, so called {\it branes}.
The name {\it brane} for higher-dimensional hyperplanes is a generalization of the two-dimensional mem-{\it brane}.
As a heuristic picture one may imagine an open string to be similar to a guitar string, being able to carry different 
excitations. Just like each excitation of the guitar string corresponds to a distinct tone,
each excitation of a string can be identified with a distinct particle. The excitations of a closed string 
correspond to different particles. For example the {\it graviton} which is the massless spin 2 gauge
boson mediating the gravitational force emerges as the quadrupole oscillation of a closed string. Since other 
exchange particles such as the photon emerge in the same way as a distinct string excitation, this theory
provides a unified concept from which the gauge interactions arise, including gravity.
Therefore string theory is capable of giving conceptual explanations for the structure of matter and its 
interactions in terms of just one string tension parameter.
For its consistency string theory requires ten dimensions~(six of which need to be compactified), supersymmetry and
it is reasonable to give dynamics to the branes, as well.
We will learn a bit more about string theory in section~\ref{sec:dualitiesNString}
but a full treatment is beyond the scope of this thesis and the reader is referred to 
textbooks~\cite[for example]{Polchinski:1998rq,Polchinski:1998rr}.

Also string theory rises many problems. First of all it is not known how to obtain the standard model from string
theory and since that is the experimentally verified theory any conceptual extension has to incorporate it. 
A pending theoretical problem is the full quantization of string theory. And finally we stress again the lack 
of experimental predictions which could distinguish string theory from others, confirm it or rule it out. Without
a way to connect to reality and to verify string theory or at least the concepts derived from it, 
it is unfortunately useless for physics.

{\bf Current state of AdS/CFT} 
How does the gauge/gravity correspondence called {\it AdS/CFT} provide tools to connect string theory and 
possibly the standard model? 
{\it AdS/CFT} is the name originally given to a correspondence between a certain gauge theory with 
conformal symmetry~(i.e. it is scale-invariant) in four flat space-time dimensions on one side 
and supergravity in a five-dimensional space with constant negative curvature called anti de Sitter space-time~(AdS)
on the other side~\cite{Maldacena:1997re,Aharony:1999ti}. Due to the mismatch in dimensions which is 
reminiscent of {\it holography} in classical optics, the correspondence is sometimes called holography. 
This correspondence arises from a string theory setup taking intricate limits which we describe in detail 
in chapter~\ref{sec:adsCft}. Originally the conformal field theory considered
on the gauge theory side of the correspondence has been~$\N =4$ Super-Yang-Mills theory~(SYM). Today 
{\it gauge/gravity correspondence}~(sometimes loosely called AdS/CFT)
is also used to refer to the extended correspondence involving non-conformal, non-supersymmetric gauge 
theories with various features modeling standard model behavior such as chiral symmetry breaking, 
matter fields in the fundamental representation of the gauge group and confinement~(to name only a few). Introducing these 
features on the gauge theory side of the correspondence requires deformation of the anti de Sitter background on
the gravity side. In other words changing the geometry on the gravity side from AdS to something else 
changes the phenomenology on the gauge theory side. Unfortunately there is no version
of the correspondence available which realizes QCD or even the whole standard model to date. At the moment
one relies on the fact that studying other strongly coupled gauge theories one still learns something 
about strongly coupled dynamics in general and maybe even of QCD in particular if one studies features
with a sufficient generality or universality, such as meson mass ratios~\cite{Erdmenger:2007cm} 
or the shear viscosity to entropy ratio of a strongly coupled thermal plasma~\cite{Son:2007vk}. 

The phenomenological virtue of this setup is that we gain a conceptual understanding of strong coupling 
physics taking the detour via AdS/CFT. That is because AdS/CFT is not only a correspondence between a 
gauge theory and a gravity theory but rather a {\it duality } 
between them. This means in particular that a gauge theory at strong coupling corresponds to a 
gravity theory at weak coupling. Thus we can formulate a problem in the gauge theory at strong coupling, 
translate the problem to the dual weakly coupled gravity theory, use perturbative methods in order to solve
this gravity problem and afterwards we can translate the result back to the strongly coupled gauge theory.
As a specific example of this we will compute flavor current correlation functions at strong coupling in
a thermal gauge theory with a finite chemical isospin potential in section~\ref{sec:anaHydroIso}, 
using the methods reviewed in chapter~\ref{sec:holoMethods}.

Recently AdS/CFT also uncovered a connection between hydrodynamics of the gauge theory 
and black hole physics~\cite{Policastro:2001yc} 
which attracted broad attention~\cite[for example]{Son:2002sd,Policastro:2002se,Policastro:2002tn,
Herzog:2002pc,Kovtun:2003wp,Kovtun:2004de,Teaney:2006nc,Kovtun:2006pf,Son:2006em,Son:2007vk}. 
Here the main motivation is the so-called {\it viscosity bound}
\begin{equation}
\frac{\eta}{s} \ge \frac{\hbar}{4\pi} \, ,
\end{equation}
which was derived from AdS/CFT for all strongly coupled gauge theories with a gravity dual. Here the 
shear viscosity~$\eta$~(measuring the momentum transfer in transverse direction) is divided by the 
entropy density~$s$. Due to its universal validity in all calculated cases one hopes that this bound is 
a generic feature of strongly coupled gauge theories which is also valid in QCD. Indeed the measurements
at the RHIC collider confirm the prediction in that the viscosity of the plasma formed there is the smallest 
that has ever been measured. This phenomenological success of AdS/CFT motivated many extensions
in order to come closer to QCD and the real world.

One particularly important extension to the original correspondence~\cite{Maldacena:1997re} 
was the introduction of flavor and matter in the fundamental representation
of the gauge group, i.e. quarks and their bound states, the mesons~\cite{Karch:2002sh} further studied
in~\cite{Babington:2003vm,Kruczenski:2003be,Kruczenski:2003uq,Kirsch:2004km,Mateos:2006nu,Kobayashi:2006sb}.
In particular in~\cite{Babington:2003vm} it was found that a gravity black hole background induces 
a phase transition in the dual gauge theory. Further studies have shown that on the gravity side
a geometric transition~(see section~\ref{sec:dualitiesNString}) corresponds to a deconfinement transition for the fundamental matter
in the thermal gauge theory. At the moment the flavored extension of the relation between hydrodynamics
and black hole physics is under intense investigation~\cite[incomplete list of closely related work]{Ghoroku:2005kg,
Maeda:2006by,Peeters:2006iu,Mateos:2006yd, Nakamura:2006xk,Hoyos:2006gb,Amado:2007yr,Nakamura:2007nx,
Parnachev:2007bc,Mateos:2007vc,Karch:2007br, Ghoroku:2007re,Erdmenger:2007bn,Mateos:2007vn,Mateos:2007yp,
Aharony:2007uu,Myers:2007we, Evans:2008tv,Myers:2008cj}. So far the effect of finite chemical baryon potential
in the gauge theory and the structure of the phase diagram of these theories have been explored. 
For a review of the field the reader is referred to~\cite{Erdmenger:2007cm}, while a brief introduction
can also be found here in section~\ref{sec:mesons}. 
This connection between introducing fundamental matter and the exploration
of its thermodynamic an hydrodynamic properties in the strongly coupled thermal gauge theory as
well as the extension to more general chemical potentials is central
to my work partly published in~\cite{Erdmenger:2007ap,Erdmenger:2007ja}. This and other extensions 
to the thermal AdS/CFT framework are also the central goal of this thesis.

In the light of the reasonable hydrodynamics findings agreeing with observations, the bridge between 
string theory and phenomenologically relevant gauge 
theories starts to become illuminated: Since AdS/CFT is a concept derived from string theory it is by construction
connected to that side of the gulch. If on the other hand we can experimentally confirm the strong coupling predictions made 
using this concept, then we have found a way to ascribe phenomenological relevance to a concept of string theory.
This is by far no proof that string theory is the fundamental theory which describes nature, but certainly it
would confirm that these concepts in question correctly capture the workings of nature. One could be even 
more brave and take such a confirmation as the motivation to take the correspondence not just as a phenomenological
tool but to take it seriously in its strongest formulation and assume that the full quantized string theory
can be related to the gauge theory fully describing nature~(this would have to be a somewhat extended standard model).

{\bf The mission for this thesis}
\label{page:theMission}
The general question I wish to answer in this thesis is: What is the impact of finite baryon and isospin chemical potentials
or densities on the thermal phenomenology of a strongly coupled flavored plasma? The gauge/gravity duality shall be 
used to obtain strong coupling results. Since no gravity dual to QCD has been found yet, we work in
a supersymmetric model theory which is similar to QCD in the properties of interest. To be more precise we consider the 
gravity setup of a stack of~$N_c$ D3-branes which produce the asymptotically AdS black hole background and we 
add~$N_f$ probe D7-branes which introduce quark probes on the gauge dual side. The AdS black hole background 
places the dual gauge theory at a finite temperature~$T$ related to the black hole horizon~$\varrho_H=\pi T R^2$, where
$R$ is the radius of the AdS space. The chemical potential is a measure for the energy which is needed in 
order to increase the thermodynamically conjugate charge density inside the plasma. On the gravity side a chemical
potential is introduced by choosing a non-vanishing background field in time direction~$A_0(\varrho)\not =0$. The chemical 
potential then arises as its boundary value~$\lim\limits_{\varrho\to\varrho_{\text{bdy}}} A_0(\varrho) = \mu$. Depending on the 
gauge group from which the flavor gauge field~$A_0$ arises, the chemical potential can give the baryon chemical
potential for the $U(1)$-part of the gauge group, the isospin chemical potential for~$SU(2)$ or other 
chemical potentials for~$SU(N_f)$.

In order to study the phenomenology of the plasma with chemical potentials dual to the gravity setup, which we have just 
described, we gradually approach the construction of the phase diagram by computing all relevant thermodynamic 
quantities. We shall also study thermal spectral functions describing the plasma as well as transport properties, in 
particular the diffusion coefficients of quarks and mesons inside the plasma.

Note, that in the previously discussed sense we confirm the AdS/CFT concept with each reasonable
thermal result that we produce. Furthermore, tracing the relation between the thermal gauge theory
and the dual gravity in detail using specific examples will also lead to a deeper understanding of the
inner workings of the AdS/CFT correspondence in general. Therefore we can aim for the additional
goal of finding out something about string concepts from our studies, rather than restricting ourselves
to the opposite direction of reasoning.

{\bf Summary of results}
We can generally answer the main question of this thesis with the statement that introducing baryon
and isospin chemical potentials into the thermal gauge theory at strong coupling has a significant effect on the 
thermodynamical quantities, on the correlation functions, spectral functions and on transport processes.
Studying both the canonical and grandcanonical ensemble, we find an enriched thermodynamics at finite 
baryon and isospin density, or chemical potential respectively. 
In particular we construct the phase diagram of the strongly coupled plasma at finite isospin and baryon
densities or chemical potentials, respectively. We compute the free energy, grandcanonical potential, entropy,
internal energy, quark condensate and chemical potentials or densities, depending on the ensemble.
Discontinuities in the quark condensate and in the baryon and isospin densities or potentials indicate a phase
transition at equal chemical potentials or densities, respectively. This newly discovered phase transition
appears to be analogous to that found for 2-flavor QCD in~\cite{Splittorff:2000mm}.
Conceptually we have also achieved the generalization to $U(N_f)$-chemical potentials with arbitrary~$N_f$
and we provide the formulae to study the effect of these higher flavor gauge groups.

As an analytical result we find thermal correlators of $SU(2)$-flavor currents at strong coupling and a non-zero
chemical isospin potential in the hydrodynamic approximation~(small frequency and momentum). 
In particular we find that the isospin potential changes the location of the correlator poles in the 
complex frequency plane. The poles we examine are the diffusion poles formerly appearing at imaginary
frequencies. Increasing the isospin potential these poles acquire a growing positive or negative real part
depending on the flavor current combination. The result is a triplet-splitting of the original pole into three 
distinct poles in the complex frequency plane each corresponding to one particular flavor combination.

From a numerical study we derive thermal spectral functions of $U(1)$-flavor currents in a thermal
plasma at strong coupling and finite baryon density. We find mesonic quasi-particle resonances
which become stable as the temperature is decreased. In this low temperature regime these 
resonance peaks are also found to follow the vector meson mass formula~\cite{Kruczenski:2003be}
\begin{equation}
M= \frac{L_\infty}{R^2}\,\sqrt{2 (n+1)(n+2)}\, ,
\end{equation}
where~$L_\infty$ and~$R$ are geometric parameters of the gravity setup described in section~\ref{sec:mesonSpectraB}.
The radial gravity excitation number~$n$ is related to the peak considered in the spectral function,
starting with the lowest frequency peak at~$n=0$. This fact and the fact that the peaks become very narrow
confirm that stable mesonic states form in the plasma at sufficiently low temperature~(or equivalently at 
large quark mass). We identify these resonances with stable mesons having survived the deconfinement
transition of the theory in agreement with the lattice results given in~\cite{Asakawa:2003xj} and the 
findings of~\cite{Shuryak:2004tx}. However, the interpretation of the small mass/high temperature regime is still controversial.
In that particular regime we observe very broad resonances which move first to lower frequencies as the temperature is decreased.
Then we discover a turning point at a certain temperature after which the mesonic behavior described above
sets in. We ascribe the turning behavior to the dissipative character of the excitations at high temperature and
argue that these resonances can not be interpreted as quasi-particles and therefore their frequency can
not be identified with a vector meson mass. The concise treatment of these speculations we delay to future
work using quasinormal modes. Nevertheless, we already record our observations in section~\ref{sec:peakTurning} 
also providing interesting insight in the gauge/gravity correspondence in terms of a bulk/boundary solution correspondence.

The spectral functions at finite isospin density show similar resonance peaks with a similar behavior.
Additionally the spectral functions for the three different flavor directions show a triplet splitting in
the resonance peaks which results from the isospin potential breaking the~$SU(2)$-symmetry in
flavor space.

Studying transport properties we find that the quark diffusion in the thermal plasma shows a vanishing phase transition 
as the baryon density is increased. This transition is smoothened to a crossover which appears
as a minimum in the diffusion coefficient versus quark mass or temperature. A similar picture 
arises when simultaneously a finite isospin density is introduced. 
For the case of quarkonium transport in the plasma we find a systematic agreement between the 
AdS/CFT calculation and the corresponding field theory calculation confirming the correspondence
on a more than empirical level. 

All these effects are caused by significant changes on the gravity side such as: the embeddings having a spike and
being only of black hole type. For a finite chemical potential there has to be a finite gauge field on the brane and the 
field lines 'end' at the horizon. Also the resonance peaks in the spectral function are shifted by both baryon and isospin densities.
We primarily find that by the presence of a baryon and/or isospin chemical potential the gravity
solutions which for example generate the peak in the spectral function are changed considerably.
The same is true for those solutions with vanishing boundary condition called quasinormal modes.
Their frequencies, called quasinormal frequencies are shifted in the complex frequency plane 
by the introduction of finite potentials. Since these quasinormal frequencies correspond to
poles in the correlation function, this result agrees with our analytically found pole shift in the case of the 
diffusion pole mentioned above. Especially the triplet-splitting of the poles upon introduction of isospin appears in
both results.

{\bf How to read this}
New results extending significantly beyond those published until now are reported in 
sections~\ref{sec:anaHydroIso},~\ref{sec:thermoB&I},~\ref{sec:hiNf},~\ref{sec:peakTurning},~\ref{sec:isospinDiffusion},
and~\ref{sec:charmDiffusion}. Completely new ideas are developed in the three outlook 
sections~\ref{sec:molecular},~\ref{sec:mesonSpectraB&I} and~\ref{sec:diffusionMatrix}.

This thesis is structured as follows: For improved readability and overview each of the main chapters
contains a small summary section at its end. After the non-technical introduction just given in the 
present introduction chapter, we establish the AdS/CFT correspondence
in chapter~\ref{sec:adsCft} on a technical level. The first three chapters~(including this introduction) are written
such that they may serve as a directed introduction to the field addressed to graduate students or researchers
who are not experts on string theory or AdS/CFT. The basic concepts needed from string theory such as branes
and duality relations are briefly introduced in section~\ref{sec:dualitiesNString}, then put together with those of conformal field theory 
considered in section~\ref{sec:CFT} in order to merge these frameworks to the statement of the AdS/CFT 
correspondence~\ref{sec:stateAdsCft}. With chapter~\ref{sec:holoMethods} we develop the mathematical methods
which we use to compute correlation functions and transport coefficients from AdS/CFT at finite temperature. 
Section~\ref{sec:chemPots} shows how chemical potentials are implemented and in section~\ref{sec:qnm} 
the concept of quasinormal modes is reviewed. This directed introduction is not designed to cover string theory at any rate
(for a concise introduction the reader is referred to reviews, e.g.~\cite{Mohaupt:2002py}, or 
books, e.g.~\cite{Polchinski:1998rq,Polchinski:1998rr}).

The last four chapters collect all my calculations and results which are relevant for the aim of this thesis.
Each of the chapters~\ref{sec:holoThermoHydro},~\ref{sec:thermalSpecFunc} and~\ref{sec:transport} contains an 
outlook section which is that one before the summary section. These outlook sections give explain some ideas
how the investigation of the present topic in that chapter can be continued. If available also initial calculations
are presented as a starting point. Chapter~\ref{sec:holoThermoHydro} shows the calculation and results of 
correlation functions for thermal flavor currents obtained analytically and the thermodynamics of the 
thermal gauge theory at finite baryon or isospin or both potentials or densities. Chapter~\ref{sec:thermalSpecFunc}
shows the numerical calculation and the results and conclusions derived from thermal spectral functions of 
flavor currents in a strongly coupled plasma. Finally the transport properties of quarks and mesons are 
studied in chapter~\ref{sec:transport}. In chapter~\ref{sec:conclusion} we will conclude this thesis putting 
stress on the interrelations between our results and on their relation to experiments, lattice and other QCD results.

\section{The AdS/CFT correspondence} \label{sec:adsCft}
In this chapter we briefly review the gauge/gravity correspondence from its origins in string theory to its
application aiming for phenomenological predictions in collider experiments. The AdS/CFT correspondence, which 
carries the properties of {\it holography}~(in analogy to holography in optics) and a {\it duality} as well, 
states that string theory in the near-horizon limit of~$N_c$ coincident
M- or D-branes is equivalent to the world-volume theory on these branes. In the first section we develop the string theory framework
in order to state the correspondence more precisely and discuss the existing evidence for this conjectured 
correspondence in the second section. The third section then introduces fundamental matter, i.e. quarks into 
the duality. Section four includes a study of the AdS/CFT correspondence at finite temperature introducing
the concepts and notation upon which this present work is based. A brief overview of other deformations of the 
original correspondence and their implications for phenomenology is given in the last section. We discuss
the role of AdS/CFT as a phenomenological tool and contrast this to ascribing a more fundamental character to it.

\subsection{String theory and AdS/CFT} \label{sec:generalAdsCft}
The AdS/CFT correspondence is a gauge theory / gravity theory duality appearing in string theory. 
We will see that it is special because it relates strongly coupled quantized gauge theories to weakly coupled classical
supergravity and therefore makes it possible to study strong coupling effects non-perturbatively. It
may also be turned around and used to study gravity at strong coupling by computations in the 
weakly coupled field theory dual. 
Nevertheless, from the string point of view this correspondence is one duality among many others.
In order to understand its role in string theory, we start out examining the general concept of 
dualities in string theory and M-theory.

\subsubsection{Dualities and string theory}
\label{sec:dualitiesNString}
The AdS/CFT correspondence is heavily used in this work and since it carries the character of a
duality relating one theory at strong coupling to a different theory at weak coupling,
in this section we explore other dualities appearing in string theory in order to understand the role 
of AdS/CFT in string theory.

Up to the early 1990s five different kinds of superstring theories had been discovered~\cite{Polchinski:1998rr}:
type I, type IIA, type IIB, heterotic~$SO(32)$, heterotic~$E_8\times E_8$. This was a dilemma 
to string theory as the unique theory of everything. But in 1995~\cite{Witten:1995ex,Horava:1995qa} this dilemma was 
resolved to great extend by virtue of dualities. All five string theories had been related to each other by 
so-called S-, T-dualities, by compactification and by taking certain limits. Let us pick T-duality 
as a representative example to study in more detail. 

{\bf A brief T-duality calculation} 
T-duality in the simplest example of bosonic string theory compactified on a circle with radius~$R$
in the 25${}^{\text{th}}$ dimension is a symmetry of the bosonic string solution under the
transformation of the compactification radius~$R \to \tilde R = {l_s}^2/R$ and simultaneous
interchange of the winding number~$W$ with the Kaluza-Klein excitation number~$K$. 
This means that bosonic string theory compactified on a circle with radius~$R$ with $W$ windings
around that circle and with momentum~$p^{\text{25}}=K/R$ is equivalent to a bosonic string theory 
compactified on a circle with radius~${l_s}^2/R$ with winding number~$K$ and momentum~$p^{\text{25}}=W/R$.
To see this in more detail, consider the closed bosonic string 
action in 25-dimensional bosonic string theory with target space coordinates~$X^\mu$~\cite{Becker:2007zj}
\begin{equation}
\label{eq:bosonicStringAction}
S_{\text{bosonic}}= - T \int \dd \sigma \dd \tau
 \sqrt{-\det{g_{\mu\nu}\partial_\alpha X^\mu\partial_\beta X^\nu}} \, ,
\end{equation}
with the metric~$g$, the string tension~$T$ and a $1+1$-dimensional
parametrization~$(\sigma^0=\tau,\sigma^1=\sigma)$ of the brane world volume where~$\alpha,\,\beta=0,1$.
Here the parameters are the world-sheet time~$\tau=0,\dots ,2\pi$ and spatial coordinate~$\sigma=0,\dots,\pi$.
Note, that we could generalize this action~\eqref{eq:bosonicStringAction} to the case of a simple 
p-dimensional object, a {\it Dp-brane} as we will learn below. The most general solution is given by the sum of one solution
in which the modes travel in one direction on the closed string~({\it left-movers}) and the   
second solution where the modes travel in the opposite direction~({\it right-movers})
\begin{equation}
\label{eq:bosonicStringSolution}
X^\mu =X^\mu_{L}+X^\mu_{R} \, ,
\end{equation}
which for closed strings are given by
\begin{eqnarray}
X^\mu_{L} &= \frac{1}{2} x^\mu+ \frac{1}{2} l_s^2 p^\mu (\tau-\sigma) +\frac{i}{2} l_s
  \sum \limits_{n\not = 0} \frac{1}{n}\alpha_n^\mu e^{-2 i n (\tau-\sigma)} \\  
X^\mu_{R} &=\frac{1}{2} x^\mu+ \frac{1}{2} l_s^2 p^\mu (\tau+\sigma) +\frac{i}{2} l_s
  \sum \limits_{n\not = 0} \frac{1}{n}\tilde \alpha_n^\mu e^{-2 i n (\tau+\sigma)} \, .
\end{eqnarray}
These solutions each consist of three parts: the center of mass position term, the total
string momentum or {\it zero mode} term and the string excitations given by the sum.
If we compactify the 25$^{th}$ dimension on a circle with radius~$R$, we get 
\begin{eqnarray}
\label{eq:compactifiedBosonicStringSolutionL}
X^{25}_{L} &= \frac{1}{2} (x^{25}+\tilde x^{25})+ (\alpha' p^{25}+ W R) (\tau+\sigma) +  \dots \\  
\label{eq:compactifiedBosonicStringSolutionR}
X^{25}_{R} &= \frac{1}{2} (x^{25}-\tilde x^{25})+ (\alpha' p^{25}- W R) (\tau-\sigma) +  \dots \, ,
\end{eqnarray}
We leave out the sum over excitation modes~(denoted by~$\dots$) since it is invariant under compactification.
The constant~$\tilde x^{25}$ is arbitrary since it cancels in the whole solution~\eqref{eq:bosonicStringSolution25}.
Only the {\it zero mode} is affected by the compactification since the momentum 
becomes~$p^{25}=K/R$ with~$K$ labeling the levels of the Kaluza-Klein tower of excitations 
becoming massive upon compactification. An extra winding term is added as well.
So the the sum of both solutions in 25-direction reads
\begin{eqnarray}
\label{eq:bosonicStringSolution25}
X^{25} &= x^{25}+ 2\alpha' \frac{K}{R} \tau+2 W R \sigma +  \dots \, .
\end{eqnarray}
We now see explicitly that the transformation~$W\leftrightarrow K, \, R\to \alpha/R$ applied to
equations~\eqref{eq:compactifiedBosonicStringSolutionL} and~\eqref{eq:compactifiedBosonicStringSolutionR} 
is a symmetry of this theory because the {\it zero mode} changes 
as~$(\alpha' K/R \pm W R) \to (\alpha' W R / \alpha' \pm K \alpha'/R) = (W R \pm \alpha' K/R)$. 
So we get the transformed solution 
\begin{eqnarray}
\label{eq:tDualBosonicStringSolution25}
\tilde X^{25} &= \tilde x^{25}+ 2W R \tau +2 \alpha'\frac{K}{R}\sigma +  \dots \, .
\end{eqnarray}
Comparing the solutions~\eqref{eq:tDualBosonicStringSolution25} and~\eqref{eq:bosonicStringSolution25}
we note that the transformed solution is equal to the original one except for the fact \
that~$\sigma$ and~$\tau$ are interchanged.
However, the bosonic string action is reparametrization invariant~\footnote{
S-duality exchanges the fundamental strings (i.e. the NS-NS or the Ramond-Ramond two-forms)
with the D1-branes.
So, roughly speaking the string behaves like a D1-brane. Generalizing the case~$p=1$ to
arbitrary~$p$ we would find that the D$p$-brane action is reparametrization invariant under a change 
of the $p+1$ world-volume coordinates given by~$\sigma^\alpha\to\sigma^\alpha (\tilde\sigma)$.} 
under~$(\tau,\sigma)\to (\tilde \tau,\tilde \sigma)$.
Therefore we see that physical quantities like correlation functions are invariant under the 
T-duality tranformation.

From this duality we learn how we may start from one string theory and by different ways of 
compactification we arrive at two distinct but equivalent formulations of the same physics. Another 
important feature is that certain quantities change their roles as we go from one compactification
to the other~(winding modes turn into Kaluza-Klein modes as~$K\leftrightarrow W$). Finally we
realize that T-duality relates a theory compactified on a large circle~$R$ to a theory 
compactified on a small circle~$\alpha'/R$. 

By virtue of T-duality another important ingredient for the gauge/gravity correspondence 
was introduced into string theory: D$p$-branes. Introducing open strings into the bosonic 
theory of closed strings, we need to specify boundary conditions at the string end points.
A natural criterion for these boundary conditions is to preserve Poincar\'e invariance. 
So we would choose Neumann boundary conditions~$\partial_\sigma X^\mu=0$ at
the end points~$\sigma=0,\,\pi$. Evaluating this condition for the general solution
given in~\eqref{eq:bosonicStringSolution25}, we see that the Neumann condition
turns into a Dirichlet boundary condition~$\partial_\tau X^\mu=0$.
This condition explicitly breaks Poincar\'e invariance by fixing~$p$ of the 
spatial coordinates of open string ends to $\tau$-independent hypersurfaces. These
surfaces are called Dirichlet- or D$p$-branes and have to be considered as dynamical 
objects in addition to the fundamental strings.  We will see below that $AdS/CFT$ is a duality 
arising from two distinct ways of describing these D$p$-branes in open string theory. 

Analogous to T-duality, S-duality relates a string theory with coupling constant~$g_s$ 
to a string theory with coupling~$1/g_s$. In this respect S-duality is very similar to the 
AdS/CFT duality which relates a gauge theory at strong coupling to a gravity theory at
weak coupling or vice versa. A particularly interesting example of S-duality is the electric/magnetic duality
(which is also present in $\N=4$~Super-Yang-Mills theory).

{\bf Gauge/gravity dualities}
We have seen in the last subsection that there exists a variety of string dualities and it is
time now to narrow our view to the subset of gauge/gravity dualities including the AdS/CFT correspondence.

As for the important special case of gauge/string dualities there are three kinds relating 
conventional (nongravitational) QFT to string or M-theory: {\it matrix theory}, {\it AdS/CFT} and {\it geometric transitions}.
It is remarkable that quantum mechanical theories are dual to~(i.e. may be replaced by) a gravity theory.

{\it Matrix theory} is a quantum description of M-theory in a flat 11-dimensional space-time 
background. So this gives an M-theory approximation beyond 11-d SUGRA limit.
In matrix theory the dilaton is not massless and therefore there is no dimensionless coupling that could
be used to define a perturbation theory.
The fundamental degrees of freedom are D0-branes and it is written down in a non-covariant formulation.

Let us briefly consider a second gauge/gravity duality called  {\it geometric transition}. It is a duality
relating open strings to closed strings, and this is a property which it shares with AdS/CFT.~\footnote{
The basic idea of a geometric transition is that a gauge theory describing
an open string sector, i.e. a gauge theory on D-branes, is dual to a 
{\it flux compactification} of a particular string theory in which no D-branes
are present, but {\it fluxes} are present instead. In other words, as a modulus is
varied, there is a transition connecting the two descriptions~\cite{Becker:2007zj}.}
One setup in which the geometric transition takes place is given by an $\N=1$-supersymmetric 
confining gauge theory obtained by wrapping D5-branes around topologically non-trivial 
two-cycles of a {\it Calabi-Yau manifold}~(determining the structure of the internal space). The
remaining four directions of the D5 span the four Minkowski directions.
On the D5-branes open string excitations form a supersymmetric
gauge theory. The shape of the Calabi-Yau manifold~(of internal space) is parametrized by {\it moduli}. 
These are scalars appearing in the theory having a constant potential which can thus take arbitrary values. 
One may now shrink the two-cycles by varying the {\it moduli} of the theory
in an appropriate way. At the limit of a zero-size two-cycle the system undergoes a geometric
transition to a (sector of  the) theory in which closed strings are the fundamental objects to be excited.
With the vanishing two-cycles also the D-branes disappear from the description of the system.
In section~\ref{sec:mesons} we will meet another particularly interesting example for a geometric
transition. That is the transition from Minkowski to black hole embeddings in the D3/D7-brane setup. 
In that case the D7-brane wraps an~$S^3$ inside the~$S^5$ of the~$AdS_5\times S^5$ background
geometry.

In order to find the {\it AdS/CFT correspondence} we have to consider collections of coincident M- or D-branes.
These branes source flux and curvature. Examples of theories on these branes with maximal 
supersymmetry~(32 supercharges) are M2-, D3- and M5-branes corresponding
to 3-, 4- and 6-dimensional world-volume theories being superconformal (SCFT):\\
\begin{center}
\begin{tabular}{l c r}
SCFT on $N_c$ M2-branes & $\leftrightarrow$ & M-theory on $AdS_4\times S^7$\\
SCFT on $N_c$ M5-branes & $\leftrightarrow$ & M-theory on $AdS_7\times S^4$\\
$\N=4$ SYM on $N_c$ D3-branes & $\leftrightarrow$ & type IIB on $AdS_5\times S^5$
\end{tabular} .
\end{center}

Note that also dS/CFT relating a gauge theory to gravity in positively curved de Sitter space is 
interesting because of the experimental observation that our universe is accelerated. If this acceleration
is due to a positive cosmological constant, the matter and radiation densities approach zero in the infinite
future and our universe approaches de Sitter space in future. On the other dS/CFT might be interesting for 
the early universe. Nevertheless it is less explored than AdS/CFT since it features no supersymmetry.
Instead of D-/M-branes, Euclidean S-branes are used.

\subsubsection{Black branes}
\label{sec:blackBranes}
The gauge/gravity correspondence we explain in this section originated from the study 
of black $p$-branes in $10$-dimensional string theory and $11$-dimensional M-theory. It turned out 
that one can describe branes in two ways which are different limits of string theory: 
a $p$-brane is a solitonic solution to classical supergravity and at the same time 
a $p$-brane is the hypersurface of points where an open string is allowed to end. It was
shown that {\it Dirichlet-$p$-branes~(D$p$-branes)} give the full string theoretic description of
the $p$-branes found as classical solutions to supergravity.
Furthermore black branes are essential for the study of dual field theories at finite temperature~(as will
be seen in the next section). Because of their doubly-important role, we will expand these thoughts
on branes.

{\bf Classical solutions}
In this paragraph we examine the classical $p$-brane solutions to supergravity because these
objects and their classical description~(in Anti de Sitter space AdS) are one of the two fundamental building 
blocks of the AdS/CFT correspondence.

Black p-branes were found as solutions to classical limits of string and M-theory, like e.g. the 
bosonic part of the $11$-dimensional SUGRA action (with M$2$ and M$5$-brane 
solutions)~\cite[equations (12.3), (12.18)]{Becker:2007zj}
\begin{equation}
S = \frac{1}{2\kappa^2_{11}} \int \dd^{11}x\sqrt{-G}\left (
	  \mathcal{R}-\frac{1}{2} |F_4|^2
       \right) -
       \frac{1}{6}\int A_3\wedge F_4\wedge F_4
\end{equation}
or the 10-dimensional SUGRA action (with D$p$-brane solutions)
\begin{equation}
S = \frac{1}{2\kappa_{10}^2} \int \dd^{10}x\sqrt{-g}\left [
	 e^{-2\Phi} (\mathcal{R}+4(\partial \Phi)^2)-\frac{1}{2} |F_{p+2}|^2
       \right] \, ,
\end{equation}
which include a dilaton~$\Phi$, the curvature scalar~$\mathcal{R}$, gauge field strengths~$F_{p+1}$ and the
corresponding gauge fields~$A_p$. $\kappa_D$ denotes the gravity constant in dimension~$D=10$
or~$11$. 
Branes are $(p+1)$-dimensional objects solving the equations of motion derived from either action.
They can be viewed as higher-dimensional generalizations of a black hole in four dimensions. 
Black hole solutions in four space-time dimensions are point-like
objects, which are surrounded by an event horizon. They have an $SO(3)$ rotational symmetry 
and a symmetry associated with time-translation invariance. 
Black $p$-branes are surrounded by a higher-dimensional event horizon, they break Lorentz 
symmetry of the $D=d+1$-dimensional theory to
\begin{equation}
 SO(d,1)\to \underbrace{SO(d-p)}_{\text{rotational symmetry transverse to brane}} \times 
   \underbrace{SO(p,1)}_{\text{Lorentz symmetry along brane}} 
\end{equation}
The Lorentz-symmetry is enlarged to Poincar\'e symmetry by translation symmetries along the brane.  
There exist two classes of $p$-brane solutions: the supersymmetric ones which are called
{\it extremal} and the ones which break supersymmetry which are called {\it non-extremal}.  
The general extremal D$p$-brane solution has the metric
\begin{equation}
\label{eq:extremalDpMetric}
\dd s^2 = H_p^{-1/2}\eta_{ij}\dd x^i\dd x^j + H_p^{1/2}\xi_{mn}\dd y^m\dd y^n \, , 
\end{equation}
with the flat Lorentzian metric~$\eta$ along the brane and the Euclidean metric~$\xi$ 
perpendicular to the brane. The harmonic function~$H_p$ is
\begin{equation}
H_p(r)=1+(\frac{r_p}{r})^{7-p}\, ,
\end{equation}
and the dilaton
\begin{equation}
e^\Phi=g_s H_p^{(3-p)/4}\, .
\end{equation}

The general non-extremal solution comes with the metric
\begin{equation}
\label{eq:nonExtremalDpMetric}
\dd s^2 = -{\Delta_+} {\Delta_-}^{-1/2}- \dd t^2 +{\Delta_-}^{1/2} \dd x^i \dd x^i + 
  {\Delta_+}^{-1} {\Delta_-}^\gamma \dd r^2+r^2{\Delta_-}^{\gamma+1} \dd \Omega^2_{8-p} \, , 
\end{equation}
with~$\gamma=-\frac{1}{2}-\frac{5-p}{7-p}$ and 
\begin{equation}
\Delta_\pm=1-(\frac{r_\pm}{r})^{7-p}\, ,
\end{equation}
and the dilaton
\begin{equation}
\label{eq:dilatonE}
e^\Phi=g_s {\Delta_-}^{(p-3)/4} \, .
\end{equation}

{\it The special case~$p=3$:} Note that the $p=3$-brane solution is special in that it is the only one in which the dilaton
is constant~$e^\Phi=g_s$. We will develop the arguments for the AdS/CFT correspondence along this specific case below and
therefore include the (classical) D3-brane solution to supergravity here
\begin{eqnarray}
\label{eq:3BraneBackgMetric}
\dd s^2 = {H_3}^{-1/2} \left( {\dd t}^2+{\dd \mathbf{x}}^2 \right) + {H_3}^{1/2} \left ( {\dd r}^2 + r^2 {\dd \Omega_5}^2 \right )\, , \\
\label{eq:3BraneBackgFiveF}
F_5= (1+\star) \dd t\wedge \dd x_1\wedge\dd x_2\wedge\dd x_3\wedge\dd {H_3}^{-1}\, , \\ 
\label{eq:3BraneBackgH3}
H_3 = 1+\frac{R^4}{r^4}\, , \quad R^4:=4\pi g_s(\alpha')^2 N \, ,
\end{eqnarray}
where we call the AdS radius~$R$ in agreement with the AdS/CFT literature.

{\bf D$p$-branes and DBI-action}
\label{sec:branesNDBI}
We have already mentioned that branes, in particular D$p$-branes are the crucial objects to consider in
order to understand the AdS/CFT correspondence. Beyond this general insight into the working of the 
correspondence in this section we also include the effective action, the {\it Dirac-Born-Infeld~(DBI)}-action.
We will make use of this formulation later in order to compute brane embeddings, or in other words 
the location of the D$p$-branes in the ten-dimensional space and additionally fluctuations on these branes.

As mentioned above, T-duality implies the existence of extended dynamical objects 
in string theory which are called D$p$-branes. Roughly speaking these are the hypersurfaces 
in target space on which end points of open strings can lie. D$p$-branes are $p+1$-dimensional
objects carrying charge and thus coupling to $(p+1)$-form gauge fields. 

The Dirac-Born-Infeld~(DBI) action is the $(p+1)$-dimensional world-volume action for fields living on a D$p$-brane
embedded in ten-dimensional space-time.
For a D$p$-brane with an Abelian gauge field~$A$ in a
background of non-flat metric~$g_{\mu\nu}$, the dilaton~$\Phi$ and the two-form~$B_{\mu\nu}$
in static gauge the DBI action in string frame is given by
\begin{equation}
\label{eq:dbiAction}
S_{\text{D}p}= -T_{\text{D}p} \int \dd^{p+1}\sigma e^{-\Phi}
 \sqrt{
  -\det  \left \{ 
  P[g+B]_{\alpha\beta} + 
  (2\pi \alpha') F_{\alpha\beta}  
\right \}
 } \, .
\end{equation}
Static gauge refers to the choice of world-volume coordinates~$\sigma^\alpha$ which by diffeomorphism-invariance
of the action are set equal to~$p+1$ of the space-time coordinates~$X^\mu$, such that the pull-back is simplified. 
The remaining $(9-p)$ coordinates are relabeled as~$2\pi\alpha' \phi^i$. The~$\phi^i$ are scalar fields of 
the world-volume theory with mass dimension~$[\phi^i]=1$.
The brane tension~$T_{\text{D}p}$ is given by
\begin{equation}
\label{eq:TDp}
T_{\text{D}p} = \frac{1}{g_s (2\pi)^p(\alpha')^{(p+1)/2}} \, .
\end{equation}
Note, that the DBI-action also contains a fermionic contribution~(see e.g.~\cite{Kirsch:2006he} for details).

The geometry of a number~$N$ D-branes is more subtle. Coordinates transverse to the 
brane are T-dual to non-Abelian gauge fields.
The DBI action for this case of non-Abelian gauge fields~$A$ is given by
\begin{multline}
\label{eq:nonAbDbiAction}
S_{\text{D}p}= -T_{\text{D}p} \int \dd^{p+1}\sigma e^{-\Phi}
 \mathrm{STr}\Big\{\sqrt{\det Q^\gamma{}_\kappa}  \\ 
  \times \sqrt{-\det (E_{\alpha\beta}
  +E_{\alpha\gamma}(Q^{-1}-\delta)^{\gamma\kappa}E_{\gamma\beta}
  +(2\pi \alpha') F_{\alpha\beta}} \Big \} \, .
\end{multline}
Here~$Q^i{}_j =\delta^i{}_j+ i (2\pi \alpha') [\phi^i,\phi^k] E_{kj}$ and~$E_{kj}=g_{kj} + B_{kj}$ 
collects the antisymmetric background tensors. Choosing the transverse scalar fields such 
that~$[\phi^i,\phi^k]=0$ we obtain the general form of the Abelian DBI action~\eqref{eq:dbiAction} but for 
non-Abelian gauge fields~$A=A^a T_a$ with generators~$T_a$ and field strengths~$F=F^a T_a$.
The symmetrized trace~$\mathrm{STr}\{\dots \}$ tells us to symmetrize the expression in the 
flavor representation indices. Note, that the non-Abelian DBI-action in this form is only valid up to
order~$\O({\alpha '}^4)$. Another limitation is that we can only consider slowly varying fields.

Let us choose the special case of~$N_c$ coincident D3-branes. The world-volume action of this
stack of branes at low energy is that of a~$d=4$ dimensional $\N=4$-supersymmetric Yang-Mills theory 
with gauge group~$U(N_c)$. This theory is supersymmetric and obeys {\it conformal invariance}, 
meaning that it is a conformal field theory as explained below. The massless modes of the low energy spectrum for open strings
ending on the stack of coincident D3-branes constitute the $\N=4$ vector supermultiplet
in~$(3+1)$ dimensions.

{\it BPS states:}
In supersymmetry representations and especially branes are often classified in terms of how many supersymmetries they
break if introduced to the brane-less theory. The Bogomolny-Prasad-Sommerfeld~(BPS) 
bound distinguishes between branes which are BPS and those which are not.
Let us see what this means in the example of massive point particles in four dimensions.
The $\N$-extended supersymmetry algebra for particles of positive mass~$M>0$ at
rest is
\begin{equation}
\label{eq:bpsAlgebra}
\{ Q_\alpha^I,Q_\beta^{\dagger\, J} \} = 2 M\delta^{IJ}\delta_{\alpha\beta}+ 2 i Z^{IJ} \Gamma^0_{\alpha\beta}\, ,
\end{equation}
with the central charge matrix~$Z^{IJ}$, supersymmetry generators~$Q^I,\, I=1,\dots,\N$ and
Majorana spinor labels~$\alpha,\, \beta$. The central charge matrix can be brought in
a form such that we can identify a largest component~$Z_1$. The BPS-bound is defined
in terms of this component as a lower bound for the particle's mass
\begin{equation}
\label{eq:bpsBound}
M\ge |Z_1| \, .
\end{equation}
States that saturate the bound~$M=|Z_1|$ belong to the {\it short supermultiplet} also called
the {\it BPS representation}. In this case some relations in the algebra~\eqref{eq:bpsAlgebra}
become zero such that less combinations of supercharges~$Q$ can be used to generate states
starting from the lowest one, resulting in less possible states. States with~$M> |Z_1|$ belong to a {\it long supermultiplet}.
Depending on the number of central charges which are equal to the mass~(e.g. $M=|Z_1| =|Z_2|$)
the number of unbroken supersymmetries changes. If for example half of the supersymmetries
of a $\N=4$ theory are unbroken because $2$~of the central charges are equal to the mass, 
then the representation is called {\it half BPS}. In general for $n$~central charges being equal to the mass we
have a {\it $(n/\N)$ BPS representation}.

Since BPS states include particles with mass equal to the central charge, the mass is not changed 
as long as supersymmetry is unbroken, i.e. these states are stable and in particular we can 
examine them at strong and at weak coupling.

{\bf Identifying D-$p$-branes with classical $p$-branes}
It is believed that the extremal $p$-brane in supergravity and the D$p$-brane from string theory
are two distinct descriptions of the same physical object in two different parameter regimes. 
Here we establish a direct comparison to consolidate this statement which lies at the heart 
of the AdS/CFT correspondence.

In the case~$p=3$ it can be shown~\cite{Aharony:1999ti} that the classical $p$-solution is 
valid in the regime~$1\ll g_s N <N$ with the string coupling~$g_s$ and
the Ramond-Ramond charge~$N = \int\limits_{S^{8-p}} \star F_{p+1}$. While the validity
of the string theoretic D$p$-brane description for a stack of~$N$ D3-branes is limited to~$g_s N \ll 1$~\cite{Aharony:1999ti}. 
As discussed in section~\ref{sec:branesNDBI} D$p$-branes are the $(p+1)$-dimensional
hypersurfaces on which strings can end. On the other hand they are also sources for closed 
strings. This fact can be translated into the heuristic picture that those particular closed string
excitations identified with gravitons are sourced by the D$p$-brane. This reflects the fact that 
D$p$-branes are massive (charged) dynamical objects which also curve the space around them. 
In particular D$p$-branes can carry Ramond-Ramond charges. A stack of $N$ coincident
D$p$-branes carries~$N$ units of the $(p+1)$-form charge which can be calculated from the corresponding
action as shown in~\cite{Polchinski:1995mt}. Turning to supersymmetry we find that the Dirichlet 
boundary condition imposed on the string modes by the presence of a D$p$-brane identifies the 
left-moving and right-moving modes~(see section~\ref{sec:dualitiesNString}) on the string and therefore breaks at least half of the
supersymmetry. It turns out that in type IIB string theory branes with odd~$p$ preserve exactly
one half of the supersymmetries and hence D$p$-branes are BPS-objects. On the other hand
the classical $p$-brane solution in supergravity carries the Ramond-Ramond charge~$N$ as
well and features the same symmetries. 
A further check of the identification is the computation of gauge boson masses (which are analogs
of the W-boson masses in the standard model) in the effective
theories in both descriptions. It turns out that breaking the $U(N)$-symmetry by a scalar
vacuum expectation value in both setups generates bosons with the same masses. These bosons 
are analogs of the W-bosons in the standard model which acquire their masses by the scalar vacuum
expectation value of the Higgs field via the Higgs mechanism.

\subsection{Gauge \& gravity and gauge/gravity} \label{sec:gaugeNGravity}
This section serves to supply a detailed description of the two theories involved in the AdS/CFT correspondence:
the superconformal quantum field theory~(CFT) in flat space on one hand, and the (limit of ) string theory
in Anti de Sitter space~(AdS) on the other hand. A direct comparison of their features inevitably leads to 
the conjectured one-to-one correspondence of fields and operators, of symmetries and eventually of the 
full theories.

\subsubsection{Conformal field theory}
\label{sec:CFT}
The original formulation of the AdS/CFT correspondence involves a conformal field theory, hence~CFT, 
on the conformal boundary of anti de Sitter space. Although we will later modify the correspondence in order to 
come to more QCD-like theories breaking superconformal symmetry, we now consider the conformal
case in order to have it as a limit to check the setups deviating from the conformal case. For example
we will see that two-point functions --which are central to this work-- in the conformal case are completely determined 
by the conformal symmetry.

CFT's are invariant under the conformal group which is essentially
the Poincar\'e group extended by scale-invariance. In the context of renormalization groups it was found that 
many quantum field theories exhibit a renormalization group flow between a scale-invariant ultraviolet~(UV) fixed-point~(repelling)
and a scale-invariant infrared~(IR) fixed-point~(attracting). The quantum theory of strong interactions, QCD
is scale-invariant at it's IR fixed-point in the so-called conformal window. This fixed-point, also called the
Banks-Zaks fixed-point, appears in a distinct window of values for the number of flavors compared to 
colors~$N_f <11/2 N_c$~(for these values asymptotic freedom is guaranteed) while imposing chiral 
symmetry~(i.e. the quarks are massless) at the same time. 
So QCD itself becomes a conformal field theory in this specific limit. This is only one connection
between QCD and CFT which motivates us to believe that CFT's are a good approach to learn something about QCD
in non-perturbative regimes. 

CFT's have played a key role in understanding two-dimensional
quantum field theories since they are exactly solvable by virtue of the conformal group being infinitely
large and yielding infinitely many symmetries. If we would like to study higher dimensions 
we obtain the conformal group in $d$ dimensions by extending the Poincar\'e group with the requirement of scale invariance. In 
general the conformal group leaves the metric invariant up to an arbitrary scale factor~$g_{\mu\nu}(x)\to \Omega^2(x) g_{\mu\nu}(x)$. 
There are two types of additional transformations enhancing Poincar\'e to conformal symmetry. First, we have the 
scale transformation~$x^\mu\to \lambda x^\mu$ which is generated by~$D$ and second, there is the 
special conformal transformation~$x^\mu\to (x^\mu+a^\mu x^2)/(1+2x^\nu a_\nu+a^2x^2)$ generated by~$K_\mu$.
Denoting the Lorentz generators by~$M_{\mu\nu}$ and translations by~$P_\mu$, the conformal algebra
is given by the set of commutators  
\begin{eqnarray}
\label{eq:confAlgebra}
[M_{\mu\nu},P_\rho]=&-i (\eta_{\mu\rho}P_\nu-\eta_{\nu\rho}P_\mu)\, ,\quad 
&[M_{\mu\nu},K_\rho]=-i (\eta_{\mu\rho}K_\nu-\eta_{\nu\rho}K_\mu)\, ,\nonumber \\
\hphantom{} [M_{\mu\nu},M_{\rho\sigma}] =& -i\eta_{\mu\rho} M_{\nu\sigma} \pm \text{permutations}\, ,\quad 
&[M_{\mu\nu},D]=0\, ,\quad [D,K_\mu]=iK_\mu\, ,\nonumber \\
\hphantom{}[D, P_\mu ] =& -iP_\mu\, ,\quad &[P_\mu,K_\nu]=2iM_{\mu\nu}-2i\eta_{\mu\nu}D \, ,
\end{eqnarray}
and all other commutators vanish.
The algebra~\eqref{eq:confAlgebra} is isomorphic to the algebra of the rotation group~$SO(d,2)$
as may be seen by defining the generators of~$SO(d,2)$ in the following way
\begin{equation}
\label{eq:rotationGenerators}
J_{\mu\nu}=M_{\mu\nu}\, ,\quad J_{\mu d}=\frac{1}{2} (K_\mu-P_\mu)\, ,\quad 
J_{\mu (d+1)}=\frac{1}{2}(K_\mu+P_\mu)\, ,\quad J_{(d+1) d}= D \, .
\end{equation}
Note, that we consider all group structures in the Minkowski, not in Euclidean signature.

The conformal algebra is extended to the superconformal algebra by inclusion of fermionic
supersymmetry operators~$Q$. From the (anti)commutators we see that we need to include 
two further operators for the algebra to be closed: a fermionic generator~$S$ and the 
$R$-symmetry generator~$R$. The conformal algebra is supplemented by the relations
given schematically as follows
\begin{eqnarray}
[D,Q]=-\frac{i}{2} Q\, ,\quad [D,S]=\frac{i}{2}S\, ,\quad [K,Q]\propto S\, ,\quad [P,S]\propto Q\, ,\nonumber \\
\{Q,Q\}\propto P\, ,\quad \{S,S\} \propto K\, ,\quad \{Q,S\} \propto M+D+R
\, .
\end{eqnarray}
In~$d=4$ dimensions the $R$-symmetry group is~$SU(4)$ and the fermionic 
generators are in the~$(\mathbf{4,4})+(\mathbf{\bar 4,\bar 4})$ of $SO(4,2)\times SU(4)$.
Unitary interacting scale-invariant theories are believed to be invariant under the 
full conformal group, but this has only been proven in~$d=2$ dimensions. Given a 
classical conformally invariant field theory, conformal invariance is broken if we define
a quantum theory since this requires introduction of a cutoff breaking scale invariance.
However, the $\N=4$ supersymmetric Yang-Mills theory~(SYM) in four dimensions is special in this
sense because it is a prominent example for a superconformal quantum field theory.
It is shown in~\cite{Nahm:1977tg} that supersymmetry and conformal symmetry are sufficiently restrictive to limit
superconformal algebras to $d\le6$~dimensions.

The physically relevant representations of the conformal group are given by Eigenfunctions
of the scaling operator~$D$. Its eigenvalues are~$-i\Delta$ where~$\Delta$ is the scaling
dimension of the corresponding state~$\phi$. Its scaling transformation 
reads~$\phi(x)\to\lambda^\Delta \phi(\lambda x)$. Note that the commutators
in~\eqref{eq:confAlgebra} imply that~$P_\mu$ raises the scaling dimension of a field while~$K_\mu$
lowers it. In unitary field theories there are operator of lowest dimension, which are
called {\it primary operators}. The defining property for a primary operator~$\O_p$ is that
it has the lowest possible dimension~$[K,\O_p]=0$. Correlation functions of fields and in particular of such
primary fields are severely restricted by conformal symmetry. Two-point functions
vanish if evaluated between two fields of different dimension~$\Delta$. For a single scalar field
with dimension~$\Delta$ it was shown that
\begin{equation}
\langle \phi (0) \phi (x) \rangle \propto \frac{1}{(x^2)^\Delta} \, .
\end{equation}
Three-point functions are restricted to have the form
\begin{equation}
\langle \phi_i (x_1) \phi_j (x_2) \phi_k (x_3) \rangle = 
 \frac{c_{ijk}}{|x_1-x_2|^{\Delta_1+\Delta_2-\Delta_3} |x_1-x_3|^{\Delta_1+\Delta_3-\Delta_2}
  |x_2-x_3|^{\Delta_2+\Delta_3-\Delta_1}} \, .
\end{equation}
For~$n$-point functions with~$n\ge4$ there are more and more independent conformally invariant
functions which can appear in the correlator.
Similar expressions arise for higher-spin operators. For example the vector-vector correlator of
conserved currents~$J^a_i (x)$ (having dimension~$\Delta= d - 1$) must take the inversion covariant,
gauge invariant form
\begin{equation}
\label{eq:confJJ}
\langle J^a_i(x) J^b_j(y) \rangle
= B \frac{\delta^{ab}}{(2¹)^d} (\Box \delta_{ij} - \partial_i\partial_j) \frac{1}{(x-y)^{2 (d-2)}} \, ,
\end{equation}
where~$B$ is a positive constant, the central charge of the~$J(x)J(y)$ {\it operator
product expansion}~(OPE). The OPE of a local field theory describes the action 
of two operators~$\O_1(x)$ and~$\O_2(y)$ shifted towards each other in terms of all other operators having the 
same global quantum numbers as their product~$\O_1 \O_2$ as follows
\begin{equation}
\label{eq:ope}
\langle \O_1(x) \O_2 (y) \rangle \to 
 \langle \sum\limits_n C^n_{12}(x-y) \O_n(y) \rangle \, .
\end{equation}
In conformal field theories the energy-momentum tensor is included in the conformal algebra 
and has scaling dimension~$\Delta=d$ just as each conserved current has scaling dimension~$\Delta=d-1$.
To leading order the OPE for the energy-momentum tensor with a primary field is
\begin{equation}
T_{\mu\nu}(x) \phi (0) = \Delta \phi (0)\partial_\mu\partial_\nu x^{-2} +\dots \, ,
\end{equation}
while its two-point function turns out to be~(see e.g.~\cite{Erdmenger:1996yc})
\begin{multline}
\langle T_{\mu\nu}(x) T_{\rho\sigma} (y)\rangle = \frac{C_T}{s^{2d}}
 \mathcal{I}^T_{\mu\nu, \rho\sigma}(s)\, , \\
  \mathcal{I}^T_{\mu\nu, \rho\sigma}(s)= 
  (\delta_{\mu\alpha}-2\frac{x_\mu x_\alpha}{x^2})(\delta_{\nu\beta}-2\frac{x_\nu x_\beta}{x^2})
   \mathcal{E}^T_{\alpha\beta, \rho\sigma}\, ,
\end{multline}
where the projection operator onto the space of symmetric traceless tensors is 
given by
\begin{equation}
 \mathcal{E}^T_{\alpha\beta, \rho\sigma}= 
  \frac{1}{2}(\delta_{\alpha\rho}\delta_{\beta\sigma}+\delta_{\alpha\sigma}\delta_{\beta\rho})
    -\frac{1}{d}\delta_{\alpha\beta}\delta_{\rho\sigma} \, .
\end{equation}
The two-point function of energy momentum tensor fluctuations in a black hole background was used to compute a lower
bound on the viscosity~\cite{Policastro:2001yc} in a strongly coupled plasma as mentioned in section~\ref{sec:morePheno}.

{\bf Symmetries and conformal compactification of~$\mathbf{R}^{1,1}$}
In this paragraph we study the causal structure and symmetries of two-dimensional Minkowski space~$\mathbf{R}^{1,1}$
by a series of coordinate transformations called {\it conformal compactification} in order to generalize 
this analysis to four dimensions in the next paragraph. We will see that {\it conformally compactified}
four-dimensional Minkowski space has the same 
structure as the {\it Einstein static universe} and that it can be identified with the {\it conformal compactification}
of $AdS_5$. 

The flat space with Euclidean signature~$\mathbf{R}^d$ can be compactified to the
$d$-dimensional hypersphere~$S^n$ with isometry~$SO(d)$. A similar compactification can be obtained in
Minkowski space.
To give a specific example for the symmetry structure of globally conformal field theories in 
flat Minkowski space consider the geometry~$\mathbf{R}^{1,1}$. It can be {\it conformally}~\footnote{Here
{\it conformal} refers to a series of transformations which are demonstrated explicitly
at the end of this section.} embedded 
into the cylinder~$\mathbf{R}\times S^1$. It has the conformal isometry group structure
$SO(2,2)$, which is generated by six conformal {\it Killing vectors}. Killing vectors are 
the vectors~$X_\mu$ which leave the metric~$g_{\mu\nu}$ invariant under infinitesimal coordinate 
transformations~$x_\mu' = x_\mu + \epsilon X_\mu$. This condition can be rewritten as follows
\begin{equation}
\label{eq:killingCondition}
\mathcal{L}_{X} g_{\mu\nu} = 0 \, ,
\end{equation}
utilizing the covariant derivative~$D$ inside the {\it Lie derivative}
\begin{equation}
\label{eq:lieDerivative}
\mathcal{L}_{X} Y = [X, Y] = X Y - Y X \, .
\end{equation}
In local coordinates the Killing condition amounts to the {\it Killing equation}
\begin{equation}
\label{eq:killingEq}
\mathcal{L}_\mu X_\nu  =  D_\mu X_\nu+D_\nu X_\mu \, .
\end{equation}
In order to incorporate conformal symmetries, i.e. rescaling of the metric with a factor~$\lambda$, 
we need to generalize the condition~\eqref{eq:killingCondition} to its conformal version
\begin{equation}
\label{eq:conformalKillingCondition}
\mathcal{L}_{X} g_{\mu\nu} = \lambda g_{\mu\nu} \, .
\end{equation}
The six vectors fulfilling the Killing equation~\eqref{eq:killingEq} in~$\mathbf{R}^{1,1}$ are
given in light-cone coordinates~$r_\pm=t\pm x$ by~$\partial_\pm,\, r_\pm \partial_\pm,\, {r_\pm}^2 \partial_\pm$.
Isometries generated by the Killing vectors are related to the standard representation
for generators of the conformal group~\eqref{eq:confAlgebra}.
The two translations along the cylinder~$\mathbf{R}\times S^1$ for example are generated by the 
linear combination~$(1+{r_\pm}^2)\partial_\pm$. We identify these two generators as~$J_{03}$ and~$J_{12}$
given in the standard representation~$J_{ab}$ of the $SO(2,2)$~rotation algebra
being linear combinations of the conformal generators as given in~\eqref{eq:rotationGenerators}. 

In order to study the causal structure of this two-dimensional Minkowski space, we utilize a series
of transformations given for example in~\cite{Carroll:2004st}. This chain of transformations is often used
to draw {\it conformal diagrams} visualizing the causal structure of a specific space-time. Our aim is to
map Minkowski space into the interior of a compact space and since the 
transformations involve a conformal rescaling of the metric, this procedure is therefore often called {\it conformal
compactification}
. Beginning with
\begin{equation}
{\dd s}^2 = -{\dd t}^2 +{\dd x}^2, \quad (-\infty < t,x<\infty) \, ,
\end{equation}
we first transform to light-cone coordinates~$u_\pm=t\pm x$ giving
\begin{equation}
{\dd s}^2 = -{\dd u_+}\dd u_- \, .
\end{equation}
Now we map this into a compact region using trigonometric functions~$u_\pm = \tan\tilde u_\pm$
with~$\tilde u_\pm = (\tau\pm\theta)/2$. This gives the metric
\begin{equation}
{\dd s}^2 = \frac{1}{4\cos^2\tilde u_+ \cos^2 \tilde u_-} (-{\dd \tau}^2 +{\dd x}^2)\quad (|u_\pm| < \frac{\pi}{2}) \, ,
\end{equation}
which we simplify by a conformal rescaling to our final expression of the conformal compactification
of two-dimensional Minkowski space
\begin{equation}
\label{eq:einsteinStaticUCft2}
{\dd s}^2 =(-{\dd \tau}^2 +{\dd \theta}^2) \, .
\end{equation}
The variables are limited to the compact region~$-\pi < \theta < \pi, \, |\tau|+\theta < \pi$.   

{\bf Symmetries and conformal compactification of $\mathbf{R}^{1,p},\, p\ge 2 $}
In this paragraph we generalize the above example of~$\mathbf{R}^{1,1}$ to $(p+1)$-dimensional
Minkowski space which can be {\it conformally compactified} and then identified with the
{\it conformal compactification} of~$AdS_{p+2}$.

Note, that we can generalize the above example to~$\mathbf{R}^{1,p}$ conformally embedded 
into~$\mathbf{R}\times S^p$, which is the {\it Einstein static universe} with isometry group
$SO(2,p+1)$ as we see by an analogous series of coordinate transformations. We start from
\begin{equation}
{\dd s}^2 =-{\dd t}^2 +{\dd r}^2 + r^2 {\dd \Omega_{p-1}}^2 \, ,
\end{equation}
and transform to~$u_\pm=t\pm r$ which gives
\begin{equation}
{\dd s}^2 =-{\dd u_+ \dd u_-} +\frac{1}{4}(u_+-u_-)^2 {\dd \Omega_{p-1}}^2 \, .
\end{equation}
Then changing to~$\tilde u_\pm$ by~$u_\pm=\tan \tilde u_\pm$ leaves us with
\begin{equation}
{\dd s}^2 =\frac{1}{4 \cos^2 \tilde u_+ \cos^2\tilde u_-}(-{\dd \tilde u_+\dd \tilde u_-} 
  +\frac{1}{4}\sin^2(\tilde u_+-\tilde u_-) {\dd \Omega_{p-1}}^2) \, ,
\end{equation}
which transforms under~$\tilde u_\pm = (\tau\pm\theta)/2$ into
\begin{equation}
{\dd s}^2 =\frac{1}{4 \cos^2 \tilde u_+ \cos^2\tilde u_-}(-{\dd \tau}^2 +{\dd \theta}^2 
  + \sin^2\theta {\dd \Omega_{p-1}}^2) \, .
\end{equation}
Finally we rescale this result conformally in order to obtain
\begin{equation}
\label{eq:einsteinStatUCFT}
{\dd s}^2 =-{\dd \tau}^2 +{\dd \theta}^2 
  + \sin^2\theta {\dd \Omega_{p-1}}^2 \, ,
\end{equation}
which we extend maximally to the region~$0\le\theta \le \pi,\, -\infty<\tau<\infty$ such 
that its geometry~$\mathbf{R}\times S^p$ becomes obvious and we can identify it as 
the {\it Einstein static universe}.

To summarize these results, we state that the universal cover of the 
subgroup~$SO(2)\times SO(p+1)$ of the conformal group~$SO(2,p+1)$ examined below
equation~\eqref{eq:confAlgebra}~(take~$d=p+1$) can be identified 
with the isometry of the whole~(not only part of it) Einstein static 
universe~$\mathbf{R}\times S^p$ which we just worked out.

\subsubsection{Supergravity and Anti-de Sitter space} 
\label{sec:sugraInAds}
The AdS/CFT correspondence relates a conformal field theory~(CFT) to a supergravity in Anti de Sitter 
space~(AdS) times a compact space.
In this subsection we examine properties of supergravity in~AdS such as symmetries, geometry, 
field content and coordinate representations.

Anti de Sitter space~$AdS_d$ is a maximally symmetric $d$-dimensional {\it Lorentzian manifold} of 
constant negative curvature. It is a vacuum solution to Einstein's field equations of general relativity 
with an attractive~(negative) cosmological constant. 
A Lorentzian manifold is a pseudo-Riemann manifold with signature~$(1,d-1)$, which again is the 
generalization of a differentiable manifold equipped with a metric, called a Riemann manifold, 
on which the restriction to a positive-definite metric has been replaced by the condition
for the metric not to be degenerate. 
To be more specific consider the metric of $AdS_{p+1}$ in Poincar\'e coordinates~$(r, t,\vec x)$ given by
\begin{equation}
\label{eq:poincareMetric}
\dd s^2 = R^2 (\frac{\dd r^2}{r^2} + r^2(-\dd t^2 + {\dd \vec x}^2))\, , 
\end{equation}
where~$R$ is the radius of AdS and $r\in [0,\infty[$ is the radial AdS-coordinate.
In this form the two subgroups~$ISO(1,p)$ and~$SO(1,1)$ of the isometry group~$SO(2,p+1)$
are manifest. $ISO(1,p)$ is the Poincar\'e transformation on~$(t,\vec x)$ and $SO(1,1)$ is 
a scaling symmetry of~\eqref{eq:poincareMetric} under the transformation~$(t,\vec x, r)\to (ct, c\vec x, c^{-1} r)$.
This scaling can be identified with the dilatation~$D$~(introduced in section~\ref{sec:CFT}) 
in the AdS/CFT-dual conformal field theory. Note, that Poincar\'e coordinates do not cover the whole AdS. This
fact is easier to understand in the Euclidean version of Poincar\'e coordinates which do not cover the whole AdS,
as well. Turning the sign of
the time component of the metric~\eqref{eq:poincareMetric} we get the Euclidean analog of Poincar\'e coordinates. 
This system only covers one of the two disconnected hyperboloids of Euclidean AdS space. We will 
discuss the structure of AdS and its identification with a hyperboloid below in the Lorentzian signature case.

Rescaling~\eqref{eq:poincareMetric} by $r R^2=\varrho$ gives the standard form of the AdS-metric
\begin{equation}
\label{eq:standardMetric}
\dd s^2 = \frac{R^2}{\varrho^2}\dd \varrho^2 + \frac{\varrho^2}{R^2}(-\dd t^2 + {\dd \vec x}^2)\, , 
\end{equation}

By transformation to the inverted coordinate~$y= r^{-1},\, \dd r^2 =y^{-4} \dd y^2 $ we find another form often used in 
the literature
\begin{equation}
\dd s^2 = R^2 (\frac{\dd y^2+(-\dd t^2 + {\dd \vec x}^2)}{y^2} )\, . 
\end{equation}

{\bf Symmetries and geometry of~AdS}
In Euclidean space-time it can be shown that the $(p+1)$-dimensional hyperbolic space, 
which is the Euclidean version of~$AdS_{p+1}$, can be {\it conformally} mapped to the $(p+1)$-dimensional
disc~$D_{p+1}$ with the boundary being~$S^p$. The conformal mapping or {\it conformal
compactification} is a series of coordinate transformations used to map  a given space-time into
a compact region and study its causal structure~(see e.g.~\cite{Carroll:2004st}). One of these transformations is a conformal 
rescaling of the metric. A similar compactification is possible in Minkowski space-time as we will see
in detail in this subsection.

In order to study $AdS_{p+2}$-space, we consider the $d=p+2$-dimensional hyperboloid 
\begin{equation}
\label{eq:hyperboloid}
{X_0}^2 + {X_{p+2}}^2 - \sum\limits_{i=1}^{p+1} {X_i}^2 = R^2 \, .
\end{equation}
The hyperboloid is embedded in the flat $(p+3)$-dimensional space with one further dimension and the 
metric of the ambient space reads
\begin{equation}
\label{eq:hyperboloidMetric}
\dd s^2 = -{\dd X_0}^2 - {\dd X_{p+2}}^2 +  \sum\limits_{i=1}^{p+1} {X_i}^2 \, .
\end{equation}
This space has isometry~$SO(2,p+1)$, it is homogeneous and isotropic. A solution to~\eqref{eq:hyperboloid}
is given by the coordinate choice
\begin{eqnarray}
X_0= R \cosh \rho \cos \tau \, , \nonumber\\
X_{p+2} = R \cosh \rho \sin \tau \, , \nonumber \\
X_i = R \sinh \rho \Omega_i \, (i = 1,\dots, p+1; \sum\limits_i \Omega_i^2 = 1)\, .
\label{eq:hyperboloidSolution}
\end{eqnarray}
Note, that the radial coordinate~$\rho$ appearing here is different from the radial coordinate~$\varrho$ in the
previous section.
The metric of $AdS_{p+2}$ can be obtained by plugging this solution~\eqref{eq:hyperboloidSolution} 
into the metric~\eqref{eq:hyperboloidMetric} giving the metric in {\it global coordinates}
\begin{equation}
\label{eq:globalCoordsMetric}
\dd s^2 = R^2 (-\cosh^2\rho\, {\dd \tau}^2 + {\dd \rho}^2 + \sinh^2 \rho\, {\dd \Omega}^2) \, .
\end{equation}
In the region~$0\le \rho,\, 0\le \tau < 2\pi$, this solution covers the hyperboloid once, hence 
these coordinates are called global.
Expanding the metric~\eqref{eq:globalCoordsMetric} near the origin~$\rho=0$ 
as~${\dd s}^2\sim R^2 (-{\dd \tau}^2 + {\dd \rho}^2 +\rho^2 {\dd \Omega}^2)$, we recognize 
the cylinder-symmetry~$S^1\times \mathbf{R}^{p+1}$. The~$S^1$ represents closed time-like
curves which violate causality. In order to cure this, we unwrap the circle by taking the 
{\it universal covering} of the cylinder with~$-\infty\le \tau\le \infty$. In order to study the
causal structure of this covering space, which we will simply call AdS-space from now on, we
proceed with the conformal compactification by transforming~$\tan \theta = \sinh \rho \,(0\le\theta< \pi/2)$.
The metric becomes
\begin{equation}
{\dd s}^2 = \frac{R^2}{\cos^2\theta} (-{\dd\tau}^2 + {\dd \theta}^2 + \sin^2\theta {\dd \Omega}^2) \, ,
\end{equation}
which we then rescale conformally in order to get
\begin{equation}
\label{eq:einsteinStatUAds}
{\dd s}^2 =  (-{\dd\tau}^2 + {\dd \theta}^2 + \sin^2\theta {\dd \Omega}^2)
  \quad (0\le \theta <\pi/2,\, -\infty<\tau<\infty) \, .
\end{equation}
We have obtained the Minkowski metric of {\it Einstein's static universe}~\eqref{eq:einsteinStatUAds}.
Recall that we found the same metric with one dimension lower after conformal compactification
of Minkowski space~$\mathbf{R}^{1,p}$ in section~\ref{sec:CFT}, equation~\eqref{eq:einsteinStatUCFT}. 
Note that the range for the variable~$\theta$ is only half as big in this conformal compactification of $AdS_{p+2}$ as for the
conformal compactification of Minkowski space~$\mathbf{R}^{1,p}$. This means that the conformally compactified
$AdS_{p+2}$ only covers one half of Einstein's static universe. 

This space has topology~$\mathbf{R}\times (\text{upper half-sphere of}\,S^{p+1})$ with a boundary at the 
$S^{p+1}$-equator~$\theta=\pi/2$ which features a topology of~$\mathbf{R} \times S^p$. The boundary found here
is the analog of the boundary of the disc~$D_{n+1}$ encountered in conformally compactified Euclidean space.
Thus we find that the boundary of conformally compactified $AdS_{p+2}$ is identical to the 
conformal compactification of $(p+1)$-dimensional Minkowski space~$\mathbf{R}^{1,p}$. 
Having stated this we are now equipped with an identification of the space in which the conformal
field theory lives~(i.e. Minkowski space) with the boundary of the space on which supergravity is defined~(i.e. AdS). 
This is a fundamental building block for the AdS/CFT correspondence which we state in the
next section. Note that here the $(p+1)$-dimensional boundary of $(p+2)$-dimensional AdS is related to 
$(p+1)$-dimensional Minkowski space. This fact implies that the information given by the extra-dimension
in the gravity theory in AdS has to be encoded in the gauge theory with one dimension less in a different way.
Since this resembles the principle of holography in optics, the AdS/CFT correspondence is also called 
{AdS/CFT holography}. To be precise the AdS/CFT holography is a particular realization of the 
more general holographic principle suggested in~\cite{tHooft:1993gx,Susskind:1994vu}.

{\bf Type IIB supergravity}  
Before we state the correspondence let us review the field content, symmetries and properties of
supergravity. This examination will reveal that the symmetries of type IIB supergravity
on~$AdS_5\times S^5$ are equal
to the symmetries of the superconformal theory we examined in the preceding section~\ref{sec:CFT}.
We will further find some evidence for the fact that the classical supergravity with $p$-branes is 
suspiciously similar to the superconformal theory living on the stack of D$p$-branes.

We are specifically interested in type IIB supergravity in ten dimensions which
can be defined on~$AdS_5$ and which is the gravity theory appearing in the AdS/CFT~(gravity/gauge)
correspondence. It is the low-energy effective theory of type IIB
string theory. So both have the same massless fields: two left-handed Majorana-Weyl gravitinos, 
two right-handed Majorana-Weyl dilatinos, the metric~$g_{\mu\nu}$, the two form~$B_2$, the dilaton~$\Phi$ and
the form fields~$C_0,\, C_2,\, C_4$. the four-form~$C_4$ has a self-dual field strength~$\tilde F_5$. 
Type IIB supergravity is constructed through supersymmetry and gauge 
arguments~\cite{Pernici:1985ju, Gunaydin:1985cu} starting from the equations of motion. Further it was
shown that supergravity is stable on anti de Sitter spaces~\cite[for supergravity in 5 dimensions]{Gibbons:1983aq} 
with an appropriate set of boundary conditions. Existence of the self-dual five-form field strength obstructs the covariant 
formulation of an action, such that we need to find an action and add a self-duality constraint by hand.
The bosonic part of the action can be written as the sum of a Neveu-Schwarz~(NS), a Ramond-Ramond~(RR)
and a Chern-Simons~(CS) term
\begin{eqnarray}
\label{eq:IIBSugraAction}
S =S_{\text{NS}}+S_{\text{RR}}+S_{\text{CS}}  \\
    = \frac{1}{2\kappa^2}\int \dd^{10} x\sqrt{-g} \Bigg [ e^{-2\Phi} 
      \left ( \mathcal{R}+ 4\partial_\mu\Phi\partial^\mu\Phi -\frac{1}{2}|H_3|^2 \right ) \\
      -\frac{1}{2} \left ( |F_1|^2+|\tilde F_3|^2+\frac{1}{2}|\tilde F_5|^2 \right )\Bigg ] \\
      -\frac{1}{4\kappa^2} \int C_4\wedge H_3\wedge F_3\, ,
\end{eqnarray}
with~$F_{n+1}=\dd C_n,\, H_3 =\dd B_2, \,\tilde F_3 = F_3-C_0 H_3,\, \tilde F_5 = 
F_5-\frac{1}{2} C_2\wedge H_3 +\frac{1}{2} B_2\wedge F_3$ and the curvature scalar~$\mathcal{R}$ . 
This is the theory which we will relate to a conformal field theory through the AdS/CFT correspondence.

Note, that this supergravity can also be Kaluza-Klein-compactified on~$S^5$ and then truncated utilizing the
Freund-Rubin Ansatz choosing the five-form to be proportional to the volume form of~$S^5$.
The resulting theory is {\it gauged supergravity} on~$AdS_5$ with possible supersymmetries~$SU(2,2|\N/2),\,
\N=2,4,6,8$. Here we only mention the maximally supersymmetric case~$\N=8$ which has gauge
group~$SU(4)$. The $SO(6)$-isometry on the compactification manifold~$S^5$ becomes
the local gauge symmetry in the truncated theory. In this thesis we will not consider the gauged
supergravities.

\subsubsection{Statement of the AdS/CFT-correspondence}
\label{sec:stateAdsCft}
In this section we state the correspondence and provide a comparison of the gravity theory
with the gauge theory which leads to the conjecture. Further, we include a dictionary and 
a discussion how to translate or identify objects, e.g. operators in the gauge theory with those, e.g. 
fields in supergravity.

The AdS/CFT-conjecture states that (for the case of D3-branes) type IIB superstring
theory compactified on $AdS_5\times S^5$ background described in section~\ref{sec:sugraInAds} is
dual to $\N = 4$, $d = 4$ Super-Yang-Mills theory with gauge group~$SU(N)$~\footnote{
Or rather with gauge group~$U(N)$ according to~\cite{Maldacena:2001ss}.} as described in
section~\ref{sec:CFT}. This equivalence is called the AdS/CFT-correspondence.
The string theory background corresponds to the ground state of the gauge
theory, while excitations and interactions in one description correspond to
excitations and interactions in the dual description. There are three different levels on 
which the gauge/gravity correspondence is conjectured. The {\it strong form} conjectures
that the full quantized type IIB string theory on~$AdS_5\times S^5$ with string coupling~$g_s$ is dual to the $\N=4$ Super-Yang-Mills
theory~(SYM) in four dimensions with gauge group~$SU(N)$ and Yang-Mills coupling~$g_{\text{YM}}$
in its superconformal phase. On the string theory side the $AdS_5$ and~$S^5$ have the same radius~$R$ and the 
five-form~$F_5$ has integer flux~$\int F_5 = N$. The parameters from the string theory
are related to those on the gauge theory side by
\begin{equation}
\label{eq:adsCftParameters}
g_s={g_{\text{YM}}}^2\, , \quad R^4=4\pi g_s N (\alpha ')^2 \, .
\end{equation}

On the second level a weaker form of the conjecture utilizes the {\it 't Hooft limit}
\begin{equation}
\lambda := g_{\text{YM}}^2 N = \text{fixed}\, ,\quad N\to \infty \, .
\end{equation}
The gauge theory,~$\N=4$ SYM, in this limit can be expanded in~$1/N$ and representing a topological 
expansion of the field theory's Feynman diagrams. It is conjectured to be equivalent to type IIB string theory, 
which can be expanded in powers of the string coupling~$g_s = \lambda / N$ representing a weak 
coupling (classical) string perturbation theory, i.e. a string loop expansion.  

The third and weakest form of the conjecture is the {\it large~$\lambda$ limit}. Expanding the 
SYM theory for large~$\lambda$ in powers~$\lambda^{-1/2}$ corresponds to an~$\alpha'$ expansion
on the gravity side. On this level the AdS/CFT correspondence conjectures that type IIB
supergravity on~$AdS_5\times S^5$ is dual to the large~$\lambda$ expansion of $\N=4$~SYM theory.

\paragraph {Road map to the conjecture}
In order to put forward an argument for the AdS/CFT conjecture, consider a stack of $N$ parallel 
D3-branes in type IIB string theory on flat Minkowski space. Two kinds of string excitations exist
in this setup: the closed strings propagating through the ten-dimensional bulk and the open 
strings which end on the D3-branes describing brane excitations. At energies lower than the 
inverse string length~$1/l_s$ only massless modes are excited such that we can integrate out
massive excitations to obtain an effective action splitting into three parts~$S = S_{\text{bulk}}+
S_{\text{brane}}+S_{\text{interaction}}$. The bulk action is identical to the action of ten-dimensional
supergravity~\eqref{eq:IIBSugraAction} describing the massless closed string excitations in the bulk plus possible
higher derivative corrections. These corrections come from integrating out the massive modes and they
are suppressed since they are higher order in~$1/\text{cut-off}=\alpha'$. 
The brane action is given by the Dirac-Born-Infeld action~(DBI) on the 
stack of D3-branes already given in~\eqref{eq:nonAbDbiAction} for D$p$-branes. It contains 
the~$\N=4$ SYM action as discussed below~\eqref{eq:nonAbDbiAction} plus higher derivative 
corrections such as~${\alpha'}^2 \mathrm{tr} {F^4}$. The interaction between the bulk modes and the 
brane modes is described by~$S_{\text{interaction}}$. These are suppressed at low energies
corresponding to the fact that gravity becomes free at large distances. In the same limit the 
higher derivative terms vanish from the brane and bulk action leaving two decoupled regimes 
describing open strings ending on the brane and closed strings in the bulk, respectively.

Now let us take the same setup of $N$~D3-branes but describe its low energy behavior
in an alternative way, with supergravity. It will turn out that we can again find two decoupled
sectors of the effective low-energy theory. In supergravity D$p$-branes are massive charged
objects sourcing supergravity fields. We have seen the D3-brane solution explicitly 
in~\eqref{eq:3BraneBackgMetric},~\eqref{eq:3BraneBackgFiveF} and~\eqref{eq:3BraneBackgH3}. 
Note that the component~$g_{tt}={H_3}^{-1/2}=-(1+R^4/r^4)$ being the measure for physical time or equivalently energy
is not constant but depends on the radial AdS coordinate~$r$. For an observer at infinity~$r=\infty$ 
this means that the local energy~$E_{\text{object}}(r=\text{constant})$ of any object placed at some 
constant position~$r$ is red-shifted on the way to the observer. The observer measures
\begin{equation}
E (r=\infty) = (1+\frac{R^4}{r^4})^{-1/4} E_{\text{object}}(r)\, .
\end{equation}
Approaching the position~$r=0$ which we call the horizon, the object appears to have smaller and 
smaller energy. This means that in the low-energy limit we can have excitations with 
arbitrarily high local energy~$E_{\text{object}}$ as long as we keep them close enough to the horizon.
This regime of the theory is called the {\it near-horizon region}.
On the other hand modes that travel through the whole bulk are only excited in the low-energy limit 
if their energy is sufficiently small. These are the two regimes~(bulk and near-horizon) of the theory which decouple
from each other in analogy to the string theory approach. In the full theory bulk excitations interact
with the near-horizon region because the D$p$-brane located at the horizon absorbs the bulk excitations
with a cross section~$\sigma\sim \omega^3 R^8$~\cite{Klebanov:1997kc, Gubser:1997yh}. 
However, in the low-energy limit this cross section becomes small because the bulk excitations 
have a wave length which is much bigger than the gravitational  size of the brane~$\sim \O(R)$. 
The low-energy excitations in the near-horizon region which have an energy low enough
to travel through the whole bulk are caught near the horizon by the deep gravitational potential
produced by the massive $p$-branes at~$r=0$.
In the near-horizon region~$r\ll R$ the metric~\eqref{eq:3BraneBackgMetric} can be approximated
with~$H_3=(1+(R/r)^4)\sim (R/r)^4$ such that it becomes
\begin{equation}
{\dd s}^2 = \frac{r^2}{R^2} \left ( -{\dd t}^2 +{\dd\mathbf{x}}^2 \right) +
 R^2\frac{{\dd r}^2}{r^2} +R^2 {\dd \Omega_5}^2 \, ,
\end{equation}
which is the metric of the AdS-space~$AdS_5\times S^5$ in the same coordinates as~\eqref{eq:standardMetric}.
This means that the effective theory near the horizon is string theory~(any kind of excitations possible)
on~$AdS_5\times S^5$ and it decouples from the bulk theory which itself is supergravity~(low-energy
excitations only) in the asymptotically~($r \gg R$ and~$H_3=1$) flat space.

In both descriptions of D$p$-branes we have now found two decoupled theories in the low-energy limit:

1. For the classical supergravity solution we found supergravity on~$AdS_5\times S^5$ near the 
horizon and supergravity in the flat bulk. 

2. For the string theoretic D$p$-brane description we found the $\N=4$ SYM theory in flat Minkowski space
on the stack of D3-branes and ten-dimensional supergravity in the flat bulk.
\begin{center}
\parbox{0.7\textwidth}{
Since supergravity in the flat bulk is present in both descriptions, we are lead to identify the near-horizon
supergravity in~$AdS_5\times S^5$ and the $\N=4$~SYM brane theory, as well.}
\end{center}

{\bf The dictionary}
The natural objects to consider in a conformal field theory are operators~$\O$ since conformal 
symmetry does not allow for asymptotic states or an S-matrix. On the other side of the correspondence
we have fields~$\phi$ which have to satisfy the IIB supergravity equations of motion in~$AdS_5\times S^5$. 
AdS/CFT states that the CFT-operators~$\O$ are dual to the fields~$\phi$ on $AdS_5\times S^5$ in a specific way.

Consider as an example for a field~$\phi$ the dilaton field~$\Phi$. Its expectation value 
gives the value of the dynamical string coupling which
is constant only for the special case of D3-branes which we do not consider here~(see equation~\eqref{eq:dilatonE}).
Moreover, the dilaton expectation value in string theory is determined by boundary condition for the
dilaton field at infinity~(AdS boundary). 
By the correspondence between couplings~\eqref{eq:adsCftParameters} we know that the coupling in the gravity theory
also determines the gauge coupling~$g_{YM}$ or 't Hooft coupling~$\lambda$. Thus changing the
boundary value~$\lim\limits_{r\to r_{\text{bdy}}}\Phi (r)=\Phi_{\text{bdy}}$ of the (string theory) dilaton field from zero
to a finite value~$\Phi_{\text{bdy}}$ changes the coupling in the dual gauge theory . 

On the gauge theory side a change in the gauge coupling is achieved 
by changing the term~$\int \dd^4 x \Phi_{\text{bdy}} \O$ in the action, where
$\O$ is the operator~$\mathrm{tr} F^2$ containing the gauge field strength~$F$ of the gauge theory. 
$\O$ is a marginal operator and thus its presence changes the value of the gauge theory coupling 
compared to the case when the marginal operator~$\O$ is not included into the gauge theory. 

So we see by considering this special case of the dilaton, that changing the boundary value of the field~$\phi$ 
leads to the introduction of a marginal operator in the dual field theory.
Therefore the AdS-boundary value~$\phi_{\text{bdy}}$ of the supergravity field~$\phi$ acts as a source 
for the operator~$\O$ in the dual field theory. This statement is conjectured to hold for all fields~$\phi$ 
in the gravity theory and all dual operators~$\O$ of the gauge theory~(not only marginal ones).

Let us be a bit more precise on what we mean by the boundary value~$\phi_{\text{bdy}}$ of the supergravity
field~$\phi$. In the geometry of~$AdS_5\times S^5$ we decompose the field~$\phi$ into spherical 
harmonics on the~$S^5$ which produces Kaluza-Klein towers of excitations with different masses 
coming from the compactification. These latter excitations live on~$AdS_5$ with the metric~$g$ 
and~(neglecting interactions) they have to satisfy the free field equation of motion
\begin{equation}
\label{eq:adsEOM}
(\Box_g+m^2) \phi= 0 \, ,
\end{equation}
which has two independent asymptotic solutions near the boundary~$r=\infty$
\begin{equation}
\label{eq:adsSolution}
\phi(r) = \phi_{\text{nn}}r^{4-\Delta} + \phi_{\text{n}}  r^{\Delta}+ \dots \, .
\end{equation} 
Here the 4 is the dimension of the AdS-boundary and $\Delta$ is the conformal dimension of the field.
The first term with the coefficient~$\phi_{\text{nn}}$ is  the non-normalizable solution, the second term
with the coefficient~$\phi_{\text{n}}$ gives the normalizable one.
The two expansion coefficients~$\phi_{\text{n}}$ and~$\phi_{\text{nn}}$ are related by the AdS/CFT correspondence 
to the vacuum expectation value~$\langle\O\rangle$ of the dual operator and the external source for the operator respectively.
This means that only the non-normalizable solution acts as a source in the way we discussed above 
in the example of the dilaton field
\begin{equation}
\Phi(r) =   \Phi_{\text{bdy}} + \langle\text{tr} F^2\rangle r^{-4}  \, ,
\end{equation}
where we used that the dilaton field has conformal dimension~$\Delta=0$ and we note that the 
non-normalizable part is related to the asymptotic string coupling~$g_s=e^{\Phi_{\text{bdy}}}$.

By virtue of the operator-field duality we can also identify correlation functions in the two theories but since this
discussion is crucial for the present work it will be presented in a separate section in~\ref{sec:G}.

{\bf Symmetry matching}
Let us recall the symmetries of IIB supergravity on~$AdS_5\times S^5$~(as considered in~\ref{sec:sugraInAds}) 
and those of~$\N=4$ super-Yang-Mills~(as studied in~\ref{sec:CFT}) in order to check 
if the symmetries match on both sides and in order to use these matching symmetries
as hints which quantities are to be identified with each other in the correspondence.

The $\N=4$ Super-Yang-Mills theory on the gauge theory side of the correspondence has the following symmetries:
a $SU(2,2)$~conformal symmetry and the $SU(4)$~R-symmetry as discussed in section~\ref{sec:CFT}.
It contains the~$U(N)$ gauge vector~$A_\mu$, the fermionic fields~$\lambda^{1,2,3,4}$ and the six scalars~$X^{4,5,6,7,8,9}$.
All these fields live in the adjoint representation of the gauge group. 

On the other hand we have supergravity which in~$AdS_5$ has the {\it isometry}~(transformations
leaving the metric invariant) group~$SO(4,2)$. The $S^5$ has isometry~$SO(6)$. We consider the
covering groups of~$SO(4,2)$ and~$SO(6)$ which are~$SU(2,2)$ and~$SU(4)$, respectively. The
$AdS_5\times S^5$-background preserves as much supersymmetries as flat Minkowski space does. 
Under the spatial isometries~$SU(2,2)\times SU(4)$ the supercharges transform as~$(4,4) +(\bar 4,\bar 4)$
and so the spatial isommetries combine with the conserved supercharges to give the full symmetry
group of~$\N=4$ Super-Yang-Mills: the superconformal group~$PSU(2,2|4)$ as written out in
section~\ref{sec:CFT}.  

A direct comparison of these symmetries shows that the global R-symmetry group~$SU(4)$ of SYM can be identified 
with the isometries of~$S^5$. Finally the conformal symmetry~$SU(2,2)$ is identified with the 
isometry group of~$AdS_5$.

{\bf Holography}
The AdS/CFT duality carries also the character of a holography. This understanding arises from 
the observation that a four-dimensional gauge theory is related to an effectively five-dimensional gravity
theory. 
The gauge theory lives on the boundary of the Anti de Sitter space. We already saw this in section~\ref{sec:sugraInAds} 
comparing the conformal compactifications of AdS on one hand and of four-dimensional Minkowski 
space on the other. There we found that the $(p+1)$-dimensional boundary of~$AdS_{p+2}$ can be conformally
mapped to one half of the Einstein static universe. In $p$ dimensions this is a whole Einstein static universe.
Minkowski space was mapped to exactly the same $p$-dimensional whole Einstein static universe. 
Since the first four coordinates in both theories are identified as the common
The $p$-dimensional Minkowski space, the extra coordinate in the gravity theory is the radial AdS coordinate. On the 
gauge theory side this coordinate translates into an energy or renormalization scale at which the gauge 
theory is defined. Excitations with energies higher than this scale are integrated out on the gauge theory side.
So placing the gauge theory on the AdS boundary corresponds to setting the renormalization scale
to infinity and therefore not integrating out any fields. As we decrease the energy scale, we integrate out
more and more fields moving the gauge theory to finite values of the radial AdS coordinate. Note, that this
picture is an incomplete heuristic view on the topic which can for example not answer why the correspondence
should still be valid at a finite radius which is not the boundary of AdS.

{\bf Evidence}
Although still a conjecture the AdS/CFT correspondence has passed a convincing number of tests of its validity.
The first check of the conjecture is the matching of all global symmetries. These are independent 
of the couplings and agree exactly as discussed in the above paragraph.

Generic objects to compute both on the AdS side and then also on the CFT side are correlation functions.
It was found in several cases that the $n$-point functions of operators~$\O$ in the gauge theory match 
exactly the $n$-point functions of the supergravity field~\cite{Lee:1998bxa} conjectured to be dual to~$\O$.

Since the correspondence is a duality relating one theory at strong coupling to another one at weak coupling, 
it is not in general possible to compute correlation functions on both sides perturbatively. However, there
are correlation functions which do not depend on the coupling~$\lambda$. $\N=4$ SYM theory is superconformal
and therefore scale-invariant. The superconformal group~$PSU(2,2|4)$ remains exact up to one-loop exact
anomalies appearing upon quantization. These one-loop diagrams appear when the theory is coupled to 
gravitational or external $SU(4)_R$ gauge fields. All higher order contributions vanish. The one-loop contributions can be calculated and 
so correlation functions of e.g. global R-currents can be calculated even at strong coupling. Thus it is possible 
to compare these correlation functions to those of the dual fields in supergravity which are computed perturbatively.
Since we do not know how a specific normalization in the gauge theory translates into a normalization of 
the gravity theory, we use the two-point functions in each theory to normalize the R-current~$J \to \tilde J$ such that
\begin{equation}
\langle \tilde J^a(x) \tilde J^b(y) \rangle  = \frac{\delta^{a b}}{|x-y|^{2 \Delta}}\, ,
\end{equation}
where~$\Delta$ is the conformal dimension of the operator~$J$.
The three-point correlator of R-currents normalized to the two-point correlator was computed in SYM 
and it was found to agree with the three-point correlation function computed from the dual supergravity 
vector field~$A^\mu$ normalized to its two-point correlator
\begin{equation}
\langle \tilde J_\mu^a(x) \tilde J_\nu^b(y) \tilde J_\rho^c(z) \rangle_{\text{Sugra}}  = 
  \langle \tilde J_\mu^a(x) \tilde J_\nu^b(y) \tilde J_\rho^c(z) \rangle_{\text{CFT}}  \, .
\end{equation}
In~\cite{Lee:1998bxa} all three-point functions of normalized chiral operators in four-dimensional
$\N=4$~SYM computed perturbatively were shown to agree with the correlators obtained from AdS/CFT
in the limit of large number of colors~$N$.
Similar results were obtained for other correlators and no counter example has been found yet.

Also the spectrum of chiral operators does not change with any coupling and has for example been compared
in the review~\cite{Aharony:1999ti}. 
The moduli space of the theories and the behavior of the theories under deformation by relevant
or marginal operators was also reviewed in~\cite{Aharony:1999ti}. These examinations have not yielded any contradiction.

After having motivated the conjecture in its original form featuring adjoint matter fields only,
we now expand the correspondence in order to include fundamental matter.

\subsection{Generalizations of~AdS/CFT: Quarks and mesons} \label{sec:mesons}
The original AdS/CFT conjecture does not include matter in the fundamental representation of the gauge 
group but only adjoint matter. In order to come closer to a QCD-like behavior we therefore investigate how
to incorporate quarks and their bound states in this section. We focus on the main results of~\cite{Karch:2002sh}
and~\cite{Kruczenski:2003be}, however for a concise review the reader is referred to~\cite{Erdmenger:2007cm}.

Since AdS/CFT has been discovered a lot of modifications of the original conjecture have 
been proposed and analyzed. This is always achieved by modifying the gravity theory in an appropriate way. 
For example the metric on which the gravity theory is defined may be changed to produce chiral symmetry breaking
in the dual gauge theory~\cite{Constable:1999ch,Babington:2003vm}. Other modifications put the 
gauge theory at finite temperature and produce 
confinement~\cite{Witten:1998zw}. Besides the introduction of finite temperature the inclusion of fundamental
matter, i.e. quarks, is the most relevant extension for us since we are aiming at a qualitative description
of strongly coupled QCD effects at finite temperature. This kind of effects are the ones observed at the
RHIC~heavy ion collider.

\begin{figure}
\includegraphics[width=0.9\textwidth]{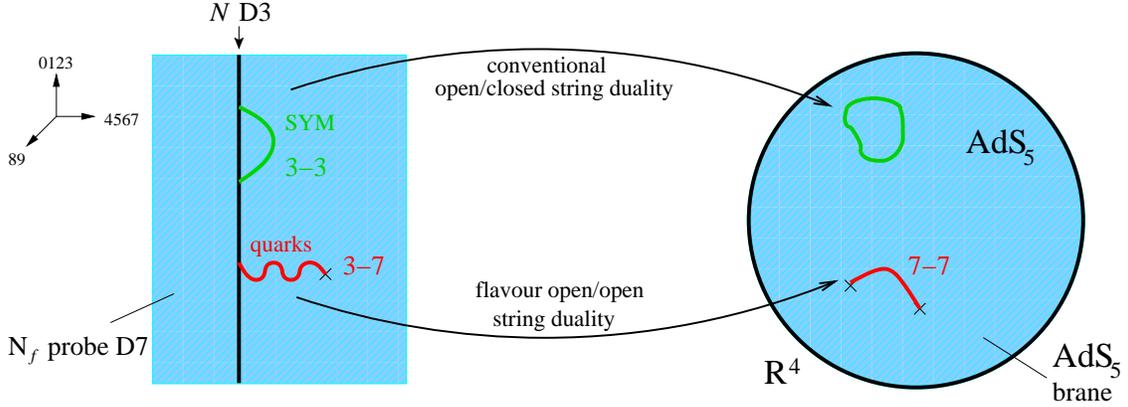}
\caption{ \label{fig:d3D7Setup}
The figure sketches the original AdS/CFT correspondence between open and closed strings and its
extension to fundamental matter relating open strings to each other. On the left side the geometry of 
a stack of coincident $N$~D3-branes~(represented by the thick vertical line) and a small number of 
coincident $N_f$~D7-branes is shown. This is the setup within which the full string theory 
description is reduced to the effective Dirac-Born-Infeld description on the world volume of the D7-branes.
On the left side of the figure the geometry of~$AdS_5\times S^5$ is outlined on which the classical
supergravity description is defined. At each point on the disc representing~$AdS_5$ an~$S^5$ exists
but is not drawn for simplicity. The curved lines with labels~$p-q$ represent strings starting at the stack of
D$p$ branes and ending on the stack of~D$q$-branes. 
This figure has been kindly provided by the authors of~\cite{Erdmenger:2007cm}.
}
\end{figure}

{\bf Adding flavor to AdS/CFT}
The change we have to make on the gravity side in order to produce fundamental matter on the gauge
theory side is the introduction of a small number~$N_f$ of D7-branes. These are also called {\it probe
branes} since their backreaction on the geometry originally produced by the stack of $N$ D3-branes
is neglected. Strings within this D3/D7-setup now have the choice of starting~(ending) on the
D3- or alternatively on the D7-brane as visualized by figure~\ref{fig:d3D7Setup}. 
Note that the two types of branes share the four Minkowski directions~$0,1,2,3$ in which also the dual gauge theory
will extend on the boundary of AdS as visualized in figure~\ref{tab:d3D7Coords}.

\begin{figure}
\begin{center}
\begin{tabular}[c]{|c|c|c|c|c|c|c|c|c|c|c|}
\hline
 & 0 & 1 & 2 & 3 & 4 & 5 & 6 & 7 & 8 & 9 \\
 \hline
 D3 & x & x & x & x  & & & & & & \\
 \hline
 D7 & x & x & x & x & x & x & x & x&  & \\
 \hline
\end{tabular} 
\end{center}
 \caption{ \label{tab:d3D7Coords}
 Coordinate directions in which the D$p$-branes extend are marked by 'x'. D3- and D7-branes 
 always share the four Minkowski directions and may be separated in the $8,9$-directions 
 which are orthogonal to both brane types.
 }
\end{figure}

The configuration of one string ending on $N$ coincident D3-branes produces an $SU(N)$~gauge symmetry of 
rotations in color space. Similarly the $N_f$ D7-branes generate a $U(N_f)$~flavor gauge
symmetry. We will call the strings starting on the stack of D$p$-branes 
and ending on the stack of D$q$-branes $p-q$ strings. The original $3-3$ strings are unchanged while
the $3-7$- or equivalently $7-3$ strings are interpreted as quarks on the gauge theory side of the correspondence.
This can be understood by looking at the $3-3$ strings again. They come in the adjoint representation of the
gauge group which can be interpreted as the decomposition of a bifundamental 
representation~$(N^2 -1)\oplus 1 = N \otimes \bar N$. So the two string ends on the D3-brane are interpreted
as one giving the fundamental, the other giving the anti-fundamental representation in the gauge theory.
In contrast to this the $3-7$ string has only one end on the D3-brane stack corresponding to a single 
fundamental representation which we interpret as a single quark in the gauge theory. 

We can also give mass to these quarks by seperating the stack of D3-branes from the D7-branes
in a direction orthogonal to both branes. Now $3-7$ strings are forced to have a finite length~$L$
which is the minimum distance between the two brane stacks. On the other hand a string is an
object with tension and if it assumes a minimum length, it needs to have a minimum energy being
the product of its length and tension. The dual gauge theory object is the quark and it now also 
has a minimum energy which we interpret as its mass~$M_q= L/(2\pi \alpha')$.

The $7-7$ strings decouple from the rest of the theory since their effective coupling is suppressed 
by~$N_f/N$. In the dual gauge theory this limit corresponds to neglecting quark loops which is often 
called {\it the quenched approximation}. Nevertheless, they are important for the description of 
mesons as we will see below.

Let us be a bit more precise about the fundamental matter introduced by $3-7$ strings. The gauge
theory introduced by these strings (in addition to the original setup) gives a $\N=2$ supersymmetric
$U(N)$ gauge theory containing~$N_f$ fundamental hypermultiplets.

{\bf D7 embeddings \& meson excitations}
Mesons correspond to fluctuations of the D7-branes~\footnote{To be precise the fluctuations 
correspond to the mesons with spins 0, 1/2 and 1~\cite{Kruczenski:2003be,Kirsch:2006he}.} embedded 
in the $AdS_5\times S^5$-background generated by the D3-branes. 
From the string-point of view these fluctuations are
fluctuations of the hypersurface on which the $7-7$ strings can end, hence these are small 
oscillations of the $7-7$ string ends. The $7-7$ strings again lie in the adjoint representation of
the flavor gauge group for the same reason which we employed above to argue that $3-3$ strings
are in the adjoint of the (color) gauge group. Mesons are the natural objects in the adjoint flavor representation.
Vector mesons correspond to fluctuations of the gauge field on the D7-branes.

Before we can examine mesons as D7-fluctuations we need to find out how the D7-branes are 
embedded into the 10-dimensional geometry without any fluctuations. Such a stable configuration needs to
minimize the effective action. The effective action to consider is the world volume action of the D7-branes
which is composed of a Dirac-Born-Infeld as given in~\eqref{eq:dbiAction} and a topological Chern-Simons part
\begin{equation}
\label{eq:d7MesonAction}
S_{\text{D}7}= -T_{\text{D}7} \int \dd^{8}\sigma e^{-\Phi}
 \sqrt{
  -\det  \left \{ 
  P[g+B]_{\alpha\beta} + 
  (2\pi \alpha') F_{\alpha\beta}  
\right \}
 }  + \frac{(2\pi \alpha')^2}{2} T_{\text{D7}} \int P[C_4]\wedge F \wedge F\, .
\end{equation}
The preferred coordinates to examine the fluctuations of the D7 are obtained from the coordinates
given in~\eqref{eq:standardMetric} by the transformation~$\varrho^2 = {w_1}^2 + \dots + {w_4}^2 , \,
r^2=\varrho^2+{w_5}^2+{w_6}^2$.
Then the metric reads
\begin{equation}
\label{eq:metricWithW}
{\dd s}^2= \frac{r^2}{R^2}{\dd \vec x}^2 + \frac{R^2}{r^2} ({\dd \varrho}^2 + \varrho^2 {\dd \Omega_3}^2
 + {\dd w_5}^2 + {\dd w_6}^2) \, , 
\end{equation}
where~$\vec x$ is a four vector in Minkowski directions~$0,1,2,3$ and $R$ is the AdS radius.
The coordinate~$r$ is the radial AdS coordinate while~$\varrho$ is the radial coordinate on the 
coincident D7-branes. For a static D7 embedding with vanishing field strength~$F$ on the D7 world volume
the equations of motion are
\begin{equation}
\label{eq:w56EOM}
0 = \frac{\dd}{\dd\varrho} \left ( \frac{\varrho^3}{\sqrt{1+{w_5'}^2+{w_6'}^2}}\frac{\dd w_{5,6}}{\dd \varrho} \right ) \, ,
\end{equation}
where~$w_{5,6}$ denotes that these are two equations for the two possible directions of fluctuation.
Since~\eqref{eq:w56EOM} is the same type of equation as for the motion of a supergravity field in the bulk which
was considered in~\eqref{eq:adsEOM}, also the solution takes a form resembling~\eqref{eq:adsSolution} 
near the boundary
\begin{equation}
w_{5,6} = L + \frac{c}{\varrho^2} +\dots \, ,
\end{equation}
with $L$~being the quark mass acting as a source and $c$~being the expectation value of the operator 
which is dual to the field~$w_{5,6}$. While $c$ can be related to the scaled quark 
condensate~$c\propto \langle\bar q q\rangle(2\pi\alpha')^3$.

If we now separate the D7-branes from the stack of D3-branes the quarks become massive and the 
radius of the~$S^3$ on which the D7 is wrapped becomes a function of the radial AdS coordinate~$r$.
The separation of stacks by a distance~$L$ modifies the metric induced on the D7~$P[g]$ such that it contains  
the term~$R^2 \varrho^2/(\varrho^2+L^2){\dd\Omega_3}^2$. This expression vanishes at a 
radius~$\varrho^2=r^2-L^2=0$ such that the~$S^3$ shrinks to zero size at a finite AdS radius. 

Fluctuations about these~$w_{5}$ and~$w_6$ embeddings give scalar and pseudoscalar mesons.
We take
\begin{equation}
w_5 = 0 + 2\pi\alpha' \chi\, , \quad w_6 = L + 2\pi\alpha' \varphi
\end{equation}
After plugging these into the effective action~\eqref{eq:d7MesonAction} and expanding to quadratic
order in fluctuations we can derive the equations of motion for~$\varphi$ and~$\chi$. As an example 
we consider scalar fluctuations using an Ansatz
\begin{equation}
\label{eq:mesonAnsatz}
\varphi = \phi (\varrho) e^{i \vec k\cdot \vec x} \mathcal{Y}_l(S^3)\, ,
\end{equation}
where~$\mathcal{Y}_l(S^3)$ are the scalar spherical harmonics on the~$S^3$, $\phi$ solves the radial
part of the equation and the exponential represents propagating waves with real momentum~$\vec k$.
We additionally have to assume that the mass-shell condition 
\begin{equation}
M^2 = -{\vec k}^2
\end{equation}
is valid. Solving the radial part of the equation we get the 
hypergeometric function~$\phi \propto F(-\alpha,\, -\alpha+l+1;\, (l+2);\, \frac{-\varrho^2}{L^2})$ and
the parameter
\begin{equation}
\alpha=-\frac{1-\sqrt{1-{\vec k}^2 R^4/L^2}}{2}
\end{equation}
summarizes a factor 
appearing in the equation of motion. In general this
hypergeometric function may diverge if we take~$\varrho\to \infty$. But since this is not compatible with our 
linearization of the equation of motion in small fluctuations, we further demand normalizability of the 
solution. This restricts the sum of parameters appearing in the hypergeometric function to take the integer values
\begin{equation}
n = \alpha - l -1  \, , \quad n=0,\, 1,\, 2, \dots \, .
\end{equation}
With this quantization condition we determine the scalar meson mass spectrum to be
\begin{equation}
\label{eq:Ms}
M_{\text{s}} =  \frac{2 L}{R^2} \sqrt{(n+l+1)(n+l+2)}\, ,
\end{equation}
where~$n$ is the radial excitation number found for the hypergeometric function.
Similarly we can determine pseudoscalar masses
\begin{equation}
M_{\text{ps}} =  \frac{2 L}{R^2} \sqrt{(n + l + 1)(n+l+2)}\, .
\end{equation}
For vector meson masses we need to consider fluctuations of the gauge field~$A$ appearing in
the field strength~$F$ in equation~\eqref{eq:d7MesonAction}. The formula for vector mesons
(corresponding to e.g. the $\varrho$-meson of QCD) is 
\begin{equation}
\label{eq:susyVectorMesonM}
M_{\text{v}} = \frac{2 L}{R^2}\sqrt{(n+l+1)(n+l+2)} \, .
\end{equation}
Note that the scalar, pseudoscalar and vector mesons computed within this framework show 
identical mass spectra. Further fluctuations corresponding to other mesonic excitations can be 
found in~\cite{Kruczenski:2003be,Kirsch:2006he}.

\subsection{AdS/CFT at finite temperature} \label{sec:finiteTAdsCft}
This present work aims at a qualitative understanding of the finite temperature effects inside a plasma governed by
QCD at strong coupling. Our focus will mainly be on the fundamental matter, the quarks and their bound states, the mesons.
In this section we describe how to construct a gravity dual to a finite temperature gauge theory with flavor 
degrees of freedom, i.e. fundamental matter.

{\bf A thermodynamics reminder}
Within this paragraph we remind ourselves of some basic concepts of thermodynamics which will be 
important for our desired study of a thermal quantum field theory at strong coupling. 

The first thing to note is that quantum field theory in its application to collider physics is a theory at zero temperature.
However, in order to study heavy-ion collision experiments, neutron stars and cosmological setups in which 
there are high enough particle number and energy densities in order to justify the thermodynamic limit, 
thermal quantum field theories have been developed in great detail~\cite[as an example]{Das:1997gg}. There
are two formalisms which can be used to introduce a notion of temperature into quantum field theory.
The simpler method is the {\it imaginary-time formalism} which basically Euclideanizes the time-coordinate~$t$
by Wick rotatation~$t\to -i \tau_{\text{Euclid}}$ and afterwards compactifies it on a circle with 
period~$\beta = 1/T$ such that~$\tau+\beta \sim \tau$. Any correlation function defined on this 
periodic Euclidean space-time can be Fourier-transformed to the four momentum coordinates~$\vec k$.
Because of the periodicity and limited range in the time-coordinate~$0\le\tau\le \beta$ the Fourier frequency~$k_0$ is 
discrete~$k_0 = 2\pi T n, \, n=0,\,1,\dots$. These are the real-valued {\it Matsubara frequencies}. 
The disadvantage here is that we basically trade the time coordinate for temperature and therefore
loose any notion of temporal evolution of our system. Therefore we can only describe equilibrium states 
with this formalism. In order to incorporate time and temperature at equal footing we need to employ the more complicated
{\it real-time formalism}. We will come back to this issue when discussing correlation functions in section~\ref{sec:G}.

If we have the notion of a temperature in our quantum field theory, we can also define a chemical potential~$\mu$
for a conserved total charge~$Q = \int_{\text{volume}}\! J^0$ with a charge density~$J^0$. Here we assume that
the chemical potential~$\mu$ is constant with respect to the four Minkowski directions~$\vec x$.
The chemical potential is a measure for the energy needed to add one unit of charge~$Q$ to the  thermal system and it is 
given in terms of the grandcanonical potential in the grandcanonical ensemble as
\begin{equation}
\label{eq:defMu}
\mu =  -\partial_{J^0} \Omega \, .
\end{equation}
In order to prove this recall also that a system in contact only with a heat bath is described by the 
canonical ensemble with the partition function 
\begin{equation}
Z_{\text{canonical}} = e^{-\beta \int \mathcal{H}} \, ,
\end{equation}
with the Hamiltonian density~$\mathcal{H}$ giving the energy of the system after integrating over the volume.
If we would like to work at a finite chemical potential, in addition we need to put our system into contact 
with a particle bath. Then the relevant ensemble is the grandcanonical one with the partition function
\begin{equation}
Z_{\text{grand}} = e^{-\beta \int (\mathcal{H} -\mu J^0)} \, .
\end{equation}
The finite charge density~$J^0$ is the thermodynamically conjugate variable to the chemical potential.
Introducing a finite charge density will also change the chemical potential while changing the 
chemical potential will in general also change the charge density.
In the grand canonical ensemble the grandcanonical potential is defined by
\begin{equation}
\Omega = -\frac{1}{\beta}\ln Z_{\text{grand}} =  \int (\mathcal{H} -\mu J^0) \, ,
\end{equation}
which immediately confirms the chemical potential formula~\eqref{eq:defMu}.

Now a chemical potential in a thermal QFT is given by the time component of a gauge field~$A_0$. This
may be seen heuristically by comparing the partition function in the grand canonical ensemble
(including the charge density~$J^0$) on one hand
\begin{equation}
Z = e^{-\beta \int (\mathcal{H}-\mu J^0)}
\end{equation}
with the partition function at zero charge density but for a gauge theory including a gauge field~$A_\mu$
coupling to the conserved current~$J^\mu$ on the other hand
\begin{equation}
Z [A_\mu] = e^{-\beta \int (\mathcal{H}-A_\mu J^\mu)} \, .
\end{equation}
Choosing only the time component of the gauge field~$A_\mu$ non-zero and having called the 
thermodynamical charge density suggestively~$J^0$, we can now identify
\begin{equation}
A_0 =  \mu \, .
\end{equation}
Thus we have seen that introducing a finite gauge field time component in a thermal QFT is equivalent
to (and therefore may be interpreted as) the introduction of a finite chemical potential~$\mu$ for the 
charge density~$J^0$. A more formal treatment of this may be found in section~\ref{sec:chemPots}.

{\bf Introducing temperature}
In order to study thermal gauge theories through AdS/CFT we need a notion of temperature on the gravity
side. This means that we need to modify the background and in particular the background metric in order to
incorporate temperature in the dual gauge theory. The idea of using a metric describing the geometry of a 
black hole comes about quite naturally since black holes are holographic thermal objects themselves whose 
$d$-dimensional exterior physics is completely captured by their $(d-1)$-dimensional horizon surface. This 
phenomenon is studied in the field called {\it black hole thermodynamics}.
The Bekenstein-Hawking formula relates the area of the black hole horizon to the entropy of the complete 
black hole~(bulk) which has a distinct Hawking temperature depending on its mass.

It was first proposed in~\cite{Witten:1998zw} that black hole backgrounds or black branes as described in
section~\ref{sec:blackBranes} are holographically dual to a gauge theory at finite temperature. The
metric for a stack of black D3-branes can be conveniently written in the form
\begin{equation}
\label{eq:bhBackgrd}
{\dd s}^2 = \frac{1}{2} \left (\frac{\varrho}{L}\right)^2 \left (
 -\frac{f^2}{\tilde f}{\dd t}^2+\tilde f {\dd\vec x}^2 \right ) +
 \left (\frac{L}{\varrho}\right)^2 \left ( {\dd \varrho}^2 + \varrho^2 {\dd \Omega_5}^2 \right ) \, ,
\end{equation}
with
\begin{equation}
f(\varrho) = 1-\frac{r_0^4}{\varrho^4}\, ,\quad \tilde f(\varrho) = 1+\frac{r_0^4}{\varrho^4} \, .
\end{equation}
We obtain this form of the metric from~\eqref{eq:3BraneBackgMetric} by the 
transformation~$\varrho^2=r^2+\sqrt{r^4-{r_0}^4}$ where~$r_0$ is the location of the horizon.
The Hawking temperature~$T_H$ of the black hole horizon is equivalent to the temperature~$T$ in the thermal 
gauge theory on the other side of the correspondence. In order to relate the temperature~$T$ to the 
factors appearing in metric components, we make the metric Euclidean by Wick rotation. Demanding regularity
at the horizon renders the Euclidean time coordinate to be periodic with period~$\beta = 1/T = r_0/(\pi L^2)$. 
Note, that this background is confining~\cite{Witten:1998zw} and preserves all the supersymmetry, i.e. the
dual field theory is $\N=4$ SYM at finite temperature. 
Further there exist crucial differences  between the Euclideanized background and its Minkowski version. 
We will discuss this issue in section~\ref{sec:adsG}.

\begin{figure}
\includegraphics[width=0.95 \textwidth]{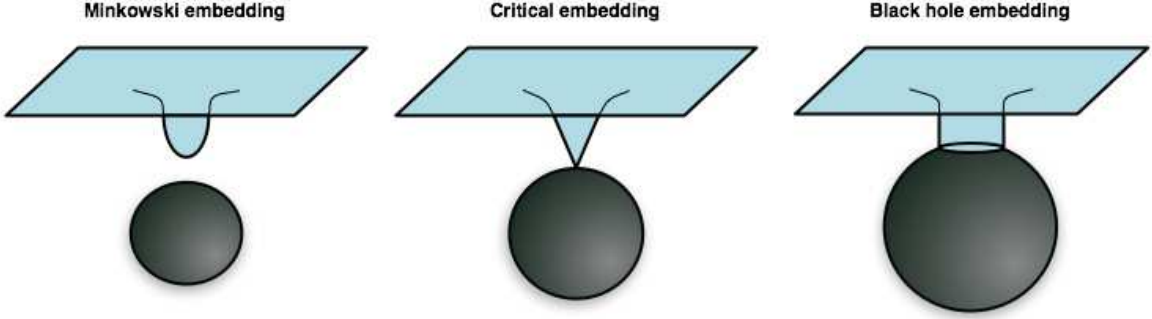}
\caption{\label{fig:geometricTrans}
Increasing the temperature from the left to the right picture we see that the black hole becomes larger. The
embedded brane is pulled towards the horizon stronger and stronger until the probe brane just touches the 
black hole horizon~(middle picture). Increasing the temperature further the brane is pulled through the horizon.
This figure has been kindly provided by the authors of~\cite{Mateos:2007vn}.
}
\end{figure}

{\bf Quarks \& chemical potential}
In order to include fundamental matter in this finite temperature setup we introduce D7-probe branes 
as described in section~\ref{sec:mesons}. At vanishing baryon density it was observed in~\cite{Babington:2003vm} 
that these thermal D7-embeddings are special because in the gauge theory a phase transition appears which is 
dual to a geometric transition on the gravity side~(see figure~\ref{fig:geometricTrans}).
A detailed study of this transition showed that it is of first order~\cite{Kirsch:2004km}.
This study has been refined and generalized to Dp/Dq-branes in~\cite{Mateos:2006nu}.
The setup is governed by a parameter~$m\propto M_q/T$ which is proportional to the quotient of quark mass~$M_q$
and temperature~$T$. At large values of~$m$ we have {\it Minkowski embeddings} which end outside the horizon. 
We write down the black hole metric in the coordinates introduced in~\eqref{eq:metricWithW}
\begin{equation}
\label{eq:metricWT}
\dd s^2 = \left ( {\tilde w}^2 +\frac{{w_H}^4}{{\tilde w}^2} \right )\dd \mathbf{x}^2 
 +\frac{({\tilde w}^4-{w_H}^4)^2}{{\tilde w}^2({\tilde w}^4-{w_H}^4)} \dd t^2 
 + \frac{1+(\partial_\varrho w_6)^2}{{\tilde w}^2} \dd \varrho^2 
 + \frac{\varrho^2}{{\tilde w}^2} \dd {\Omega_3}^2 \, ,
\end{equation}
where we define~$\tilde w^2= \varrho^2 + {w_6(\varrho)}^2$ and~$w_H$ is the location of the horizon.
In the $AdS_5\times S^5$-background the D7-brane fills the AdS wrapping an~$S^3$ inside the~$S^5$. 
Looking at the $S^3$-part of the metric~\eqref{eq:metricWT}, for~$\rho=0,\,, w_6 >w_H$ we find that 
the~$S^3$ shrinks to zero size before reaching the horizon. 
These Minkowski embeddings resemble those present at vanishing temperature at large values of~$m$.

Decreasing the parameter~$m$ we reach a critical value below which the D7-brane always reaches to the horizon.
The geometrical difference is that for these {\it black hole embeddings} now the $S^1$ in time direction collapses
as can be seen from the time component of the metric~\eqref{eq:metricWT}.

This means that the D3/D7-system in presence of a black hole undergoes a geometrical transition. That 
transition is dual to a first order phase transition in the thermal field theory dual.
The physics of this transition is discussed in greater detail in section~\ref{sec:thermoBaryon}.

An extension of the black hole background to a field theory dual with a finite baryon charge
density has been performed in~\cite{Kobayashi:2006sb}. Note, that the latter work corrected 
a similar analysis performed earlier in~\cite{Nakamura:2006xk}.

However, the central achievement of this present work is to introduce a finite baryon and isospin density
{\it simultaneously} in
the setup we have just described. We will see that this changes the embeddings and also the phase structure of the theory. 
We will further observe that the phase transition is softened. This statement 
will be explained in the discussion of this system's hydrodynamics in chapter~\ref{sec:transport}. We 
discover a further transition at equal baryon and isospin densities discussed in the thermodynamics 
section~\ref{sec:thermoB&I}.

{\bf Brane thermodynamics and holographic renormalization}
At finite temperature an extension of the standard AdS/CFT claim is the conjecture that the thermodynamics of the
thermal field theory is described by the gravity theory. In particular, on the thermal field theory side one has to Euclideanize 
by a Wick rotation~$x_0\to i \tau_E$ in order to identify the Euclidean path integral with a thermal partition function. 
On the gravity dual side one equivalently has to Euclideanize the AdS-black hole background~\eqref{eq:adsBHmetric}.
The Euclidean black hole is interpreted as a saddle-point of the Euclidean path integral such that the classical
supergravity action is conjectured to give the leading contribution to the free energy
\begin{equation}
\label{eq:actionFreeEnergy}
S_{\text{E}} = \beta F \, .
\end{equation}
Note the typographical difference between the action~$S$ and the entropy~$\S$. Recall that~$F = -\ln Z$.
In what follows we will find these thermodynamic definitions of entropy~$\S$, internal energy~$E$ and the 
speed of sound~$v_s$ useful
\begin{equation}
\label{eq:thermoRelFSE}
\S= -\frac{\partial F}{\partial T} \, , \quad E = F+ T \S\, , \quad
  {v_s}^2 = \frac{\partial P}{\partial E}= \frac{\partial P}{\partial T}\left (\frac{\partial E}{\partial T}\right )^{-1} = \frac{\S}{c_v}\, .
\end{equation}
For a stack of~$N_c$ black D3-branes such as those described by~\eqref{eq:nonExtremalDpMetric} the free energy turns out to be
\begin{equation}
F = \frac{-\pi^2}{8} {N_c}^2 T^4 \, .
\end{equation}
From this the energy and entropy are easily computed and the speed of sound is given by
\begin{equation}
{v_s}^2  = \frac{1}{3}\, .
\end{equation}

In order to obtain these finite results we had to {\it holographically renormalize} the gravity action by adding boundary terms.
Let us review the process of {\it holographic renormalization} in order to apply it to our setups later on.
In general the Euclideanized AdS-bulk action~$S^{\text{bulk}}_{E\, \text{Dp}}$ contains ultra-violet~(UV) divergences. 
The first step is to identify the divergent terms by introducing a UV-cutoff~$r_{\text{max}}$. Integrating the bulk Dp-action over
the $(p-5)$-remaining directions and evaluating the result at the cutoff~$r = r_{\text{max}}$ we obtain
the boundary action~$S^{\text{bdy}}_{E\, \text{Dp}}$. This action contains the UV-divergent terms and in order to renormalize
the bulk action we simply subtract this boundary action
\begin{equation}
\label{eq:holoRenormalization}
S^{\text{renormalized}}_{E\, \text{Dp}} = \lim\limits_{r_{\text{max}}\to r_{\text{bdy}}}
 \left (S^{\text{bulk}}_{E\, \text{Dp}} -S^{\text{bdy}}_{E\, \text{Dp}}\right ) \, .
\end{equation}
This Euclideanized and renormalized bulk action is the one we will derive all thermodynamic quantities from.
We stop at this point since we will not show the explicit applications of this method in this thesis. The 
interested reader is referred to the review on holographic renormalization~\cite{Skenderis:2002wp}.

\subsection{More Phenomenology from AdS/CFT} \label{sec:morePheno}
In this section we give a sketchy overview of the phenomenologically relevant outcomes of AdS/CFT-applications.  
Only the paragraph discussing the Sakai-Sugimoto model is a bit more detailed because that model is in many 
respects a valuable, partly complementary competitor to the D3/D7-setup which we study in this thesis.
We also briefly discuss both the model building aspect and the fundamental value of AdS/CFT. 

{\bf Low viscosity bound}
The phenomenologically most striking prediction of AdS/CFT is that the viscosity~$\eta$ to entropy density~$s$ ratio is
incredibly small
\begin{equation}
\label{eq:viscosityBound}
\frac{\eta}{s} = \frac{1}{4\pi} \, .
\end{equation}
This bound is satisfied to leading order in~$1/N_c$ in all theories with gravity duals computed up to now~\footnote{
Note, that a recent investigation~\cite{Brigante:2007nu} had claimed that higher derivative corrections violate the 
viscosity bound for a certain family of models. But the same authors also found these very theories to be inconsistent
violating microcausality~\cite{Brigante:2008gz} supporting again the idea of the universality of this bound.}.
It was observed at the RHIC heavy-ion collider that the quark gluon plasma supposedly formed 
in this experiment has an extremely low viscosity~(well below any viscosity measured before) numerically 
comparable with the AdS/CFT value. Most of the models used to analyze the RHIC data are consistent
with ratios in a range of~$\eta/s \approx \frac{4/3}{4\pi}\dots \frac{2}{4\pi}$~\cite[e.g.]{Adare:2006nq,Romatschke:2007mq}. 
This discovery was even celebrated as an experimental possibility of testing the AdS/CFT correspondence. One has to be 
careful though since no QCD-dual gravity theory has been discovered yet and thus one has to rely on the
{\it universality} of the observables to be measured. 
In the context of these viscosity investigations many different backgrounds have been employed in order to find
out what makes this bound so universal. All investigated gauge theories  with gravity duals show this universal behavior
no matter if one breaks conformal symmetry, supersymmetry or if one introduces flavor or a finite chemical potential.
Three interrelated proofs of the universality of the viscosity ratio have been provided 
successively~\cite{Buchel:2004qq,Kovtun:2004de,Starinets:2008fb}.
It is still under lively investigation, which underlying principle is the origin of the viscosity universality and if it 
applies to QCD as well.

In a series of papers~\cite{Policastro:2001yc,Policastro:2002se,Policastro:2002tn,Kovtun:2003wp,Kovtun:2004de,
Kovtun:2006pf,Son:2006em,Son:2007vk} an identification of hydrodynamic modes with gravity objects was achieved leading
to a detailed gravity description of the hydrodynamics in a strongly coupled fluid.
Recently this framework has been extended to second order hydrodynamics~\cite{Baier:2007ix,Natsuume:2007ty,
Natsuume:2007tz,Natsuume:2008iy}. Here also a
correction of the widely used {\it Mueller-Israel-Stewart theory} is proposed based on gravity consistency arguments.
It is well known that hydrodynamics violates causality. Mueller-Israel-Stewart theory is a relativistic generalization
of second order hydrodynamics which the authors of~\cite{Baier:2007ix,Natsuume:2007ty,Natsuume:2007tz,Natsuume:2008iy}
claim to be incomplete.

{\bf D3/D7-setup}
A particularly promising setup is the D3/D7-brane configuration described in~\ref{sec:mesons}. Its gauge dual contains 
massive quarks and a chemical potential can be consistently introduced. Further it exhibits confinement and
thus a first order phase transition of the fundamental matter in the spectrum. We will study this particular
system in most of this thesis.

The calculation of meson spectra~\cite{Kruczenski:2003be} in this system was one of the first phenomenological 
applications of AdS/CFT. Also ratios of B-meson masses were recently given~\cite{Erdmenger:2006bg}.

A topic under ongoing investigation is that of heavy-light mesons~\cite{Erdmenger:2006bg,Erdmenger:2007vj, Herzog:2008bp} 
modeled by strings spanning from one D7-brane to another after having separated the D7-branes from each other.

Recently the hadron multiplicities after hadronization of the final state in a particle-antiparticle 
annihilation~\cite{Evans:2007sf} have been modelled to surprising accuracy~(see also~\cite{Castorina:2008gf}).

Interesting effects such as mass shift analogous to the Stark effect and chiral symmetry breaking are also observed
in gauge/gravity duals with flavor for which pure-gauge Kalb-Ramond B fields are turned on in the background, 
into which a D7 brane probe is embedded~\cite{Erdmenger:2007bn, Albash:2006ew,Albash:2006bs}.

{\bf QCD duals?}
Although some aspects of the D3/D7-brane configuration mirror QCD quite well one main point of criticism
is that the dual gauge theory has too much symmetry. Various deformations of the background metric have 
been devised to break symmetries in a controlled way for the gauge dual to come closer to QCD behavior. 
Remember that for a D3/D7-configuration, on the gauge theory side we have
$\N=2$ supersymmetric Yang-Mills theory coupled to~$\N=4$ SYM and the conformal symmetry
is broken if the quarks become massive by seperating the D3 from the D7-branes. Also a finite temperature,
i.e. a black hole background metric breaks conformal symmetry. 

In a different background, the Constable-Myers background, all of the supersymmetry is broken and the 
theory turns out to be confining~\cite{Constable:1999ch}. In contrast to the black hole background 
which has a singularity covered by the black hole horizon, the Constable-Myers background contains
a naked singularity. This singularity was found to be screened by a condensate in~\cite{Babington:2003vm}.
Furthermore the first occurance of spontaneous breaking of a chiral symmetry, in this
case the axial~${U(1)}_A$-symmetry, in the context of the gauge/gravity correspondence
has been achieved in~\cite{Babington:2003vm}. The breaking of the axial part of 
the chiral symmetry is achieved by the formation of the same condensate screening the singularity.
Thus, in the configuration considered in~\cite{Babington:2003vm} all the supersymmetries, the chiral symmetry,
and the conformal symmetry are broken.

A great advantage of the D3/D7-setup is that the gauge theory living on the boundary of AdS is four-dimensional, 
matching the observed dimensionality of our real world. Nevertheless, an explicit QCD-dual has not been found, yet.
Therefore it is worthwhile to also study different brane configurations with possibly different dimensionality.

{\bf Another QCD model: Sakai-Sugimoto model}
The {\it Sakai-Sugimoto model} is an alternative D4/D8 anti-D8 brane system with $N_c$ D4-branes and $N_f$ 
pairs of D8/anti-D8-branes. Here the D4-branes generate the geometry very much like the D3 branes do in D3/D7 setups and the
D8 and anti-D8 branes are the flavor branes corresponding to the D7. Since this model is the second most studied
model~(after the D3/D7-setup) introducing fundamental matter, we discuss also a few technical points here. 
This setup features no quark masses but two distinct phase transitions corresponding to the {\it chiral symmetry breaking and
deconfinement transition}, respectively. Supersymmetry is explicitly broken. 
In contrast to the D3-setup, there is one extra-dimension~$x_4$ in the worldvolume of the gauge theory. In order to 
come down to four space-time dimensions this extra coordinate needs to be compactified. There is also a geometrical
argument for this coordinate to be periodic: together with the "radial" coordinate $u$ it forms a cigar-shaped submanifold, which
has a tip at $u=u_T$. To avoid a singularity at this tip, $x_4$ needs to be periodic with period $2\pi R$.
The metric of the background at low temperature is
\begin{equation}
\label{eq:ssMetric}
\dd s^2 = (\frac{u}{R_{D4}})^{3/2} (\dd t^2+\delta_{ij} \dd x^i \dd x^j +f(u) \dd x_4^2) +
(\frac{R_{D4}}{u})^{3/2}(\frac{\dd u^2}{f(u)} +u^2 \dd\Omega_4^2)
\end{equation}
The $x_4$-circle shrinks to zero at $u=u_\Lambda$ and the $D8$ and their antibranes have nowhere to end thus 
staying connected. So the chiral $U(N_f)_L\times U(N_f)_R$ is broken to a diagonal $U(N_f)_V$ in the low temperature phase.

At finite temperature there always exist two solutions of which one is preferred at low temperature and the other at high temperature. 
Connected to this an asymptotic symmetry among the two circles (time-direction and $x_4$) exists. In the high temperature
phase $t$ and $x_4$ interchange roles (the $f(u)$ in the metric is shifted from one to the other), so that the $x_4$-circle
now does not shrink to zero, but the $t$-circle does. Chiral symmetry is restored as the flavor branes may be
parallel now. 

The biggest advantage of this model over the D3/D7-setup is that chiral symmetry breaking can be achieved
quite naturally. On the contrary, the quark masses are not incorporated from the start but also arise dynamically. 
Mesons have also been studied in the Sakai-Sugimoto model. For example quark bound states
which play the role of QCD pions arise as Goldstone bosons from the spontaneous symmetry
breaking generated upon introducing the probe branes giving fundamental degrees of freedom.
Recent developments of mesons at finite temperature may be found in~\cite{Mazu:2007tp}.
One recent approach generating quark masses dynamically can be found in~\cite{Dhar:2008um}. 

{\bf Fundamentalism \& phenomenology}
Let us briefly discuss the phenomenological versus fundamental value of AdS/CFT. Although still only a conjecture,
AdS/CFT has failed no comparative test so far and it succeeds in describing strong coupling phenomena. The perturbative
or geometric understanding on the gravity side can be translated to an understanding of the strongly coupled gauge theory
on the other side of the correspondence. In this way AdS/CFT makes it possible to get a qualitative understanding of 
strong coupling phenomena. At the present level where we do not have an explicit QCD / gravity dual, the qualitative 
understanding AdS/CFT supplies us with, should be seen as being complementary to for example lattice data 
providing exact QCD data but also hiding the inner workings of the strongly coupled theory. In some cases such
as for the viscosity bound the quantities involved may even be protected by universality and thus solely depend 
on the fact that the gauge theory is strongly coupled. If this is the case then AdS/CFT results may even continue 
to be valid for QCD or the real world. All these results justify the duality at least as a valid phenomenological tool.

Turning around the argument, the phenomenological success of AdS/CFT may be seen as a hint that the gauge gravity 
correspondence and the principles from which it was derived come indeed close to the principles governing nature.
Studying explicit instances of the correspondence, for example studying correlators in the D3/D7-setup, could also provide us 
with a detailed understanding of how the duality works in general and it might even suggest a way to prove AdS/CFT.

\subsection{Summary} \label{sec:sumAdsCft}
In this technical introduction chapter we have developed the concepts of the AdS/CFT correspondence and we 
investigated how these ideas emerged from the careful study in rather formal areas of string theory~(see \ref{sec:generalAdsCft}).  We
have shown how modifications of the original correspondence give rise to temperature and fundamental matter in the gauge theory.
Temperature in the gauge theory is generated by a black hole background such as~\eqref{eq:bhBackgrd} on the gravity side. 
Fundamental matter alias quarks is introduced by embedding a stack of~$N_f$ probe D7-branes into the ten-dimensional setup
in addition to the $N$~D3 branes, which determine the gravity geometry~(as explained in~\ref{sec:mesons}). 
Finally, we discussed the phenomenological picture which can be drawn by putting the results in different gravity duals 
together and extrapolating from it what the phenomenology of a QCD-dual at strong coupling might look like.
We are now ready to develop the methods which we will apply to investigate the promising D3/D7-configuration.

\section{Holographic methods at finite temperature} \label{sec:holoMethods}
The goal of this work is to develop a qualitative description of thermal QCD-plasma at strong coupling as it is claimed to be
seen at the RHIC heavy ion collider. In order to compute observables and study qualitative features of this class
of systems we utilize the AdS/CFT duality in order to overcome the difficulty that the system is governed by
QCD at strong coupling. In this present chapter we develop the methods which are needed to derive correlation
functions~(section~\ref{sec:G}) in the strongly coupled field theory by computations on the weakly coupled gravity side. Furthermore
we review how to obtain non-equilibrium observables such as diffusion coefficients and shear viscosity by
the formulation of a gravity dual to relativistic hydrodynamics~(section~\ref{sec:holoHydro}). Finally in section~\ref{sec:qnm} 
we elucidate the connection between quasinormal modes known from general relativity in presence of a black hole on the 
gravity side and distinct hydrodynamic modes. Note that as stated in the previous chapter no gravity dual for QCD 
has been found, yet. Thus we will apply our holographic methods to quantum field theories which are similar to 
QCD in the properties of interest.

\subsection{Holographic correlation functions} \label{sec:G}
Since we are interested in the spectral functions~$\R$ and in particular in the resonances appearing therein which 
correspond to mesons due to AdS/CFT~(as will be argued in section~\ref{sec:thermalSpecFunc}), our 
motivation to compute retarded correlators~$G^R$ is sourced by the formula~$\R = -2\, \mathrm{Im} G^R$.
Correlation functions in AdS/CFT have been under intensive examination during the past ten years. They are useful quantities
to compare the conjectured gauge/gravity results to results directly obtained in the quantum field theory as outlined
in section~\ref{sec:stateAdsCft}~(paragraph 'Evidence'). Moreover retarded two point correlators in Minkowski space 
are needed to compute non-equilibrium observables such as transport coefficients~(shear viscosity, diffusion coefficient,
heat conductivity,\dots). We briefly distinguish Euclidean formulation from the Minkowski formulation of correlation functions
in AdS/CFT in section~\ref{sec:adsG}. Afterwards we develop analytical~(\ref{sec:anaAdsG}) and numerical~(\ref{sec:numAdsG}) 
recipes by which correlation functions may be obtained. 

\subsubsection{Correlation functions in AdS/CFT} \label{sec:adsG}
In the beginning of AdS/CFT the correspondence for correlation functions was formulated in 
Euclidean space-time for simplicity. The idea was to obtain Euclidean correlators from a conjectured generating functional identity
\begin{equation} 
\label{eq:euclideanGenFId}
\left < e^{\int_{\partial M} \, \phi^{\text{bdy}} \hat\O}\right > = e^{-S_{\text{classical}}[\phi]} \, ,
\end{equation}
and to analytically continue them afterwards. 
In this section we review this Euclidean procedure and the subtleties which make it fail in general if naively extended 
to the gravity dual of finite temperature field theories on Minkowski space-time. Finally we justify the correct prescription 
to get thermal Minkowski space correlators from a conjectured AdS/CFT identity similar to~\eqref{eq:euclideanGenFId}. 

The left hand side of~\eqref{eq:euclideanGenFId} is the Euclidean space-time
generating functional for correlators of operators~$\hat\O$ in the boundary field theory. In order to 
Euclideanize the originally Minkowskian space-time we had to perform a Wick rotation~$t\to\tau_E= i t$. On the right
hand side we find the action for the classical solution to the equation of motion for the bulk field~$\phi$
in the bulk metric obeying a boundary condition of the form~$\lim\limits_{r\to r^{\text{bdy}}}\phi = \phi^{\text{bdy}}$. 
Either side may be functionally derived with respect to the boundary field~$\phi^{\text{bdy}}$ in order to get Euclidean correlators
of the dual operator~$\O$, such that for the two-point function we have
\begin{equation}
\label{eq:euclideanTwoPointFunc}
\left < \O(x) \O(y) \right >  =  \frac{\delta^2 e^{-S_{\text{classical}}[\phi]}}{\delta \phi^{\text{bdy}}(x) \phi^{\text{bdy}}(y)} \, .
\end{equation}
Note that this implies that on the right hand side we know the explicit form of the field~$\phi$
in terms of its boundary value~$\phi^{\text{bdy}}$, i.e. we need to solve the equations of motion for the field~$\phi$ first. The use of 
the identity~\eqref{eq:euclideanGenFId} has proven very useful and was confirmed by the results for correlators at zero temperature 
and for extremal metrics on the gravity side, respectively. 

At finite temperature however this prescription fails. It should be clear that the Euclideanization is only a tool for simplification and in principle
the correlators should be obtainable from the full Minkowskian description in AdS/CFT. In practice
it will be necessary to derive Minkowski correlators directly since in order to get them from their 
Euclidean versions, one would need to know all the {\it Matsubara frequencies}~$\omega_n$. Matsubara frequencies 
are the discrete values which arise in finite temperature field theory from the compactification~$\tau_E\sim \tau_E+T^{-1}$ of the 
Euclidean time coordinate~$\tau_\text{E}$ on a circle with the period~$T$ being identified as the temperature
in the field theory. Only at these particular frequencies the Euclidean correlators are defined. 
The compactification of the Euclidean time appearing in the black hole background 
on the gravity side is dual to the {\it imaginary time formalism} in the dual thermal field theory. In many 
applications for correlation functions such as the derivation of hydrodynamic transport coefficients an 
approximation of some sort is needed during the calculation. For example in hydrodynamics the frequencies have to 
be small such that we can not work with all Matsubara frequencies as would be required to analytically 
continue Euclidean correlators to Minkowski correlators. Due to this fact we need the full Minkowski prescription.

As shown in~\cite{Son:2002sd} also a naive formulation of~\eqref{eq:euclideanGenFId} in 
Minkowski space-time given by
\begin{equation} 
\label{eq:naiveMinkGenFId}
\left < e^{i \int_{\partial M} \, \phi^{\text{bdy}} \hat\O}\right > = e^{i S_{\text{classical}}[\phi]} \, ,
\end{equation}
fails since it produces only real valued correlators. In the same work the authors propose a working recipe
to obtain two-point Minkowski correlators. This is the recipe which we will make heavy use of and we explain
it in the next section~\ref{sec:anaAdsG}. Finally~\cite{Herzog:2002pc} developed a general prescription involving an analog 
of~\eqref{eq:euclideanGenFId} which can be used to obtain~$n$-point correlators and which we briefly review here in 
order to clarify the limits of the two-point correlator recipe we will use here.

\begin{figure}[h]
\begin{center}
\begin{picture}(350,75)(-15,-5)
\put(0,55){\circle*{5}}
\put(0,55){\vector(1,0){150}}
\put(150,55){\line(1,0){150}}
\put(300,55){\line(0,-1){20}}
\put(300,35){\vector(-1,0){150}}
\put(150,35){\line(-1,0){150}}
\put(0,35){\vector(0,-1){18}}
\put(0,17){\line(0,-1){17}}
\put(0,0){\circle*{5}}
\put(5,58){\makebox(0,0)[bl]{$t_i$}}
\put(-5,55){\makebox(0,0)[r]{$A$}}
\put(-5,0){\makebox(0,0)[r]{$B$}}
\put(300,58){\makebox(0,0)[bl]{$t_f$}}
\put(190,10){\makebox(0,0)[b]{$\mathcal{C}$}}
\put(140,60){\makebox(0,0)[b]{$1$}}
\put(140,20){\makebox(0,0)[b]{$2$}}
\put(300,34){\makebox(0,0)[tl]{$t_f-i\sigma$}}
\put(5,0){\makebox(0,0)[bl]{$t_i{-}i\beta$}}
\end{picture}
\end{center}
\caption{The Schwinger-Keldysh contour is a time contour~$\mathcal{C}$ where points A and B are identified~(this figure
is a slightly modified version of that shown in~\cite{Herzog:2002pc}).}
\label{fig:sKContour}
\end{figure}
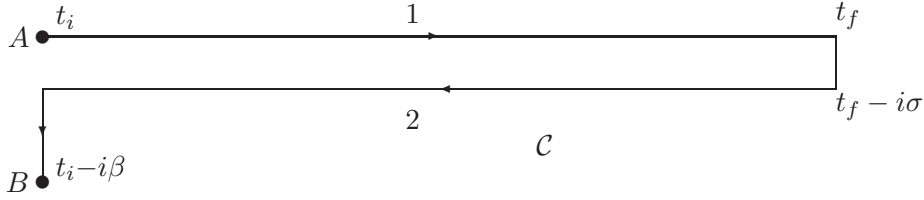

{\bf Schwinger-Keldysh formalism for thermal QFT}
In general the authors of~\cite{Herzog:2002pc} developed a detailed gravity dual to the real-time formalism of thermal
quantum field theory. For a detailed review of the real-time or Schwinger-Keldysh formalism the reader is
referred to~\cite{Das:1997gg,Herzog:2002pc} but let us work out the rough ideas here in order to understand the
equivalent features on the gravity side. In this formalism the operators (or
fields)~$\O$ live on the time contour~$\mathcal{C}$ shown in figure~\ref{fig:sKContour}. 

The starting point~$A$ and the end~$B$ are identified with the condition~$\O|_A =-\O|_B$ for fermionic~$\O$ and~$\O|_A =\O|_B$ for bosonic~$\O$. 
Now one introduces sources~$\phi_{1,2}$ for the operator~$\O_{1,2}$ on the upper~($1$), respectively lower~($2$)
part of the contour along the original real Minkowski time direction. Defining an appropriate generating functional~$Z$
one can then define the matrix valued Schwinger-Keldysh propagator which correlates operators on the
upper and lower parts of the time contour in figure~\ref{fig:sKContour}
\begin{equation}
  iG_{ab}(x-y) = 
  \frac 1{i^2}\,\frac{\delta^2\ln Z[\phi_1,\phi_2]}
  {\delta\phi_a(x)\,\delta\phi_b(y)}
   = i \left(\begin{array}{cc} G_{11} & -G_{12}\\-G_{21} & G_{22}
  \end{array}\right) \ .
\end{equation} 

Transforming to momentum space by~$  G(k) = \int\!\dd x\,e^{-ik\cdot x}G(x)$ we can write down the relations
between the components of the Schwinger-Keldysh propagator and the ordinary retarded two-point function~$G^R$
\begin{eqnarray}
\label{sKRelations}
  G_{11}(k) &=& \mathrm{Re}\, G^R(k) + i\coth\frac\omega{2T}\,\mathrm{Im}\, G^R(k)\,,
  \qquad \omega \equiv k^0\,, \nonumber \\
  G_{12}(k) &=& \frac{2i e^{-(\beta-\sigma)\omega}}{e^{\beta\omega}-1}\,\mathrm{Im}\, G^R(k)\,,
   \nonumber \\
  G_{21}(k) &=& \frac{2i e^{-\sigma\omega}}{1-e^{-\beta\omega}}\,\mathrm{Im}\, G^R(k)\,,
   \nonumber \\
  G_{22}(k) &=& -\mathrm{Re}\, G^R(k) + i\coth\frac\omega{2T}\,\mathrm{Im}\, G^R(k)\,.
\end{eqnarray}
For the choice of the arbitrary length parameter~$\sigma=\beta/2$ we see that the Schwinger-Keldysh correlator
is symmetric~$G_{12}=G_{21}$. 

{\bf Holographic Schwinger-Keldysh formulation}
Let us now turn to the gravity dual description of the Schwinger-Keldysh formalism reviewed in the 
previous paragraph. For the asymptotically AdS spaces containing a 
black hole which we consider here, there exists an analog of {\it Kruskal coordinates}. Kruskal coordinates in
general relativity cover the entire space-time manifold of the maximally extended Schwarzschild solution and they 
are well-behaved everywhere outside the physical singularity, i.e. they show no coordinate singularities as other 
coordinates do, e.g. at the horizon. The identity~\eqref{eq:euclideanGenFId} suggests, that one has to know 
the explicit form of the classical action including the solution of the equation of motion for the field~$\phi$
in terms of boundary values for the field in order to take derivatives of the expression on the left hand side as
shown in~\eqref{eq:euclideanTwoPointFunc} 
and get an explicit expression for the correlation function. Now the main idea is to use this standard AdS/CFT prescription to get the 
correlation functions but to carefully impose boundary conditions on the gravity fields in the analog of the 
Kruskal time coordinate and not in the ordinary Minkowski time. These boundary conditions on the gravity 
fields will be subject of a detailed discussion on the level of two point correlators in the next section~\ref{sec:anaAdsG}. 
Let us note here only that these boundary conditions are the point where the naive Minkowski formulation of the AdS/CFT 
correlator prescription fails. The reason for this is the fact that in ordinary coordinates the boundary conditions
in Euclidean space-time are completely fixed by the requirement of regularity but this is not the case in 
the Minkowski version. For example a scalar gravity field has to fulfill a second order equation of motion and 
therefore one needs to fix two boundary conditions. One of these is fixed by the boundary 
data~$\phi|_\text{bdy} = \phi^\text{bdy}$. The other condition is imposed at the horizon where the scalar 
locally behaves like~$(1-u)^\beta$ with the radial AdS-coordinate~$u\in [0,1]$ which is defined in the
context of the black hole metric
\begin{equation}
\dd s^2=\frac{(\pi T R)^2}{u} [-f(u){\dd t}^2 +{\dd \mathbf{x}}^2] 
 + \frac{R^2}{4 u^2 f(u)}{\dd u}^2+R^2 {\dd\Omega_5}^2 \, .
\end{equation}
Here the horizon is located at~$u=1$, spatial infinity at~$u=0$ and 
the function~$f$ is defined by~$f(u)=1-u^2$. This metric is obtained
from the standard AdS~black hole metric with radial coordinate~$r$ by the transformation~$u=(r_0/r)^2$. 
The temperature~$T=r_0/(\pi R^2)$ is a function of the AdS-radius~$R$ and the black hole horizon~$r_0$.
In Euclidean space-time we 
have~$\beta=\pm\omega/(4\pi T)$ and only one of the two signs produces a regular solution. In Minkowski
signature this is completely different since there we compute~$\beta=\pm i\omega/(4\pi T)$ and both
signs can produce regular solutions, thus leaving an ambiguity which needs to be fixed by another requirement.
\begin{figure}[ht]
\begin{center}
\includegraphics[width=0.4\textwidth]{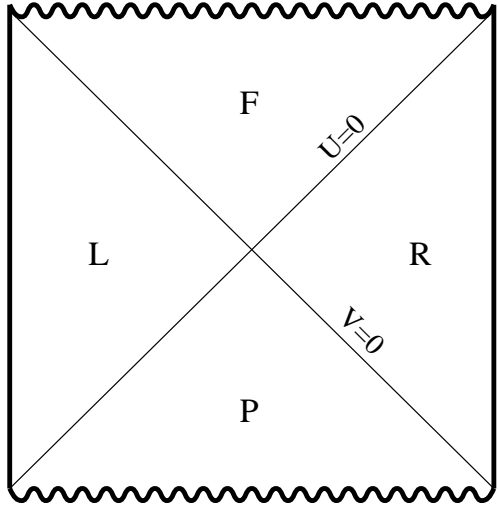}
\end{center}
\caption{ \label{fig:adsPenrose}
The Penrose diagram for AdS containing a black hole.
This figure has been kindly provided by the authors of~\cite{Herzog:2002pc}.
}
\end{figure}
Now the main achievement in~\cite{Herzog:2002pc} was to single out such a requirement which is in
general applicable to any $n$-point correlator. This requirement involves applying boundary conditions at
the boundaries of different quadrants of the Penrose diagram shown in figure~\ref{fig:adsPenrose} and 
forming a superposition of those. The diagram shows the causal structure of our asymptotically AdS space
(which contains a black hole) in Kruskal coordinates. In our earlier attempt to fix boundary conditions we
only considered the R-quadrant and its boundaries. The prescription of~\cite{Herzog:2002pc} takes into account
that the full space-time contains four quadrants. 

Nevertheless, in what follows it will be sufficient to use a simplified 
boundary condition, the {\it incoming wave boundary condition} which allows us to restrict ourselves to
the R-quadrant, to use the original Minkowski coordinates and it finally enables us to calculate two-point functions
as discussed in the following section~\ref{sec:anaAdsG}. It is argued in~\cite{Herzog:2002pc} that the 
general but also more complicated prescription involving Kruskal coordinates in the case of two-point
correlators reduces to the (simple) prescription that we are about to use.

\subsubsection{Analytical methods: correlators and dispersion relations} \label{sec:anaAdsG}
In order to obtain correlation functions for an operator~$\O$ in~AdS/CFT one usually has to solve a second
order differential equation~(as we have already mentioned in the previous section), 
the equation of motion for the particular field~$\phi$ which is dual to the 
operator~$\O$. Often that equation of motion can only be solved numerically.\footnote{ Especially if
we consider massive quarks, which implies that we embed a~$D7$-brane. The embedding functions 
can in general only be obtained numerically. In this case already the metric components~$g^{\mu\nu}$ appearing
in the equation of motion for our field~$\phi$ are only given numerically since they contain the
embedding functions.} Thus it is remarkable that in~\cite{Policastro:2002se,Policastro:2002tn} a method has been developed to find the
correlators analytically for a field theory at finite temperature and without quarks. The main idea of
this approach is to use the ratio of four-momentum and temperature~$\vec k/(2\pi T):= (\wn,0,0,\qn)$~\footnote{This choice 
for the four-momentum is adapted to the symmetries of the problem we will consider in this section.} as an expansion parameter.
Then the fields are expanded in a perturbation series in orders of~$\wn$ and~$\qn^2$ and exact solutions 
to the equations of motion can be obtained up to the desired order in~$\wn$ and~$\qn^2$. This kind of
expansion is known from statistical mechanics and goes by the name of {\it hydrodynamic expansion}.
Note that we only consider the diffusive modes with this choice. In order to find for example the sound modes
and their damping we would have to consider an expansion in~$\wn,\,qn$~\cite{Kovtun:2005ev}. 
From the solutions expanded in~$\wn$ and~$\qn^2$ we will obtain the correlators of the operator~$\O$.
The poles of these correlators can be read off directly from the analytical expressions giving the dispersion
relations~$\wn(\qn)$. Note, that we work in the geometry described in~\cite{Policastro:2002se} where the
fluctuations are chosen along the $x_3$-direction such that~$\vec x = (x_0=t,\,0,\, 0,\, x_3 = z)$. Furthermore we 
choose the gauge in which~$A_4\equiv 0$ and we assume that the remaining space-time directions have 
already been compactified such that we have to consider a five-dimensional theory only.
 
{\bf The correlator recipe}
Let us review the three-step recipe to obtain two-point correlation functions motivated and developed in~\cite{Son:2002sd}.
We calculate the retarded two-point correlator~$G^R$ of the operator~$\O$ in Minkowski space. 
The operator~$\O$ is dual to a field which we denote by~$\phi$, where~$\phi$ can be a scalar~$\Phi$, vector~$A_\mu$
or tensor field~$T_{\mu\nu}$ merely changing the index structure.
Step number one is to find the part of the action which is quadratic in the field~$\phi$ dual to~$\O$
\begin{equation}
\label{eq:recipeAction}
S^{(2)} = \int \dd u \dd^4 x \mathfrak{B}(u) (\partial_u \phi)^2 + \dots \, ,
\end{equation}
where the factor~$\mathfrak{B}$ depends on~$u$ and the momenta only, collecting metric components and all other factors
in front of the derivatives~$(\partial_u \phi)^2$.
Now the second step is to solve the equation of motion for the field~$\phi$. We rewrite the
space-time equation of motion in Fourier space such that all derivatives except~$\partial_u\phi =:\phi'$
can be expressed in terms of four-momenta~$\vec k$
\begin{equation}
0 = \phi '' + a(\vec k,u) \phi ' +b(\vec k,u) \phi \, .
\end{equation} 
This second order differential equation in special cases can be solved analytically in the hydrodynamic limit 
of small~$\wn,\qn^2\ll 1$.~\footnote{If the coefficients~$a,\,b$ are sufficiently complicated~(they might be given only
numerically) we have to reside to numerical methods, two of which are explained 
in~\cite{Kovtun:2006pf,Teaney:2006nc} and~\cite{Myers:2007we} reviewed in section~\ref{sec:numAdsG} of this work.}
The general solution can be split into the field's boundary value~$\phi^{\text{bdy}}(\vec k)$ and 
the bulk function~$\mathfrak{F}(u,\vec k)$
\begin{equation}
\phi (u, \vec k) = \phi^{\text{bdy}} (\vec k) \mathfrak{F}(u,\vec k)\, .
\end{equation}
To clearly illustrate this step, we will consider details of this general procedure in the specific example below.
In step three we finally assemble the solution~$\mathfrak{F}(u,\vec k)$ obtained in step two and the 
coefficient~$\mathfrak{B}(u)$ from step one to obtain the retarded correlator in Fourier space
\begin{equation}
\label{eq:anaGFormula}
G^R (\vec k)=  -2 \mathfrak{B}(u) \mathfrak{F}(u,-\vec k) \partial_u \mathfrak{F}(u,\vec k)\big |_{u\to 0} \, .
\end{equation}

{\bf An example}
To illustrate the three steps in more detail we consider the example of~$\N=4$ supersymmetric Yang-Mills
theory with an $R$-charge current~$J^\mu$ dual to the vector field~$A_\mu$ in five-dimensional
supergravity. The part of the action quadratic in the gauge field~$A$ is given by
\begin{equation}
\label{eq:quadraticSuperMaxwellAction}
S^{(2)}= -\frac{N^2}{16\pi^2} \int \dd u \dd ^4 x \sqrt{-g(u)} F_{\mu\nu} F^{\mu\nu} \, .
\end{equation} 
In order to place our field theory at finite temperature, we will work in the dual AdS~black hole background
\begin{equation}
 \label{eq:adsBHmetric}
\dd s^2=\frac{(\pi T R)^2}{u} [-f(u){\dd t}^2 +{\dd \mathbf{x}}^2] 
 + \frac{R^2}{4 u^2 f(u)}{\dd u}^2+R^2 {\dd\Omega_5}^2 \, , 
\end{equation}
with the radial AdS-coordinate~$u\in [0,1]$, the horizon at~$u=1$, spatial infinity at~$u=0$ and 
the function~$f(u)=1-u^2$. This metric is obtained
from the standard AdS~black hole metric with radial coordinate~$r$ by the transformation~$u=(r_0/r)^2$. 
The temperature~$T=r_0/(\pi R^2)$ is a function of the AdS-radius~$R$ and the black hole horizon~$r_0$.

Applying step one of our recipe to the quadratic super-Maxwell action~\eqref{eq:quadraticSuperMaxwellAction},
we find the coefficient 
\begin{equation}
\mathfrak{B}(u)= -\frac{N^2}{16\pi^2} \sqrt{-g(u)} g^{uu} g^{\nu\nu'} \, ,
\end{equation}
(hiding the index structure on the left hand side). 

{\bf Hydrodynamic expansion and equation of motion}
Now in step two of the recipe we take a closer look on the method for solving the equation of motion 
for our field~$A_\mu$. Using~\eqref{eq:quadraticSuperMaxwellAction} in the 
Euler-Lagrange equation, we get the equation of motion
\begin{equation}
\label{eq:superMaxwellEom}
0 = \partial_\nu [\sqrt{-g(u)} g^{\mu\varrho} g^{\nu\sigma} (\partial_\varrho A_\sigma-\partial_\sigma A_\varrho)] \, .
\end{equation}
We make use of the Fourier transformation
\begin{equation}
\label{eq:fourierTrafo}
A_i (u, \vec x)= \int \frac{\dd^4 k}{(2\pi)^4} e^{-i\omega t+i \mathbf{k \cdot x}} A_i (u, \vec k) \, .
\end{equation}
Rewritten in Fourier space we may split the equation of motion~\eqref{eq:superMaxwellEom} 
into five separate equations labeled by the free index~$\mu=0,1,2,3,4$
\begin{eqnarray}
A_t''-\frac{1}{uf}(\qn^2 A_t +\wn\qn A_z) =0\, , \label{eq:eomSMFSt} \\
A_{x,y}'' + \frac{f'}{f} A_{x,y}' + \frac{1}{uf} (\frac{\wn^2}{f} -\qn^2) A_t =0\, , \label{eq:eomSMFSxy}\\
A_z'' + \frac{f'}{f} A_z' + \frac{1}{u f^2} (\wn^2 A_z+\wn\qn A_t) = 0\, , \label{eq:eomSMFSz}\\
\wn A_t'+\qn f A_z' = 0 \, . \label{eq:eomSMFSu}
\end{eqnarray}
Note that~$A_t$ and~$A_z$ need to satisfy the coupled set of three equations~\eqref{eq:eomSMFSt},
\eqref{eq:eomSMFSz} and \eqref{eq:eomSMFSu} while the transversal~$A_{x,y}$ decouple and
merely have to satisfy the stand-alone equation~\eqref{eq:eomSMFSxy} separately.
However, we can decouple the system for~$A_t, \, A_z$ rewriting~\eqref{eq:eomSMFSt} as 
\begin{equation}
\label{eq:sMFSAz}
A_z = \frac{u f}{\wn \qn} A_t '' - \frac{\qn}{\wn} A_t \, ,
\end{equation}
and use it to substitute~$A_z$ in~\eqref{eq:eomSMFSu} yielding a single second order equation
for~$A_t'$
\begin{equation}
\label{eq:eomSMAtp}
A_t ''' +\frac{(u f)'}{uf} A_t '' + \frac{\wn^2 -\qn^2 f}{u f^2} A_t' =0 \, .
\end{equation}
Note that the appearance of the third derivative~$A_t'''$ is a generic feature of this
particular example and has nothing to do with the general method.
Since this equation does not depend on any of the other field components we will solve  it
separately and impose conditions for the other components afterwards. Note that~\eqref{eq:eomSMAtp}
has singular coefficients at the horizon~$u=1$ (and at the boundary as well). We have to
invoke the {\it indicial procedure} in order to split the singular behavior~$(1-u)^\beta$ from the regular part~$F(u)$
of the solution
\begin{equation}
\label{eq:asymptoticHorSol}
A_t' = (1-u)^\beta F(u) \, .
\end{equation}
The {\it indicial exponent}~$\beta$ characterizing the singular behavior is
determined by setting~$A_t' \to (1- u)^\beta$, expanding the singular coefficients of~\eqref{eq:eomSMAtp}
around the horizon~$u=1$ keeping only the leading order term and evaluating~\eqref{eq:eomSMAtp} with
these restrictions. The result is a quadratic equation for~$\beta$ giving
\begin{equation}
\label{eq:SMIndices}
\beta = \pm \frac{i\wn}{2} \, .
\end{equation}
By the variable change to a radial~$\xi=-\ln(1-u)$ with~$0<\xi<\infty$ we see that the positive sign in~$\beta$ describes
an outgoing wave at the horizon~$A_t'(\xi)\propto e^{-i\wn \xi/2}$ while the negative sign gives an
incoming wave~$A_t'(\xi)\propto e^{i\wn \xi/2}$. We select the latter solution to be the physical one since
no radiation should come out of the black hole. This is often referred to as the {\it incoming wave boundary condition}. 

Now we are ready to write down the hydrodynamic expansion in momentum-temperature ratios~$\wn, \qn^2\ll 1$ for 
the regular part~$F(u)$ of the solution
\begin{eqnarray}
\label{eq:hydroExpansion}
F(u) = &F_0 + \wn F_1 + \qn^2 G_1 \\ 
&+ \wn^2 F_2 + \qn^4 G_2 + \wn\qn^2 H_1 + \dots \, .
\label{eq:hydroExpansionSecond}
\end{eqnarray}
We will refer to the first line~\eqref{eq:hydroExpansion} as the {\it leading order or first order hydrodynamics terms}, while 
we coin the second line~\eqref{eq:hydroExpansionSecond} {\it second order hydrodynamics terms}.
Substituting the leading order hydrodynamic expansion~\eqref{eq:hydroExpansion} into the equation of 
motion~\eqref{eq:eomSMAtp} with~$A_t'=(u-1)^{-i\wn/2} F(u)$ and comparing coefficients in the 
orders~$\O(1),\, \O(\wn)$ and~$\O(\qn^2)$ yields three equations for the three hydrodynamic 
functions~$F_0,\, F_1,\, G_1$  
\begin{eqnarray}
F_0'' + \frac{(u f)'}{uf} F_0 ' =0 \label{eq:eomF0} \, ,\\
F_1 '' +\frac{(u f)'}{uf} F_1 ' +\frac{i}{2}[\frac{1}{(u-1)^2}-\frac{(u f)'}{uf (u-1)}] F_0 = 0 \label{eq:eomF1}\, , \\
G_1 '' + \frac{(u f)'}{uf} G_1 ' -\frac{1}{uf} F_0 =0 \, . \label{eq:eomG1}
\end{eqnarray}
Note that we can compute higher order corrections in this hydrodynamic perturbation approach
by inclusion of higher order terms, e.g. the second order terms~\eqref{eq:hydroExpansionSecond}.
We would have to compare coefficients up to the desired order of accuracy and would end up with
e.g. three further equations added to~\eqref{eq:eomF0},~\eqref{eq:eomF1} and~\eqref{eq:eomG1} for 
three additional hydrodynamic functions~$F_2,\, G_2, \, H_1$ in the case of second order corrections.

The solutions to~\eqref{eq:eomF0},~\eqref{eq:eomF1} and~\eqref{eq:eomG1} can be obtained analytically
if we start out noting that we may set~$F_0=\text{constant} =C$. Then we get~\footnote{Note, that the
complex logarithm~$\ln z$ being a multivalued function has branch points at~$z=0,\,\infty$ and in general a branch cut is defined
to extend between these points on the negative axis. Here we define the complex logarithm on the
first Riemann sheet, such that e.g.~$\ln(-1) = + i\pi$. All the equations here should be read with this in mind.}
\begin{equation}
\label{eq:solF1withCs}
F_1 = C_2 +\frac{i C}{2} \ln (u-1) - C_1 \ln u +\frac{C_1}{2} \ln\{(u+1) (u-1)\} \,
\end{equation}
with two undetermined integration constants~$C_1,\, C_2$. These can be fixed by recalling that
we have already chosen the constant order in~$F(u)$ independent from~$u,\, \wn, \,\qn^2$ to
be given by~$C$. So we now have to impose the condition on our solution for~$F_1$ that it 
gives no corrections to this constant~$C$, meaning~$\lim\limits_{u\to 1}F_1 = 0$. In this limit 
two of the terms in~$F_1$ become divergent and the constants have to be chosen such that 
these cancel each other. After application of this procedure to~$G_1$ as well, we are left with
\begin{eqnarray}
F_1=\frac{iC}{2} \ln\frac{2 u^2}{u+1} \, , \\
G_1 =  C \ln \frac{1+u}{2u} \, .
\end{eqnarray}
Now we have a first order solution for the derivative~$A_t'$. 
We can also fix the constant~$C$ in terms of boundary values~$A^{\text{bdy}}$ for the physical fields.
This is important because~$C$ contains the $\wn$-pole structure of the solution as we will
see shortly. 
First we recall that~$\lim\limits_{u\to 0} A_t = A_t^{\text{bdy}}$  and~$\lim\limits_{u\to 0} A_z = A_z^{\text{bdy}}$.
Now substitute the solution for~$A_t'$ into equation~\eqref{eq:sMFSAz} and take the 
boundary limit of this expression. This yields
\begin{equation}
\label{eq:C}
C=\frac{\qn^2 A_t^{\text{bdy}}+\wn\qn A_z^{\text{bdy}}}{i\wn -\qn^2 +\O(\wn^2,\qn^4,\wn\qn^2)} \, . 
\end{equation}
The denominator of~\eqref{eq:C} contains the poles of the solution which are the poles of the
retarded correlator as well. 

Taking our third and final recipe-step we assemble the correlator for time components of the~$R$-charge current
\begin{equation}
G^R_{tt}=-2 \lim\limits_{u\to 0} (-\frac{N^2}{16 \pi^2})\sqrt{-g} g^{uu} 
 \frac{\delta^2}{\delta A_t^{\text{bdy}}\delta A_t^{\text{bdy}}} (g^{tt} A_t' A_t +g^{zz}A_z' A_z ) \, ,
\end{equation}
where the double functional derivative encodes the step of selecting the terms in the action which are 
relevant (meaning quadratic in the field~$A_t^{\text{bdy}}$) in order to be more illustrative here.
We finally get
\begin{equation}
G^R_{tt}=\frac{N^2 T}{16 \pi^2} \frac{q^2}{i \omega - D q^2} \, ,
\end{equation}
with the constant~$D = 1/(2\pi T)$ which is identified with the diffusion coefficient. This interpretation
is best understood by noting that the diffusion equation
\begin{equation}
\partial_t J^t = D \nabla^2 J^t
\end{equation}
can be Fourier transformed to
\begin{equation}
-i\omega J^t = D (iq)^2 J^t \, .
\end{equation}
This suggests that the retarded correlator we found has the correct pole structure to be the
Greens function for a diffusion problem, in our case this is the diffusion of $R$-charges.

{\bf Dispersion relations} 
The dispersion relation for the $R$-charge current~$J^\mu$ to first hydrodynamic order is given by
\begin{equation}
0 = i \wn -\qn^2 +\O(\wn^2,\qn^4,\wn\qn^2) \, .
\end{equation}
Computing the second order hydrodynamics corrections as described above, we obtain
the dispersion relation
\begin{equation}
\label{eq:dispersion2}
0 = i \wn -\qn^2 + \ln 2 (\frac{\wn^2}{2}+\frac{i}{2}\wn\qn^2-\qn^4) +\O(\wn^3,\qn^6,\wn^2\qn^4)\, .
\end{equation}
Since this equation is quadratic in~$\wn$ one at first suspects that two solutions exist,
but if we solve~\eqref{eq:dispersion2} and then (recalling~$\wn,\qn^2\ll 1$) expand both solutions 
in~$\wn$, we get
\begin{eqnarray}
\label{eq:goodDisp}  
\wn = - i\qn^2 - i\ln 2 \qn^4 +\O(\qn^6)  \, , \\
\label{eq:badDisp} 
\wn_{\text{discard}} =  -\frac{2i}{\ln 2} + i \ln 2 \qn^4 + \O(\qn^6) \, .
\end{eqnarray} 
Only the first~\eqref{eq:goodDisp} of these two solutions is compatible with our initial assumption that~$\wn \sim \qn^2 \ll 1$ 
since the second solution~\eqref{eq:badDisp} has a constant leading order with an absolute value of order one.

Dispersion relations and correlators of other operators~$\O$~(e.g. the energy-momentum tensor~$T^{\mu\nu}$) 
dual to other fields $\phi$ are obtained in the same way.

\subsubsection{Numerical methods} \label{sec:numAdsG}
It was already mentioned and should be stressed here again that the main difficulty in the computation of the 
two-point function for any field theory operator~$\O$ is that of solving the equation of motion for the dual gravity
field~$\phi$. This is the {\it second step} undertaken in the context of the recipe from section~\ref{sec:anaAdsG}. 
In the previous section we took the small frequency, small momentum limit~(which is called the hydrodynamic limit) in order to obtain an
analytical solution. In this present section we describe two different numerical methods to obtain the full solution
to the equation of motion for the gravity field~$\phi$ without taking the hydrodynamic limit. We consider the 
(dis)advantages of both methods.

{\bf Integrating forward}
The kinetic term in the classical gravity action for any field fluctuation~$\phi$ has the quadratic 
form~$\partial_\mu \phi \partial_\nu \phi$. Neglecting interaction terms~(since we are only interested in the
two-point functions) the Euler-Lagrange equation for any gravity field fluctuation is thus quadratic in derivatives of the
field fluctuation~$\phi$. Fourier-transforming the Minkowski-direction derivatives into four-momentum components
according to equation~\eqref{eq:fourierTrafo} and assuming no dependence on the three angular coordinates, 
only the radial AdS-derivatives~$\partial_u \phi \equiv \phi '$ has the general form
\begin{equation}
\label{eq:generalFlucEom}
0 = \phi'' + A(u) \phi' + B(u) \phi \, ,
\end{equation}
and the coefficients~$A, B$ in the backgrounds we will consider only depend on the radial AdS-coordinate~$u$ and 
on the Minkowski four-momentum~$\vec k$. Therefore we need to solve second order differential equations
with non-constant coefficients. The coefficients~$A, B$ can be singular at the boundary~$u_{\text{bdy}}$ and at the 
horizon~$u_H$. In this case one has to perform the {\it indicial procedure} described in section~\ref{sec:anaAdsG}
yielding an asymptotic form for the solution at the horizon given by~\eqref{eq:asymptoticHorSol} as
\begin{equation*}
\phi = (u_H-u)^\beta F(u) \, .
\end{equation*}
The incoming wave boundary condition determines~$\beta$ to be negative.
Now we proceed by plugging this Ansatz into the equation of motion~\eqref{eq:generalFlucEom} yielding a 'regular'
equation of motion for the regular factor~$F(u) = F(u_H) + F'(u_H) (u_H-u)+ \dots$ of the solution~$\phi$ 
\begin{equation}
\label{eq:regularFlucEom}
0 = F''(u) + \tilde A(u) F'(u) + \tilde B(u) F(u) \, .
\end{equation}
This has to be solved numerically with the boundary conditions
\begin{equation}
\label{eq:numericalBCs}
F(u_H) = a_0\, ,\quad F'(u_H) = a_1 \, .
\end{equation}
Explicit values for~$a_0,\, a_1$ are found by plugging the asymptotic form of the regular solution near the horizon
\begin{equation}
F(u) = a_0 + a_1 (u_H-u) + a_2 (u_H-u)^2 +\dots \, ,
\end{equation}
into the equation of motion~\eqref{eq:regularFlucEom}. This procedure yields an equation which we can expand around
$u_H$ and by matching coefficients of orders in~$(u_H-u)$ we get recursive relations for~$a_0,\, a_1,\, \dots$ 
to any desired order in~$u$. Since we are free to normalize~$F(u)$, we can choose~$a_0\equiv 1$ and 
determine~$a_1$ from the recursive relations fixing our numerical boundary conditions~\eqref{eq:numericalBCs}.
We will use this method for example in chapter~\ref{sec:thermalSpecFunc}.

This method is straightforward and easy to use. We will apply it to find the correlators giving spectral functions
in chapter~\ref{sec:thermalSpecFunc}.
 
{\bf Matching in the bulk}
There are cases~(such as the calculation of quasinormal modes) in which the numerical method described in the previous paragraph fails.
An alternative method to solve the AdS-equations of motion numerically is described in~\cite{Kovtun:2006pf,Teaney:2006nc}.
The basic idea is to use two asymptotic solutions at the horizon as starting values for numerically integrating them 
forward into the bulk, then doing the same with two asymptotic solutions at the AdS-boundary and to afterwards 
match the two boundary solutions to the particular horizon solution which satisfies the incoming wave boundary condition, 
which we already discussed in section~\ref{sec:anaAdsG}. We can not directly determine the linear combination of 
integrated boundary solutions which is compatible with the boundary conditions since the incoming wave bounary condition is 
given only at the horizon.

We again start out with a second order differential equation of motion for fluctuations~$\phi$ as given in~\eqref{eq:generalFlucEom}
in the AdS Schwarzschild black hole background~\eqref{eq:adsBHmetric}. Note that in these coordinates the black hole horizon 
is located at~$u=1$ while the AdS-boundary lies at~$u=0$. The coefficients~$A,\, B$ again depend on the dimensionless
frequency~$\wn=\omega/(2\pi T)$, momentum~$\qn = q/(2\pi T)$ and on the radial coordinate~$0\le u\le 1$. For definiteness
we work in the setup of~\cite{Kovtun:2006pf} as a specific example where~$\phi$ are fluctuations of the metric tensor. We first 
have to determine the asymptotic behavior of the solution to this equation at the boundary~$u=0$. The {\it indicial procedure}
described in~\ref{sec:anaAdsG} yields the leading order asymptotic behavior~$\phi\propto u^0$ or~$\phi\propto u^2$ with
the indicial exponents~$\beta_1=0$ or~$\beta_2=2$ 
corresponding to the two possible solutions respectively. For a second order differential equation we can expand the
asymptotic solutions~$\Phi^I,\, \Phi^{II}$ according to~\cite{Bender} into general series~$\Phi^{II}=(u-u_{\text{bdy}})^{\beta_{2}} A(u)$ 
and~$\Phi^{I}=(u-u_{\text{bdy}})^{\beta_{2}} A(u) \ln u + (u-u_{\text{bdy}})^{\beta_1}C(u)$  with the indicial 
exponents~$\beta_1,\, \beta_2$ and the functions~$A(u),\,B(u),\, C(u)$
being analytic at~$u=0$. So in our example we have the asymptotic solutions for~$\phi$ at the boundary~$u=0$
\begin{eqnarray}
\label{eq:boundarySolsI}
&\Phi^{I} =  &u^{0} \left ( b^{(0)}_I+b^{(1)}_I u +b^{(2)}_I u^2 + \dots \right  )+ h Z^{II} \ln u \, , \\
\label{eq:boundarySolsII}
&\Phi^{II} = &u^{2} \left ( b^{(0)}_{II} + b^{(1)}_{II} u + b^{(2)}_{II} u^2 + \dots \right) \, .
\end{eqnarray}
We obtain recursive relations for the coefficients~$b_I,\, b_{II},\, h$ by plugging each expansion~\eqref{eq:boundarySolsI}
and~\eqref{eq:boundarySolsII} separately into the equation of motion, expanding in~$u$ around~$u=0$ and by then 
comparing coefficients in orders of~$u$.
The most general solution of the equation of motion is a linear combination~$\Phi(u) = a \Phi^{I} + b \Phi^{II}$ of the two solutions
given in~\eqref{eq:boundarySolsI},~\eqref{eq:boundarySolsI} with coefficients~$a,\,b$.
But since we have two boundary conditions our solution is fully determined and we give special names to the 
coefficients~$a,\,b$ which satisfy the two boundary conditions:~$a\to\A$ and~$b\to\B$, such that
\begin{equation}
\phi(u) = \A(\wn,\qn) Z^{I} + \B(\wn,\qn) Z^{II} \, .
\end{equation}
But how do we find~$\A,\, \B$ explicitly? In order to see this we also need the two asymptotic solutions at the 
AdS-horizon~$u=1$, where we calculate the indices~$\gamma_1=i\wn/2$ and~$\gamma_2 = -i\wn/2$. Just as we did
on the boundary, we now have to use the general expansion at the horizon~$\phi^I = (1-u)^{\gamma_1} D(u)$ 
and~$\phi^{II}=(1-u)^{\gamma_2} \bar{D(u)}$ giving
\begin{eqnarray}
\label{eq:horizonSolsI}
&\phi^I = &(1-u)^{-i\wn/2} \left ( a^{(0)}_{I}+a^{(1)}_{I} u +a^{(2)}_{I} u^2 + \dots \right) \, , \\
&\phi^{II} = &\bar{\phi^I} \, .
\end{eqnarray} 
The first thing we note is that only the first solution~$\phi^I$ is compatible with the incoming wave boundary condition
as described below equation~\eqref{eq:SMIndices}. We again obtain recursive relations for the coefficients~$a_{I}$
by plugging~\eqref{eq:horizonSolsI} into the equation of motion and comparing coefficients. 

Now the idea is that we determine the first two coefficients in the asymptotic horizon-expansion of the one solution~$\phi^I$ satisfying
the incoming wave boundary condition~(i.e.~$a^{(0)}_{I}$ and~$a^{(1)}_{I}$ 
in~\eqref{eq:horizonSolsI}). Then we use these two values to numerically integrate~$\phi^I$ forward into the bulk. 
We repeat this procedure with the two solutions~$\Phi^I$ and~$\Phi^{II}$ at the boundary. Then we find that linear combination
of integrated boundary solutions~$\Phi^I$ and~$\Phi^{II}$ which equals the incoming horizon solution~$\phi^I$
\begin{equation}
\A \Phi^I + \B \Phi^{II}=\phi^I \, .
\end{equation}
The values of coefficients~$a_I,\, b_I, \, b_{II}, \, h$ are all fixed by recursive relations, with the exception of~$b^{(0)}_I,\,b^{(0)}_{II}$
and~$b^{(2)}_I$. Note, that we are free to normalize the solutions~$\Phi^{I,II}$ such that~$b^{(0)}_I=1$ and~$b^{(0)}_{II}=1$.
Our freedom to choose~$b^{(2)}_I$ arbitrarily reflects the fact that the solution~$\Phi_I$ is still a solution if one adds a multiple of
the other solution~$\Phi_{II}$. We choose~$b^{(2)}_I= 0$ for convenience. This fixes all the asymptotic expansions.

This particular procedure is more complicated and involves a few more steps than the forward integration but in some cases
such as the search for the quasinormal modes~(QNMs) it is necessary to employ a matching in the 
bulk see e.g.~\cite[section 7.2]{Erdmenger:2007cm} or~\cite{Hoyos:2006gb}. The problem there is that one has to satisfy
the incoming wave boundary condition, which implies that the solution near the horizon is heavily oscillating as~$(1-u)^{-i\wn/2}$
and on the other side at the boundary~$u=0$ the solution is required to be normalizable. Numerically it would be very difficult to
for example start at the boundary with a normalizable solution and try to match a highly oscillating solution at the horizon
by directly integrating forward. Thus the method of matching integrated solutions in the bulk is preferred here.
  
\subsection{Holographic hydrodynamics} \label{sec:holoHydro}
There is convincing evidence~\cite[for a review]{Son:2007vk} that the AdS/CFT correspondence maps relativistic hydrodynamics on the 
(thermal) field theory side to black hole physics on the gravity side. In this section we remind ourselves of some facts about relativistic
hydrodynamics~(\ref{sec:relativisticHydro}), we review how to introduce a chemical potential in thermal quantum field 
theory~(\ref{sec:chemPots}) and we rederive a method to compute (non-equilibrium) transport coefficients like the heat 
conductivity or shear viscosity~(\ref{sec:transportCoeffs}). The understanding we gain here on the field theory side
will help us substantially interpreting the results from gravity calculations we perform in the AdS/CFT context in the
coming chapters. 

\subsubsection{Relativistic hydrodynamics} \label{sec:relativisticHydro}
Relativistic hydrodynamics~\cite{Landau,Forster} is an effective theory which describes the dynamics of a fluid at
long wave length and small frequency for fluctuations. Since this theory historically includes dissipative effects it
is formulated in terms of equations of motion and not in terms of an action principle. These hydrodynamic equations
are mostly obtained from a system of conservation equations an so-called {\it constitutive equations}. These
constitutive equations express the conserved quantities~(e.g. tensor, vector current) in terms of {\it hydrodynamic variables},
such as temperature and four-velocity of a fluid element.
The thermal system is assumed to be in local thermal equilibrium but globally the hydrodynamic variables may
vary. We can define the local temperature~$T(\vec x)$ and the local four-velocity~$u^\mu (\vec x)$ of a fluid element
in the system, where~$u^\mu u_\mu = -1$.
The simplest example of a set of hydrodynamic equations is the conservation equation for energy and momentum
\begin{equation}
\partial_\mu T^{\mu\nu} = 0 \, ,
\end{equation}
together with the constitutive equation for the energy-momentum tensor
\begin{equation}
\label{eq:constitutiveTLO}
T^{\mu\nu} = (\epsilon +P) u^\mu u^\nu + P g^{\mu\nu} \, ,
\end{equation}
with the (internal) energy density~$\epsilon$, the pressure~$P$
The constitutive equation~\eqref{eq:constitutiveTLO} is obtained by writing down all possible terms in an expansion
in powers of spatial derivatives of hydrodynamic variables to leading order. We can also include the next to leading order
yielding
\begin{equation}
\label{eq:constitutiveTNLO}
T^{\mu\nu} = (\epsilon +P) u^\mu u^\nu + P g^{\mu\nu} - \sigma^{\mu\nu} \, .
\end{equation}
While the leading order~\eqref{eq:constitutiveTLO} conserves entropy, the next to leading order~\eqref{eq:constitutiveTNLO}
contains the dissipative part~$\sigma^{\mu\nu}$ containing first derivatives of~$T(\vec x)$ and~$u^\mu(\vec x)$. 

In systems with a conserved current~$J^\mu$ satisfying
\begin{equation}
\partial_\mu J^\mu = 0 \, .
\end{equation}
And this current can be expressed in terms of the hydrodynamic variables by the constitutive equation
\begin{equation}
J^\mu = d u^\mu - D (g^{\mu\nu} + u^\mu u^\nu) \partial_\nu d \, ,
\end{equation}
with the charge density~$d$ in the fluid rest frame and the constant~$D$. The terms correspond to the processes
of convection and diffusion respectively and~$D$ is the diffusion coefficient. In the fluid rest frame this reduces to
\begin{equation}
\mathbf{J} = - D  \nabla d \, ,
\end{equation}
which is Fick's diffusion law. 

There is an intimate relation between the poles of thermal field theory correlators and the hydrodynamic modes
like for example the diffusion mode governed by the diffusion equation
\begin{equation}
\label{eq:diffusionEquation}
0 = \partial_t d  - D \nabla^2 d \, .
\end{equation}
Transformed to Fourier space this equation reads
\begin{equation}
\label{eq:diffusionEquationFT}
0 = (\omega + i D \mathbf{k}^2) d \, .
\end{equation}
The corresponding field theory two point correlator of a conserved current~$J^\mu$ is given  in Fourier space by
\begin{equation}
G(\omega, \mathbf{k}) \propto \frac{1}{i\omega - D\mathbf{k}^2}\, .
\end{equation}
We easily verify that this two point current correlator is a Green function for the diffusion equation or in other words
a solution to the diffusion problem. Such identifications are also possible for other hydrodynamic modes like 
the shear and sound modes of the energy-momentum tensor which are identified with poles in the metric fluctuation
correlators~(for details the reader is referred to~\cite[and references therein]{Son:2007vk}). 

Let us also include the relation between the thermal spectral function~$\R$ and the retarded correlation function~$G^R$ here 
for completeness.
\begin{equation}
\label{eq:spectralFunction}
\R = -2\, \mathrm{Im}\, G^R \, .
\end{equation}
Heuristically the thermal spectral function gives the thermal spectrum of the system at finite temperature. Resonances
appearing in this spectral function are analogous to the spectral lines one gets when analyzing light with a prism. The 
resonances are interpreted as quasi-particles produced in the plasma. Just as it is the case for e.g. the visible light spectrum, 
the resonances here have a finite width corresponding to the lifetime of the quasi-particle excitation since the thermal system features 
dissipative processes. 
Let us write the energy $\omega$ and spatial momentum $\bm q$ in a four
vector $\vec k=(\omega,\bm q)$ while the Green function $G^R$ may be written as
\begin{equation}
        G^R(\omega, \bm q) = -i \int\!\! \dd^4x\:  e^{i\:\vec{k}\vec{x}}\, \theta(x^0) \left< \left[ J(\vec{x}), J(0) \right] \right>
\end{equation}
We may find singularities of $G^R(\omega, \bm q)$ in the lower half of the complex
$\omega$-plane, including hydrodynamic poles of the retarded real-time Green function. Consider for example
\begin{equation}
        \label{eq:quasiNormalModes}
        G^R = \frac{1}{\omega-\omega_0 + i \Gamma}.
\end{equation}
These poles emerge as peaks in the spectral function,
\begin{equation}
        \R = \frac{2\, \Gamma}{(\omega-\omega_0)^2+\Gamma^2}\,,
\end{equation}
located at $\omega_0$ with a width given by $\Gamma$. These peaks are
interpreted as quasi-particles if their lifetime $1/\Gamma$ is considerably
long, i.e.\ if $\Gamma \ll \omega_0$.
We will discuss the spectral function again in chapter~\ref{sec:thermalSpecFunc}.

Another facet of the spectral function will be made use of in the diffusion chapter~\ref{sec:transport}. In its zero
frequency limit the spectral function evaluated at zero spatial momentum is related to the diffusion coefficient~$D$
of the charge~$Q$ to which the correlated~($G^R\propto\langle J J \rangle$) current~$J$ couples
\begin{equation}
\label{eq:spectralFunctionDiffusion}
\Xi D = \lim\limits_{\omega\to 0} \frac{\R (\rho,\omega,q=0)}{2\omega} = 
  -2 \lim\limits_{\omega\to 0}\, \frac{1}{2\omega}\mathrm{Im}\, G^R (\rho,\omega,q=0) \, ,
\end{equation}
where~$\rho$ is the radial AdS coordinate and the susceptibility~$\Xi$ is given by
\begin{equation}
\Xi = \left .\frac{\partial J^0(\mu)}{\partial \mu} \right |_{\mu=0} \, ,
\end{equation}
with the charge density~$J^0$ for the conserved charge~$Q$ and the thermodynamically conjugate
chemical potential~$\mu$. This provides us with a method to compute diffusion coefficients using the fluctuations
about a background. An alternative method makes use of the {\it membrane paradigm} in order to compute the 
diffusion coefficient from metric components only~(see section~\ref{sec:membranePara}).

\subsubsection{Chemical potentials in QFT} \label{sec:chemPots}
Since the introduction of a chemical potential~$\mu$ and its thermodynamic conjugate charge density~$d$
is a central point in this work, in this section we make the heuristic statements given in section~\ref{sec:finiteTAdsCft}
more precise. All the ideas explained below should be read with the QFT path integral formalism in mind.

Introducing a chemical potential~$\mu$ in a QFT at finite temperature T is formally
analogous to turning on a fictitious gauge fields time-component $A_t$~\footnote{The identification $A_t\sim \mu$ is solely a 
field theory matter and has a priori nothing to do with the AdS/CFT-correspondence.} .
Heuristically this can be seen by the comparison of terms entering
the partition function $Z$ by turning on a chemical potential on one hand
\begin{equation}
Z\propto e^{-\beta H}\to e^{-\beta (H-\mu N)} \, ,
\end{equation}
where $\beta$ is the temperature obtained from compactifying the time
coordinate in the imaginary time formalism, $\mu$ is the chemical
potential and $N$ is the number operator for a particle. We are working 
in the grand canonical ensemble. On the other hand we can turn on
a fictitious gauge field $A_\mu$ belonging to a symmetry which conserves
a certain current $J^\mu$
\begin{equation}
Z\propto e^{-\beta H}\to e^{-\beta (H-A_\mu J^\mu)} \, ,
\end{equation}
where $A_\mu=(\mu,\mathbf{0})$.
So roughly we obtain the relations
\begin{equation}
A_t\sim \mu \ ,  \ \ \ \ \ \ \ \ J^t\sim N \ . 
\end{equation}
The next paragraph
describes the above statements in greater detail.

From the Noether theorem we know that every symmetry of a theory contributes a 
conserved current
\begin{equation}
J^\mu_{\text{Noether}}=\frac{\partial L}{\partial(\partial_\mu A_\nu)}\delta A_\nu \, ,
\end{equation}   
which for QED equals the electromagnetic current that we observed formally above as $J^\mu$.
The conserved charge is obtained by integrating the first component of the current over
all space. This shows that conserved currents are intimately related to charges
and introduction of either implies existence of the other. The electromagnetic current
can by Maxwells equations be written as 
\begin{equation}
J^\mu=\partial_\mu F^{\mu\nu}=\frac{\partial L}{\partial A_\mu} \, ,
\end{equation}
and can therefore be regarded as a source of the field strength or the gauge field.

{\bf Finite temperature currents and the chemical potential}
In the context of a non-SUSY complex scalar field theory we would like to evaluate the
partition function with a chemical potential $\mu$ and show that at finite
temperatures its introduction is equal to introducing a fictitious gauge field.
In the grand canonical ensemble the partition function is~\footnote{This is a standard 
finite temperature QFT result. See e.g.~\cite{Kapusta:1989tk} for reference.} 
\begin{eqnarray}
Z &=& tr[\ e^{-\beta(H-\mu N)}] \nonumber \\
&=& C \int \mathcal{D}\pi^{\dagger}\mathcal{D}\pi
\int \mathcal{D} \phi^{\dagger} \mathcal{D}\phi\
e^{\int_0^\beta d\tau \int \! d^3 x \,\left[
    i\pi \frac{\partial\phi}{\partial\tau}
    +i\pi^{\dagger} \frac{\partial\phi^{\dagger}}{\partial\tau}
    -\mathcal{H}(\pi,\phi) +
    \mu \mathcal{N}(\pi,\phi) \right]} \, ,
\end{eqnarray}
with the Hamiltonian
\begin{equation}
\mathcal{H}=\pi^\dagger \pi + \nabla\phi^\dagger \cdot
\nabla\phi + m^2 \phi^\dagger \phi  \, .
\end{equation}
The conjugate momenta are defined to be $\pi(\vec x)=
\frac{\partial L}{\partial \dot\phi(\vec x)}=\dot\phi^\dagger$.  
By the Noether formula the conserved current is 
$J_\mu = i(\phi^\dagger \partial_\mu \phi - \phi \partial_\mu \phi^\dagger)$.
The first component of this is the conserved charge density
$\rho=i(\phi^\dagger \pi^\dagger - \phi\pi)$.
The integrand of the exponent in the path integral can be rewritten
\begin{eqnarray}
&i\left(\pi^\dagger \partial_\tau \phi^\dagger+\pi
\partial_\tau \phi\right) -\left(\pi^\dagger \pi +
\nabla\phi^\dagger \cdot
\nabla\phi + m^2 \phi^\dagger \phi\right) + i\mu\left(\pi^\dagger \phi^\dagger -
\pi\phi \right) \nonumber 
&\\ = &-\left( \pi^\dagger -i(\partial_\tau -
\mu)\phi\right)\left(\pi-i(\partial_\tau + \mu)\phi^\dagger\right) -\\ 
&-(\partial_\tau+\mu)\phi^\dagger(\partial_\tau-\mu)\phi\nonumber 
 -\nabla\phi^\dagger \cdot \nabla\phi - m^2 \phi^\dagger \phi\, . &  
\end{eqnarray}
This shows that using the Euclidean time $\tau$ we can redefine 
the time derivative $\partial_\tau\to\partial_\tau-\mu\equiv D_0$ (and
equivalently $(\partial_\tau\phi)^\dagger
\to[(\partial_\tau+\mu)\phi]^\dagger\equiv (D_0\phi)^\dagger$ ).
From standard gauge theory we know that this is the same as introducing
a covariant derivative. But in the case at hand this gauge field has only one
non-zero, constant component, the time component $A_\tau$.
Therefore this gauge field is non dynamical having no kinetic term.   

Performing the functional integration over $\pi$ and $\pi^\dagger$ leads
to the following expression for the partition function~\footnote{Note
that only the first term in the integrand of the path integral 
is depending on $\pi$ and $\pi^\dagger$. Considering $-i(\partial_\tau-\mu)\phi$
and $-i(\partial_\tau+\mu)\phi^\dagger$ as shifts of the integration variables
$\pi$ and $\pi^\dagger$ (these shifts do not depend on either of the integration
variables), only the second and third term survive the integration to yield
the partition function as an integral over fields $\phi$ and $\phi^\dagger$ 
only}:
\begin{equation}
Z = C' \int \mathcal{D}\phi^\dagger \mathcal{D} \phi\ e^{-\int_0^\beta
d\tau \int \! d^3 x \,\left[
(\partial_\tau+\mu)\phi^\dagger(\partial_\tau-\mu)\phi + \nabla\phi^\dagger \cdot
\nabla\phi + m^2 \phi^\dagger \phi \right]} \nonumber \, ,
\end{equation}
which can be analytically continued
to Minkowski space to yield an effective Lagrangian:
\begin{eqnarray}
&& C' \int
\mathcal{D}\phi^\dagger \mathcal{D} \phi\ e^{i\int
dt \int \! d^3 x \,\left[
(\partial_t+i\mu)\phi^\dagger(\partial_t-i\mu)\phi - \nabla\phi^\dagger \cdot
\nabla\phi - m^2 \phi^\dagger \phi \right]} \nonumber  \\ &&\equiv C' \int
\mathcal{D}\phi^\dagger \mathcal{D} \phi\ e^{i\int
dt \int \! d^3 x \,\mathcal{L}_\mathrm{eff} } \, .
\end{eqnarray}

It is important to note that $\mathcal{L}_\mathrm{eff}$ is not simply
$\mathcal{L}+\mu\mathcal{N}$, since $\mathcal{N}$ is a function of
$\pi$ in addition to $\phi$. Instead,
\begin{equation}
\label{eqn:Lscalar}
\mathcal{L}_\mathrm{eff} = \partial^\nu \phi^\dagger \partial_\nu \phi
+ i\mu \left(\phi^\dagger \partial_t \phi-\phi \partial_t
\phi^\dagger \right) - (m^2-\mu^2)
\phi^\dagger \phi \, .
\end{equation}
The term linear in $\mu$ is the expected $\mu \mathcal{N}$
contribution. The term quadratic in $\mu$ arises from the modification
of the conjugate momenta $\pi=\dot{\phi}^\dagger+i\mu \phi$.   

The symmetry under which the current coupling to the chemical potential
is conserved could for example be the $U(1)_B$ symmetry. In this case
the conserved current $\mathcal{N}$ is the baryon-number-operator density.
The conserved charge is then the baryon number. 

\subsubsection{Transport coefficients: Kubo formula} \label{sec:transportCoeffs}
Kubo formulae relate transport coefficients~$\Lambda$ in non-equilibrium 
thermodynamics with retarded Green functions of the associated
thermodynamical current~$J$. Symbolically we can write 
\begin{equation}
\Lambda \propto \langle [J,J]\rangle_{\mathrm{ret.}} \, .
\end{equation}
These relations hold up to linear order expanding in thermodynamical
forces. This is called the linear response approximation. In the 
following subsection the Kubo formula will be derived for 
a system with energy-momentum conservation only. The principle
is extended to an additional conserved current in the next-to-next
subsection.

The derivation of Kubo formulae assumes that the system under
consideration has established a local thermal equilibrium in 
order to define meaningful state variables like temperature~$T$,
 mass density~$\rho$ and others locally. On the other hand
globally there exists a non-equilibrium, that means there are
gradients of thermodynamical state variables or potentials 
across the whole system.

{\bf Nonequilibrium Kubo formulae in theory with $\partial_\mu T^{\mu\nu}=0$}
This subsection generally follows~\cite{Hosoya:1983id}.
Let's imagine a thermodynamical system in which a gradient~(e.g. a
temperature gradient $\partial_{\mu}T$) exists. This gradient 
will cause the system to respond by forming a current $J$~(e.g. a 
heat current). With this current the system tries to equilibrate
the gradient~(e.g. heat flow diminishes temperature gradient
by levelling out temperatures in the whole system). This 
kind of gradient is synonymously called a thermodynamical force~$F$.

We can expand the response of the system~(namely the current) to 
a gradient as a series of powers of the gradient:
\begin{equation}
J=a_0\ (F)^0+a_1\ (F)^1+a_2\ (F)^2 +\dots  \, .
\end{equation} 
Since the current should vanish with vanishing gradient, the
constant term has to vanish and the linear one is the lowest order
contribution with respect to the gradient-expansion. This contribution
gives the linear response of the system to the gradient~$F$. The
proportionality-factor~$a_1$ is called a transport coefficient, 
which we will generally denote by $\Lambda$. 
\begin{equation}
J=\Lambda\ F \, .
\end{equation}
We now would like to establish this connection between currents $J$ and the 
gradients $F$ driving them. For this reason we will compute the
thermal non-equilibrium average of the energy-momentum tensor~$T_{\mu\nu}$
containing several~(scalar, vector and tensor) currents. In order to be able to use
thermodynamics, we assume from now on that a local equilibrium is established and 
we can thus define the state variables such as temperature and pressure locally. We
can also compute local averages and denote them by~$\langle .\rangle$.  

Remember that in equilibrium we can define the probability density matrix
$\varrho_{\mathrm{eq}}=\mathrm{e}^{\beta H}$ to be used as a calculational
tool determining thermal averages of operators~$\mathcal{O}$
\begin{equation}
\langle \mathcal{O} \rangle_{\mathrm{eq}} =\mathrm{tr}[\varrho_{\mathrm{eq}}\mathcal{O}] \, .
\end{equation}
This has an analog in nonequilibrium
\begin{equation}
\langle \mathcal{O} \rangle_{\mathrm{non-eq}} =
\mathrm{tr}[\varrho_{\mathrm{non-eq}}\mathcal{O}] \, .
\end{equation}
We are interested in the operator $\mathcal{O}\equiv T_{\mu\nu}$ and how to relate its 
non-equilibrium expectation value back to quantities which are in local equilibrium (averaged by~$\langle .\rangle$).
Zubarev already proposed the following construction for non-equilibrium~\cite{Zubarev:1974}:
\begin{equation}
\varrho_{\mathrm{non-eq}}= {
\exp}\{
\underbrace{-
 \int d^3xF^\nu T_{0\nu}}_{=\beta H\mathrm{\  equilibrium}}
 +\int d^3x\int\limits^t_\infty dt_1 e^{\epsilon(t_1-t)}
T_{\mu\nu}
\partial^\mu F^\nu \}
\, ,
\end{equation}
where~$T_{\mu\nu}$ is the thermodynamical 'current', $\partial^\mu F^\nu$~is the thermodynamical
force~(gradient), while 
\begin{equation}
F^\nu=\beta u^\nu \, .
\end{equation}
Here we have $\beta\equiv 1/T$ with the temperature~$T$, $u^\nu$ is the four-velocity component of the fluid and the 
parameter~$\epsilon$ is a small number which will be sent to zero in the end. Note that we set the Boltzmann
constant~$k_B=1$ throughout this work.
So the average over the energy-momentum tensor can be written
\begin{equation}
\langle 
T_{\mu\nu}\rangle_{\mathrm{non-eq}} = 
\mathrm{tr}[\exp\{
\underbrace{-
 \int d^3xF^\nu T_{0\nu}}_{=\beta H\mathrm{\  equilibrium}}
 +\int d^3x\int\limits^t_\infty dt_1 e^{\epsilon(t_1-t)}
T_{\rho\sigma}
\partial^\rho F^\sigma \}\ T_{\mu\nu}]  \, .
\end{equation}
Expanding the exponential to linear order in the gradient~$\partial^\rho F^\sigma$
we get
\begin{equation}
\label{eq:<T>}
\langle {T_{\mu\nu}}\rangle_{\mathrm{non-eq}}{\approx}
\langle T_{\mu\nu}\rangle_{\mathrm{eq}}
 +\int d^3x'\int\limits_{-\infty}^t dt' \mathrm{e}^{\epsilon(t'-t)}
 \underbrace{
  (T_{\mu\nu}(\vec x,t),T_{\rho\sigma}(\vec x',t'))
 }_{\propto
 \langle T_{\mu\nu},T_{\rho\sigma}\rangle_{\mbox{retarded}}}
 {\partial^\rho F^\sigma(\vec x',t')}  \, .
\end{equation}
The currents collected in $T_{\mu\nu}$ may be separated into
tensor, vector and scalar currents by means of the 
thermodynamical standard form
\begin{equation}
\label{eq:thermoT}
T_{\mu\nu}={\large\sigma_{\mu\nu}}+\epsilon u_\mu u_\nu-p g_{\mu\nu}+pu_\mu u_\nu
 +P_\mu u_\nu+P_\nu u_\mu \, ,
\end{equation}
where the expansion coefficients are the energy density~$\epsilon$,~$p$ the pressure, the 
tensor structure~$\sigma_{\mu\nu} = [(g_{\mu\rho} -u_\mu u_\rho )(g_{\nu\sigma} -u_\nu u_\sigma )
-\frac{1}{3}(g_{\mu\nu} -u_\mu u_\nu )(g_{\rho\sigma} -u_\rho u_\sigma )] T^{\rho\sigma}$ and the heat 
current~$P_\mu=(g_{\mu\nu}-u_\mu u_\nu) u_\sigma T^{\nu\sigma}$. Note, that the latter will be absent
if there is no quantity~(such as a charge density) relative to which that current could be measured. This
is a consequence of relativity since the flow of mass and the flow of heat, i.e. energy becomes indistinguishable.
The bracket~$(.,.)$ denotes the quantum time correlation functions defined by
\begin{equation}
(T_{\mu\nu}(\vec x,t),T_{\rho\sigma}(\vec x',t') ) = \int\limits_0^1 \dd \tau \langle T_{\mu\nu}(\vec x,t) 
( e^{-A\tau} T_{\rho\sigma}(\vec x',t') e^{A\tau} -\langle T_{\rho\sigma}(\vec x',t') \rangle ) \rangle \, ,
\end{equation}
where we used the abbreviation~$A = \int \dd^3 x F^\nu(\vec x, t) T_{0\nu}(\vec x, t)$.
Due to Curie's theorem tensor currents like~$\sigma_{\mu\nu}$ are only driven
by 'tensor gradients'. The scalar and vector processes as well are 
only related to scalar and vector gradients. Correlation functions 
between currents of different tensor rank vanish. In other words
to linear order in the gradients the scalar, vector and tensor
processes have nothing to do with each other. We now use this fact
picking out the tensor process to replace $T_{\mu\nu}$ by the tensor current $\sigma_{\mu\nu}$
But which tensor gradient drives this current? To answer this question,
we compute
\begin{equation}
\label{eq:currForc}
T_{\rho\sigma}\partial^\rho F^\sigma=
\sigma_{\rho\sigma}
\ {
\beta\partial^\rho u^\sigma}
 +\beta P_\rho(\beta^{-1}\partial^\rho \beta+u^\kappa\partial_\kappa u^\rho)
 -\beta p'\partial_\rho u^\rho  \, .
\end{equation}
Now using equation~(\ref{eq:thermoT}) on the left hand side of (\ref{eq:<T>}) 
and (\ref{eq:currForc})
on the right, we get
\begin{equation}
\langle{ \sigma_{\mu\nu}}\rangle_{\mathrm{non-eq}}
{\approx} \underbrace{
\langle \sigma_{\mu\nu}\rangle_{\mathrm{eq}}}_{\equiv 0}
 +\int d^3x'\int\limits_{-\infty}^t dt' \mathrm{e}^{\epsilon(t'-t)}
 \underbrace{(\sigma_{\mu\nu}(\vec x,t),\sigma_{\rho\sigma}(\vec x',t'))}_{
 \propto(\mbox{tensor structures}_{\mu\nu\rho\sigma})
 \times(\sigma_{\alpha\beta},\sigma^{\alpha\beta})}
 {\beta \partial^\rho u^\sigma} \, .
\end{equation}
Analogous separation works for vector and scalar currents.
If now the gradient~${\beta \partial^\rho u^\sigma}$ varies only slowly
compared to the correlation lenght of the rest of this integral,
we can pull it in front and get an integral expression for the 
transport coefficient associated with tensor processes, namely
the shear viscosity 
\begin{equation}
\eta \equiv \frac{\beta}{5}
 \int d^3x'\int\limits_{-\infty}^t dt' \mathrm{e}^{\epsilon(t'-t)}
\underbrace{(\sigma_{\alpha\beta}(\vec x,t),\sigma^{\alpha\beta}(\vec x',t'))}_{
\langle \sigma_{\alpha\beta},\sigma^{\alpha\beta}\rangle_{\mbox{retarded}}}  \, .
\end{equation}
This is called a Kubo formula. The coefficients connected to
scalar and vector processes are called bulk viscosity~$\xi$ and 
the heat conductivity~$\kappa$ respectively.

{\bf Kubo formulae in theory with additional conserved current $\partial_\mu J^{\mu}=0$}
This subsection is oriented along the application to R-charged black holes discussed in~\cite{Son:2006em}.
Following Landau and Lifshitz~\cite{Landau} one can imagine a thermodynamical
system with an additionally conserved current
\begin{equation}
\partial_\mu T^{\mu\nu}=0 \ \ \ \ , \ \ \ \ 
\partial_\mu J^\mu =0    \, .
\end{equation}
In such a relativistic hydrodynamic system the energy-momentum tensor
and the conserved current take the form
\begin{equation}
\label{eq:TJ}
T_{\mu\nu}=p g_{\mu\nu}+\omega u_\mu u_\nu + \tau_{\mu\nu},\ \ \ \ \ J_\mu=\rho_C u_\mu+\nu_\mu \, ,
\end{equation}
where p is the pressure, with the heat function $\omega=\epsilon+p$.  
$\rho_C$ is the charge density of the conserved charge given by the
first component of the conserved current $J_0=\rho_C$. The dissipative
part of the current is denoted by $\nu_\mu$. A contemporary application in the context of 
the gauge/gravity correspondence is the calculation of the heat conductivity in a R-charged black hole
background~\cite{Son:2006em}.  

According to~\cite{Zubarev:1974} the method described in the previous subsection
is completely general and we can replace $T_{\mu\nu}$ by any 
conserved current. So for the non-equilibrium density matrix
\begin{equation}\label{eq:varrhoX}
\varrho_{\mathrm{non-eq}}= {
\exp}\{
\underbrace{-
 \int d^3xF^\nu T_{0\nu}}_{=\beta H\mathrm{\  equilibrium}}
 +\int d^3x\int\limits^t_\infty dt_1 e^{\epsilon(t_1-t)}
 \underbrace{{
J_{\mu}}}_{\mbox{current   }}
 \underbrace{{
 X^\mu \}}}_{\mbox{associated gradient}} \, .
\end{equation}
The dissipative part~$\nu_\mu$ of the conserved current 
is driven by a gradient in the corresponding
chemical potential~$\mu$. From thermodynamical relations~\footnote{To be more precise: from the fact that dissipative processes like the current or the shear-processes will produce entropy. $\nu_\mu$ and $\tau_{\mu\nu}$ can be related to the entropy they produce.} one can obtain
\begin{equation}\label{eq:nu}
\nu^\mu=-\varkappa\left (g^{\mu\lambda}+u^\mu u^\lambda \right )\partial_\lambda \frac{\mu}{T}  \, ,
\end{equation}
which tells us that $\varkappa$ is the corresponding transport
coefficient. By anology we conclude its Kubo formula to read
\begin{equation}
\varkappa \equiv \frac{\beta}{3}
 \int d^3x'\int\limits_{-\infty}^t dt' \mathrm{e}^{\epsilon(t'-t)}
{\langle \nu_{\alpha}(\vec x,t),\nu^{\alpha}(\vec x',t'))\rangle_{\mbox{retarded}}
} \, .
\end{equation} 
But now that we know how to compute $\varkappa$ we should also
give a physical interpretation of it. $\varkappa$ certainly
tells us how big the dissipative current $\nu_\mu$ will be,
given a certain gradient in the chemical potential and 
temperature $\partial_\lambda\frac{\mu}{T}$. Like in the previous
subsection $\tau_{\mu\nu}$ had been the dissipative current connected
to the velocity gradient $\partial_\mu u_\nu$, now $\nu_\mu$ is
the dissipative current connected to the chemical potentials and
the temperatures gradient. So we also know
that $X_\lambda\equiv \partial_\lambda\frac{\mu}{T}$ in (\ref{eq:varrhoX})

\textit{Interpretation of the dissipative transport coefficient~$\varkappa$:}
The idea here is to relate the new current~$\nu_\mu$ to the 
energy-momentum tensor by expressing part of $T_{\mu\nu}$ 
by $\nu_\mu$. This is done in two steps. First the new current 
needs to be translated into a current we know. Second, the 
new gradient of the chemical potential needs to be translated.
 
The authors of~\cite{Son:2006em} choose to set the charge currents to zero $J^i=0$ for
the sake of interpretation. Using the form of $J^\mu$ from~(\ref{eq:TJ}) this immediately
tells us 
\begin{equation}
J_0=\rho_C,\ \ \ \ 0=J_i=\rho_C u_i+\nu_i
\end{equation}
which we will use to get an expression for the velocity $u_i=-\nu_i/\rho_C$.
We assume the local velocity to be small.
From equation~(\ref{eq:nu}) it is known, that neglecting terms quadratic in 
the velocity:
\begin{equation}
\nu^\lambda=-\varkappa \partial^\lambda \frac{\mu}{T}
\end{equation}
So we derive
\begin{equation}
u_i=\frac{1}{\rho_C}\varkappa \partial_i \frac{\mu}{T} 
\end{equation}
Also from (\ref{eq:TJ}) it is seen that the part~$T_{0i}$ of the 
energy-momentum tensor (usually interpreted as the heat current) 
now amounts to 
\begin{equation}
T_{0i}=p g_{0i}+\omega u_0 u_i + \tau_{0i}=(\epsilon +p)u_0 u_i 
\end{equation}  
if we can assume the stress tensor to have vanishing components
here by its general interpretation measuring spatial shear effects
only.
Plugging in our expression for the velocity yields 
\begin{equation}\label{eq:uT0i}
T_{0i}=(\epsilon +p)u_0\frac{1}{\rho_C}\varkappa \partial_i \frac{\mu}{T}  
\end{equation}  
This completes the first step being the sought after translation
of $\nu_\mu$ into $T_{0i}$ via $u_i$.

The second step uses $\dd\mu=\dd p/\rho_C-s\dd T/\rho_C$ in order to translate
the gradient:
\begin{equation}
\partial_i\mu=\frac{1}{\rho_C}(\partial_i p- s\partial_i T)\, .
\end{equation}    
Putting the two steps together gives a well known relation
\begin{equation}\label{eq:T0i}
T_{0i}=-\underbrace{\left (\frac{\epsilon+p }{T\rho_C}\right)^2 \varkappa
 }_{\equiv\kappa}\left[\partial_i T-\frac{T}{\epsilon + p}\partial_i p \right]
\end{equation}
As described in Landau and Lifshitz~\cite{Landau}, this expression gives 
the relativistic hydrodynamics heat current. Compared to the
non-relativistic one it gets an extra contribution from the 
pressure gradient throughout the system. The transport coefficient
related to heat flow is the heat conductivity $\kappa$.

We now have two interpretations of the new $\varkappa$: \\ 
1. It relates the dissipative current with the temperature and
chemical potential gradient by~(\ref{eq:nu}). This is true
for general currents~$J_\mu$. \\
2. It also relates the heat current with the temperature and 
pressure gradient by~(\ref{eq:T0i}). This interpretation though
only holds if the charge current vanishes, so $J_i=0$. 
 
In the application to R-charged black holes~\cite{Son:2006em} the authors conclude by 
examining the limit of 
vanishing charge current, that the dissipative part of the charge current
contributes to the heat current. Thus they identify~$\varkappa$ to be proportional to the 
heat conductivity~$\kappa = \varkappa (\epsilon+p)^2/(\rho T)^2$. 

\subsection{Quasinormal modes} \label{sec:qnm}
Quasinormal modes of fields on the gravity side of AdS/CFT are intimately related to the retarded two-point correlation 
functions~$G^R$ of the dual operators~$\O$ in the thermal field theory. To be more precise the poles appearing in the
correlator~$G^R$ are exactly located at the frequency values~$\omega_{qn}$ of the quasinormal modes belonging
to the dual gravity field. In order to understand this relation on a technical level, we here review the concept of quasinormal
modes in gravity and explore their relation to thermal correlators through AdS/CFT.

{\bf Quasinormal modes in gravity}
This paragraph follows closely the work of~\cite{Horowitz:1999jd} and details may be obtained from that original work.
Normal modes are the preferred time harmonic states~$e^{-i\omega_n t}$ of compact classical linear oscillating systems such as
finite strings or cavities filled with electromagnetic radiation. The normal frequencies~$\omega_n$ of these systems are 
real~$\omega_n\in \mathbf{R}$ and the general solution can be written as a linear superposition of all possible eigenmodes~$n$. 
Quasinormal modes in classical supergravity are the analog of normal modes but for a non-conservative system. The quasinormal
frequencies assume complex values~$\omega_{qn}\in \mathbf{C}$ where the imaginary part is associated with the dissipation. 
In the case of a black hole background excitations dissipate energy into the black hole and are therefore damped when traveling 
through the bulk. Since we would like to utilize AdS/CFT, we are interested in quasinormal modes in the $d$ dimensional AdS Schwarzschild metric
\begin{equation}
\label{eq:rMetric}
\dd s^2=-h(r)\dd t^2+h(r)^{-1} \dd r^2 +r^2 \dd \Omega^2_{d-2}\, ,
\end{equation}
with 
\begin{equation}
h(r)=\frac{r^2}{R^2}+1-\left ( \frac{r_0}{r} \right )^{d-3}\, .
\end{equation}
This factor for large black holes with~$r_0\gg R$ in~$AdS^5$ becomes~$h(r)=\frac{r^2}{R^2}-\left ( \frac{r_0}{r} \right )^{2}$
\footnote{Which is identical to the form used e.g. by Myers et al. in~\cite{Myers:2007we} up to a scaling with~$R^2$}.
Quasinormal modes are the (quasi) Eigenmodes of fluctuations of fields in presence of a black
hole (or black brane) background, also referred to as the {\it ringing of the black hole}.
As a simple example let us follow~\cite{Horowitz:1999jd} and consider 
the wave equation of a minimally coupled scalar~$\Phi$
\begin{equation}
\nabla^2\Phi = 0\, .
\end{equation}
Assuming spherical symmetry we may use the product Ansatz
\begin{equation}
\Phi(t,r,\theta)=r^\frac{2-d}{2} \psi (r) Y(\theta) e^{-i\omega t}\, ,
\end{equation}
with the spherical harmonics~$Y$ on $S^{d-2}$.
Splitting the radial from the spherical equation of motion we obtain
\begin{equation}
[\partial_{r^*}^2+\omega^2-\tilde V(r^*)]\psi(r) = 0\, ,
\end{equation}
where the tortoise coordinate~$r^*$ is given by
\begin{equation}
r^* = \int \frac{\dd r}{h(r)+1} \, .
\end{equation}
The potential~$\tilde V(r^*)$ vanishes at the horizon~$r^*=-\infty$ and diverges at~$r=\infty$. In
general this equation has solutions for arbitrary~$\omega$. The solutions which are 
called quasinormal modes are defined to be purely incoming at the horizon~$\Phi\sim e^{-i\omega(t+r^*)}$ 
(and purely outgoing at infinity~$\Phi\sim e^{-i\omega(t-r^*)}$~, where the boundary of AdS is located in these coordinates).
This condition can only be satisfied at discrete complex values of~$\omega$ called quasinormal frequencies.
In the AdS black hole case the potential~$\tilde V$ diverges at infinity~$r=\infty$, such that we 
require the solution to vanish at this location. 

In order to have a finite variable range we invert the radial coordinate~$r\to1/x$. 
The radial equation of motion for the minimally coupled scalar then reads
\begin{equation}
\label{eq:qnmEom}
s(x) \frac{\dd^2}{\dd x^2}\psi (x)+ \frac{t(x)}{x-x_+}\frac{\dd}{\dd x}\psi (x)+\frac{u(x)}{(x-x_+)^2}\psi (x) = 0\, .
\end{equation}
In our $AdS_5$-case the coefficients are given by~\cite{Horowitz:1999jd}
\begin{eqnarray}
s(x) = \frac{({x_+}^2+1) x^5}{{x_+}^4}+\frac{({x_+}^2+1) x^4}{{x_+}^3}+\frac{x^3}{{x_+}^2}+\frac{x^2}{{x_+}^2}\, , \\
t(x) = 4 {r_0}^2 x^5-2 x^3-2x^2 i \omega \, , \\
u(x) = (x-x_+) V(x)\, , \\
V(x) = \frac{15}{4}+\frac{3+4 l (l+2)}{4} x^2 +\frac{9 {r_0}^2}{4} x^4\, \\
{r_0}^2 = \frac{{x_+}^2+1}{{x_+}^4} \, ,
\end{eqnarray}
where~$l(l+2)$ is the Eigenvalue of the Laplacian on~$S^3$. Note that we do not rewrite~\eqref{eq:qnmEom} such
that the factor in front of the second derivative becomes one. That is because the coefficients~$s, t, u$ have 
finite expansions in~$(x-x_+)$ and thus are more tractable.

We compute the quasinormal modes numerically by expanding the solution in
a power series about the horizon at~$x=x_+$. In order to find the near-horizon behavior
we determine the indices (as explained in section~\ref{sec:anaAdsG})~$\alpha =0$ and~$\alpha=i\omega/(2\pi T)$. 
Only the first index describes ingoing modes at the horizon and we discard the second one. This 
fixes the leading order~$(x-x_+)^{0}$ for our solution and we expand the remaining analytic part of it in a Taylor series
about the horizon~\cite{Bender}
\begin{equation}
\label{eq:qnmExp}
\psi (x) = (x-x_+)^\alpha \sum\limits_{n=0}^{\infty} a_n (x-x_+)^n \, ,
\end{equation}
Then we demand this series to vanish at infinity~$r=\infty$ equivalent to~$x=0$. The expansion~\eqref{eq:qnmExp} is substituted in
the equation of motion~\eqref{eq:qnmEom} in order to compare coefficients of~$(x-x_+)^n$ in each order~$n$.
From this we find the recursion relations
\begin{equation}
a_n=-\frac{1}{P_n} \sum\limits_{k=0}^{n-1}[k (k-1) s_{n-k}+k t_{n-k}+u_{n-k}] a_k\, ,
\end{equation}
with the expression~$P_n=n (n-1) s_0+n t_0=2 x_+^2 n (n\kappa-i \omega)$.
Only the coefficient~$a_0$ remains undetermined as expected for a linear equation. 
  
Together with the condition that the solution~$\psi$ should be normalizable and therefore has to vanish
at spatial infinity~$\psi(x=0)=0$, we have mapped the problem of finding quasinormal frequencies to the  problem of
finding the zeroes of
\begin{equation}
\label{eq:gravityZeroes}
\sum\limits_{n=0}^{\infty} a_n(\omega) (-x_+)^n = 0 \, ,
\end{equation} 
in the complex~$\omega$ plane. Equation~\eqref{eq:gravityZeroes} can only be satisfied for 
discrete  values of complex~$\omega$. We approach the exact result by truncating the series at~$n$ and finding the 
zeroes of the partial sum~$\psi_N(\omega, x=0)=\sum\limits_{n=0}^N a_n (-x_+)^n$. To be more specific, we really 
find the minima of the absolute value squared~$|\psi_N(\omega, x=0)|^2$
of the partial sum and then check if the value at that minimum is (numerically) zero. 
The accuracy can be increased by going to larger~$n$ and the error is estimated
from the change of~$w (n)$ as $n$ is increased.

{\bf Alternative QNM computations}
In more complicated backgrounds (such as the D3/D7-setup) it is hard or even impossible to write down analytical
expressions as those used in the previous paragraph, especially if some factors like the embedding function in the 
metric components are only given numerically. In this case one has to reside to numerical methods. 

Numerically we can compute~$|\phi|^2$ directly starting with two boundary conditions at the horizon and search 
its minimum. In some cases~(especially if the solution is oscillating heavily on one boundary) the numerical
method of matching in the bulk~\cite[section 7.2]{Erdmenger:2007cm} has proven more adequate to find solutions~$\phi$.
Numerics may also be improved by a coordinate transformation to more tractable~(non-singular) 
coordinates. An application of this latter method is given in~\cite{Evans:2008tv}.

{\bf Quasinormal modes in AdS/CFT}
In the context of AdS/CFT it has been shown~\cite{Son:2002sd,Kovtun:2005ev} 
that the lowest lying~(i.e. those with the smallest absolute value) quasinormal frequency of the perturbation of a
distinct gravity field~$\phi$ coincides with the pole of the two-point function for the operator~$\O$ dual to this distinct field. 
We can see this by approaching the problem with the question: what is the two-point correlator of 
two gauge-invariant operators? As described above, the correlator is given by
\begin{equation}
\label{eq:minkCorrRecipe}
\langle \O \O \rangle = \lim\limits_{r\to r_\text{bdy}} B(r) \phi (r) \partial_r \phi (r) \, ,
\end{equation}
where~$\phi (r)$ is the solution to the gravity equation of motion~(ordinary differential equation ODE) 
for the field~$\phi$ dual to the operator~$\O$. Here we use the same radial coordinate~$r$
defined above in equation~\eqref{eq:rMetric}.
At the boundary the solution can be written as linear combination of two local solutions
\begin{equation}
\label{eq:solAB}
\phi (r) = \A \phi_1 (r) +\B \phi_2 (r)\, ,
\end{equation}
with~$\A$ and~$\B$ being determined by the coefficients in the differential equation for~$\phi$.
The coefficients~$\A$ and~$\B$ give that particular linear combination which satisfies the 
incoming wave boundary condition at the horizon.
Near the boundary the solution~\eqref{eq:solAB} splits into the normalizable and non-normalizable parts
\begin{equation}
\label{eq:solABNearBdy}
\phi (r) = \A r^{-\Delta_-}(1+\dots) +\B r^{-\Delta_+} (1+\dots ) \, ,
\end{equation}
The action quadratic in field fluctuations~$\phi$ reduces to the boundary term
\begin{equation}
S^{(2)}\propto  \lim\limits_{r\to r_\text{bdy}} \int\dd \omega\, \dd^p q\, B(r,\omega,\mathbf{q}) 
 \phi(r) \partial_r\phi (r) +\text{contact terms} \, .
\end{equation} 
Applying~\eqref{eq:minkCorrRecipe} and assuming~$\Delta_+>\Delta_-,\, \Delta_+>0$ 
we obtain the two-point function of operators~$\O$ by an expansion
in the radial coordinate~$r$ and taking the boundary limit afterwards
\begin{equation}
\label{eq:corrAB}
\langle \O \O \rangle \propto \frac{\B}{\A} +\text{contact terms} \, .
\end{equation}
The poles of the retarded correlator thus correspond to the zeroes of the connection
coefficient~$\A$. On the other hand~$\A$ is determined by the coefficients of the equation of motion for
the field fluctuation~$\phi$ and therefore~$\A=0$ is a particular choice of boundary condition for that
field fluctuation~$\phi$. 
As an example consider~$\Delta_- =0,\, \Delta_+ =2$ and~$B(r,\omega,\mathbf{q}) \propto r^3$ and 
$r_\text{bdy}=\infty$. Then
\begin{equation}
\langle \O \O \rangle \propto \lim\limits_{r\to r_{\text{bdy}}} r^3 \frac{-2\B r^{-3} }{\A + \B r^{-2}} +\text{contact terms} \, .
\end{equation}

Now we are ready to connect our holographic considerations back to
the gravity definition of quasinormal modes given above~\eqref{eq:gravityZeroes}. Comparing the two 
approaches we conclude that the condition for having quasinormal modes coming from 
gravity~\eqref{eq:gravityZeroes} and the boundary condition for the field fluctuation in~AdS/CFT~$\A=0$ are identical.
For this reason the quasinormal frequencies of black hole excitations are identical to the 
poles of the retarded two-point correlator of their~AdS/CFT-dual operators.

\subsection{Summary} \label{sec:sumHoloMethods}
In this chapter we have reviewed some thermodynamics and hydrodynamics in the context of thermal quantum
field theories and we have developed holographic tools to calculate thermal field theory quantities at strong coupling.
The formulation of the gauge/gravity correspondence in the Euclidean version has been contrasted to the 
Minkowski version. In particular we found out that the Euclidean prescription is not sufficient to describe 
non-equilibrium processes at finite temperature. Motivated by this fact we went on to develop a recipe to 
retrieve two-point correlation functions in Minkowski space, which is dual to the {\it real time formalism}
frequently used in thermal quantum field theory.
We have especially seen that correlation functions may be obtained by analytical or numerical methods. The analytical
recipe~\ref{sec:anaAdsG} relies on the hydrodynamic approximation of perturbations with only small frequency and 
momentum. In this case we can extract the relevant boundary term of the on-shell action~(first step), solve the equation of 
motion for the field which is dual to the operator which we would like to find correlations of~(second step), and finally we can use the
formula~$G^R (\vec k)=  -2 \mathfrak{B}(u) \mathfrak{F}(u,-\vec k) \partial_u \mathfrak{F}(u,\vec k)\big |_{u\to 0}$ 
given in~\eqref{eq:anaGFormula}~(third step). Beyond this hydrodynamic limit we have seen in~\ref{sec:numAdsG} that we can
employ two different numerical methods to take the second step in the prescription and solve the equation of motion 
for the gravity field numerically. Furthermore we have derived the Kubo formula which relates transport coefficients
to the retarded two-point correlation function. Finally the poles in the thermal field theory two-point correlators of
an operator~$\O$ have been identified with the quasinormal frequencies of the dual gravity field~$\phi$.

\section{Holographic thermo- and hydrodynamics} \label{sec:holoThermoHydro}
In this chapter I present my~(partly unpublished) own work on introducing a chemical isospin and baryon 
potential~(cf.~section~\ref{sec:chemPots}) into the thermal $\N=4$~Super-Yang-Mills theory coupled to fundamental matter as 
described in section~\ref{sec:mesons}. We were the first to consider the non-Abelian part of the flavor gauge group in the 
context of AdS/CFT with a finite charge density~\cite{Erdmenger:2007ap, Erdmenger:2007ja} and the results are 
summarized and considerably enhanced especially in sections~\ref{sec:anaHydroIso} and~\ref{sec:thermoB&I}. 

In the upcoming section~\ref{sec:appKubo} we will start out with an application of 
the Kubo formula for heat conductivity derived in the previous chapter~\ref{sec:transportCoeffs}. 
The rest of this chapter considers the D3/D7-brane setup with a background flavor gauge field introduced on the~D7-brane
as described in section~\ref{sec:mesons}, section~\ref{sec:finiteTAdsCft} and section~\ref{sec:chemPots}. 
In section~\ref{sec:anaHydroIso} we first take an analytical approach to get some exact results for massless quarks,
while in chapter~\ref{sec:thermalSpecFunc} we will use numerical techniques. In order to do so 
we have to employ a small-frequency/small-momentum approximation coined the {\it hydrodynamic expansion}~(
cf.~equation~\eqref{eq:hydroExpansion}, and those following it). These requirements are then relaxed and in 
section~\ref{sec:thermoBaryon} the background and its thermodynamics are generalized to non-zero quark masses
in a setup where also arbitrary frequencies / momenta of the perturbations~(cf.~chapter~\ref{sec:thermalSpecFunc}) 
are treatable. The price for this generalization is that we have to use
numerical techniques in order to find the (massive) D7-brane embeddings as analyzed in~\cite{Mateos:2007vn}. 

In this context we will review the thermodynamics at finite $U(1)$~baryon density~\cite{Kobayashi:2006sb} or finite baryon chemical 
potential~\cite{Mateos:2007vc} in section~\ref{sec:thermoBaryon}. Investigating the effects of isospin and the non-Abelian
part of the flavor group we will develop the thermodynamics for the non-Abelian part~$SU(N_f)$ of 
the full flavor gauge symmetry~$U(N_f)$. We find a significant impact of isospin on the hydrodynamics as well as on thermodynamics.

\subsection{Application of the Kubo formula} \label{sec:appKubo}
The purpose of the calculation ahead is to understand and check the non-equilibrium methods introduced
in the previous chapter~\ref{sec:transportCoeffs}. This general understanding is needed in the coming chapter~\ref{sec:transport}
and all our asides on diffusion or other non-equilibrium phenomena. We will keep the computation as general as possible 
and only in the very end we apply the result to a conformal field theory in order to check it. The present computation
may be seen as a preparation to apply similar calculations to more QCD-like theories in order to find their
transport coefficients.

In~\cite{Hosoya:1983id} a relatively general treatment of the
problem ahead is given. The problem is simply how to (carefully) 
carry out the integrals inside a Kubo formula. Hosoya does this
for the Kubo formula giving the shear viscosity; whereas we are
actually interested in the heat conductivity. But having this
sample calculation at hand let's follow it and we will see that
the steps for our Kubo formula will walk exactly the same path (up
to some constant factors).  

The viscosity Kubo formula is~\cite{Hosoya:1983id}
\begin{equation}
\label{eq:kuboSVis}
\eta=-\frac{1}{5}\mathrm{lim}_{\epsilon\to 0}\int\limits^0_{-\infty}dt_1 
      \mathrm{e}^{\epsilon t_1}
      \int\limits^{t_1}_{-\infty}dt' \int\limits^{\infty}_{-\infty}
      \frac{dk^0}{2\pi} \mathrm{e}^{i k^0 t'} \tilde{\Pi}(k^0)   \, ,
\end{equation} 
where $\tilde{\Pi}(k^0)$ is a 2-point correlation function only 
depending on $k^0$ out of the integration variables. For the
shear viscosity this correlator is the energy-momentum tensor
2-point function $\langle T_{ij} T_{ij}\rangle$. 
The $\epsilon$ appearing here comes from the non-equilibrium
thermodynamics formalism and it parametrizes the (small) deviation
from thermal equilibrium. Since we will see that it formally has
exactly the same effect as an ordinary QFT regulator, I will call
it the \textit{thermal regulator}. Speaking about the ordinary
QFT regulators, as is common habit, in (\ref{eq:kuboSVis})
the field theory regulator is not explicitly written. We put
it back in by $k^0\to (1-i\epsilon_0) k^0$ in order to keep 
track of all the poles appearing. 
\begin{equation}
\label{eq:kuboSVisEps}
\eta=-\frac{1}{5}\mathrm{lim}_{\epsilon,\epsilon_0\to 0}
     \int\limits^0_{-\infty}dt_1 \mathrm{e}^{\epsilon t_1}
     \int\limits^{\infty}_{-\infty}\frac{dk^0}{2\pi} 
     \tilde{\Pi}(k^0(1-i\epsilon_0)) \underbrace{\int\limits^{t_1}_{-\infty}dt'
     \mathrm{e}^{i k^0(1-i\epsilon_0) t'}}_{
       \left [\frac{1}{ik^0(1-i\epsilon_0)}\mathrm{e}^{ik^0(1-i\epsilon_0)t'}
       \right ]^{t_1}_{-\infty}} \, .
\end{equation}   
Of the $t'$-integral only the upper limit~($t_1$) remains because for the
lower bound~($-\infty$) we get
\begin{equation}
\mathrm{lim}_{t'\to -\infty}\mathrm{e}^{ip^0(1-i\epsilon_0)t'}=
 \mathrm{lim}_{t'\to -\infty}\underbrace{\mathrm{e}^{ik^0t'}}_{oscillating}
 \underbrace{\mathrm{e}^{i(-i)k^0\epsilon_0)t'}}_{\to 0} \, .
\end{equation}      
So from this integral we are left with
\begin{equation}
\frac{1}{ik^0(1-i\epsilon_0)}\mathrm{e}^{ik^0(1-i\epsilon_0)t_1} . 
\end{equation}
Note that the use of the regulator $\epsilon_0$ together with 
the integral gives us a new pole for the $k^0$-integration at
$k^0=0$. We will see that subsequent integration of this over 
$t_1$ together with
the thermal regulator $\epsilon$ will give us yet a different
pole structure in the complex $k^0$-plane.  
Explicitly carrying out the same procedure as before with this new expression
we are left with
\begin{equation}
\eta=-\frac{1}{5}\mathrm{lim}_{\epsilon,\epsilon_0\to 0}
     \int\limits^{\infty}_{-\infty}\frac{dk^0}{2\pi}
     \frac{\tilde{\Pi}(k^0(1-i\epsilon_0))}{ik^0(1-i\epsilon_0)}   
     \underbrace{\int\limits^0_{-\infty}dt_1 
       \mathrm{e}^{\epsilon t_1+ik^0(1-i\epsilon_0)t_1}}_{
        \frac{1}{ik^0(1-i\epsilon_0)-i\epsilon}
         \left [1-\mathrm{e}^{(\epsilon+\epsilon_0 k^0)(-\infty)}
         \mathrm{e}^{-i k^0)(-\infty)}
       \right ]}  \, .
\end{equation}
This leaves us with the $k^0$-integration and an integrand
having two poles~\footnote{We have to assume that the
function $\tilde{\Pi}(k^0(1-i\epsilon_0))$ introduces no
additional poles.}:
\begin{equation}
\frac{1}{5(2\pi)(1-2i\epsilon_0)}\int\limits^\infty_{-\infty} dk^0
 \frac{\tilde{\Pi}(k^0(1-i\epsilon_0))}
 {\underbrace{k^0}_{\equiv A}
  \underbrace{(k^0-i\frac{\epsilon}{1-i\epsilon_0})}_{\equiv B}}  \, . 
\end{equation}
To integrate a function like this the Cauchy-Riemann formula
\begin{equation}
\int\limits_{\textrm{closed contour}} \frac{f(z)dz}{(z-z_0)^2}=
   \left . (2\pi i)\partial_z f(z)\right |_{z=z_0}   \, ,
\end{equation}
is usually of great help. But to apply it we first need to
turn the integrand with two different poles into one with
two poles at the same position to match the form of the 
integrand  the Cauchy-Riemann formula. This can be
done by introducing Feynman parameters $a,b$ making use
of the formula 
\begin{equation}
\frac{1}{A B}= \int\limits^1_0 da\, db\, \delta(a+b-1) \frac{1}{(aA+bB)^2}  \, ,
\end{equation}
which can be verified by carrying out the integrals on 
the right hand side. Plugging in $A=k^0$ and 
$B=k^0-i\frac{\epsilon}{1-i\epsilon_0}$ we get
\begin{equation}
\frac{1}{A B}= \int\limits^1_0 da db \frac{\delta(a+b-1)}{(a+b)^2}
 \frac{1}{(k^0-i\frac{\epsilon}{1-i\epsilon_0}\frac{b}{a+b})^2} \, ,
\end{equation}
 which displays the sought-after second order pole at 
$k^0=i\frac{\epsilon}{1-i\epsilon_0}\frac{b}{a+b}$. 
Use of the Cauchy formula and integration over the Feynman parameter 
$b$ yields
\begin{equation}
\eta=\frac{i}{5}\mathrm{lim}_{\epsilon_0\to 0} \mathrm{lim}_{\epsilon\to 0}
     \frac{1}{1-2i\epsilon_0} \int\limits^1_0 da \left . \partial_{k^0}
     \tilde{\Pi} \right |_{k^0=i\frac{\epsilon}{1-i\epsilon_0}(1-a)}   \, . 
\end{equation}   
Now first taking the $\epsilon_0$ ordinary field theory limit gives
\begin{equation}
\label{eq:hosoyaRegKubo}
\frac{i}{5} \mathrm{lim}_{\epsilon\to 0}
     \int\limits^1_0 da \left . \partial_{k^0}
     \tilde{\Pi} \right |_{k^0=i\epsilon (1-a)} \, ,
\end{equation}
 and afterwards the thermal regulator limit $\epsilon\to 0$ produces
\begin{equation}
\label{eq:kuboPIofk0}
\eta=\frac{i}{5}\left . \partial_{k^0}\tilde{\Pi} \right |_{k^0=0} \, . 
\end{equation} 
This formula is true for any correlator $\tilde{\Pi}$ which 
introduces no new poles in $k^0$ and which does not depend
on any of the time-variables~($t', t_1$).
The Kubo formula for thermal conductivity will only have a
different numerical factor and it will contain the 
current correlator $\langle J^a_i J^b_i\rangle(\vec k)$ 
instead of the energy-momentum correlator
$\langle T_{ij} T_{ij}\rangle(\vec k)$. 
But both are only functions of $k^0$ as required. And
from the Fourier-transformation of~(\ref{eq:confJJ}) essentially given by
\begin{equation}
\label{eq:confJJFT}
\langle J^a_i J^b_i \rangle = -\delta^{ab} \lim\limits_{\epsilon_d\to 0} C(\epsilon_d) \vec{k}^{2+\epsilon_d} \, ,
\end{equation}
with the dimensional regularization parameter~$\epsilon_d$ and the coefficient~$C(\epsilon_d)$
we can see, that the conformal flavor current correlator contains no poles in $k^0$. 

Simply applying formula (\ref{eq:hosoyaRegKubo}) to the 
conformal flavor current correlator~\eqref{eq:confJJFT}~(see also~\cite{Freedman:1998tz}) we get
the transport coefficient 
\begin{equation}
\eta=\mathrm{lim}_{\epsilon_d\to 0}\frac{i}{M}
\underbrace{
\left . 
\partial_{k^0}
 \{
  \delta^{ab}
  \tilde{C}^{4-\epsilon_d} {\vec k}^{2+\epsilon_d}
 \}
\right |_{k^0=0}}_{
\tilde{C}^{4-\epsilon_d}(1+\epsilon_d/2)\vec k^{\epsilon_d} 2k^0|_{k^0=0}
} =0  \, ,
\end{equation} 
where $M$ stands for the factors different from the viscosity
case. The vanishing of this transport coefficient can be traced
to the thermal regulator by plugging it in before taking any
of the limits $\epsilon_d, \epsilon_0, \epsilon\to 0$. Carrying
out all integrations and derivatives before taking these three
limits, the coefficient vanishes exactly when taking the 
'thermal' limit $\epsilon\to 0$.  

Our interpretation of this fact is that the conformal symmetry
realized in the correlator does not allow any scale in the
theory. In particular conformal symmetry does not allow introduction
of an energy scale like the temperature. Plugging in the conformal
correlator essentially amounts to setting the temperature $T=0$ in the 
non-equilibrium theory from which the Kubo formula is 
derived. 

\subsection{Analytical Hydrodymamics at finite isospin potential} \label{sec:anaHydroIso}
In this section I present the first available analytical approach towards incorporating a non-Abelian chemical potential into the
context of the AdS/CFT correspondence. The solution of this problem is a central point in this thesis and we published first
results in~\cite{Erdmenger:2007ap}. I have extended these calculations considerably for this thesis. 
In particular we will study AdS/CFT-predictions about the hydrodynamics on the field theory side of the
duality. The calculation presented in this section builds on the achievements in the case without any chemical potential
which is presented in~\cite{Policastro:2002se}. Nevertheless, this study is the first one to take the non-Abelian effects 
into account. All earlier approaches have been restricted to the~$U(1)$ baryonic part of the full flavor group~$U(N_f)$. 
Also in order to incorporate the non-Abelian structure we need  to develop some new methods and ideas. 
These mainly unpublished results are interpreted and compared with the published results~\cite{Erdmenger:2007ap}
involving an additional approximation.

We need to write down the Dirac-Born-Infeld
action in this background and derive the non-Abelian equations of motion which will be differential equations 
coupled through the space-time indices~$\nu$ in field components~$A^a_\nu$ and also through the flavor indices~$a$.
We need to find the {\it flavor transformation}~\eqref{eq:flavorTrafo} from flavor gauge fields~$A^a$ to combinations of those, which
decouple the equations of motion in the flavor indices~$a$. Then we have to find the on-shell action to apply the correlator
prescription studied in~\ref{sec:anaAdsG}. Furthermore, we need to develop a modified understanding of how the
incoming wave boundary condition fixes the singular behavior of the gauge field fluctuations at the horizon. This idea
amounts to a distinction of cases for the {\it indices}~\eqref{eq:indexX1}. In the next four subsections I present my 
calculations in some detail. A comprehensive discussion and interpretation is given in subsection~\ref{sec:discAnaHydro}.

\subsubsection{Calculation of transversal fluctuations} \label{sec:transFlucs}
We will work in the D3/D7-setup described in section~\ref{sec:finiteTAdsCft} at vanishing quark mass, i.e. 
with flat D7-brane embeddings. The coordinates we use are those introduced in equation~\eqref{eq:adsBHmetric}.
In order to find the effective action which suffices to describe small gauge field fluctuations we start from the
Dirac-Born-Infeld action~\eqref{eq:dbiAction} for a D7-brane, constant dilaton field~$e^\Phi=g_s$ and vanishing
transversal scalars~$\phi_i\equiv 0$ so that we get
\begin{equation}
\label{eq:dbiSD7}
S_{\text{D7}} = -T_{\text{D7}} \int \dd^8 \xi\, \mathrm{Str}\sqrt{\det\{g +(2\pi\alpha ')\hat F\}} \, 
\end{equation}
where~$g$ is the pull-back of the originally ten-dimensional metric to the eight-dimensional brane 
and~$\hat F$ is the non-Abelian field strength on the brane for a field~$\hat A$. Making use of the determinant expansion formula
for small values of~$|M|$
\begin{equation}
\label{eq:detExpansion}
\sqrt{\det (1 + M)} = e^{\frac{1}{2} \tr (M -\frac{1}{2} M^2 +\frac{1}{3} M^3 +\dots)} = 1+\frac{1}{2} \tr M -\frac{1}{4} \tr M^2 
  +\frac{1}{8}(\tr M)^2+ \dots \, ,
\end{equation}
we expand the action in gauge field fluctuations~$A$ up to quadratic order in~$A$, which are contained in~$\hat F$.
The non-Abelian field strength tensor~$F$ consists of flavor components~$F^a$ and representation matrices~$T^a$ 
as follows
\begin{equation}
\hat{F}_{\mu\nu}=\hat{F}^a_{\mu\nu} T^a = 2 \partial_{[\mu}\hat{A}^a_{\nu]} T^a + f^{abc} \hat{A}^b_\mu \hat{A}^c_\nu\, T^a ,
\end{equation}
and the field~$\hat A$ is comprised of a background gauge field and fluctuations~$A$ in the context of the
background field method of quantum field theory
\begin{equation}
\hat{A}^a_\nu = \delta_{\nu 0} \delta^{a 3} \mu + A^a_\nu \, ,
\end{equation}
where~$\mu$ is the constant time-component which is interpreted as the chemical potential at the AdS-boundary.
Using~\eqref{eq:detExpansion}, and noting that~$M= g^{-1}\hat F$ so that~$\tr (g^{-1}\hat F) =0$ by tracing the symmetric~$g$
together with the antisymmetric~$\hat F$, we obtain
\begin{eqnarray}
S_{\text{D7}} &=& -T_{\text{D7}} \int \dd^8 \xi\, \mathrm{Str}\{ \sqrt{-g} \sqrt{\det [1 +g^{-1} (2\pi\alpha ')\hat F] }\} \nonumber \\
  &=& -T_{\text{D7}} \int \dd^8 \xi\, \mathrm{Str} \{ \sqrt{-g} [1 + 
   \frac{1}{2} (2\pi\alpha ')\tr ( g^{\Sigma\Sigma'}\hat F_{\Sigma'\Omega})  \nonumber \\
   &&-\frac{1}{4} (2\pi\alpha ')^2 \tr ( g^{\Sigma\Sigma'}\hat F_{\Sigma'\Omega} g^{\Omega\Omega'}\hat F_{\Omega'\Delta}) + \dots ]\} \\
   &=& -T_{\text{D7}} \int \dd^8 \xi\, \mathrm{Str}\{\sqrt{-g} [ 1 
    -\frac{ (2\pi\alpha ')^2}{4} g^{\Sigma\Sigma'} g^{\Omega\Omega'} \hat F_{\Sigma\Omega} \hat F_{\Sigma'\Omega'} + \dots ]\} \\
    &=&-T_{\text{D7}} \int \dd^8 \xi\, \mathrm{Str}\{\sqrt{-g} [ \mathbbm{1}_{N_f\times N_f} 
    -\frac{ (2\pi\alpha ')^2}{4} g^{\Sigma\Sigma'} g^{\Omega\Omega'} \hat F^a_{\Sigma\Omega} \hat F^b_{\Sigma'\Omega'} T^a T^b + \dots]\} 
     \, .
\end{eqnarray}
The symmetrized flavor trace~$\text{Str}\{\dots\}$ applied to the first two terms in the action merely gives a 
factor of~$N_f$ for the trace over unity while in the second term it gives
\begin{eqnarray}
\text{Str}\{ T^a T^b \} &=& \tr_{\text{fund}(N_f)} \{ T^aT^b + T^bT^a \}  \nonumber \\ 
  &=&  \tr_{\text{fund}(N_f)}\underbrace{[T^a,T^b]}_{i f^{abc}T^c}
  + 2  \tr_{\text{fund}(N_f)}(T^b T^a) \nonumber \\ 
  &=& 2 \tr_{\text{fund}(N_f)}(T^b T^a) \, ,
\end{eqnarray}
where we have used that our flavor group generators~$T^c$ are traceless. Furthermore the generators are Hermitian~${T^a}^\dagger = T^a$
and they live in the fundamental representation of the flavor gauge group~$SU(N_f)$. It is in general possible to choose 
linear combinations of a given set~$\{T^a\}$ such that the trace property
\begin{equation}
 \tr_{\text{fund}(N_f)} (T^a T^b) = k_a \delta^{ab}\, \text{(no sum)} \, ,
\end{equation}
is satisfied~\cite[equation (II.7)]{Georgi:1982jb}. The standard conventions~\cite{Bohm:2001yx} fix the 
factor~$k_a = T_R$ for all~$a=1,2,\dots, (N_f^2-1)$, where the {\it Dynkin index}~$T_R$ only depends on the 
representation. For the fundamental representation we  have~$T_R = 1/2$ as we can check explicitly in the example 
with Pauli matrices~$\sigma^a$ for the~$SU(2)$ isospin generators~$T^a_{\text{Iso}} = \sigma^a/2$
\begin{equation}
\tr_{\text{fund}(N_f)} (T^a_{\text{Iso}}T^b_{\text{Iso}} ) =\frac{1}{4} \tr_{\text{fund}(N_f)} (\delta^{ab} \mathbbm{1} + i \epsilon^{abc}\sigma^c) 
  = \frac{2 \delta^{ab}}{4} = \frac{1}{2} \delta^{ab} \, .
\end{equation}
In the hypothetical case that our flavor generators~$T^a$ would live in the adjoint representation the Dynkin index~$T_R$ 
would equal the value of the Casimir operator~$C_A = T_{\text{adj}} = N_f$. 

As mentioned before, we work in the fundamental representation of the flavor group~$SU(N_f)$, 
therefore we find the D7-brane action in quadratic order of gauge field fluctuations~$A$
\begin{eqnarray}
\label{eq:constIsoSD7}
S^{(2)}_{\text{D7}} = T_{\text{D7}}\frac{(2\pi\alpha')^2}{4} T_R (2\pi^2 R^3) \, \int\, \dd u \dd^4 x\, 
  \sqrt{-g} g^{\mu\mu'} g^{\nu\nu'} \hat{F}^a_{\mu\nu}\hat{F}^a_{\mu'\nu'}\, ,
\end{eqnarray}
where we have already integrated over the three angular directions~$5,6$ and~$7$~(on which none of the fields
depends) giving the factor~$(2\pi^2 R^3)$. With the help of equation~\eqref{eq:TDp} the factor in front of the 
action integral in~\eqref{eq:constIsoSD7} can be written as
\begin{equation}
T_{\text{D7}}\frac{(2\pi\alpha')^2}{4} T_R (2\pi^2 R^3) = 2^{-7}\pi^{-3}{g_s}^{-1} (\alpha')^{-2} R^3 \, .
\end{equation}
Note, that equation~\eqref{eq:constIsoSD7} still contains cubic and quartic terms in the fluctuations but we have 
deliberately chosen this covariant form since it is more compact. We will neglect cubic and quartic contributions in a later step.
The fluctuations~$A^a_\mu (t,x=0,y=0,z,u)$ without loss of generality are assumed to depend on time~$t$, the third direction
$x_3=z$ and the radial AdS coordinate~$u$ only while we choose a gauge such that the field has components in the 
Minkowski directions only, i.e.~$\nu=0,\,1,\,2,\,3$.
\begin{eqnarray}
\hat{F}^a_{\mu\nu} \hat{F}^{a\, \mu\nu}
&=& 4 \partial_{[\mu}\hat{A}^a_{\nu]} \partial^{[\mu}\hat{A}^{\nu]\, a} 
+4 f^{abc}\partial_{[\mu}\hat{A}^a_{\nu]} \hat{A}^{\mu\, b}\hat{A}^{\nu\,c}
+f^{abc} f^{ab'c'}\hat{A}_\mu^{b}\hat{A}_{\nu}^b\hat{A}^{\mu\, b'}\hat{A}^{\nu\,c'}\nonumber\\
&=& 4 \partial_{[\mu}(\delta^{a3}\delta_{\nu]0}\mu+A^a_{\nu]}) \partial^{[\mu}(\delta^{a3}\delta^{\nu]0}\mu+A^{\nu]\, a}) \nonumber \\
&&+4 f^{abc}\partial_{[\mu}(\delta^{a3}\delta_{\nu]0}\mu+A^a_{\nu]}) (\delta^{b 3}\delta^{\mu]0}\mu+A^{\mu]\, b})
(\delta^{c3}\delta^{\nu]0}\mu+A^{\nu]\, c}) \nonumber\\  
&&+f^{abc} f^{ab'c'}(\delta^{b 3}\delta_{\mu]0}\mu+A^b_{\mu]})(\delta^{c 3}\delta_{\nu]0}\mu+A^c_{\nu]})
(\delta^{b' 3}\delta^{\mu]0}\mu+A^{\mu]\, b'})(\delta^{c' 3}\delta^{\nu]0}\mu+A^{\nu]\, c'}) \, .
\label{eq:ggFF1}
\end{eqnarray}
This expression simplifies considerably by noting that derivatives acting on the constant~$\mu$ vanish. Furthermore
the terms including more than two background fields~$\mu$ vanish because of the antisymmetrization. For example
\begin{equation}
f^{abc} \partial_{[\mu} A^a_{\nu]} \delta^{\mu 0}\delta^{b 3}\mu \delta^{\nu 0}\delta^{c 3}\mu =0\, .
\end{equation}
The mathematical reason for this to vanish is that more than one background gauge field term is contracted with
one single structure constant. Since every term including the background gauge field~$\mu$ by our choice
always has to contain the factor~$\delta^{3 a}$, it is clear that more than one such factor forces two of the flavor
indices in~$f^{abc}$ to be equal to 3:~$f^{33c}=0$. Since there are at most two different structure constants in
one single term such as~$f f \mu \mu A A$~(schematically), we can have at most two background gauge fields in
one term. One of the two~$\mu$ has to be contracted with the first structure constant~$f$ the other has to be
contracted with the other one. Since we are interested in two-point functions we are also free to neglect all terms
that are cubic or higher order in the field fluctuations~$\O(AAA,AAAA,\dots)$. After these considerations the
action factor~\eqref{eq:ggFF1} becomes
\begin{eqnarray}
\hat{F}^a_{\mu\nu} \hat{F}^{a\, \mu\nu}
&=& 4 \partial_{[\mu}A^a_{\nu]} \partial^{[\mu}A^{\nu]\, a} 
     +4 f^{a3c} g^{00}\partial_{[0}A^a_{\nu]} A^{\nu\, c} \mu + 4 f^{ab3} g^{00}\partial_{[\mu}A^a_{0]} A^{\mu\, b} \mu \nonumber \\
 &&+f^{a3c} f^{a3c'} \mu^2 A_\nu^{c} A^{\nu\, c'}  +f^{a3c} f^{ab'3}\mu^2 g^{00} A_0^{c} A_0^{b'}  \nonumber \\
 &&+f^{ab3} f^{a3c'} \mu^2 g^{00} A_0^{b} A_0^{c'} + f^{ab3} f^{ab'3}\mu^2 A_\mu^{b} A^{\mu\, b'}\\
&=& 4 \partial_{[\mu}A^a_{\nu]} \partial^{[\mu}A^{\nu]\, a} +8 f^{ab3} \mu g^{00}\partial_{[\nu}A^a_{0]} A^{\nu\, b}\nonumber\\ 
\label{eq:ggFF2}
 &&+2 \mu^2  (g^{\mu\mu'}A^1_\mu A^{ 1}_{\mu'} + g^{\mu\mu'}A^2_\mu A^{2}_{\mu'} - g^{00}A^1_0 A^1_0 - g^{00}A^2_0 A^{2}_0)  \, .
\end{eqnarray}
Using this simplified factor~\eqref{eq:ggFF2} in the quadratic action~\eqref{eq:constIsoSD7} we derive the equations of motion for the
gauge field components~$A_\mu^a$ using the Euler-Lagrange equation
\begin{eqnarray}
0&=& \partial_\kappa\left [ \frac{\delta S^{(2)}_{\text{D7}}}{\delta(\partial_\kappa A^d_\sigma)} \right] 
 -\frac{\delta S^{(2)}_{\text{D7}}}{\delta A^d_\sigma} \\
  &=& \partial_\kappa\left [ \frac{\delta }
   {\delta(\partial_\kappa A^d_\sigma)} (\sqrt{-g} g^{\mu\mu'} g^{\nu\nu'} \hat{F}^a_{\mu\nu}\hat{F}^a_{\mu'\nu'}) \right] 
 -\frac{\delta }{\delta A^d_\sigma}(\sqrt{-g} g^{\mu\mu'} g^{\nu\nu'} \hat{F}^a_{\mu\nu}\hat{F}^a_{\mu'\nu'}) \\
\end{eqnarray}
After a few simplifications by interchanging indices the equations of motion can be written as
\begin{eqnarray}
0&=& 2 \partial_\kappa \left [\sqrt{-g} g^{\kappa\kappa'} g^{\sigma\sigma'} \left( \partial_{[\kappa'} A^d_{\sigma']} \right )\right ]\nonumber \\
  &&+ \mu f^{db3} \left [ \delta_{\sigma 0} \partial_\kappa (\sqrt{-g} g^{00} g^{\kappa\kappa'} A^b_{\kappa'})
    +\sqrt{-g} g^{00} g^{\sigma\mu}\partial_\mu A^b_0 - 2 \sqrt{-g} g^{00} g^{\sigma\mu}\partial_0 A^b_\mu \right ] \nonumber \\
  &&-\mu^2 \sqrt{-g}g^{00}g^{\sigma\sigma'}\left [ \delta^{d1}(A^1_{\sigma'} - A^1_0 \delta_{0\sigma'}) 
    + \delta^{d2} (A^2_{\sigma'} - A^2_0\delta_{0\sigma'}) \right ] \, .
\label{eq:anaEom}
\end{eqnarray}
There is one free space-time index~$\sigma$ which can take values in the four Minkowski 
directions~$(x_0=t,x_1=x,x_2=y,x_3=z)$ and in the radial
AdS-direction~$x_4=u$ as well. Therefore we can split equation~\eqref{eq:anaEom} into five distinct differential equations which
are coupled with each other. There is also one free flavor index~$d$ which we will consider in detail shortly. 
Let us start choosing the free index~$\sigma=1$ 
\begin{eqnarray}
0&=& 2 \partial_\kappa \left [\sqrt{-g} g^{\kappa\kappa'} g^{1\sigma'} \left( \partial_{[\kappa'} A^d_{\sigma']} \right )\right ]\nonumber \\
  &&+ \mu f^{db3} \left [ \delta_{1 0} \partial_\kappa (\sqrt{-g} g^{00} g^{\kappa\kappa'} A^b_{\kappa'})
    +\sqrt{-g} g^{00} g^{1\mu}\partial_\mu A^b_0 - 2 \sqrt{-g} g^{00} g^{1\mu}\partial_0 A^b_\mu \right ] \nonumber \\
  &&-\mu^2 \sqrt{-g}g^{00}g^{1\sigma'}\left [ \delta^{d1}(A^1_{\sigma'} - A^1_0 \delta_{0\sigma'}) 
    + \delta^{d2} (A^2_{\sigma'} - A^2_0\delta_{0\sigma'}) \right ] \, .
\label{eq:anaEomSigma1}
\end{eqnarray}
This equation only involves the gauge field components in the~$x_1$-direction and writing down the other four equations
we will see later, that this equation decouples from all of them and is therefore the simplest one to solve.
We note here that the inverse metric is diagonal such  that~$g^{\mu\mu'}=g^{\mu'\mu}$ and it vanishes for~$\mu\not=\mu'$, 
so we get
\begin{eqnarray}
0&=& 2 \partial_\kappa \left [\sqrt{-g} g^{\kappa\kappa'} g^{11} \left( \partial_{[\kappa'} A^d_{1]} \right )\right ]\nonumber \\
  &&+ \mu f^{db3} \left [\sqrt{-g} g^{00} g^{11}\partial_1 A^b_0 - 2 \sqrt{-g} g^{00} g^{11}\partial_0 A^b_1 \right ] \nonumber \\
  &&-\mu^2 \sqrt{-g}g^{00}g^{11}\left [ \delta^{d1}A^1_{1} + \delta^{d2} A^2_{1} \right ] \, .
\end{eqnarray}
Now recall that we have chosen the geometry such that~$A^a_\mu(x_0,x_1=0,x_2=0,x_3,x_4)$, which implies that the derivatives
of fluctuations in all other than~$x_0,\,x_3,\,x_4$-directions vanish
\begin{equation}
\partial_{1,2} A^a_\mu \equiv 0\, , \quad \partial_{5,6,7} A^a_\mu \equiv 0\, .
\end{equation}
Considering this gives
\begin{eqnarray}
0&=& \partial_\kappa \left [\sqrt{-g} g^{\kappa\kappa'} g^{11} \left( \partial_{\kappa'} A^d_{1} \right )\right ]
  -2 \mu f^{db3} \sqrt{-g} g^{00} g^{11}\partial_0 A^b_1 \nonumber \\
  &&-\mu^2 \sqrt{-g}g^{00}g^{11}\left [ \delta^{d1}A^1_{1} + \delta^{d2} A^2_{1} \right ] \, .
\end{eqnarray}
Now we transform to Fourier space with conventions given in equation~\eqref{eq:fourierTrafo}
\begin{eqnarray}
0&=& -i\omega \left [\sqrt{-g} g^{00} g^{11} \left( -i\omega A^d_{1} \right )\right ]
    +i q \left [\sqrt{-g} g^{33} g^{11} \left( i q A^d_{1} \right )\right ]
    +\partial_u \left [\sqrt{-g} g^{44} g^{11} \left( \partial_u A^d_{1} \right )\right ]\nonumber \\
  &&-2 \mu f^{db3} \sqrt{-g} g^{00} g^{11}\partial_0 A^b_1 
    -\mu^2 \sqrt{-g}g^{00}g^{11}\left [ \delta^{d1}A^1_{1} + \delta^{d2} A^2_{1} \right ] \, .
\end{eqnarray}
We abbreviate derivatives in radial AdS-direction~$\partial_u A = A'$ and sort the equation in derivatives of the gauge
field fluctuations~$A,\, A',\, A''$ and normalize it such that the second derivative has the coefficient one
\begin{equation}
\label{eq:anaEomSigma1d}
0 = {A^d_1}'' + \frac{\partial_u\left ( \sqrt{-g} g^{11} g^{44} \right )}{\sqrt{-g} g^{11} g^{44}} {A^d_1}' 
  - \frac{g^{00}\left[ \omega^2 A_1^d -2 i f^{db3}\omega\mu A_1^b
  +\mu^2 (\delta^{d1}A_1^1+\delta^{d2}A_1^2)\right ] + g^{33} q^2 A_1^d}{g^{44}} \, .
\end{equation}
Turning to the free flavor index~$d$ we recall that it can take the values~$1,2,3$ corresponding to the three
flavor directions we introduced by assuming an~$SU(N_f=2)$-isospin flavor symmetry. We split~\eqref{eq:anaEomSigma1d}
into three equations
\begin{eqnarray*}
0 &=& {A^1_1}'' + \frac{\partial_u\left ( \sqrt{-g} g^{11} g^{44} \right )}{\sqrt{-g} g^{11} g^{44}} {A^1_1}' 
  - \frac{g^{00}\left[ \omega^2 A_1^1 -2 i f^{1b3}\omega\mu A_1^b
  +\mu^2 (\delta^{11}A_1^1+\delta^{12}A_1^2)\right ] + g^{33} q^2 A_1^1}{g^{44}}\, , \\
0 &=& {A^2_1}'' + \frac{\partial_u\left ( \sqrt{-g} g^{11} g^{44} \right )}{\sqrt{-g} g^{11} g^{44}} {A^2_1}' 
  - \frac{g^{00}\left[ \omega^2 A_2^2 -2 i f^{2b3}\omega\mu A_1^b
  +\mu^2 (\delta^{21}A_1^1+\delta^{22}A_1^2)\right ] + g^{33} q^2 A_1^2}{g^{44}}\, , \\
0 &=& {A^3_1}'' + \frac{\partial_u\left ( \sqrt{-g} g^{11} g^{44} \right )}{\sqrt{-g} g^{11} g^{44}} {A^3_1}' 
  - \frac{g^{00}\omega^2 + g^{33} q^2 }{g^{44}}A_1^3 \, .
\end{eqnarray*}
By using the antisymmetry of the structure constants~$f^{3b3}=0,\,f^{abc}=-f^{bac}$ we arrive at
\begin{eqnarray}
0 &=& {A^1_1}'' + \frac{\partial_u\left ( \sqrt{-g} g^{11} g^{44} \right )}{\sqrt{-g} g^{11} g^{44}} {A^1_1}' 
  - \frac{g^{00}\left[ \omega^2 A_1^1 -2 i \omega\mu A_1^2
  +\mu^2 A_1^1\right ] + g^{33} q^2 A_1^1}{g^{44}}\, , \\
0 &=& {A^2_1}'' + \frac{\partial_u\left ( \sqrt{-g} g^{11} g^{44} \right )}{\sqrt{-g} g^{11} g^{44}} {A^2_1}' 
  - \frac{g^{00}\left[ \omega^2 A_2^2 + 2 i \omega\mu A_1^1
  +\mu^2 A_1^2\right ] + g^{33} q^2 A_1^2}{g^{44}}\, , \\
0 &=& {A^3_1}'' + \frac{\partial_u\left ( \sqrt{-g} g^{11} g^{44} \right )}{\sqrt{-g} g^{11} g^{44}} {A^3_1}' 
  - \frac{g^{00}\omega^2 + g^{33} q^2 }{g^{44}}A_1^3 \, .
\end{eqnarray}

{\bf Decoupling transformation}
These three differential equations for flavor components of the gauge field in~$x_1$-direction are coupled
in the first two flavor directions while the third equation for the component~$A^3_1$ decouples from all
others. We decouple the first two equations as well by a field transformation
\begin{equation}
\label{eq:flavorTrafo}
X_1 = A_1^1 + i A_1^2\, , \quad Y_1 = A_1^1 - i A_1^2 \, .
\end{equation}
After this transformation the equations of motion for the three new fields~$X_1,\,Y_1,\,A^3_1$ are given by
\begin{eqnarray}
0 &=& {X_1}'' + \frac{\partial_u\left ( \sqrt{-g} g^{11} g^{44} \right )}{\sqrt{-g} g^{11} g^{44}} {X_1}' 
  - \frac{g^{00}(\mu-\omega)^2 + g^{33} q^2 }{g^{44}} X_1\, ,\nonumber \\
0 &=& {Y_1}'' + \frac{\partial_u\left ( \sqrt{-g} g^{11} g^{44} \right )}{\sqrt{-g} g^{11} g^{44}} {Y_1}' 
  - \frac{g^{00} (\mu + \omega)^2 + g^{33} q^2}{g^{44}} Y_1\, ,\nonumber \\
0 &=& {A^3_1}'' + \frac{\partial_u\left ( \sqrt{-g} g^{11} g^{44} \right )}{\sqrt{-g} g^{11} g^{44}} {A^3_1}' 
  - \frac{g^{00}\omega^2 + g^{33} q^2 }{g^{44}}A_1^3 \, .
\label{eq:anaEomXY}
\end{eqnarray}
We are working in the background given by the metric~\eqref{eq:adsBHmetric} with the inverse components and determinant
\begin{eqnarray}
\label{eq:metricCoeffs}
g^{00} &=& -\frac{u}{b^2 R^2 f(u)} \, ,\quad g^{11}=g^{22} = g^{33} = \frac{u}{b^2 R^2}\, ,\quad
 g^{44}=g^{uu} = \frac{4 u^2 f(u)}{R^2} \, , \nonumber\\
\sqrt{-g} &=& \frac{b^4 R^5}{2 u^3}\, , \quad b = \pi T \, ,
\end{eqnarray}
so that the coefficients can be evaluated to
\begin{eqnarray}
\label{eq:anaEomCoefficients}
\frac{\partial_u\left ( \sqrt{-g} g^{11} g^{44} \right )}{\sqrt{-g} g^{11} g^{44}} = \frac{f'(u)}{f(u)}\, , \quad
 - \frac{g^{00}(\mu \mp \omega)^2 + g^{33} q^2 }{g^{44}} = \frac{(\mn \mp \wn)^2-\qn^2 f(u)}{u f(u)^2} \, ,
\end{eqnarray}
where we used the dimensionless frequency, momentum and chemical potential
\begin{equation}
\wn=\omega/(2\pi T)\, , \quad \qn=q/(2\pi T)\, , \quad \mn=\mu/(2\pi T) \, ,
\end{equation}
respectively, which have already been introduced at the beginning of section~\ref{sec:anaAdsG}.
These coefficients~\eqref{eq:anaEomCoefficients} are singular at the horizon~$u=1$ and at the boundary~$u=0$
just like in the example given in section~\ref{sec:anaAdsG}. Therefore we apply exactly the same steps in
order to gain the indices at the horizon
\begin{equation}
\label{eq:anaGHorIndices}
\beta = \mp \frac{i}{2} (\wn\mp\mn) \, ,
\end{equation}
where the upper sign inside the bracket belongs to the index for the field~$X_1$ and the lower one gives the
index for~$Y_1$. The indices at the boundary for both fields are given by
\begin{equation}
\label{eq:anaGBdyIndices}
\alpha_1 = 0\, ,\quad \alpha_2 = 1 \, .
\end{equation}
Now the question which index produces the solution that satisfies the {\it incoming wave condition}~(which 
tells us to choose only those solutions which propagate into the black hole horizon, see section~\ref{sec:anaAdsG}
for a detailed discussion) is a bit more subtle than in the previous example. Let us assume for definiteness that
both~$\mn,\, \wn \ge 0$. So in the rest of this thesis we assume that the chemical potential~$\mu$ or~$\mn$ is real and 
writing~$\wn$ in order relations we mean only the real part of~$\wn$. In this case there is only one index 
choice for the field~$Y_1$ since~$\wn+\mn\ge 0$ and we
know that the negative index~$\beta = - i/2 (\wn+\mn)$ corresponds to the incoming wave. In contrast to this we
have to distinguish four cases for the index of~$X_1$
\begin{equation}
\label{eq:indexX1}
\beta= \left \{
\begin{array}{c}
-\frac{i}{2}(\wn-\mn)  \,\text{for}\, \wn\ge\mn : \text{incoming}\\
-\frac{i}{2}(\wn-\mn)  \,\text{for}\, \wn<\mn : \text{outgoing}\\
+\frac{i}{2}(\wn-\mn)  \,\text{for}\, \wn\ge\mn : \text{outgoing}\\
+\frac{i}{2}(\wn-\mn)  \,\text{for}\, \wn<\mn : \text{incoming}\\
\end{array} 
\right . \, ,
\end{equation}
so fixing~$\mn$ we choose the incoming solution by choosing the first index if the frequency~$\wn$ is greater or equal to
the chemical potential~$\mn$, and we choose the last index if~$\wn$ is smaller.
Let us carry on considering~$X_1$ first. We also need to modify the hydrodynamic expansion Ansatz~\eqref{eq:hydroExpansion}.
Recall that our approach is to split the singular from the regular behavior in the solution according to
\begin{equation}
\label{eq:splitSingRegX1}
{X_1} = (1-u)^\beta F (u) \, ,
\end{equation}
where~$F$ is a regular function of~$u$.
Our first choice is that the chemical potential is of the same order as the frequency~$\wn\sim\mn$ and therefore
the small quantities to expand the solution in are~$(\wn-\mn)$ and~$\qn^2$. In other words we expand in the spatial
momentum~$\qn^2$ around zero while we expand in the frequency~$\wn$ around the fixed value of the chemical potential~$\mn$.
\begin{eqnarray}
\label{eq:hydroExpX}
X_1(u) &=& (1-u)^{\beta} \left ( F_0+(\wn-\mn) F_1 + \qn^2 G_1+\dots \right ) \, , \\
{X_1}'(u) &=& -\beta (1-u)^{\beta-1} \left ( F_0+(\wn-\mn) F_1 + \qn^2 G_1+\dots \right )\nonumber \\
&& +(1-u)^{\beta} \left ( {F_0}'+(\wn-\mn) {F_1}' + \qn^2 {G_1}'+\dots \right ) \, , \\
{X_1}''(u) &=& \beta (\beta -1)(1-u)^{\beta-2} \left ( F_0+(\wn-\mn) F_1 + \qn^2 G_1+\dots \right )\nonumber \\
&&-2\beta (1-u)^{\beta-1} \left ( {F_0}'+(\wn-\mn) {F_1}' + \qn^2 {G_1}'+\dots \right ) \nonumber \\
&&+ (1-u)^\beta \left ( {F_0}''+(\wn-\mn) {F_1}'' + \qn^2 {G_1}''+\dots \right ) \, .
\end{eqnarray}
For definiteness let us consider only the case~$\beta = -i(\wn-\mn)/2$ where~$\wn\ge\mn$.
Plugging this expansion into the equation of motion~\eqref{eq:anaEomXY} and seperating orders~$\O(1),\,\O(\qn^2)$
and~$\O(\wn-\mn)$ from each other gives
\begin{eqnarray}
\O(1): \, 0&=& {F_0}'' - \frac{2u}{1-u^2} {F_0}' \, , \nonumber \\
\O(\wn-\mn): \, 0&=& \frac{i}{2 (1-u)^{2}} F_0 + \frac{i}{1-u} {F_0}' + {F_1}'' -\frac{i u}{(1-u^2)(1-u)} F_0
  -\frac{2u}{(1-u)^2}{F_1}' \, , \nonumber \\
\O(\qn^2): \, 0&=& {G_1}'' -\frac{2 u}{1-u^2} {G_1}' - \frac{1}{u (1-u^2)} F_0 \, .
\end{eqnarray}

{\bf Alternative hydrodynamic expansion}
By choosing the hydrodynamic Ansatz~\eqref{eq:hydroExpX} we assumed from the beginning that the
frequency and chemical potential have to be treated at equal footing. We can check this assumption by taking a 
slightly more general Ansatz
\begin{equation}
X_1(u) =(1-u)^{\beta} \left ( F_0+\wn F_1 + \mn H_1 + \qn^2 G_1+\dots \right ) \, .
\end{equation}
The key point here is that we still assume~$\wn,\, \mn,\, \qn^2$ to be of the same order but we allow an individual
expansion coefficient~$H_1$ for the chemical potential. Using this more general expansion in the equation of
motion~\eqref{eq:anaEomXY} and seperating orders~$\O(1),\,\O(\qn^2),\,\O(\wn)$ and~$\O(\mn)$ from each other 
gives
\begin{eqnarray}
\O(1): \, 0&=& {F_0}'' - \frac{2u}{1-u^2} {F_0}' \, , \nonumber \\
\O(\wn): \, 0&=& \frac{i}{2 (1-u)^{2}} F_0 + \frac{i}{1-u} {F_0}' + {F_1}'' -\frac{i u}{(1-u^2)(1-u)} F_0
  -\frac{2u}{(1-u)^2}{F_1}' \, , \nonumber \\
\O(\mn): \, 0&=& -\frac{i}{2 (1-u)^{2}} F_0 - \frac{i}{1-u} {F_0}' - {H_1}'' +\frac{i u}{(1-u^2)(1-u)} F_0
  +\frac{2u}{(1-u)^2}{H_1}' \, , \nonumber \\
\O(\qn^2): \, 0&=& {G_1}'' -\frac{2 u}{1-u^2} {G_1}' - \frac{1}{u (1-u^2)} F_0 \, .
\end{eqnarray}
Here we see that the coefficients~$H_1$ and~$F_1$ have to satisfy the same equation of motion. This is 
already clear from the start if we look at the differential equation~\eqref{eq:anaEomXY} and the Ansatz
so that we note that~$\mn$ and~$\wn$ always appear as a sum~$(\wn-\mn)$, at least at linear order
in~$\wn,\,\mn$ which we solely consider here. So there is no single~$\mn$
or~$\wn$, so both have identical factors in the equation of motion and thus their expansion coefficients
have to be identical~(provided both satisfy the same boundary conditions)
\begin{equation}
H_1 = F_1 \, .
\end{equation}
We have now learned explicitly that our first Ansatz~\eqref{eq:hydroExpX} is fully justified.

{\bf Solving the hydrodynamic differential equations}
Our efforts have recast our problem into a set of differential equations~\eqref{eq:anaEomXY} which are only
coupled through the leading order function~$F_0$. Choosing~$F_0$ to be constant~(with respect to the radial
coordinate)~$F_0=C$ is compatible with
all the equations of motion and decouples the system
\begin{eqnarray}
\O(1): \, F_0 &=& C \, , \nonumber \\
\O(\wn-\mn): \, 0&=& \frac{i C}{2 (1-u)^{2}} + {F_1}'' -\frac{i C u}{(1-u^2)(1-u)} 
  -\frac{2u}{(1-u)^2}{F_1}' \, , \nonumber \\
\O(\qn^2): \, 0&=& {G_1}'' -\frac{2 u}{1-u^2} {G_1}' - \frac{C}{u (1-u^2)} \, .
\end{eqnarray}
These are effectively first order differential equations with an inhomogeneity and we can solve them with
\begin{eqnarray}
F_0 &=& C \, , \nonumber \\
F_1 &=& \frac{iC}{2} \ln \frac{1+u}{2}\, , \nonumber \\
G_1 &=& \frac{C}{24}\left [ \pi^2 + 12\ln u \ln (1+u) + 12 \mathrm{Li}_2(1-u) + 12 \mathrm{Li}_2(-u) \right ] \, .
\label{eq:hydroSolsX}
\end{eqnarray}
The function~$\mathrm{Li_2(u)}$ is the double logarithm and the polylogarithm in general is defined as
\begin{equation}
\mathrm{Li}_n (u) = \sum\limits_{n=1}^{n=\infty}\frac{u^k}{k^n} \, .
\end{equation}
Note, that we would not get these solutions~\eqref{eq:hydroSolsX} simply using Mathematica since the boundary conditions we have 
to satisfy here are a bit tricky. Just as described in section~\ref{sec:anaAdsG} the general solutions for~$F_1$ and~$G_1$
each come with two integration constants which have to be fixed by requiring that~$\lim_{u\to 1}F_1 = 0$ 
and~$\lim_{u\to 1}F_1 = 0$. In this horzion limit two terms in each solution become divergent and one 
has to impose the condition that these cancel each other in order to get a regular solution. See also 
equation~\eqref{eq:solF1withCs} and the discussion below it. The constant~$C$ can now be determined in terms
of the boundary fields, momentum and frequency as described in section~\ref{sec:anaAdsG} and we get
\begin{equation}
\label{eq:C12}
C=\frac{8 X_{1}^{\text{bdy}}}{8-4\wn \ln 2 + \pi^2 \qn^2} \, .
\end{equation}

Now using the solutions~\eqref{eq:hydroSolsX} and the expression for~$C$ from~\eqref{eq:C12} in the hydrodynamic 
Ansatz~\eqref{eq:hydroExpX} we get the solution to the transversal field
\begin{eqnarray}
\label{eq:solX12}
X_{1,2} &=& \frac{8 X^{\text{bdy}}_{1,2} (1-u)^{-i\frac{\wn-\mn}{2}}}{8+\pi^2 \qn^2 - 4 i \ln 2 (\wn-\mn)} 
 \left [ 1+(\wn-\mn)\frac{i}{2} \ln\frac{1+u}{2} \right. \\  \nonumber
 &&\left . + \frac{\qn^2}{24} \left(\pi^2+ 12\ln u \ln (1+u) + 12 \mathrm{Li}_2(1-u) + 12 \mathrm{Li}_2(-u)\right) \right ] \, \text{for}\,\wn\ge\mn \, ,
\end{eqnarray}
while the derivative of its finite part turns out to be
\begin{equation}
{X_{1,2}}' = i(\wn-\mn) X^{\text{bdy}}_{1,2} \, \text{for}\, \wn\ge\mn\, . 
\end{equation}
We have also included the Minkowski index~$2$ here because writing down the equations of motion for the component~$X_2$ 
we discover that it is identical to the equation for~$X_1$. Now recall that we have choosen~$\wn\ge\mn$. Finding the
solution for smaller frequencies~$\wn<\mn$ amounts to redoing the above equation with replacing all the frequency
potential brackets by absolute values~$(\wn-\mn) \to |\wn-\mn|=(\mn-\wn)$ and keeping all the signs as they are. So 
we only have to switch the order in the final solution to get the small frequency solution and we can write
\begin{eqnarray}
X_{1,2} &=& \frac{8 X^{\text{bdy}}_{1,2} (1-u)^{-i\frac{\mn-\wn}{2}}}{8+\pi^2 \qn^2 - 4 i \ln 2 (\mn-\wn)} 
 \left [ 1+(\mn-\wn)\frac{i}{2} \ln\frac{1+u}{2} \right. \\  \nonumber
 &&\left . + \frac{\qn^2}{24} \left (\pi^2+ 12\ln u \ln (1+u) + 12 \mathrm{Li}_2(1-u) + 12 \mathrm{Li}_2(-u)\right) \right ] \, \text{for}\,\wn<\mn \, ,
\end{eqnarray}
while the derivative of it's finite part is given by
\begin{equation}
{X_{1,2}}' = i(\mn-\wn) X^{\text{bdy}}_{1,2} \, \text{for}\, \wn < \mn\, . 
\end{equation}

Similarly we get the solution for the other flavor combination fields~$Y_{1,2}$ by an analogous computation replacing~$(\wn-\mn)\to (\wn+\mn)$
\begin{eqnarray}
Y_{1,2} &=& \frac{8 Y^{\text{bdy}}_{1,2} (1-u)^{-i\frac{\wn+\mn}{2}}}{8+\pi^2 \qn^2 - 4 i \ln 2 (\wn+\mn)} 
 \left [ 1+(\wn+\mn)\frac{i}{2} \ln\frac{1+u}{2} \right. \\  \nonumber
 &&\left . + \frac{\qn^2}{24} \left (\pi^2+ 12\ln u \ln (1+u) + 12 \mathrm{Li}_2(1-u) + 12 \mathrm{Li}_2(-u)\right) \right ] \, \text{for any}\,\wn \, ,
\end{eqnarray}
and its derivative
\begin{equation}
{Y_{1,2}}' = i(\wn+\mn) X^{\text{bdy}}_{1,2} \, \text{for any}\, \wn\, . 
\end{equation}
Finally the third flavor direction components are obtained as in~\cite{Policastro:2002se} 
\begin{equation}
{A^3_{1,2}}' = i \wn A^{3\, \text{bdy}}_{1,2} \, \text{for any}\, \wn\, . 
\end{equation}

Comparing our solutions with those at vanishing chemical potential~$\mu\equiv 0$~\cite{Policastro:2002se} we learn
that turning on a constant chemical potential~$\mn$ results in the substitution
\begin{equation}
\label{eq:subsConstIsoMu}
\left \{
\begin{array}{c}
 \wn \rightarrow (\wn - \mn) \quad \text{for} \, \wn\ge\mn \\
 \wn \rightarrow (\mn - \wn) \quad \text{for} \, \wn < \mn 
\end{array} \right . \, .
\end{equation}
This is due to the fact that the way~$A_\nu\to \mu\delta_{0\nu} + A_\nu$ in which we introduce~$\mn$ makes~$\mu$ 
formally identical to a time derivative. The easiest way to understand this fact is to note the form of the covariant 
derivative appearing in the Lagrangian in time direction~$D_0 = \partial_0 + A_0 = \partial_0 + \mu$.

\subsubsection{Correlators of transversal components}
In this section we compute the on-shell action for transversal and longitudinal or time-like components of the gauge field.
Furthermore the correlators of transversal components are computed here.
Let us first assume~$\wn\ge\mn$ for definiteness.

{\bf The non-Abelian on-shell action}
In order to apply the correlator recipe and identify the relevant terms in the on-shell action to be evaluated at the boundary, 
we need to compute the on-shell action first. Starting from the action~\eqref{eq:constIsoSD7} together with the explicit 
expression~\eqref{eq:ggFF2} we integrate the action by parts to obtain
\begin{eqnarray}
S^{(2)}_{D7} &=& T_{D7} T_R (2\pi^2R^3) \frac{(2\pi\alpha')^2}{4} 2
\left \{ 
  \int\dd^4 x \left [ \sqrt{-g} g^{44} g^{\nu\nu'}(\partial_4 A_\nu^a) A_{\nu'}^a \right ]_{u=0}^{u=1}\right . \nonumber \\
  && \left . -\int \dd^4x \dd u \left [ 2\partial_{\mu'} (\sqrt{-g} g^{\mu\mu'} g^{\nu\nu'} \partial_{[\mu}A^a_{\nu]}) A^a_{\nu'}\right .\right. \nonumber \\
  &&\left .\left . -\mu^2 f^{db3}f^{ba3} \sqrt{-g} g^{00} g^{jj'}(A^a_{j'}-A^a_{0}\delta_{j' 0}) A_j^b \right .\right .\nonumber  \\
  &&\left .\left . +2 \mu f^{ab3}\sqrt{-g} g^{00} g^{jj'} (\partial_{j}A_{j'}^b A_{0}^a-\partial_{0}A_{j'}^b A_{j}^a) \right ]
\right \} \, .  
\label{eq:nonShellAction}
\end{eqnarray}
Note that we recover the AdS-boundary term~(the first term in equation~\eqref{eq:nonShellAction}) which is also 
present in the Abelian background. All other (Minkowski) boundary terms vanish by the standard QFT normalizability argument for
fields~$[A^\nu \nabla A_\nu]_{\vec x =-\infty}^\infty = 0$, i.e. the field~$A_\nu(\vec x)$ has to vanish at infinity in order for the action 
to be finite and for the theory to be normalizable. In addition we have three non-vanishing terms with the full integral over the 
four Minkowski directions and over the radial AdS direction. 

We now identify the second and third term of this on-shell action~\eqref{eq:nonShellAction} with parts of the equation of motion.
After multiplying the equation of motion~\eqref{eq:anaEom} with the field~$A_\mu^d$ and reordering we get
\begin{eqnarray}
&&2\partial_{\mu'} (\sqrt{-g} g^{\mu\mu'} g^{\nu\nu'} \partial_{[\mu}A^a_{\nu]}) A^a_{\nu'} 
  -\mu^2 f^{db3}f^{ba3} \sqrt{-g} g^{00} g^{jj'}(A^a_{j'}-A^a_{0}\delta_{j' 0}) A_j^b \\ \nonumber  
  &&= -\mu f^{ab3} \left \{ \partial_\kappa [\sqrt{-g} g^{00} g^{\kappa\kappa'} A_{\kappa'}^b) A_0^a
  +\sqrt{-g} g^{00} g^{\kappa\kappa'} ( \partial_\kappa A_0^b A_{\kappa'}^a -2 (\partial_0 A_\kappa^b) A_{\kappa'}^a] \right \}
\, .
\end{eqnarray}
Substituting this into the action~\eqref{eq:nonShellAction} finally yields the on-shell action
\begin{eqnarray}
S^{(2)}_{\text{on-shell}}& = &
T_{D7} T_R (2\pi^2R^3) \frac{(2\pi\alpha')^2}{4} 2 \int \dd^4 x \left \{ \left [ 
  \sqrt{-g} g^{44} g^{\nu\nu'}(\partial_4 A_\nu^a) A_{\nu'}^a \right ]_{u=0}^{u=1} 
\nonumber \right .\\ 
&& \left . -2 \mu f^{ab3} \int \dd u \sqrt{-g} g^{00} g^{33} \partial_{3} A^b_{[3} A_{0]}^a \right \} \, .
\end{eqnarray}
Since we transformed the solutions to flavor combinations~$X_\mu,\,Y_\mu$ we also need to transform the on-shell action
to obtain correlators of the new field combinations. In order to make the result obvious note the relations
\begin{equation}
A_j^1 = \frac{X_j+Y_j}{2}\, , \quad A_j^2 = \frac{X_j-Y_j}{2 i} \, ,
\end{equation}
and we get the on-shell action for flavor fields~$X,\, Y$ and~$A^3$ in momentum space after a Fourier transformation
of each gauge field fluctuation 
\begin{align}
\label{eq:onShellActionOfX}
S^{(2)}_{\text{on-shell}} =\;& T_{D7}T_R \frac{(2 \pi \alpha')^2}{4} (2\pi^2 R^3) 2 \nonumber  \\
        & \times  \int \frac{\mathrm d^4 q}{(2 \pi)^4}\:
        \Big \{\left. \sqrt{-g}\,g^{44}g^{jj'}\left[
         \frac{1}{2}\left({X_j}'Y_{j'}
         +{Y_j}'{X_{j'}}\right)  +
           {A^3_j}' A^3_{j'}
        \right]\right|_{u_b=0}^{u_h=1}  \\ \nonumber
        &\qquad\qquad\qquad +
         \mu q \int\limits_0^1
         \mathrm d u \sqrt{-g}g^{00}g^{33}
        \left(X_{[0} Y_{3]}-Y_{[0} X_{3]}
         \right) \Big\}  \, .
\end{align}
The term in the last line merely gives contact terms which we neglect here. Our on-shell action~\eqref{eq:onShellActionOfX} 
superficially suggests that the off-diagonal correlators, such as~$G^{XY}_{03}$, vanish. However, due to the fact that
some of our bulk solutions~$X_j$ and~$Y_j$ depend on more than one boundary field~(as we will see later in 
e.g.~\eqref{eq:solX0'}), the time-$x_3$-component off-diagonal correlators~$G_{03},\, G_{30}$ do not vanish.

{\bf Correlators}
Using the solutions~\eqref{eq:solX12} in the on shell action~\eqref{eq:onShellActionOfX} as prescribed by the 
recipe~\eqref{eq:anaGFormula} gives the transversal correlators
\begin{equation}
\label{eq:generalG11}
G^{XY}_{11} = G^{XY}_{22} = \left.  (-2) T_R T_{\text{D7}} \frac{(2\pi \alpha')^2}{4} (2\pi^2) R^3 \sqrt{-g} g^{uu} g^{11}
 \frac{{X_1}' Y_1}{X_1 Y_1} \right |_{u\to 0} \, .
\end{equation}
The factor can be simplified to
\begin{equation}
(-2)\, T_R T_{\text{D7}} \frac{(2\pi \alpha')^2}{4} (2 \pi^2 R^3) = -\frac{R^3 T_R}{32\pi^3 (\alpha')^2 g_s} \, ,
\end{equation}
which combines with factors from the metric components to give the overall factor~$N_c T_R T^2/4$.
Then~\eqref{eq:generalG11} yields the correlators
\begin{equation}
G^{XY}_{11} = G^{XY}_{22} = -i \frac{N_c T_R T}{8 \pi} (\omega-\mu) \frac{16+\pi^2\qn^2+4 i (\omega-\mu)\ln 2}
  {16+2\pi^2\qn^2+8 i (\omega-\mu)\ln 2} + \dots \, .
\end{equation}
Expanding the fraction in a Taylor double series in~$(\wn-\mn)$ and~$\qn$ leaves us with
\begin{eqnarray}
G^{XY}_{11} = G^{XY}_{22} &=& -i \frac{N_c T_R T}{8 \pi} (\omega-\mu)\left [ 1+\frac{\pi^2\qn^2}{16}+\text{divergent}\,\O({\qn^2})  
 + \dots \right ]  \, \\
 &=& -i\frac{N_c T_R T}{8 \pi} (\omega-\mu) + \dots \, \text{for}\, \wn\ge\mn \, ,
\label{eq:GXY11}
\end{eqnarray}
where we have renormalized all expressions in the second step~(subtracted the divergent term of order~$\qn^2$).
Recall that we have to go through the same procedure with the other index~$\beta$ for small frequencies~$\wn <\mn$.
By analogy we know that the correlator turns out to be
\begin{equation}
G^{XY}_{11} = G^{XY}_{22} = -i\frac{N_c T_R T}{8\pi} (\mu-\omega) + \dots\, \text{for}\, \wn<\mn \, .
\end{equation}

The other nonzero $YX$-flavor combination gives the correlators~$G^{YX}$ which involve a derivative of the field~$Y$ for
which we have only one index choice~$\beta=-i (\wn+\mn)/2$. 
\begin{equation}
G^{YX}_{11} = G^{YX}_{22} = -i\frac{N_c T_R T}{8\pi} (\mu+\omega) + \dots\, \text{for any}\, \wn \, .
\end{equation}
Gauge fluctuations pointing along third flavor direction and thus along the background gauge field do not feel the 
chemical potential. Their correlations turn out to be equal to those found at vanishing chemical potential~\cite{Policastro:2002se}
up to a different normalization~(the correlators from~\cite{Policastro:2002se} have to be multiplied by 4 in order to 
match the corresponding ones computed here). Our correlators read
\begin{equation}
G^{33}_{11} = G^{33}_{22} =  -i\frac{N_c T_R T}{4\pi} \omega + \dots\, \text{for any}\, \wn \, .
\end{equation}
All other flavor combinations vanish since the on-shell action~\eqref{eq:onShellActionOfX} does not contain any combination 
such as~$X'X,\, Y'Y$.  

\subsubsection{Calculation of longitudinal fluctuations}
Starting from the general equation of motion~\eqref{eq:anaEom} we choose the free index~$\sigma=0,\,3,\,4$ which gives a
system of three coupled equations of motion for the components of gauge field fluctuations~$A_0^a,\,A_3^a$
\begin{eqnarray}
\sigma=0:\,  0&=& {A_0^d}'' + \frac{\partial_4(\sqrt{-g} g^{00} g^{44})}{(\sqrt{-g} g^{00} g^{44})} {A_0^d}'
  -\frac{g^{33}}{g^{44}} \left [ q^2 {A_0^d} + \omega q {A_3^d} - i q\mu f^{db3} {A_3^b} \right ] \, , \\
\sigma=3:\, 0&=& {A_3^d}'' +  \frac{\partial_4(\sqrt{-g} g^{33} g^{44})}{(\sqrt{-g} g^{3} g^{44})} {A_3^d}' 
  -\frac{g^{00}}{g^{44}} \left [ (\omega^2 A_3^d+\omega q A_0^d) \nonumber \right . \\  
  && \left . - i\mu f^{db3} (2\omega A_3^b + q A_0^b) -\mu^2 (\delta^{d1} A_3^1+\delta^{d2} A_3^2) \right ] \, , \\
\sigma=4:\, 0&=& \omega {A_0^d}' - q\frac{g^{33}}{g^{00}} {A_3^d}' -i \mu f^{db3} {A_0^b}' \, .
\end{eqnarray}
Recall here that our gauge choice has fixed~$A_4^a\equiv 0$. Using the metric coefficients~\eqref{eq:metricCoeffs} of the black
hole background gives
\begin{eqnarray}
\label{eq:eomX0}
\sigma=0:\,  0&=& {A_0^d}'' -\frac{1}{u f(u)} \left [ \qn^2 {A_0^d} + \wn \qn {A_3^d} - i \qn\mn f^{db3} {A_3^b} \right ] \, , \\
\label{eq:eomX3}
\sigma=3:\, 0&=& {A_3^d}'' +  \frac{f'(u)}{f(u)} {A_3^d}' 
  +\frac{1}{u f(u)^2} \left [ (\wn^2 A_3^d+\wn \qn A_0^d) \nonumber \right . \\  
  && \left . - i\mn f^{db3} (2\wn A_3^b + \qn A_0^b) + \mn^2 (\delta^{d1} A_3^1+\delta^{d2} A_3^2) \right ] \, , \\
\label{eq:eomX4}
\sigma=4:\, 0&=& \wn {A_0^d}' + \qn f(u) {A_3^d}' -i \mn f^{db3} {A_0^b}' \, .
\end{eqnarray}
These three equations for the two components~$A_0^a,\,A_3^a$ are not independent. Equations~\eqref{eq:eomX0}
and~\eqref{eq:eomX4} imply~\eqref{eq:eomX3}. In order to see this we rewrite~\eqref{eq:eomX4}
\begin{equation}
\label{eq:A0fromA3}
{A_0^d}' = -\frac{\wn}{\wn} f(u) {A_3^d}' +i\frac{\mn}{\wn} f^{db3} {A_0^b}' \, ,\quad 
{A_0^d}'' = -\frac{\wn}{\wn} (f'(u){A_3^d}' +f(u) {A_3^d}'' ) +i\frac{\mn}{\wn} f^{db3} {A_0^b}'' \, .
\end{equation}
Using~\eqref{eq:A0fromA3} in~\eqref{eq:eomX0} gives
\begin{equation}
\label{eq:eomIdenticalX3}
0 = {A_3^d}'' + \frac{f'(u)}{f(u)} {A_3^d}' + \frac{1}{u f(u)^2} \left ( 
  \wn \qn A_0^d + \wn^2 A_3^d \right ) -i f^{db3}\left ( \frac{\mn}{\qn f(u)} {A_0^b}'' + \frac{\wn\mn}{u f(u)^2} A_3^b \right ) \, .
\end{equation}
We compare this expression to the third equation in the system~\eqref{eq:eomX3} and conclude that, if these two 
expressions ought to be identical, the following equation has to be satisfied
\begin{equation}
\label{eq:eomX0X4ImplyX3}
-i f^{db3}\left ( \frac{\mn}{\qn f(u)} {A_0^b}'' + \frac{\wn\mn}{u f(u)^2} A_3^b \right ) = 
  \frac{1}{u f(u)^2} \left [ - i\mn f^{db3} (2\wn A_3^b + \qn A_0^b) + \mn^2 (\delta^{d1} A_3^1+\delta^{d2} A_3^2) \right ] \, .
\end{equation}
In order to verify this relation we go one step back from the general equation of motion~\eqref{eq:anaEom} and rewrite
the term quadratic in the chemical potential~$\mn$ in terms of structure constants~$f^{db3}$ of the flavor group
\begin{equation}
\mn^2 \left ( \delta^{d1} A_3^1+\delta^{d2} A_3^2 \right ) = - \mn^2  f^{db3} f^{b a 3} {A_3^{a}} \, .
\end{equation}
Using this identity in~\eqref{eq:eomX0X4ImplyX3} we get
\begin{equation}
0= \frac{i\mn f^{db3}}{\qn f(u)} \left [ {A_0^b}'' -\frac{1}{u f(u)} (\qn\wn A_3^b + \qn^2 A_0^b
  - i\qn\mn f^{b a 3} A_3^{a}) \right ] \, ,\\
\end{equation}
and comparing to~\eqref{eq:eomX0} we find that the expression in brackets is identical to the right hand side of the
equation of motion~\eqref{eq:eomX0} and therefore has to vanish. In this way we verified that equation~\eqref{eq:eomIdenticalX3} 
implied by~\eqref{eq:eomX0} and~\eqref{eq:eomX4}
is equivalent to~\eqref{eq:eomX3}. We thus effectively have two coupled second order differential equations for two 
components. These we can decouple as far as the Minkowski indices are concerned by rewriting~\eqref{eq:eomX0} 
\begin{equation}
A_3^d = \frac{u f(u)}{\wn\qn} {A_0^d}'' -\frac{\qn}{\wn} A_0^d + i \frac{\mn}{\wn} f^{db3} A_3^b \, ,
\end{equation}
and using it in~\eqref{eq:eomX4} gives
\begin{equation}
0 = {A_0^d}''' + \frac{(uf(u))'}{u f(u)} {A_0^d}'' +\frac{1}{u f(u)^2}\left ( \wn^2{A_0^d}' - \qn^2 f(u) {A_0^d}'
  -\mn^2 f^{db3} f^{ba3} {A_0^a}' - 2 i \wn\mn f^{db3} {A_0^b}' \right ) \, ,
\end{equation}
which depends only on gauge fluctuation components~$A_0^d$ in time direction. This equation can be split into
three equations, one for each flavor~$d=1,\,2,\,3$ and we note that the flavor structure couples these three equations
\begin{eqnarray}
0&=& {A_0^1}''' + \frac{(u f(u))'}{u f(u)} {A_0^1}'' + \frac{1}{u f(u)^2} \left [
(\wn^2 - f(u)\qn^2 +\mn^2){A_0^1}' -2 i\wn\mn {A_0^2}'
\right ] \, ,\\
0&=& {A_0^2}''' + \frac{(u f(u))'}{u f(u)} {A_0^2}'' + \frac{1}{u f(u)^2} \left [
(\wn^2 - f(u)\qn^2 + \mn^2){A_0^2}' +2 i\wn\mn {A_0^1}'
\right ] \, ,\\
0&=& {A_0^3}''' + \frac{(u f(u))'}{u f(u)} {A_0^3}'' + \frac{1}{u f(u)^2} (\wn^2 - f(u)\qn^2){A_0^3}' \, .
\end{eqnarray}
The flavor coupling can be resolved as in the transversal case by use of a flavor transformation
\begin{equation}
\label{eq:flavorTrafoA0}
X_0 = A_0^1+ i A_0^2 \, , \quad Y_0 = A_0^1 - i A_0^2 \, ,
\end{equation}
which has the same structure as~\eqref{eq:flavorTrafo}, and we are left with
\begin{eqnarray}
0&=& {X_0}''' + \frac{(u f(u))'}{u f(u)} {X_0}'' + \frac{(\wn-\mn)^2-f(u)\qn^2}{u f(u)^2} {X_0}'  \, ,\\
0&=& {Y_0}''' + \frac{(u f(u))'}{u f(u)} {Y_0}'' + \frac{(\wn+\mn)^2-f(u)\qn^2}{u f(u)^2} {Y_0}'  \, ,\\
0&=& {A_0^3}''' + \frac{(u f(u))'}{u f(u)} {A_0^3}'' + \frac{1}{u f(u)^2} (\wn^2 - f(u)\qn^2){A_0^3}' \, .
\end{eqnarray}

From this point on the solution of this decoupled system of equations almost concurs with the method applied in the
transversal case~\ref{sec:transFlucs}. The only substantial difference is that because of the equations being second
order equations for the derivatives~${X_0}',\, {Y_0}'$, we have to choose the Ansatz
\begin{equation}
\label{eq:splitSingRegX03}
{X_{0,3}}' = (1-u)^\beta F (u) \, ,
\end{equation}
where~$F$ is a regular function of~$u$ which is different for~$X_0$ and~$X_3$. We have chosen an 
Ansatz for the derivative of the field instead of choosing this Ansatz for the field~$X$ itself as in~\eqref{eq:splitSingRegX1}. 
Proceeding analogously to the transversal case we obtain solutions for the derivatives directly as
\begin{eqnarray}
\label{eq:solX0'}
{X_0}' &=& 
\left \{\begin{array}{c}
  \frac{\qn^2 X_0^{\text{bdy}}+(\wn-\mn)\qn X_3^{\text{bdy}}}{i(\wn-\mn) -\qn^2} 
    +\lim\limits_{\epsilon\to 0}\ln \epsilon \left [ \qn^2 X_0^{\text{bdy}}+(\wn-\mn)\qn X_3^{\text{bdy}} \right ]\, \text{for} \, 
    \wn\ge\mn \\
  \frac{\qn^2 X_0^{\text{bdy}}+(\mn-\wn)\qn X_3^{\text{bdy}}}{i(\mn-\wn) -\qn^2} 
    +\lim\limits_{\epsilon\to 0}\ln \epsilon \left [ \qn^2 X_0^{\text{bdy}}+(\mn-\wn)\qn X_3^{\text{bdy}} \right ]\, \text{for} \,
    \wn < \mn
\end{array} \right .  \, , \nonumber \\ \\
{Y_0}' &=& \frac{\qn^2 Y_0^{\text{bdy}}+(\wn+\mn)\qn Y_3^{\text{bdy}}}{i(\wn+\mn) -\qn^2}
  +\lim\limits_{\epsilon\to 0}\ln \epsilon \left [ \qn^2 Y_0^{\text{bdy}}+(\wn+\mn)\qn Y_3^{\text{bdy}} \right ] \, ,\\
{A_0^3}' &=& \frac{\qn^2 A_0^{3\,\text{bdy}}+\wn\qn A_3^{3\,\text{bdy}}}{i\wn -\qn^2}
  + \lim\limits_{\epsilon\to 0}\ln \epsilon \left [\qn^2 A_0^{3\,\text{bdy}}+\wn\qn A_3^{3\,\text{bdy}}\right ] \, ,
\end{eqnarray}
for the time components and similarly for the spatial components 
\begin{eqnarray}
{X_3}' &=& 
\left \{\begin{array}{c}
  -\frac{(\wn-\mn)\qn X_0^{\text{bdy}}+(\wn-\mn)^2 X_3^{\text{bdy}}}{i(\wn-\mn) -\qn^2}
 { -\lim\limits_{\epsilon\to 0}\ln \epsilon \left [ (\wn-\mn)\qn X_0^{\text{bdy}}+(\wn-\mn)^2 X_3^{\text{bdy}} \right ]}
   \, \text{for}\, \wn\ge\mn \\
   -\frac{(\mn-\wn)\qn X_0^{\text{bdy}}+(\mn-\wn)^2 X_3^{\text{bdy}}}{i(\mn-\wn) -\qn^2} 
  -\lim\limits_{\epsilon\to 0}\ln \epsilon \left [ (\mn-\wn)\qn X_0^{\text{bdy}}+(\mn-\wn)^2 X_3^{\text{bdy}} \right ]
  \, \text{for} \, \wn < \mn
\end{array} \right . , \nonumber \\ \\
{Y_3}' &=& -\frac{(\wn+\mn)\qn Y_0^{\text{bdy}}+(\wn+\mn)^2 Y_3^{\text{bdy}}}{i(\wn+\mn) -\qn^2}\nonumber \\
   &&- \lim\limits_{\epsilon\to 0}\ln \epsilon \left [ (\wn+\mn)\qn Y_0^{\text{bdy}}+(\wn+\mn)^2 Y_3^{\text{bdy}} \right ] \, , \\
{A_3^3}' &=& -\frac{\wn\qn A_3^{3\,\text{bdy}}+\wn^2 A_3^{3\,\text{bdy}}}{i\wn -\qn^2}
   - \lim\limits_{\epsilon\to 0}\ln \epsilon \left [ \wn\qn A_3^{3\,\text{bdy}}+\wn^2 A_3^{3\,\text{bdy}} \right ] \, .
\end{eqnarray}
Here just as in the case for transversal fluctuations we need to choose the appropriate signs for the solutions
to the fields~$X_{0,3}$ in order for the index to be negative such that the incoming wave boundary condition is
satisfied as described in the tranversal case below~\eqref{eq:indexX1}.

\subsubsection{Correlators of longitudinal components}
The longitudinal and time component correlators are evaluated in analogy to the previous section
and we obtain
\begin{eqnarray}
\label{eq:GXY00}
G^{XY}_{00} &=& \frac{N_c T_R T q^2}{8 \pi [i(\omega-\mu) - D q^2]} \, , \\
\label{eq:GXY33}
G^{XY}_{33} &=&  \frac{N_c T_R T (\omega-\mu)^2}{8\pi [i(\omega-\mu) - D q^2]} \, , \\
\label{eq:GXY03}
G^{XY}_{03} &=&  -\frac{N_c T_R T (\omega-\mu) q}{8 \pi [i(\omega-\mu) - D q^2]}= G^{XY}_{30}  \, , \\
\label{eq:GYX00}
G^{YX}_{00} &=&  \frac{N_c T_R T q^2}{8 \pi [i(\omega+\mu) - D q^2]} \, , \\
\label{eq:GYX33}
G^{YX}_{33} &=&  \frac{N_c T_R T (\omega+\mu)^2}{8\pi [i(\omega+\mu) - D q^2]}  \, , \\
\label{eq:GYX03}
G^{YX}_{03} &=&  -\frac{N_c T_R T (\omega+\mu) q}{8 \pi [i(\omega+\mu) - D q^2]} = G^{YX}_{30}  \, , \\
\label{eq:G3300}
G^{33}_{00} &=&  \frac{N_c T_R T q^2}{4 \pi [i\omega - D q^2]} \, , \\
\label{eq:G3333}
G^{33}_{33} &=&  \frac{N_c T_R T \omega^2}{4\pi [i\omega - D q^2]}  \, , \\
\label{eq:G3303}
G^{33}_{03} &=&  -\frac{N_c T_R T \omega q}{4 \pi [i\omega - D q^2]}= G^{33}_{30}  \, ,
\end{eqnarray}
where we have introduced the coefficient
\begin{equation}
D = \frac{1}{2\pi T} \, .
\end{equation}
We have not written this out here but the above correlators are understood to change sign in the same way
the transversal ones did. This means we have above correlators for~$\omega\ge\mu$ but we need to replace
$(\omega - \mu)\to(\mu-\omega)$ for~$\omega < \mu$ for the same reasons discussed below~\eqref{eq:anaGBdyIndices}.

\subsubsection{Discussion} \label{sec:discAnaHydro}
This section gives a physical interpretation of the effects coming from adding a finite constant isospin chemical potential
to the~$\N=4$ SYM theory coupled to a fundamental~$\N=2$ hypermultiplet. As seen in the previous 
sections on the gravity side this
addition amounts to adding a background gauge field time component in the AdS-Schwarzschild black hole
background. Furthermore, we compare the approach presented here to the approach taken 
in~\cite{Erdmenger:2007ap} which neglects more terms, in particular those of order~$\O(\mn^2)$, 
in the action than the present approach. We will see that the results of~\cite{Erdmenger:2007ap} which appear rather 
cumbersome undergo a natural completion by taking into account the neglected terms of order~$\O(\mn^2)$. 
The keypoint to note is that the additional approximation in~\cite{Erdmenger:2007ap} lead to a misidentification
of the leading order term.

{\bf Discussion of~\cite{Erdmenger:2007ap}}
The approach taken in~\cite{Erdmenger:2007ap} is identical with the one presented in the previous sections up to 
one additional approximation. In that earlier work~\cite{Erdmenger:2007ap} it was assumed that the chemical potential 
is small~$\mn \ll 1$. Therefore we expanded the action to quadratic order in fluctuations to arrive at an equation 
identical to~\eqref{eq:ggFF2}. But then we went on also neglecting the terms of order~$\O(\mn^2)$ in that action 
which leads to the equations of motion
\begin{eqnarray}
0&=& 2 \partial_\kappa \left [\sqrt{-g} g^{\kappa\kappa'} g^{\sigma\sigma'} \left( \partial_{[\kappa'} A^d_{\sigma']} \right )\right ]\nonumber \\
  &&+ \mu f^{db3} \left [ \delta_{\sigma 0} \partial_\kappa (\sqrt{-g} g^{00} g^{\kappa\kappa'} A^b_{\kappa'})
    +\sqrt{-g} g^{00} g^{\sigma\mu}\partial_\mu A^b_0 - 2 \sqrt{-g} g^{00} g^{\sigma\mu}\partial_0 A^b_\mu \right ] \nonumber \, .
\end{eqnarray}
The approximations taken here imply~$\mn \sim {A_\mu}^2,\, {(\partial_\nu A_\mu)}^2\ll 1$. 

Following the standard procedure to study the singular behavior of the solutions at the horizon, we essentially find the 
same indices as before in e.g.~\eqref{eq:anaGHorIndices}, but with the order~$\O(\mn^2)$ missing
\begin{equation}
\label{eq:fullIndex}
\beta = \mp \sqrt{ -\frac{1}{4} (\wn\mp\mn)^2} = \mp \sqrt{ -\frac{1}{4} (\wn^2 \mp 2\wn\mn \underbrace{+\mn^2}_{\text{set to 0}})} \, .
\end{equation}
As a result of this the index obtains a non-analytic structure
\begin{equation}
\beta =  \mp \sqrt{ -\frac{1}{4} (\wn^2 \mp 2\wn\mn)} = \mp\frac{i \wn}{2} \sqrt{1\mp \frac{2\mn}{\wn}}  \, ,
\end{equation}
inheriting this non-analytic structure to all the solutions. At this point in the earlier approach we had to take a further 
approximation in order to carry out the {\it indicial procedure} and the {\it hydrodynamic expansion} properly. The index
containing the square root mixes different orders of the hydrodynamic expansion parameters~$\wn,\, \qn^2$. Therefore
we approximate the index through~$\mn\gg \wn \ll 1$ by
\begin{equation}
\label{eq:badIndex}
\beta = \mp\sqrt{\frac{ \wn\mn}{2}}\quad \text{or} \,  \pm i\sqrt{\frac{\wn\mn}{2}}  \, .
\end{equation}
At this point an intricate contradiction with the first approximation~$\O(\mn^2)\sim 0$ taken in~\cite{Erdmenger:2007ap} 
emerges~\footnote{The author 
thanks Laurence G. Yaffe for drawing his attention to this point and especially for all helpful discussions of this.}.
As we know from our full calculation including terms of order~$\O(\mn^2)$ yields analytic indices and no second
approximation is needed. Nevertheless, if we would like to we can simply take the full index~\eqref{eq:fullIndex} 
without setting~$\mn^2\sim 0$, take the full equations of motion at this 
point and try to neglect the order~$\O(\mn^2)$ by~$\mn\gg\wn\ll 1$. Doing so we are forced to conclude that~$\mn^2 \gg \wn^2$.
Therefore it becomes clear now from the full calculation that we should have included the order~$\O(\mn^2)$ rather
than the order~$\O(\wn^2)$. We also see that the term quadratic in chemical potential is even larger than the 
mixed term which we considered in~\eqref{eq:badIndex}. Neglecting the terms quadratic in the chemical potential~$\O(\mn^2)$
right from the beginning in~\cite{Erdmenger:2007ap} has obstructed the clear view of the situation that our full
calculation now admits.

As a result the cumbersome combination of approximations~$\mn \gg \wn$ and~$\wn$ produced non-analytic structures in
the correlators which we misidentified as frequency-dependent diffusion coefficients.

\begin{figure}
\psfrag{-0.005}{\tiny$-0.005$}
\psfrag{-0.015}{\tiny$-0.015$}
\psfrag{-0.1}{\tiny$-0.1$}
\psfrag{-0.01}{\tiny$-0.01$}
\psfrag{-0.02}{\tiny$-0.02$}
\psfrag{-0.04}{\tiny$-0.04$}
\psfrag{0}{\tiny$0$}
\psfrag{0.025}{\tiny$0.025$}
\psfrag{0.05}{\tiny$0.05$}
\psfrag{0.075}{\tiny$0.075$}
\psfrag{0.125}{\tiny$0.125$}
\psfrag{0.15}{\tiny$0.15$}
\psfrag{0.175}{\tiny$0.175$}
\psfrag{0.01}{\tiny$0.01$}
\psfrag{0.02}{\tiny$0.02$}
\psfrag{0.03}{\tiny$0.03$}
\psfrag{0.04}{\tiny$0.04$}
\psfrag{0.08}{\tiny$0.08$}
\psfrag{0.09}{\tiny$0.09$}
\psfrag{0.1}{\tiny$0.1$}
\psfrag{0.2}{\tiny$0.2$}
\psfrag{0.4}{\tiny$0.4$}
\psfrag{0.6}{\tiny$0.6$}
\psfrag{0.8}{\tiny$0.8$}
\psfrag{1}{\tiny$1$}
\psfrag{0.11}{\tiny$0.11$}
\psfrag{0.12}{\tiny$0.12$}
\psfrag{w}{\small$\wn$}
\psfrag{ReW}{\small$\mathrm{Re}\,\wn$}
\psfrag{ImW}{\small$\mathrm{Im}\,\wn$}
\psfrag{R}{\small$\R/ (N_c T_R T)$}
\psfrag{ImR}{\small$\R / (N_c T_R T)$}
\includegraphics[width=0.45\textwidth]{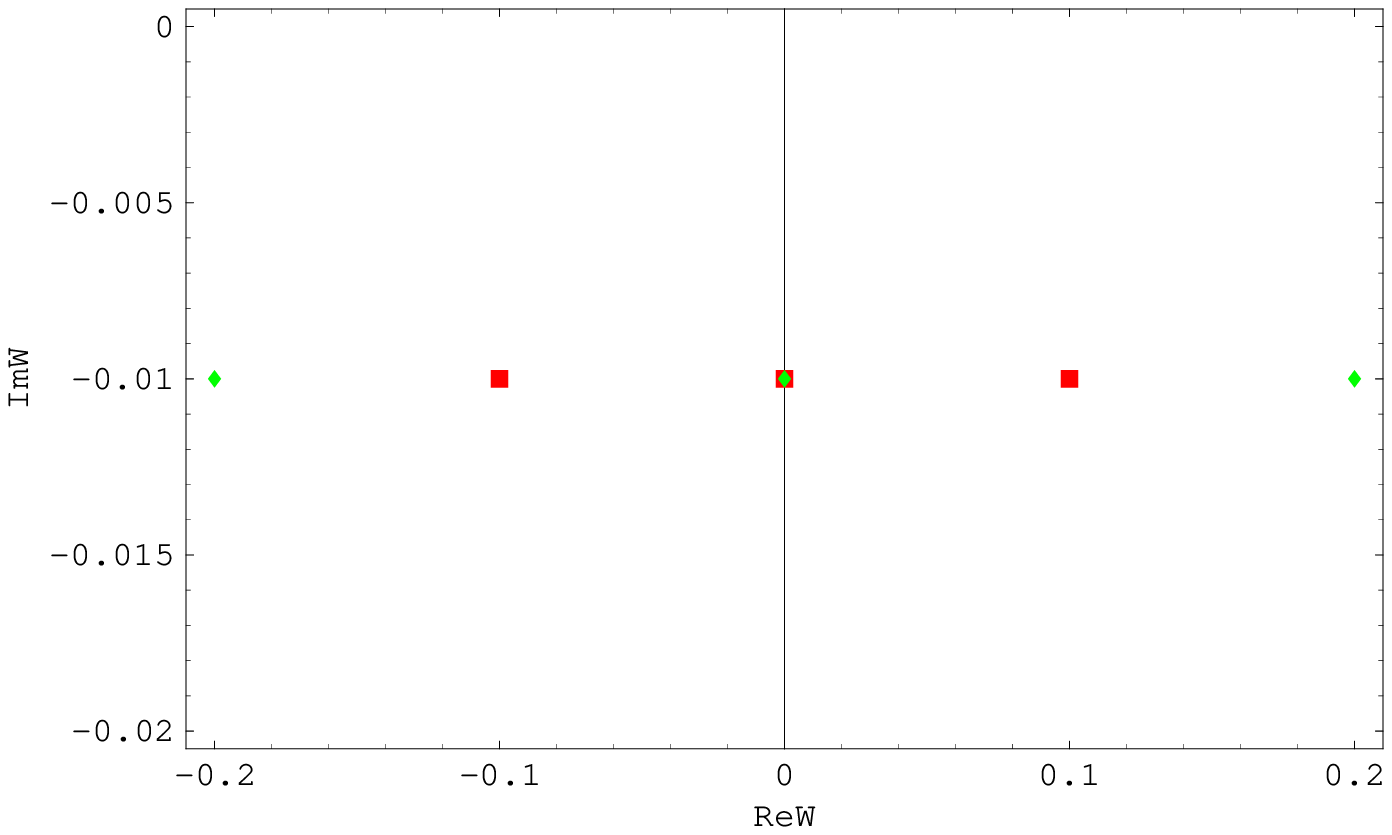}
\hfill
\includegraphics[width=0.45\textwidth]{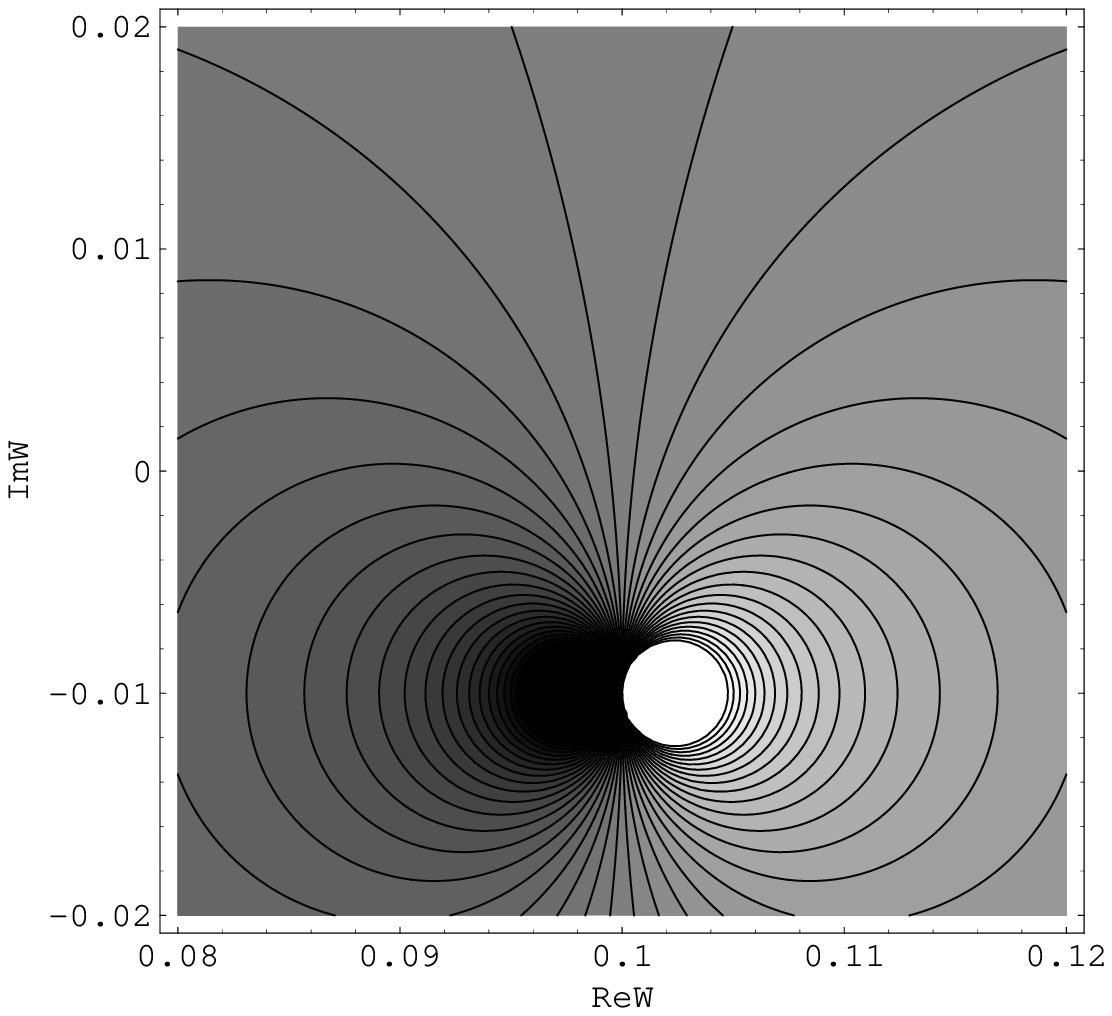}
\caption{\label{fig:anaQNM}
Left plot: The analytically computed location of the poles in the flavor-transverse correlation
functions~$G^{XY}$ and~$G^{YX}$ at finite chemical potentials~$\mu=0.1$~(red squares) and
at~$\mu=0.2$~(green diamonds). The left most pole corresponds to the combination~$YX$, 
the one in the middle to~$33$ and the right most one to~$XY$.
Right plot: The contour plot shows the value of the spectral function near the pole for~$\mu=0.1$
in the complex frequency plane.
}
\end{figure}

{\bf Technical interpretation and quasinormal modes}
We can use the intuition we have gained from our hydrodynamic considerations in section~\ref{sec:relativisticHydro} and 
from the example calculation in~\ref{sec:anaAdsG} to identify the coefficient~$D$ appearing in the 
correlators~\eqref{eq:GXY00} to~\eqref{eq:G3303} on the gauge theory side with the diffusion coefficient
for the isospin charge we have introduced. Comparing our correlators to those at vanishing chemical potential we learn 
that the main effect of an isospin chemical potential is to shift the location of poles in the correlators 
by~$\pm \mu$. In particular this can be seen from the dispersion relation which we read off the longitudinal
correlation functions
\begin{eqnarray}
\omega = -i D q^2 \pm \mu \, \text{for} \, \wn\ge\mn \, , \\
\omega =  i D q^2 + \mu \, \text{for} \, \wn < \mn\, \text{and only in} \, G^{XY}\, ,
\end{eqnarray}
where the positive sign of~$\mu$ corresponds to the dispersion of the flavor combination~$G^{XY}$ and the negative
sign of~$\mu$ corresponds to~$G^{YX}$. For the third flavor direction correlators~$G^{33}$ there is no chemical potential 
contribution in the dispersion relation. Looking at the transversal flavor directions with~$\wn\ge\mn$ we note that the imaginary part
of the pole location is unchanged while the real part is changed from zero to the value of the chemical potential~$\mu$. 
So the diffusion pole is shifted from its position on the imaginary axis to the left and right into the complex frequency plane. 
According to the AdS/CFT hydrodynamics interpretation this corresponds to shifting the hydrodynamic 
modes~(poles in the retarded gauge theory correlator are identified with the quasinormal frequencies as discussed
in section~\ref{sec:qnm}) or equivalently on the gravity dual side to shifting the quasinormal modes in the complex frequency plane 
as shown in figure~\ref{fig:anaQNM} for the two examples~$\mu=0.1,\, 0.2$. To be more precise we observe 
a shift in the frequency or energy~$(\wn\pm\mn)$ of the $SU(2)$-flavor gauge field fluctuations. Note that the other
solution for the case~$\wn < \mn$ would produce a pole/ quasinormal frequency in the upper complex frequency
plane corresponding to an enhanced mode. This solution is unphysical since if we have the finite chemical potential~$\mn$
then any perturbation introduced into the system has to have this minimum energy at least, i.e. only 
perturbations with~$\wn\ge\mn$ can form inside the plasma. Now since we are working
at finite spatial momentum~$\qn$ for that perturbation, the energy of that excitation needs to be even larger than~$\mn$.

\begin{figure}
\psfrag{-0.005}{\tiny$-0.005$}
\psfrag{-0.015}{\tiny$-0.015$}
\psfrag{-0.1}{\tiny$-0.1$}
\psfrag{-0.01}{\tiny$-0.01$}
\psfrag{-0.02}{\tiny$-0.02$}
\psfrag{-0.04}{\tiny$-0.04$}
\psfrag{0}{\tiny$0$}
\psfrag{0.025}{\tiny$0.025$}
\psfrag{0.05}{\tiny$0.05$}
\psfrag{0.075}{\tiny$0.075$}
\psfrag{0.125}{\tiny$0.125$}
\psfrag{0.15}{\tiny$0.15$}
\psfrag{0.175}{\tiny$0.175$}
\psfrag{0.01}{\tiny$0.01$}
\psfrag{0.02}{\tiny$0.02$}
\psfrag{0.03}{\tiny$0.03$}
\psfrag{0.04}{\tiny$0.04$}
\psfrag{0.08}{\tiny$0.08$}
\psfrag{0.09}{\tiny$0.09$}
\psfrag{0.1}{\tiny$0.1$}
\psfrag{0.2}{\tiny$0.2$}
\psfrag{0.4}{\tiny$0.4$}
\psfrag{0.6}{\tiny$0.6$}
\psfrag{0.8}{\tiny$0.8$}
\psfrag{1}{\tiny$1$}
\psfrag{0.11}{\tiny$0.11$}
\psfrag{0.12}{\tiny$0.12$}
\psfrag{w}{\small$\wn$}
\psfrag{ReW}{\small$\mathrm{Re}\,\wn$}
\psfrag{ImW}{\small$\mathrm{Im}\,\wn$}
\psfrag{R}{\small$\R/ (N_c T_R T)$}
\psfrag{ImR}{\small$\R / (N_c T_R T)$}
\includegraphics[width=0.4\textwidth]{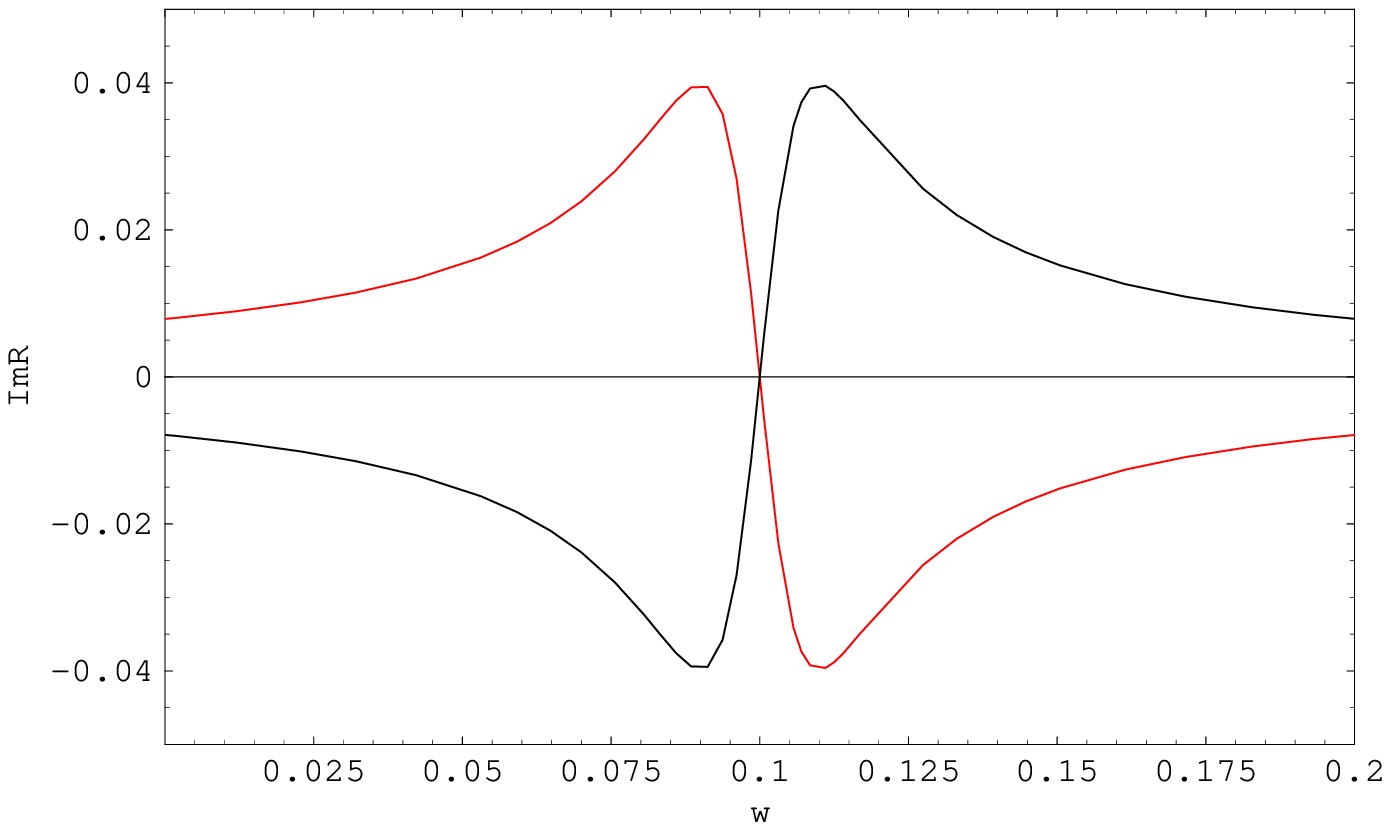}
\hfill
\includegraphics[width=0.4\textwidth]{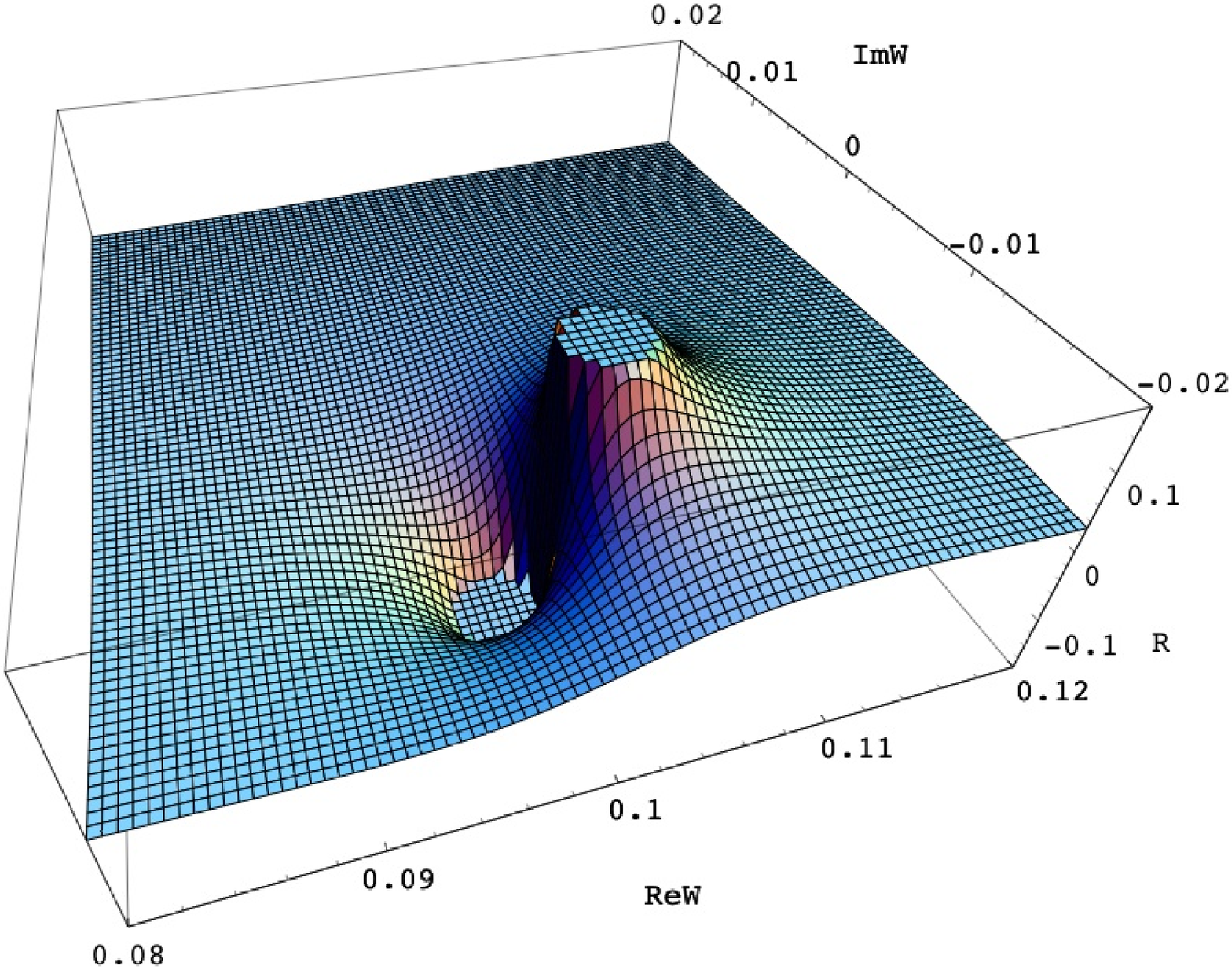}
\caption{\label{fig:anaSpecFunc}
Left plot: The spectral function computed from the two correlators is shown versus only real frequencies~$\wn\in \mathbf{R}$
for the chemical potential~$\mn=0.1$.
We have chosen to include the negative branches  for completeness but note that the incoming wave boundary condition
always selects the positive branch such that the spectral function is always positive.
Right plot: The spectral function surface is shown over complex values of the frequency. This plot shows the structure 
of the spectral function around the diffusion pole shifted to~$\mathrm{Re}\,\wn=\mn=0.1$. Note that the left plot is a vertical cut
through the right plot along the plane~$\mathrm{Im}\,\wn =0$.
}
\end{figure}

In figure~\ref{fig:anaSpecFunc} we see as an example the two spectral 
functions~$\R^{XY}_{00}=-2 \mathrm{Im} G^{XY}_{00}$~(from equation~\eqref{eq:spectralFunction})
valid in different regions~(see section~\ref{sec:relativisticHydro} for a discussion of the spectral function).
The red curve is the spectral function for the case~$\wn < \mn$ while the black curve shows the case~$\wn\ge\mn$.
In any case it is true that the spectral function is non-negative since the negative parts are cut off because they lie
outside the region of validity for that particular solution. Moreover, only the one which is cut off below~$\wn=\mn$~(black curve
in figure~\ref{fig:anaSpecFunc} for~$\wn\ge\mn$) is physical, i.e. the red curve is discarded entirely.

\begin{figure}
\psfrag{-0.005}{\tiny$-0.005$}
\psfrag{-0.015}{\tiny$-0.015$}
\psfrag{-0.1}{\tiny$-0.1$}
\psfrag{-0.01}{\tiny$-0.01$}
\psfrag{-0.02}{\tiny$-0.02$}
\psfrag{-0.04}{\tiny$-0.04$}
\psfrag{0}{\tiny$0$}
\psfrag{0.025}{\tiny$0.025$}
\psfrag{0.05}{\tiny$0.05$}
\psfrag{0.075}{\tiny$0.075$}
\psfrag{0.125}{\tiny$0.125$}
\psfrag{0.15}{\tiny$0.15$}
\psfrag{0.175}{\tiny$0.175$}
\psfrag{0.01}{\tiny$0.01$}
\psfrag{0.02}{\tiny$0.02$}
\psfrag{0.03}{\tiny$0.03$}
\psfrag{0.04}{\tiny$0.04$}
\psfrag{0.08}{\tiny$0.08$}
\psfrag{0.09}{\tiny$0.09$}
\psfrag{0.1}{\tiny$0.1$}
\psfrag{0.2}{\tiny$0.2$}
\psfrag{0.4}{\tiny$0.4$}
\psfrag{0.6}{\tiny$0.6$}
\psfrag{0.8}{\tiny$0.8$}
\psfrag{1}{\tiny$1$}
\psfrag{0.11}{\tiny$0.11$}
\psfrag{0.12}{\tiny$0.12$}
\psfrag{w}{\small$\wn$}
\psfrag{ReW}{\small$\mathrm{Re}\,\wn$}
\psfrag{ImW}{\small$\mathrm{Im}\,\wn$}
\psfrag{R}{\small$\R/ (N_c T_R T)$}
\psfrag{ImR}{\small$\R / (N_c T_R T)$}
\includegraphics[width=0.4\textwidth]{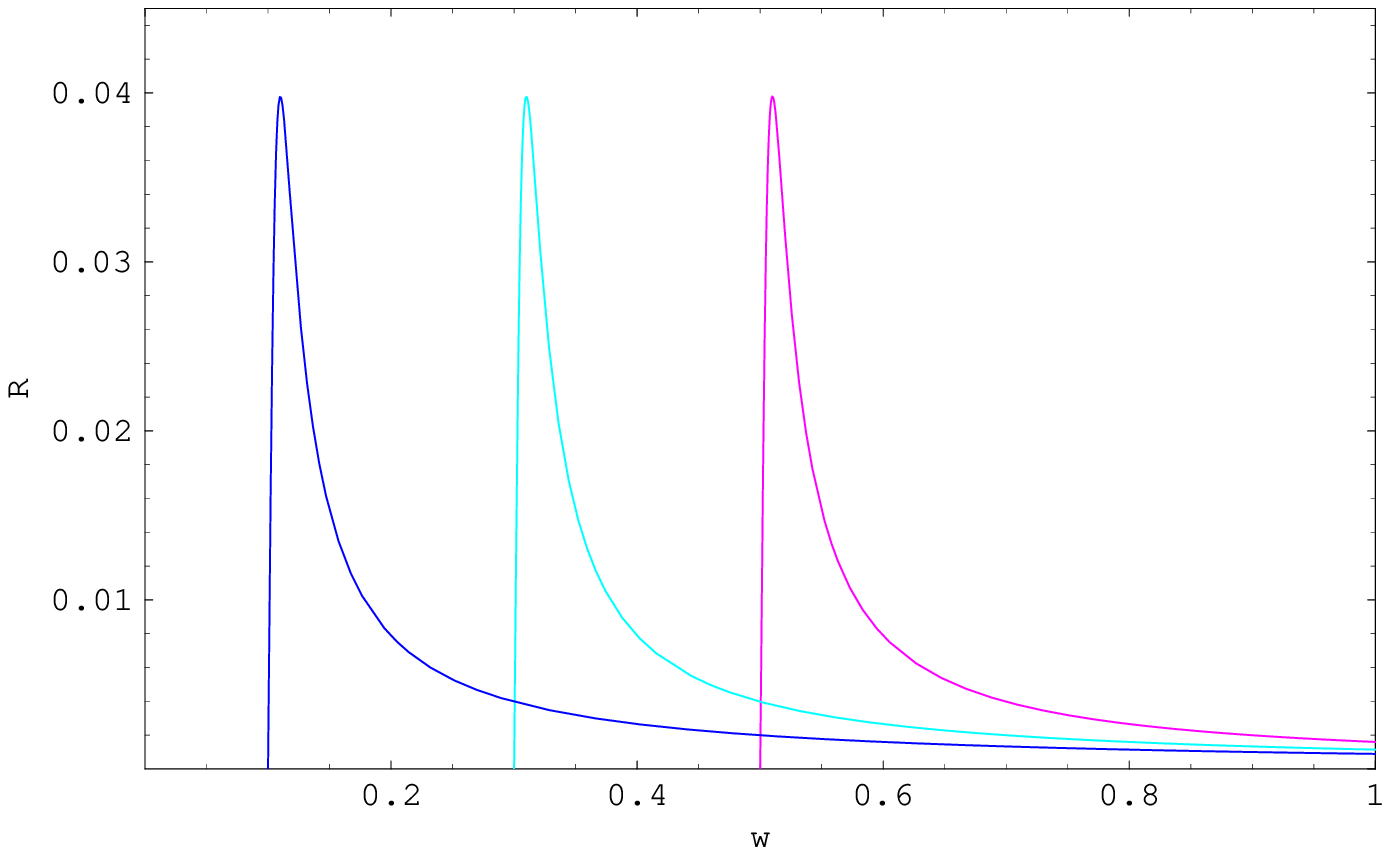}
\hfill
\includegraphics[width=0.4\textwidth]{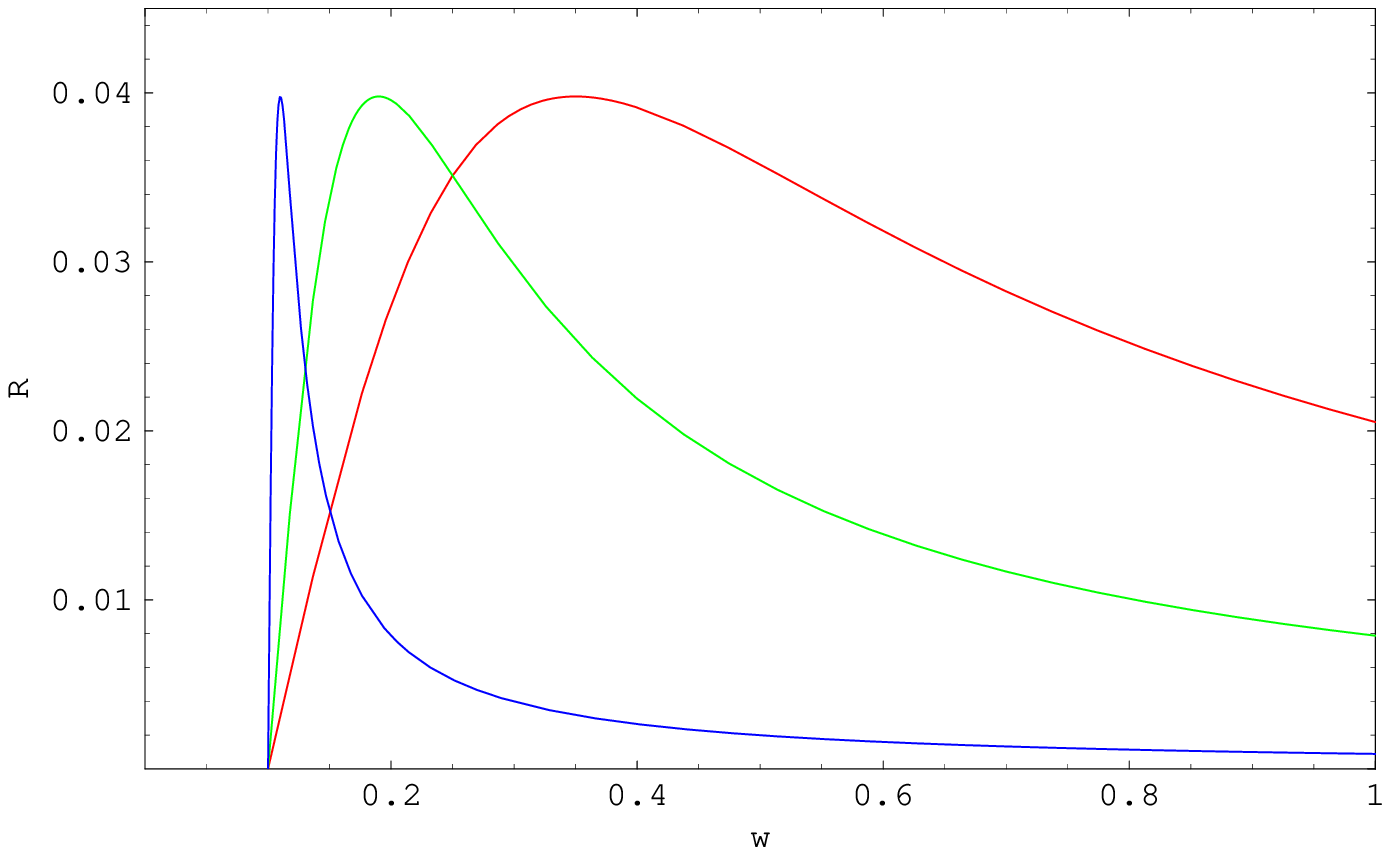}
\caption{\label{fig:anaSpecFuncMuQ}
Left plot: The spectral function in transversal flavor direction and longitudinal space-time direction~$\R^{XY}_{00}$
for different values of the chemical potential~$\mu = 0.1$~(blue), $0.3$~(light-blue), $0.5$~(purple). 
For simplicity we have chosen~$D=1/(2\pi T)=1,\, q=0.1$~(this means that we set the temperature to $T =1/(2\pi) $).
Right plot: This is the same picture as the left plot with the blue curve being identical to the blue curve in the left plot
but the other curves correspond to a fixed~$\mu=0.1$ and changing momentum~$q=0.1$~(blue), $q=0.3$~(green),
$q=0.5$~(red). 
}
\end{figure}

The right plot in figure~\ref{fig:anaSpecFuncMuQ} shows the dependence of the peak in the spectral function 
on spatial momentum~$q=0.1,\,0.3,\,0.5$~(in units of~$2\pi T$). Increasing the momentum shifts the 
peak in the spectral function to larger frequencies while in the limit~$q\to0$ the peak approaches~$\wn=\mn$.
This behavior confirms the interpretation given above of an excitation having to have at least the 
energy~$\wn=\mn$ in order to be produced in the plasma.  
The dependence on the chemical potential is shown in the left plot of figure~\ref{fig:anaSpecFuncMuQ}. 
The peaks and the frequency cut-off at~$\wn=\mn$, even the whole spectral function is shifted to a higher 
frequency by the amount of the chemical potential. The peak appearing here is the lowest lying one in a series 
of resonance peaks which under certain circumstances we will identify with quasi-particle excitations in 
section~\ref{sec:mesonSpectraB}. It is important to note that this particular diffusion peak is not 
contained in the spectra computed in section~\ref{sec:mesonSpectraB} because in that section
we set~$\qn=0$ for simplicity. Nevertheless the higher peaks and quasinormal modes show similar
behavior. In the present setup the peak is just interpreted as a resonance in the plasma which corresponds to the diffusive 
hydrodynamic mode at small~$\wn,\,\qn,\,\mn\ll 1$. Note that the high frequency tail for values~$\wn\not \ll 1$
is not physical since this is the region where our hydrodynamic expansion breaks down. 

The most striking feature here is that the peak
in the spectral function does not appear directly below the pole in the complex frequency plane but slightly shifted to 
a higher~$\mathrm{Re}\, \wn$. Looking at the contour plot this behavior can be traced back to the antisymmetric structure
of the pole. The spectral function surface~$\R (\mathrm{Re}\, \wn,\mathrm{Im}\, \wn)$ over the complex frequency plane
as shown in figure~\ref{fig:anaSpecFunc} is antisymmetric around the pole with the high~$\mathrm{Re}\, \wn$ side
being positive  showing a pole at~$+\infty$ and the low~$\mathrm{Re}\, \wn$ side being negative showing a pole at~$-\infty$.
From figure~\ref{fig:anaSpecFunc} it is also obvious that the poles in the spectral function deform the spectral function
surface antisymetrically such that the spectral function at~$\mathrm{Im}\,\wn =0$ is deformed antisymmetrically accordingly
receiving the structure shown as the black (physical) curve above~$\wn=\mn$ in the left plot of figure~\ref{fig:anaSpecFunc}.
Note that this behavior is still present if we set~$\mu=0$ such that the diffusion pole lies on the imaginary frequency axis, 
but the peak of the spectral function appears at a shifted position~$\omega \propto \pm D q^2$. A computation of the 
residues~(see also~\cite{Amado:2007yr}) at~$\mu=0$ confirms this behavior for the correlators~$G_{00}$ and~$G_{33}$ while 
the mixed correlator~$G_{03}$ gives a peak in the spectral function centered at~$\wn=0$.

{\bf Physical interpretation}
The physical interpretation of this frequency or energy shift leads us into the internal flavor space. Switching on a
background gauge field in the third flavor direction only and letting the $SU(2)$-fluctuations about it point into an arbitrary
internal direction is completely analog to the case of {\it Larmor precession} in external space-time. Larmor precession 
of a particle with spin, i.e. with a finite magnetic moment 
in external space~(Minkowski space-time) occurs if for example an electron~(spin~$|\mathbf{s}|= 1/2$) is placed in an external 
magnetic field~$\mathbf{B}$. If the {\it magnetic moment}~$\mathbf{m}$ of the electron points along the external 
field~$\mathbf{m}||\mathbf{B}$ then
the electron does not feel the field and nothing is changed. In contrast to that the transversal spin-components or equivalently
spins entirely orthogonal to the magnetic field feel a torque~$\mathbf{m} \times \mathbf{B}$ leading to the precession
of the spin around the magnetic field~$\mathbf{B}$. The frequency of this precession depends on the strength of the external field
as well as on the {\it gyromagnetic moment} taking into account quantum effects and is called Larmor frequency. 
Let us choose the geometry with the magnetic field pointing along the third space direction, then the torque on
the magnetic moment becomes
\begin{equation}
\mathbf{m}\times \mathbf{B} =
\left (\begin{array}{c}
m_2 B_3 \\
- m_1 B_3 \\
0
\end{array} 
\right ) \, .
\end{equation} 
Our situation for the flavor field fluctuations is completely analogous except for the 
fact that our precession takes place in the internal flavor space rather than in space-time~\footnote{The author is grateful to Dam T. Son for suggesting this interpretation.}. 
We have the torque on the flavor field fluctuations inside flavor space
\begin{equation}
\left(\begin{array}{c} X\\ Y\\ A^3\end{array}\right)\times \left (\begin{array}{c} 0\\0\\ \mu \end{array}\right ) =
\left (\begin{array}{c}
\mu Y \\
-\mu X \\
0
\end{array} 
\right ) \, ,
\end{equation} 
where the components correspond to the three flavor directions~$\{T^1+i T^2,T^1-i T^2, T^3 \}$ in the case of~$SU(2)$-flavor.
Assuming that components~$X, Y, A^3$ and~$\mu$ are positive, we conclude that~$X$ and~$Y$ are precessing with
opposite sense of rotation.
The flavor field Larmor frequency is given by the chemical potential~$\omega_{\text{L}}=\mu$. The chemical potential~$\mu$
is the minimum energy which an excitation has to have in order to be produced and propagated in the 
plasma~$\wn_{\text{min}}=\mn$.

{\bf Problem at the horizon}
We have introduced the chemical potential in our D3/D7-setup in the simplest possible way by choosing
the corresponding gravity background gauge field component~$A_0 = \mu + c/\rho +\dots$ to be constant
throughout the whole AdS bulk. This includes the special case that this gravity field does not vanish at the
black hole horizon.
Unfortunately there remains a conceptual problem with this simple constant potential apporach. Studying the 
AdS black hole metric~\eqref{eq:adsBHmetric}, we see that in these coordinates at the horizon~$u=1$ the time component of the metric
vanishes~$\lim\limits_{u\to 1}g_{00}=0$. Therefore a vector in time direction such as~$\partial_0$ is not well-defined in 
these coordinates. One possible solution to this problem is to claim that the background flavor gauge field should vanish 
at the horizon~\footnote{The author is grateful to Robert Myers and David Mateos for pointing this out and suggesting
to work with a non-constant background flavor gauge field.}. Nevertheless we can argue that the constant background field 
approach is still justified as a qualitative estimate. Comparing to figure~\ref{fig:backgAt} in the next section where we choose a 
non-constant background field~$\tilde A_0$ which vanishes at the horizon, we notice that the background gauge field solving the
equations of motion is constant almost everywhere. Only in a small region near the horizon it has a non-zero derivative which
drops quickly to approach zero in the bulk as seen from the slope of~$\tilde A_0$ in figure~\ref{fig:backgAt}. 
Since we are interested in the boundary theory only, we can argue that the 
constant background field is a good approximation in that region. Taking in account the non-constant behavior of the flavor
background near the horizon merely influences the equation of motion~(not the on-shell action for correlation functions). 
Solving the equations of motion for gauge field fluctuations we see that the difference is only a shift of values at the boundary
coming from integrating the peak near the horizon in~$\partial_u \tilde A_0$. 

In order to incorporate both the simplicity of a constant background field in the bulk and the vanishing boundary condition at the 
horizon we could use the theta function~$\mu (u) = \Theta(u-u_H) \mu$ with a constant~$\mu$. Nevertheless, the derivative 
of this potential has a delta peak at the horizon and we have not studied yet how this influences our computation. 
Finally we should note that there may be other background field configurations solving this setup which might not have to
vanish at the horizon. In order to study this point we would have to go to non-singular coordinates such as Kruskal coordinates.

\subsection{Thermodynamics at finite baryon density or potential} \label{sec:thermoBaryon}
In this section we will review the thermodynamics of the strongly coupled thermal field theory dual to a D3/D7-brane
configuration in the AdS black hole background~\eqref{eq:adsBHmetric}. This section summarizes the results
of the work of Myers et al. on this topic~\cite{Kobayashi:2006sb,Mateos:2007vc,Mateos:2007vn,Mateos:2006nu} 
and provides a few additional remarks. This will help us to interpret our own results within this and similar setups
that follow in the next sections and chapters. 

Except from changing the radial coordinate
from~$u$ to~$\varrho$ we also have to be careful with the definition of the thermodynamic ensemble in which
we are working. It is crucial for the understanding of all brane thermodynamics to understand that we can work 
either in the {\it canonical ensemble} or in the {\it grandcanonical ensemble}. The canonical ensemble is in contact
with a heat bath only and we work at an arbitrary but fixed charge density~$n_B$. In contrast to this situation the
grandcanonical ensemble additionally is in contact with a particle bath such that the chemical potential is fixed at 
an arbitrary value. In the thermodynamic limit both ensembles are equivalent but we will see that there are 
phase space regions in one ensemble which we can not reach in the other. Therefore it is instructive to consider both.

{\bf Brane configuration and background}
Let us describe the gravity dual of the canonical ensemble first, i.e. we fix the charge density which in 
our case ist the baryon charge density~$n_B$.
We consider asymptotically $AdS_5\times S^5$ space-time which arises 
as the near horizon
limit of a stack of $N_c$ coincident D3-branes. More precisely, our background
is an $AdS$ black hole, which is the geometry dual to a field theory at finite
temperature (see e.g.~\cite{Policastro:2002se}).
We make use of the coordinates of \cite{Kobayashi:2006sb} to write this
background in Minkowski signature as
\begin{equation}
\begin{split}
\label{eq:AdSBHmetric}
\dd s^2 =\; &\frac{1}{2} \left(\frac{\varrho}{R}\right)^2
\left( -\frac{f^2}{\tilde f}\dd t^2 + \tilde{f} \dd \bm{x}^2 \right)\\
 & + \left(\frac{R}{\varrho}\right)^2\left( \dd\varrho^2 +\varrho^2 \dd\Omega_5^2
  \right)  ,
\end{split}
\end{equation}
with the metric $\mathrm d \Omega_5^2$ of the unit $5$-sphere, where
\begin{equation}
\begin{split}
\label{eq:metricDefinitions}
f(\varrho)=1-\frac{\varrho_H^4}{\varrho^{4}},\qquad\tilde f(\varrho)=1+\frac{\varrho_H^4}{\rho^{4}},\\
R^4=4\pi g_s N_c  {\alpha'}^2,\qquad \varrho_H = T \pi R^2.
\end{split}
\end{equation}
Here $R$ is the $AdS$ radius, $g_s$ is the string coupling constant, $T$ the
temperature, $N_c$ the number of colors. In the following some equations may be
written more conviniently in terms of the dimensionless radial coordinate
$\rho=\varrho/\varrho_H$, which covers a range from $\rho=1$ at the event
horizon to $\rho \to \infty$, representing the boundary of $AdS$ space.

Into this ten-dimensional space-time we embed $N_f$ coinciding D7-branes,
hosting flavor gauge fields $A_\mu$. The embedding we choose lets the D7-branes
extend in all directions of $AdS$ space and, in the limit $\rho \rightarrow
\infty$, wraps an $S^3$ on the $S^5$. It is convenient to write the D7-brane
action in coordinates where
\begin{equation}
 \dd \varrho^2 +\varrho^2 \dd \Omega_5^2 = \dd \varrho^2 + \varrho^2 ( \dd \theta^2 
 + \cos^2 \theta \dd \phi^2 + \sin^2 \theta \dd \Omega_3^2) ,
\end{equation}
with $0\leq\theta<\pi/2$. From the viewpoint of ten dimensional Cartesian $AdS_5
\times S^5$, $\theta$ is the angle between the subspace spanned by
the 4,5,6,7-directions, into which the D7-branes extend perpendicular to the
D3-branes, and the subspace spanned by the 8,9-directions, which are transverse
to all branes.

Due to the symmetries of this background, the embeddings depend only on the
radial coordinate $\rho$. Defining $\chi \equiv \cos \theta$, the embeddings of
the D7-branes are parametrized by the functions~$\chi(\rho)$. They describe
the location of the D7-branes in $8,9$-directions. Due to our choice of the
gauge field fluctuations in the next subsection, the remaining three-sphere
in this metric will not play a prominent role.

The metric induced on the D7-brane probe is then given by
\begin{equation}
\label{eq:inducedMetric} 
\begin{split}
\dd s^2 =\; & \hphantom{\:\,+}\frac{1}{2} \left(\frac{\varrho}{R}\right)^2 \left(-\frac{f^2}{\tilde f}\,\dd t^2+
  \tilde f\, \dd \bm{x}^2 \right) \\
  & +  
  \frac{1}{2} \left(\frac{R}{\varrho}\right)^2  \frac{1-\chi^2+\varrho^2{\chi'}^2}{1-\chi^2}\, \dd \varrho^2 
  \\ & +R^2(1-\chi^2) \dd \Omega_3^2.  
\end{split}
\end{equation}
Here and in what follows we use a prime to denote a derivative with respect to
$\varrho$ (resp.\ to $\rho$ in dimensionless equations). The symbol $\sqrt{-g}$
denotes the square root of the determinant of the induced metric on the
D7-brane, which is given by
\begin{equation}
        \sqrt{-g} = \varrho^3 \frac{f \tilde f}{4}\, (1-\chi^2) \sqrt{1-\chi^2 + \varrho^2 {\chi'}^2}.
\end{equation}

The table below gives an overview of the indices we use to refer to
certain directions and subspaces.

\begin{center}
       \includegraphics[height=5.5\baselineskip]{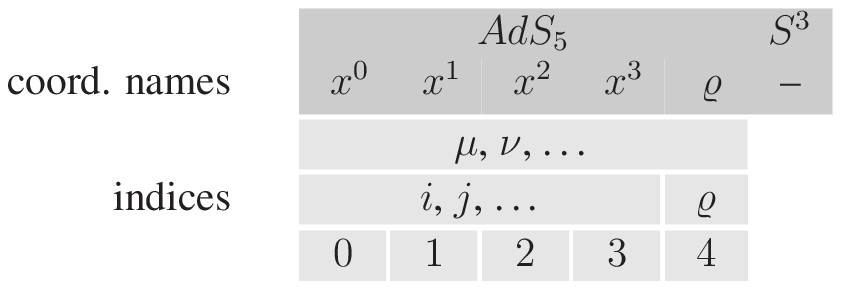}
\end{center}

The background geometry described so far is dual to thermal $\mathcal N = 4$
supersymmetric $SU(N_c)$ Yang-Mills theory with $N_f$ additional $\mathcal N =
2$ hypermultiplets. These hypermultiplets arise from the lowest excitations of
the strings stretching between the D7-branes and the background-generating
D3-branes. The particles represented by the fundamental fields of the $\mathcal
N = 2$ hypermultiplets model the quarks in our system. Their mass $M_q$ is given
by the asymptotic value of the separation of the D3- and D7-branes. In the
coordinates used here we write \cite{Myers:2007we}
\begin{equation}
\label{eq:quarkMass}
\frac{2M_q}{\sqrt{\lambda} T} = \frac{\bar M}{T} = \lim_{\rho \to \infty} \rho\, \chi(\rho) = m,
\end{equation}
where we introduced the dimensionless scaled quark mass $m$.

In addition to the parameters incorporated so far, we aim for a description of
the system at finite chemical potential $\mu$ and baryon density $n_B$. In field
theory, a chemical potential is given by a nondynamical time component of the
gauge field. In the gravity dual, this is obtained by introducing a
$\rho$-dependent gauge field component $\bar A_0(\rho)$ on the D7 brane probe.
For now we consider a baryon chemical potential which is obtained from the
$U(1)$ subgroup of the flavor symmetry group. The sum over flavors then yields a
factor of $N_f$ in front of the DBI~action written down below.

The value of the chemical potential $\mu$ in the dual field theory is then
given by
\begin{equation}
\label{eq:chemPotLimit}   
\mu = \lim_{\rho\to\infty} \bar A_0(\rho) = \frac{\varrho_H}{2\pi\alpha'} \tilde\mu,
\end{equation}
where we introduced the dimensionless quantity $\tilde\mu$ for convenience. We
apply the same normalization to the gauge field and distinguish the
dimensionful quantity $\bar A$ from the dimensionless $\tilde A_0=\bar
A_0\,(2\pi\alpha')/\varrho_H$.

The action for the probe branes' embedding function
and gauge fields on the branes is
\begin{equation}
        \label{eq:dbiActionD7}
        S_{\text{DBI}} = -N_f\, T_{\text{D7}}\int\!\! \dd^8 \xi\; \sqrt{| \det ( g + \tilde F )|}.
\end{equation}
Here $g$ is the induced metric \eqref{eq:inducedMetric} on the brane,
$\tilde F$ is the field strength tensor of the gauge fields on the brane and $\xi$ are
the branes' worldvolume coordinates. $T_{\text{D7}}$ is the brane tension and
the factor $N_f$ arises from the trace over the generators of the
symmetry group under consideration. For finite baryon density, this factor will be different 
from that at finite isospin density.

In~\cite{Kobayashi:2006sb}, the dynamics of this system of branes and gauge
fields was analyzed in view of describing phase transitions at finite baryon
density. Here we use these results as a starting point which gives the
background configuration of the brane embedding and the gauge field values at
finite baryon density. To examine vector meson spectra, we will then investigate
the dynamics of fluctuations in this gauge field background.

In the coordinates introduced above, the action $S_{\text{DBI}}$ for the
embedding $\chi(\rho)$ and the gauge fields' field strength $\tilde F$ is obtained by
inserting the induced metric and the field strength tensor into
\eqref{eq:dbiActionD7}. As in \cite{Kobayashi:2006sb}, we get
\begin{multline}
\label{eq:actionEmbeddingsAt} 
S_{\text{DBI}}=-N_f T_\mathrm{D7} \varrho_H^3 \int\!\! \dd^8 \xi\; \frac{\rho^3}{4} f
\tilde{f} (1-\chi^2)\\ 
\times\sqrt{1-\chi^2+\rho^2 
{\chi'}^2-2 \frac{\tilde{f}}{f^2}(1-\chi^2) \tilde F_{\rho 0}^2} \; , 
\end{multline}
where $\tilde F_{\rho 0}=\partial_\rho \tilde A_0$ is the field strength on the
brane. $\tilde A_0$ depends solely on $\rho$.

\begin{figure}
	\includegraphics[width=0.9\linewidth]{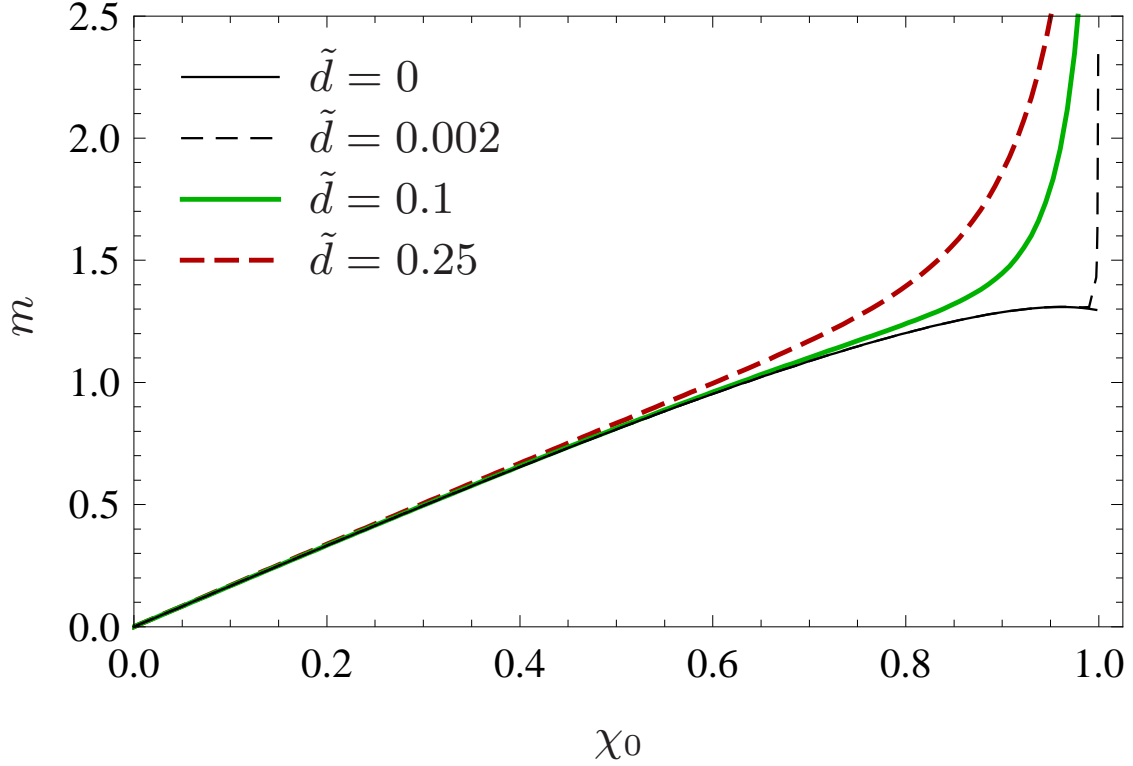}   
	\caption{The dependence of the scaled quark mass~$m=2 M_q/\sqrt{\lambda} T$
		on the horizon value~$\chi_0=\lim_{\rho\to1}\chi$ of the embedding.}
	\label{fig:mOfChi0}   
\end{figure}

According to \cite{Kobayashi:2006sb}, the equations of motion for the background
fields are obtained after Legendre transforming the
action~\eqref{eq:actionEmbeddingsAt}. Varying this Legendre transformed action
with respect to the field~$\chi$ gives the equation of motion for the
embeddings~$\chi(\rho)$,
\begin{equation}
\begin{aligned}
\label{eq:eomChi}
 & \partial_\rho\left[\frac{\rho^5 f \tilde{f} (1-\chi^2)
{\chi'}}{\sqrt{1-\chi^2+\rho^2{\chi'}^2}} \sqrt{1 +
\frac{8 \tilde{d}^2}{\rho^6 \tilde{f}^3 (1-\chi^2)^3}}\right] \\
= & - \frac{\rho^3 f \tilde{f} \chi }{\sqrt{1-\chi^2+\rho^2{\chi'}^2}}
\sqrt{1 +\frac{8 \tilde{d}^2}{\rho^6 \tilde{f}^3 (1-\chi^2)^3}}\\
 & \times\left[3 (1-\chi^2) +2 \rho^2 {\chi'}^2 -24 \tilde{d}^2
\frac{1-\chi^2+\rho^2{\chi'}^2}{\rho^6 \tilde{f}^3 (1-\chi^2)^3+8 \tilde{d}^2}
\right] .
\end{aligned}
\end{equation}
The dimensionless quantity $\tilde d$ is a constant of motion. It is related to
the baryon number density $n_B$ by \cite{Kobayashi:2006sb}
\begin{equation}
        \label{eq:nbfromdtilde}
        n_B = \frac{1}{2^{5/2}} N_f N \sqrt{\lambda} T^3 \tilde d.
\end{equation}
Below,  equation \eqref{eq:eomChi} will be solved numerically for different
initial values~$\chi_0$ and~$\tilde d$. The boundary conditions used are
\begin{equation}
\chi(\rho=1)=\chi_0,\qquad  \partial_\rho \chi(\rho) \Big|_{\rho=1}=0 .
\end{equation}
The quark mass $m$ is determined by $\chi_0$. It is zero for $\chi_0=0$ and
tends to infinity for $\chi_0 \to 1$. Figure~\ref{fig:mOfChi0} shows the
dependence of the scaled quark mass~$m=2 M_q/\sqrt{\lambda}T$ on the starting value~$\chi_0$
for different values of the baryon density parametrized by~$\tilde d\propto n_B$.
In general, a small (large)~$\chi_0$ is equivalent to a small (large) quark mass.
For~$\chi_0<0.5$, $\chi_0$ can be viewed as being proportional
 to the large quark masses. At
larger~$\chi_0$ for vanishing~$\tilde d$, the quark mass reaches a finite value.
In contrast, at finite baryon density, if~$\chi_0$ is close to
$1$, the mass rapidly increases when increasing~$\chi_0$ further. At small densities 
there exists a black hole to black hole phase transition which we will discuss in section~\ref{sec:thermoBaryon}.
In embeddings where this phase transition is present, there exist more than one embedding for one
specific mass value. In a small regime close to~$\chi_0=1$, there are more than
one possible value of~$\chi_0$ for a given~$m$. So in this small region,
$\chi_0$ is not proportional to~$M_q$.   
\begin{figure}
\centering
\includegraphics[width=0.49\linewidth]{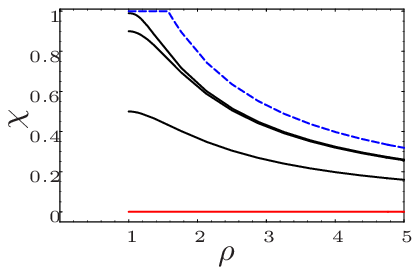}
\hfill
\includegraphics[width=0.49\linewidth]{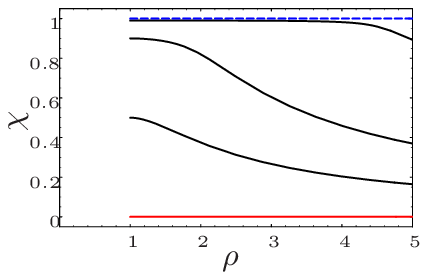}
\vfill\vspace{5pt}  
\includegraphics[width=0.49\linewidth]{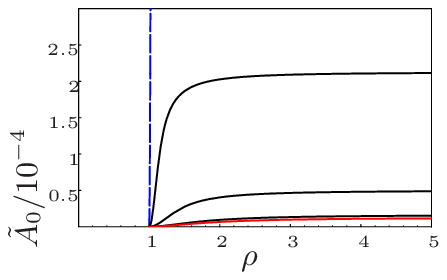}
\hfill
\includegraphics[width=0.49\linewidth]{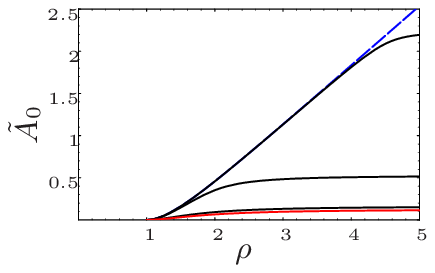}
\vfill\vspace{5pt}  
\includegraphics[width=0.49\linewidth]{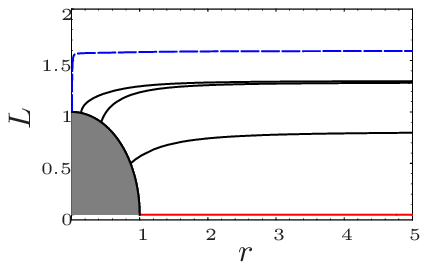}
\hfill
\includegraphics[width=0.49\linewidth]{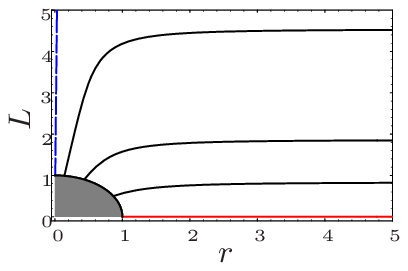}
\caption{\label{fig:backgAt}
        The three figures of the left column show the embedding function~$\chi$
        versus the radial coordinate~$\rho$, the corresponding background gauge 
        fields~$\tilde A_0$ and the distance~$L=\rho\,\chi$ between the D3 and
		the D7-branes at $\tilde d=10^{-4}/4$.
		$L$ is plotted versus $r$, given by $\rho^2 = r^2 + L^2$.
		In the right column, the same three quantities are
        depicted for~$\tilde d=0.25$. 
        The five curves in each plot correspond to parametrizations of the 
        quark mass to temperature ratio with~$\chi_0=\chi(1)=0,\,0.5,\,0.9,\,0.99$~(all solid)  
	and $0.99998$~(dashed) from bottom up.
        These correspond to scaled quark 
        masses~$m=2 M_q/T\sqrt{\lambda}=0,\,0.8089,\,1.2886,\,1.3030,\,1.5943$ in
        the left plot and to~$m=0,\,0.8342,\,1.8614,\,4.5365,\,36.4028$ on the right.  
	The curves on the left
        exhibit~$\mu\approx 10^{-4}$. Only the upper most curve    
        on the left at~$\chi_0=0.99998$ develops 
        a large chemical potential of~$\mu=0.107049$. In the right column 
	curves correspond to chemical potential values~$\mu=0.1241,\,0.1606,\,
	0.5261,\,2.2473,\,25.3810$~from bottom up.  
        }
\end{figure}
The equation of motion for the background gauge field $\tilde A$ is 
\begin{equation}
\label{eq:eomD}
 \partial_\rho\tilde{A}_0 = 2 \tilde{d}
\frac{f^2 \sqrt{1-\chi^2+\rho^2 {\chi'}^2}}
{\sqrt{\tilde{f}(1-\chi^2) [\rho^6 \tilde{f}^3 (1-\chi^2)^3+8 \tilde{d}^2]}}. 
\end{equation}
Integrating both sides of the equation of
motion from
$\rho_{\text{H}}$ to some $\rho$, and respecting the boundary condition $\tilde
A_0(\rho=1)=0$ \cite{Kobayashi:2006sb}, we obtain the full
background gauge field
\begin{equation}
\label{eq:backgroundAt}
 \tilde{A}_0(\rho) = 2\tilde{d}
\int\limits_{\rho_0}^\rho\!\! \dd\rho\;
\frac{f\sqrt{1-\chi^2+\rho^2{\chi'}^2}}{\sqrt{\tilde{f}
(1-\chi^2)[\rho^6 \tilde{f}^3 (1-\chi^2)^3 +8 \tilde{d}^2]}}\,.
\end{equation}
Recall that the chemical potential of the field theory is given by
$\lim_{\rho\to\infty}\tilde A_0(\rho)$ and thus can be obtained from the formula
above.  Examples for the functional behavior of $A_0(\rho)$ are shown in
figure~\ref{fig:backgAt}. Note that at a given baryon density $n_B\neq 0$ there
exists a minimal chemical potential which is reached in the limit of massles
quarks.

The asymptotic form of the fields $\chi(\rho)$ and $A_0(\rho)$ can be found from
the equations of motion in the boundary limit~$\rho\to\infty$,
\begin{align}
\label{eq:boundaryAt}
\bar A_0 & =\mu - \frac{1}{\rho^2} \frac{\tilde d}{2\pi\alpha'} + \cdots  \, ,  \\
\label{eq:boundaryChi}
\chi & =\frac{m}{\rho}+\frac{c}{\rho^3}+ \cdots .
\end{align}
Here $\mu$ is the chemical potential, $m$ is the dimensionless quark mass
parameter given in \eqref{eq:quarkMass}, $c$ is related to the quark condensate by
\begin{equation}
\label{eq:quarkCondensate}
\langle\bar \psi \psi \rangle = -\frac{1}{8} \sqrt{\lambda} N_f N_c T^3 c  \, ,
\end{equation}
and $\tilde d$ is related to the baryon number
density as stated in \eqref{eq:nbfromdtilde}. See also figure~\ref{fig:backgAt}
for this asymptotic behavior.
The $\rho$-coordinate runs from the horizon value~$\rho=1$ to the boundary
at~$\rho=\infty$. In most of this range, the gauge field is almost constant
and reaches its asymptotic value, the chemical potential~$\mu$, at
$\rho\to\infty$. Only near the horizon the field drops rapidly to zero. For
small~$\chi_0\to 0$, the curves asymptote to the lowest~(red) curve. So
there is a minimal chemical potential for fixed baryon density in this
setup. At small baryon density~($\tilde d \ll 0.00315$) the embeddings
resemble the Minkowski and black hole embeddings known from the case 
without a chemical potential. Only a thin spike always reaches down to 
the horizon. 

\begin{figure}[h!]
 \includegraphics[width=0.8\textwidth]{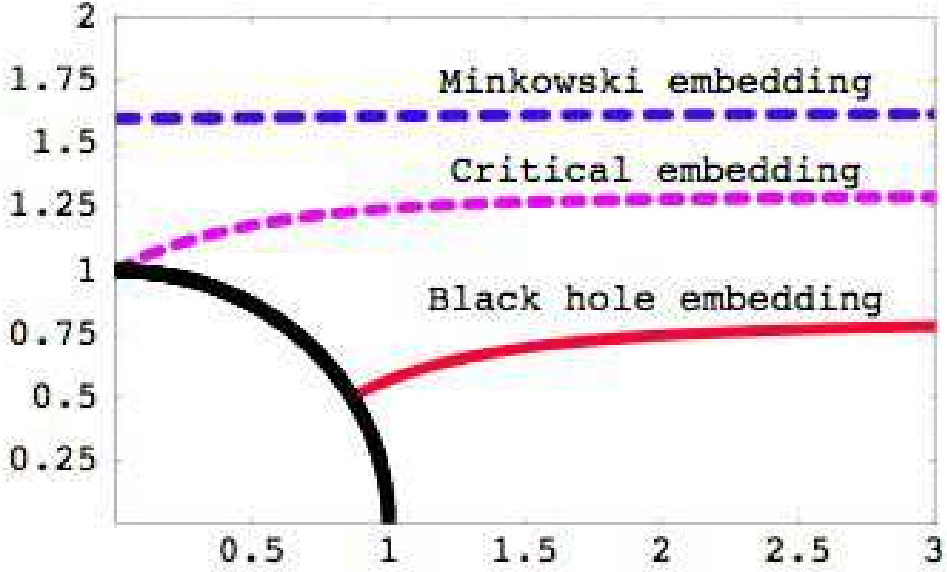}
 \caption{
   \label{fig:schematicEmbeddings}
   Schematic sketch of the three different kinds of embeddings which solve the background equations of motion
   at vanishing charge density and potential.
   This figure has been kindly provided by the authors of~\cite{Mateos:2007vn}.
 }
\end{figure}

\begin{figure}[h!]
 \includegraphics[width=0.95\textwidth]{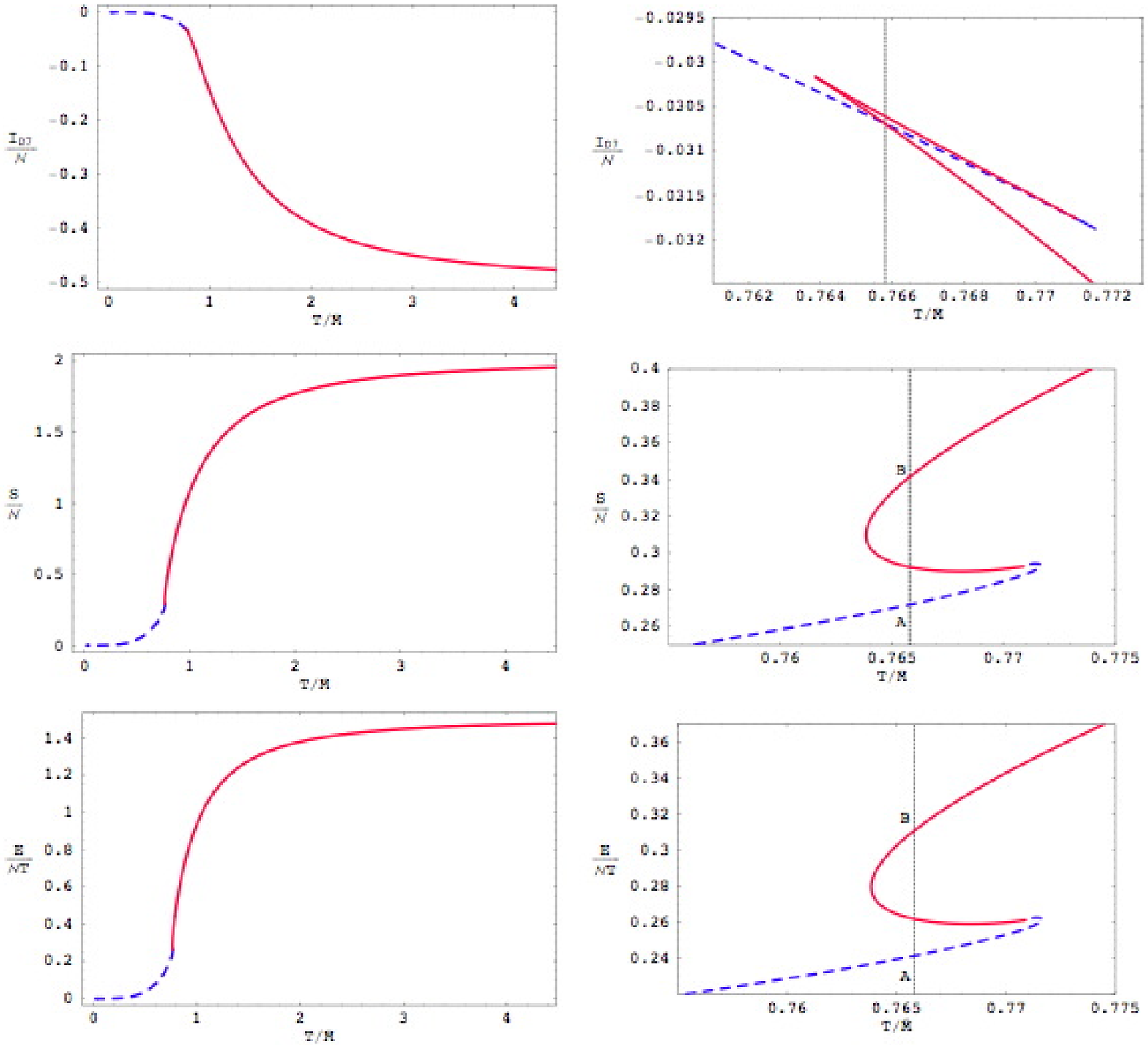}
 \caption{
   \label{fig:pureD7Thermo}
   The free energy, entropy and internal energy are shown as functions of the scaled temperature at vanishing charge 
   density and potential.
   This figure has been kindly provided by the authors of~\cite{Mateos:2007vn}.
 }
\end{figure}

\begin{figure}[h!]
 \includegraphics[width=0.95\textwidth]{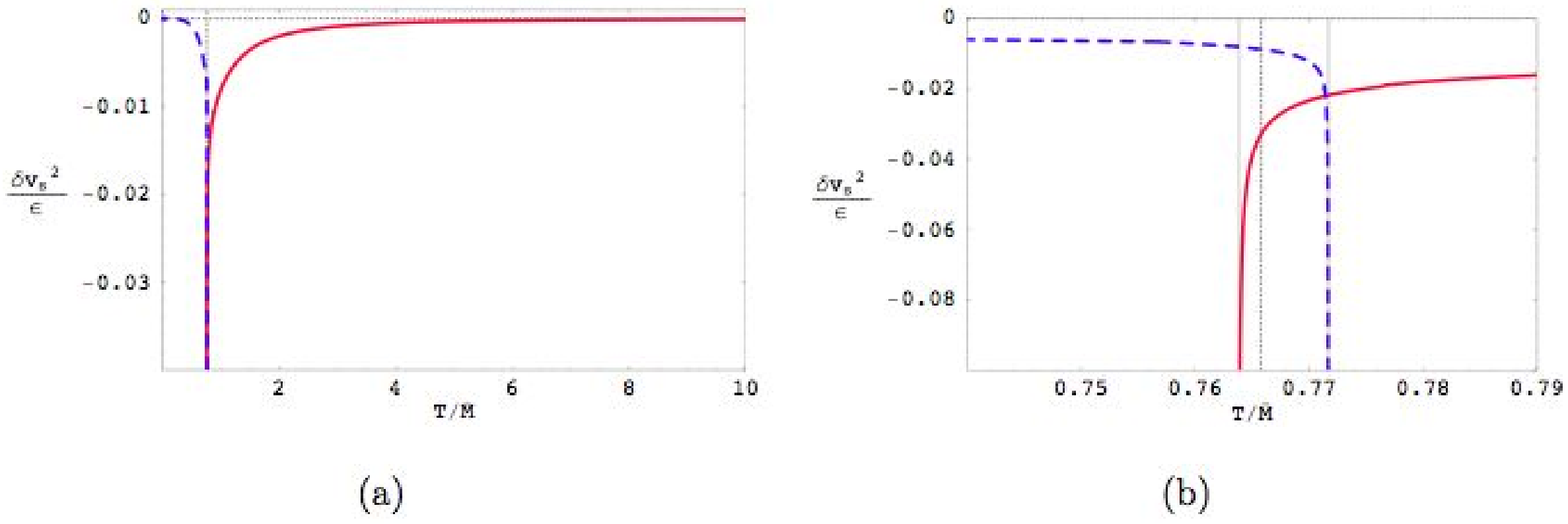}
 \caption{
   \label{fig:pureD7Vs}
   The speed of sound shown as a function of the scaled temperature at vanishing charge density and potential. 
   This figure has been kindly provided by the authors of~\cite{Mateos:2007vn}.
 }
\end{figure}

{\bf Brane thermodynamics at vanishing charge density and potential}
In order to understand the dual gauge theory thermodynamics of this gravity setup we have just built up,
let us take one step back and choose the baryon density to vanish, i.e.~$\tilde d =0$. This setup was analyzed 
in~\cite{Mateos:2007vn} and we briefly review the results. 
The most prominent thermodynamic feature of the D3/D7-setup at vanishing charge densities is a phase transition
for the fundamental matter between a confined and a deconfined phase taking place at the temperature~$T_{\text{fund}}$. 
Dual to this we have a geometric transition as shown in figure~\ref{fig:schematicEmbeddings} on the gravity side of the 
correspondence from a {\it Minkowski phase} to a {\it black hole phase}, respectively. This means that at vanishing
density and potential depending on the parameter~$m$ Minkowski embeddings and black hole embeddings are
both present. Looking at the free energy~(cf.~figure~\ref{fig:pureD7Thermo}) of these 
configurations reveals that there are actually three different 
regions: one low-temperature region where only Minkowski-embeddings~(blue dotted line in figure~\ref{fig:pureD7Thermo}) 
are possible, one intermediate
region where both embeddings are possible but one is thermodynamically favored, and finally one
high-temperature region~($m > 0.92$) where only black hole embeddings~(red line in figure~\ref{fig:pureD7Thermo}) are present.
The intersection point of the branches with lowest free energy marks the phase transition near~$\bar M/T = 0.766$.
This transition of course is reflected in discontinuities and multi-valued regions 
in thermodynamic quantities such as the free energy~$F$, the entropy~$S$, the internal energy~$E$ and 
the speed of sound~$v_s$. The free energy, entropy and internal energy are shown for the D3/D7-setup in 
figure~\ref{fig:pureD7Thermo}. These quantities are computed using equations~\eqref{eq:actionFreeEnergy}
and~\eqref{eq:thermoRelFSE} as well as the {\it holographically renormalized}~(see section~\ref{sec:finiteTAdsCft}) 
D7-brane action.
Furthermore the speed of sound can be written as a sum of contributions from the D3 and D7-branes which we
expand in~$N_f/N_c$ keeping only the leading order 
\begin{equation}
{v_s}^2 = \frac{\S}{c_v} = \frac{\S_3+\S_7}{c_{v\,3}+c_{v\,7}} = \frac{1}{3} + \frac{\lambda N_f}{(12\pi) 2\pi N_c} 
   \left (m c + \frac{1}{3} mT\frac{\partial c}{\partial T}\right) + \dots \, ,
\end{equation}
with the parameter~$m$ which is related to the quark mass by~\eqref{eq:quarkMass} and the parameter~$c$ being related to the 
quark condensate by~\eqref{eq:quarkCondensate}.
The numerical result is shown in figure~\ref{fig:pureD7Vs}.

\begin{figure}[h!]
 \includegraphics[width=0.95\textwidth]{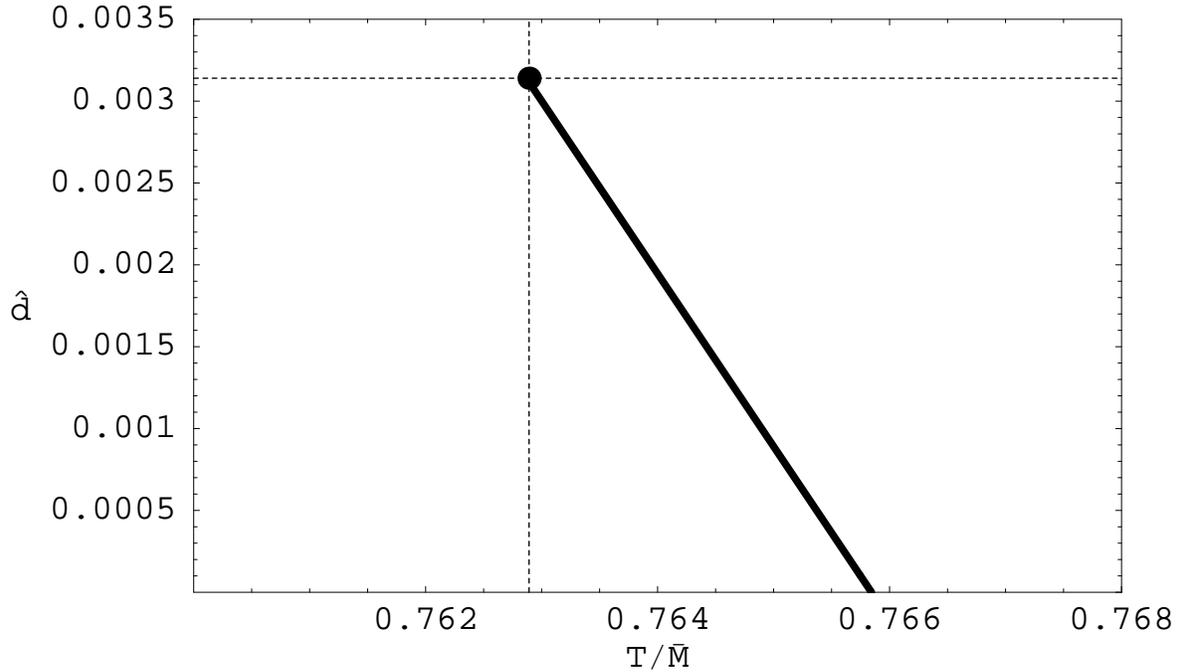}
 \caption{
   \label{fig:canPhaseDiagB}
   The phase diagram in the canonical ensemble for a system at finite baryon density. On the axes the scaled
   baryon density~$\tilde d$ is shown versus the scaled temperature~$T/\bar M$. 
   This figure has been kindly provided by the authors of~\cite{Kobayashi:2006sb}.
 }
\end{figure}

{\bf Brane thermodynamics at finite baryon density}
Now we consider a finite baryon density setup as described at the beginning of this section as was done in~\cite{Kobayashi:2006sb}.
This paragraph's title already states clearly that we are working in the canonical ensemble here fixing the baryon
density to a finite value and having the chemical potential as a thermodynamic variable. Looking at the
embeddings we find numerically in figure~\ref{fig:backgAt}, we observe that no Minkowski embeddings exist at finite
baryon density. In other words: there is always a thin spike reaching from the D7-branes down to the 
black hole horizon. This spike can be characterized more closely looking at the Legendre transformed
D7-action for embeddings with a very thin spike, i.e. in the limit~$\chi\to 1$ we find
\begin{equation}
S_{D7} \sim - n_q V_x \frac{1}{2\pi\alpha'}\int \dd t\, \dd \varrho\, \sqrt{-g_{00} (g_{44}+g_{\Theta\Theta}(\partial_4\Theta )^2))}\, ,
\end{equation}
which is the Nambu-Goto action for a bundle of fundamental strings with a density~$n_q$ stretching from the D7-brane
to the horizon. This means that in the canonical setup for non-zero baryon density we only have access to black hole embeddings.
We can only reach Minkowski embeddings in the case of vanishing baryon density~$n_q= 0$~(equivalently~$\tilde d =0$) 
while the chemical potential may be chosen arbitrarily. In contrast to this vanishing density case, in our setup developed
for finite baryon density,
a vanishing density also implies that the chemical potential vanishes~$\tilde \mu =0$ as seen from~\eqref{eq:backgroundAt}.
Note, that Minkowski embeddings are still possible but these always imply vanishing density. 
The system at finite baryon density features an apparent phase transition. The transition takes place from black hole embedings to 
other black hole embeddings which is different from the Minkowski to black hole transition at vanishing density. Furthermore
the black hole to black hole transition ceases to exist at a critical point in the phase diagram~\ref{fig:canPhaseDiagB} which
lies at~$(\tilde d^* = 0.00315,\, T^*_{\text{fund}}/\bar M = 0.7629)$. Later examinations in the grandcanonical ensemble
have shown that this black hole to black hole transition is not the thermodynamic process taking place in this region.
That is because there actually exists a mixed~(Minkowski and black hole) phase in the region around the transition line in 
figure~\ref{fig:canPhaseDiagB} and the mere black hole embeddings considered here do not give the thermodynamic ground 
state of the system. Therefore the transition takes place between a black hole and a (possibly) mixed phase.

\begin{figure}[h!]
 \includegraphics[width=0.95\textwidth]{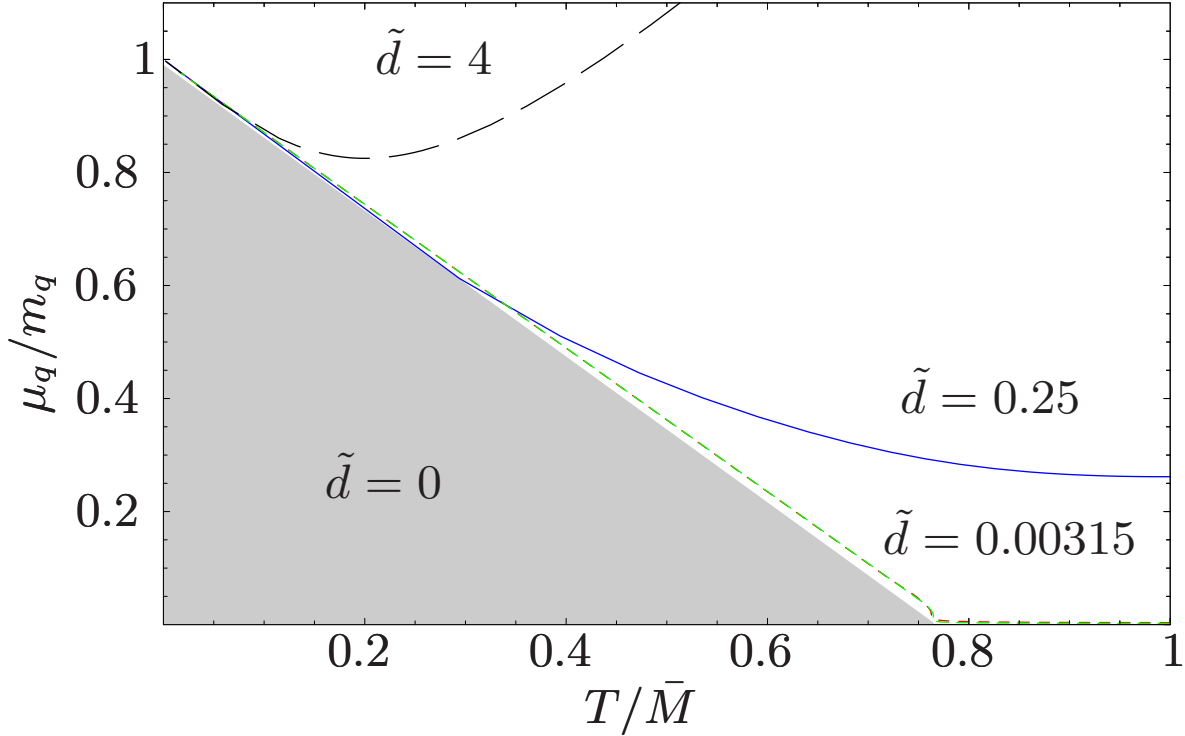}
 \caption{
   \label{fig:grandcanPhaseDiagB}
   The phase diagram in the canonical ensemble plotted against the variables of the grandcanonical ensemble. 
   On the axes the scaled chemical potential~$\mu_q/M_q$, with the quark mass~$M_q$
   is shown versus the scaled temperature~$T/\bar M$. This figure is taken from our work~\cite{Erdmenger:2007ja}.
 }
\end{figure}

\begin{figure}[h!]
 \includegraphics[width=0.95\textwidth]{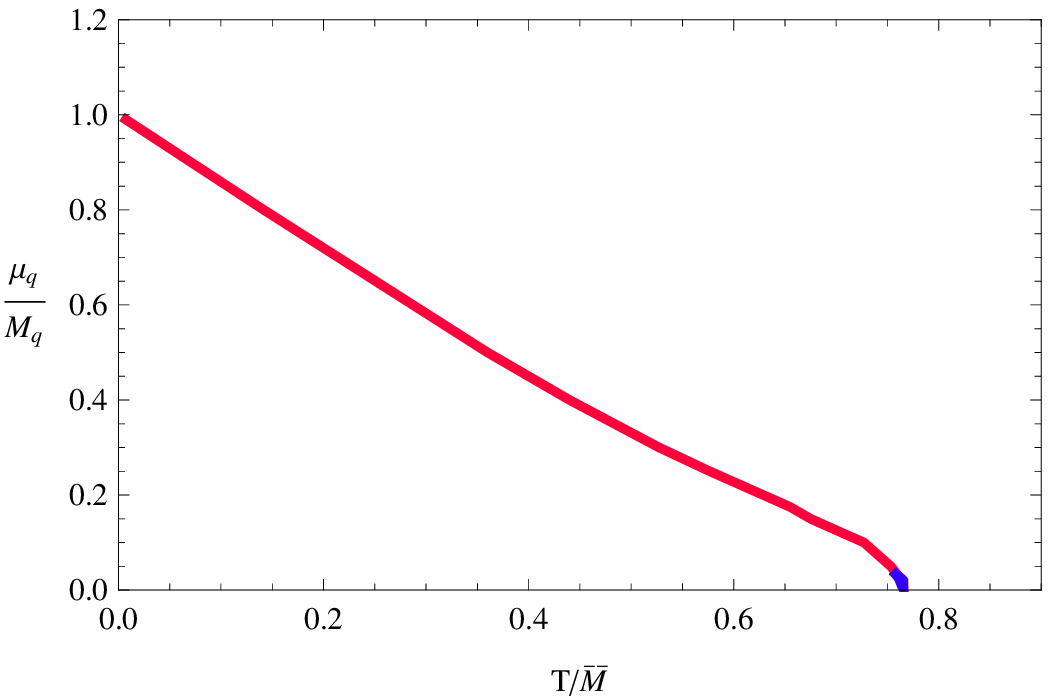}
 \caption{
   \label{fig:grandcanTransLine}
   The line of phase transitions in the grandcanonical ensemble for a system at finite baryon chemical potential. 
   On the axes the scaled chemical potential~$\mu_q/M_q$, with the quark mass~$M_q$ 
   is shown versus the scaled temperature~$T/\bar M$. 
   This figure has been kindly provided by the authors of~\cite{Mateos:2007vc}.
 }
\end{figure}

\begin{figure}[h!]
 \includegraphics[width=0.95\textwidth]{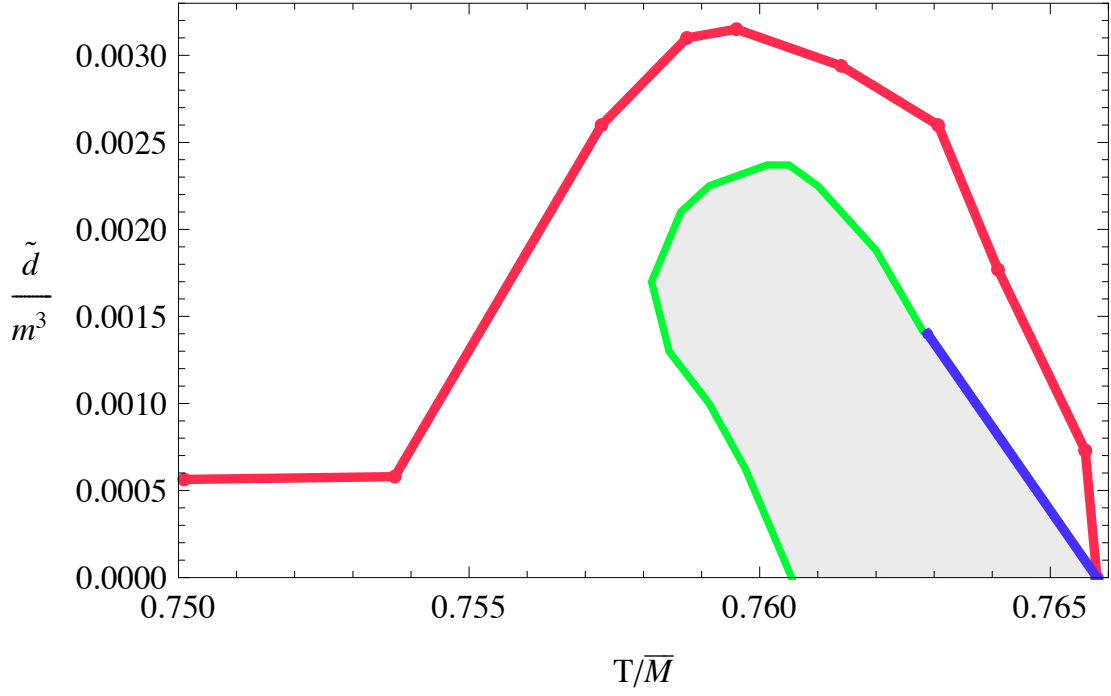}
 \caption{
   \label{fig:grandcanDVsTZoom}
   The phase diagram in the grandcanonical ensemble for a system at finite chemical potential. On the axes the scaled
   baryon density~$\tilde d$ is shown versus the scaled temperature~$T/\bar M$. 
   This figure has been kindly provided by the authors of~\cite{Mateos:2007vc}.
 }
\end{figure}

{\bf Brane thermodynamics at finite baryon chemical potential}
In order to understand the statements about the correct ground state and how to find the valid phase transition, let us
now turn to the grandcanonical ensemble. We fix the chemical potential to a finite value and consider the baryon
density as our thermodynamic variable. In figure~\ref{fig:grandcanPhaseDiagB} we have sketched the Minkowski with
vanishing density as a grey shaded region at small temperature and chemical potential. Meanwhile the black hole phase
with finite baryon density is shown in white. It is important to note here that the separation line between these
two grey and white regions does in principal not have to be identical with the line of phase transitions. 
Recall that in the canonical ensemble we have found, at least apparently, a black hole to black hole
transition, so this would be a white region to white region transition in the diagram~\ref{fig:grandcanPhaseDiagB}. The
line of phase transitions is not shown in figure~\ref{fig:grandcanPhaseDiagB} and
one has to determine it from looking at the free energy of all configurations that are possible at a given point~$(T, \mu)$ 
in the phase diagram. The resulting grandcanonical phase transition line is shown 
as the red line in figure~\ref{fig:grandcanTransLine}. 
In figure~\ref{fig:grandcanPhaseDiagB} we merely show some exemplary equal-density lines in
order to illustrate what region we are able to scan in the canonical ensemble. 
Figure~\ref{fig:grandcanDVsTZoom} shows the density-temperature phase diagram which follows from a
thorough examination of the system in the grandcanonical ensemble. The red line in figure~\ref{fig:grandcanDVsTZoom} 
shows the charge density which is computed along the line of transitions in the grandcanonical ensemble which
again is given by the red line in figure~\ref{fig:grandcanTransLine}.
Note that on the other side of the phase transition the density is zero and so in the grandcanonical ensemble the 
charge density jumps from zero to a finite density in this region and the intermediate densities under the red
curve in figure~\ref{fig:grandcanDVsTZoom} are not accessible. The blue line shows the line of 
black hole to black hole phase transitions which were found in the canonical ensemble~(cf.~figure~\ref{fig:canPhaseDiagB}).
The grey shaded region enclosed by blue and green lines shows a region where the present black hole embeddings
are unstable against fluctuations of baryon charge, i.e. the condition~$\partial n_q / (\partial \mu_q) |_T > 0$ is not
satisfied for these embeddings. Since both ensembles in the infinite volume limit are equivalent, we need to 
explain why there seem to be regions which one can only enter in the canonical ensemble but not in the grandcanonical
one. The idea here is that for the density-temperature values under the red curve in figure~\ref{fig:grandcanDVsTZoom}
the system stays in a mixed phase where both Minkowski and black hole phase are present. As an analog to this
we may recall that for example water features such a mixed phase in the transition from its liquid to its gaseous phase.
Note that the region of the mixed phase~(under the red curve in figure~\ref{fig:grandcanDVsTZoom}) is not identical with 
the region where unstable embeddings exist~(grey shaded region in figure~\ref{fig:grandcanDVsTZoom}).

Now we understand the statement that the black hole to black hole phase transition found in the canonical ensemble
is not realized. This is because that transition~(blue line in figure~\ref{fig:grandcanDVsTZoom}) lies entirely in the
mixed phase. Since in the canonical setup we considered the pure black hole phase to be the thermodynamic ground
state, those results can not be trusted in this particular region of the mixed phase. We would have to carry out our 
thermodynamic analysis with that mixed phase.

\subsection{Thermodynamics at isospin \& baryon density or potential} \label{sec:thermoB&I}
Here we consider an extension of the previous section where we worked at finite baryon backgrounds, i.e. we considered
only the~$U(1)$-part of the full $U(N_f)$~flavor group. Now we supplement this setup by switching on
a finite isospin background, i.e. the $SU(N_f)$-part of the flavor group, at the same time. The results presented here 
are my work in collaboration with Patrick Kerner~(cf.~\cite{Kerner:2008diploma}) and the results presented here 
are currently  to be published~\cite{Erdmenger:2008yj}. We have to develop a few 
new concepts and interpretations but the resulting calculations are analogous to those in section~\ref{sec:thermoBaryon}.

The main point of the previous section was to understand the phase diagram and thermodynamics of the gauge theory
with finite baryon density or potential which is dual to the D3/D7-brane setup in a non-extremal AdS-black hole background 
on the gravity side. We have learned in that simple example that we need to carry out holographic 
renormalization~(cf.~section~\ref{sec:finiteTAdsCft})
in order to get finite thermodynamics and we experienced that there may be unstable configurations or mixed phases
which force us to make use of the thermodynamic ensembles in a complementary way. 
That is so important because now we are going to use very similar embeddings and carry out the same thermodynamic
analysis for a thermal gauge theory when an isospin and baryon chemical potential~(or equivalently their conjugate densities) 
are switched on simultaneously. In principle we are free to compute thermodynamic quantities for any~$N_f$ but
since we will work numerically, we need to plug in definite numbers and for this purpose let
us confine our examination to the special isospin case~$N_f=2$. The generalization to arbitrary number of
flavors is accomplished in the next section~\ref{sec:hiNf}.

\subsubsection{Introducing baryon and isospin chemical potentials and densities}
Starting from the Dirac-Born-Infeld action
\begin{equation}
\label{eq:SBI}
S_{\text{B\&I}} = - T_{\text{D7}} \int \dd^8\xi \mathrm{Str}\{ \sqrt{\det (g + (2\pi\alpha') F_{\text{B\&I}})} \}\, ,
\end{equation}
with the baryon and isospin background gauge field
\begin{equation}
\label{eq:FBI}
{F_{\text{B\&I}}}_{\mu\nu} = \delta_{\mu 4}\delta_{\nu 0} 
  \left [ F^0_{40}T^0 +F^1_{40}T^1 +F^2_{40}T^2 +F^3_{40}T^3  \right]  \, ,
\end{equation}
with flavor group generators~$T^a, \, a = 1, 2, \dots , ({N_f}^2 -1)$.
Here we have assumed that the background gauge field~$A$ has its only component in time direction~$A_0$ and
that it only depends on the radial AdS-coordinate~$x_4=\varrho$. Therefore the only non-vanishing derivative
acting on the background gauge field is the radial one~$\partial_4 A_0\not = 0$, while~$\partial_{0,1,2,3,5,6,7} A_0 \equiv 0$.
In general the background field strength would be~$F^a_{\mu\nu} = 2\partial_{[\mu} A^a_{\nu]} + f^{abc} A^b_\mu A^c_\nu$, 
which with our assumptions becomes~$F^a_{\mu\nu} = \delta_{\mu 4}\delta_{\nu 0}\partial_{4} A^a_{0} + 
\delta_{\mu 0}\delta_{\nu 0}f^{abc} A^b_0 A^c_0$ and the second term vanishes because of the antisymmetry in
indices~$b,\,c$. The first term in~\eqref{eq:FBI} is the baryonic background already considered in the previous section.
The remaining three terms correspond to the three flavor directions~$a = 1,\, 2,\, 3$ in flavor space and the 
generators are~$T^a =  \sigma^a /2$ with the Pauli matrices which we complete by the identity~$\sigma^0$ in order to have 
a complete basis 
\begin{equation}
\sigma^0 = \left ( \begin{array}{c c} 1 & 0 \\ 0 & 1  \end{array} \right ) \, ,\quad 
\sigma^1 = \left ( \begin{array}{c c} 0 & 1 \\ 1 & 0  \end{array} \right ) \, ,\quad 
\sigma^2 = \left ( \begin{array}{c c} 0 & -i \\ i & 0  \end{array} \right ) \, ,\quad 
\sigma^3 = \left ( \begin{array}{c c} 1 & 0 \\ 0 & -1  \end{array} \right ) \, .
\end{equation}

Now we would like to find an exact solution for the background field and the D7-brane embedding and thus we rewrite the 
action~\eqref{eq:SBI}
\begin{eqnarray}
\label{eq:explicitSBI}
S_{\text{B\&I}}& = &- T_{\text{D7}} \int \dd^8\xi \mathrm{Str}\{ \sqrt{-\det g}\sqrt{\det ({1} + g^{-1} (2\pi\alpha') F_{\text{B\&I}})} \}\, , \\
  &=& -T_{\text{D7}} \int \dd^8\xi \mathrm{Str}\{ \sqrt{-\det g}\sqrt{\det [{1} + g^{00}g^{44} (2\pi\alpha')^2 ({F_{\text{B\&I}}}_{40})^2]} \}\, , \\
\end{eqnarray}
and we have performed the second step by calculating the determinant
\begin{eqnarray}
\det (g + (2\pi\alpha') F_{\text{B\&I}})&=&\left | \begin{array} {c c c c c c c c} 
g_{00}& 0 & 0 & 0 & (2\pi\alpha') {F_{\text{B\&I}}}_{40} & 0 & 0 & 0 \\
0 & g_{11} & 0 & 0 & 0 & 0 & 0 & 0 \\
0 & 0 & g_{22} & 0 & 0 & 0 & 0 & 0 \\
0 & 0 & 0 & g_{33} & 0 & 0 & 0 & 0 \\
-(2\pi\alpha'){F_{\text{B\&I}}}_{40}& 0 & 0 & 0 & g_{44} & 0 & 0 & 0 \\
0 & 0 & 0 & 0 & 0 & g_{55} & 0 & 0 \\
0 & 0 & 0 & 0 & 0 & 0 & g_{66} & 0 \\
0 & 0 & 0 & 0 & 0 & 0 & 0 & g_{77} \\
\end{array} \right |  \nonumber \\ 
  & = & g_{00} g_{11} g_{22} g_{33} g_{44} g_{55} g_{66} g_{77}  
  + g_{11} g_{22} g_{33}  g_{55} g_{66} g_{77} (2\pi\alpha')^2 ({F_{\text{B\&I}}}_{40})^2\, , \nonumber \\
  & = & \det g \left [ {1} +g^{00} g^{44} (2\pi\alpha')^2 ({F_{\text{B\&I}}}_{40})^2\right ] \, .
\end{eqnarray}

Making use of the spin-representation property~(Clifford algebra) of Pauli matrices
\begin{equation}
\{ \sigma^a, \sigma^b \} = 2 \delta^{ab} \, , 
\end{equation}
we evaluate the square of non-Abelian background gauge field strengths appearing in~\eqref{eq:explicitSBI} 
\begin{eqnarray}
\label{eq:FBI2}
({F_{\text{B\&I}}}_{40})^2 & = & 
  \left [ (F^0_{40})^2 + (F^1_{40})^2 +(F^2_{40})^2 +(F^3_{40})^2 \right ](\sigma^0 /2)^2 + \nonumber \\
  &&2 F^0_{40}\left [ F^1_{40}\sigma^1/2 + F^2_{40}\sigma^2/2 + F^3_{40}\sigma^3/2 \right ] (\sigma^0/2) \, .
\end{eqnarray}
Recall that we have~$F_{40} = -F_{04} = \partial_4 A_0$, so we do not have to take care of the structure constant term
or any commutator. Now we observe that all terms coupling different flavor representations~$\sigma^i \sigma^j\, ,\, i\not = j\, ,
i,j = 0,1,2,3$ are  proportional to the baryonic piece~$F^0$ and thus have the form~$F^0 \sigma^0 F^a \sigma^a\, ,\, a=1,2,3$. 
Thus the determinant simplifies to a sum in which the flavors are decoupled if we set the baryonic field to zero~$F^0_{40}\equiv 0$. 
Then for pure isospin background we have the action
\begin{eqnarray}
\label{eq:SI}
S_{\text{I}}& = & 
  -T_{\text{D7}} \int \dd^8\xi \mathrm{Str}\{ \sqrt{-\det g}\sqrt{\mathbbm{1}_{N_f\times N_f} 
  + (2\pi\alpha')^2 g^{00}g^{44}  \left [ (F^a_{40})^2 \right ](\sigma^0/2)^2  } \}\, , \nonumber \\
  & = & 
  -T_{\text{D7}} \int \dd^8\xi \mathrm{Str}\{ \mathbbm{1}_{N_f\times N_f} \}\sqrt{-\det g}  
  \sqrt{1 + \frac{(2\pi\alpha')^2}{4} g^{00}g^{44} [(F^a_{40})^2 ] }\, , \nonumber \\
  & = & 
  -T_{\text{D7}} N_f \int \dd^8\xi \sqrt{-\det g} \sqrt{1 + \frac{(2\pi\alpha')^2}{4} g^{00}g^{44}  
  [(F^1_{40})^2 + (F^2_{40})^2  + (F^3_{40})^2] }\, . \nonumber \\  
\end{eqnarray}
In this setup we can study how the three different charge densities or equivalently how the three components of the
chemical potential in flavor directions influence each other. We will elaborate on this in section~\ref{sec:diffusionMatrix}.

A slightly more complicated case emerges if none of the field-strengths vanishes~$F^i_{40}\not =0\, \forall \, i=0,1,2,3$
\begin{eqnarray}
S_{\text{B\&I}}& = & 
  -T_{\text{D7}} \int \dd^8\xi \mathrm{Str}\left\{ \sqrt{-\det g} \right . \nonumber \\
  &&\times \left. \sqrt{\mathbbm{1}_{N_f\times N_f} 
  + (2\pi\alpha')^2 g^{00}g^{44}  \left [ \frac{1}{4}((F^0_{40})^2 + (F^a_{40})^2)(\sigma^0)^2 
  + \frac{1}{2} F^0_{40} F^a_{40}\sigma^a \sigma^0 \right ] } \right \}\, . \nonumber \\ 
\end{eqnarray}
The complicating feature here is that one has to evaluate the square root of a sum of partly non-diagonal flavor
representations.
In order to simplify taking the square root inside this action we only consider the diagonal gauge representations~$\sigma^0$
which gives the baryonic part and~$\sigma^3$ which gives the isospin piece.
This is equivalent to turning the flavor coordinate system until our chemical potential points along the third isospin direction.
In this case we get the action
\begin{eqnarray}
S_{\text{B\&I3}}& = & 
  -T_{\text{D7}} \int \dd^8\xi \mathrm{Str}\left \{ \sqrt{-\det g} \right . \nonumber \\
  &&\times \left .\sqrt{\mathbbm{1}_{N_f\times N_f} 
  + \frac{ (2\pi\alpha')^2 g^{00}g^{44}}{4}  \left [ \underbrace{((F^0_{40})^2 + (F^3_{40})^2)}_{(F^{03})^2}(\sigma^0)^2 
  + {2} F^0_{40} F^3_{40}\sigma^3 \sigma^0 \right ] } \right \}\, . \nonumber \\ 
  &=& 
  -T_{\text{D7}} \int \dd^8\xi \mathrm{Str}\left \{ \sqrt{-\det g} \right . \nonumber \\
  && \times \left .\sqrt{ 
  \left( \begin{array}{c c}
   1 + \frac{ (2\pi\alpha')^2 g^{00}g^{44}}{4} [(F^{03})^2 +{2}F^0_{40}F^3_{40}] & 0 \\
   0 & 1 + \frac{ (2\pi\alpha')^2 g^{00}g^{44}}{4} [(F^{03})^2 - {2}F^0_{40}F^3_{40}]
  \end{array} \right ) } \right \}\, . \nonumber \\ 
  &=&
  -T_{\text{D7}} \int \dd^8\xi \sqrt{-\det g} \nonumber   
  \times\left [
  \sqrt{1 + \frac{(2\pi\alpha')^2 g^{00}g^{44}}{4} [(F^{03})^2 +{2}F^0_{40}F^3_{40}]} \right . \\ 
  &&\left . + \sqrt{1 + \frac{(2\pi\alpha')^2 g^{00}g^{44}}{4} [(F^{03})^2 -{2}F^0_{40}F^3_{40}]}
  \right ] \, . \nonumber \\ 
\end{eqnarray}
Note, that there is a term mixing the two flavor field strengths~$F^0,\, F^3$ in each of the two square roots.
Since we are interested in the equations of motion for the gauge fields appearing as~$F_{40}^i = \partial_4 A^i_0$,
we would end up with a set of coupled equations of motion for~$A^0_0$ and~$A^3_0$ if we simply 
applied the Euler-Lagrange equation to this action. In order to decouple the dynamics right here, we introduce
the rather obvious flavor combinations
\begin{equation}
\label{eq:flavorTrafoBackg}
X_1 = A^0_0 + A^3_0 \, , \quad X_2 = A^0_0 - A^3_0 \, ,
\end{equation}
which yields the action
\begin{eqnarray}
\label{eq:SBI3}
S_{\text{B\&I3}}& = &
  -T_{\text{D7}} \int \dd^8\xi \sqrt{-\det g} \left [
  \sqrt{1 + \frac{ (2\pi\alpha')^2 g^{00}g^{44}}{4} {\partial_4 X_1}^2} + \sqrt{1 + \frac{ (2\pi\alpha')^2 g^{00}g^{44}}{4} {\partial_4 X_2}^2}
  \right ] \, . \nonumber \\ 
\end{eqnarray}
Substituting in the explicit metric induced on the D7-brane~\eqref{eq:inducedMetric} gives
\begin{eqnarray} 
\label{eq:explicitSBI3}
S_{\text{B\&I3}}&=&-T_{D7}\int\,\dd^8\xi\frac{\sqrt{h_3}}{4}\rho^3f\tilde f(1-\chi^2) \nonumber \\
  && \times\Bigg(\sqrt{1-\chi^2+\varrho^2(\partial_\varrho\chi)^2-2(2\pi\alpha')^2\frac{\tilde f}{f^2}(1-\chi^2)(\partial_\varrho X_1)^2} 
  \nonumber \\
      &&   +\sqrt{1-\chi^2+\varrho^2(\partial_\varrho\chi)^2-2(2\pi\alpha')^2\frac{\tilde f}{f^2}(1-\chi^2)(\partial_\varrho X_2)^2}
\Bigg)\,.
\end{eqnarray}
These are just two summed up copies of the Abelian action given in~\eqref{eq:actionEmbeddingsAt} and in order to solve 
for the background gauge fields and for the brane embedding~$\chi$ we have to apply the same steps as 
in~\ref{sec:thermoBaryon} to each of the two terms. This means that we find two constant of motion~$d_1,\, d_2$ each of 
which is proportional to a certain flavor charge density. Legendre transforming the action in order to eliminate the 
fields~$X_1,\,X_2$ in favor of these constants~$d_1,\, d_2$, we obtain the action
\begin{equation}
\label{eq:legendreTSBI3}
\begin{split}
\tilde{S}_{\text{B\&I3}}&=S_{\text{B\&I3}}-\int\,\dd^8\xi \left ( X_1\frac{\delta S}{\delta X_1} + X_2\frac{\delta S}{\delta X_2} \right )\\
  &=-T_{D7}\int\,\dd^8\xi\frac{\sqrt{h_3}}{4}\varrho^3f\tilde f(1-\chi^2)\sqrt{1-\chi^2+\varrho^2(\partial_\varrho\chi)^2} \\
  & \left(\sqrt{1+\frac{8{d_1}^2}{(2\pi\alpha')^2T_{D7}^2\varrho^6\tilde f^3(1-\chi^2)^3}} 
    + \sqrt{1+\frac{8{d_2}^2}{(2\pi\alpha')^2T_{D7}^2\varrho^6\tilde f^3(1-\chi^2)^3}}\right)\,.
\end{split}
\end{equation}
And from this the equation of motion for the embedding function~$\chi$ can be deduced in the following form
\begin{equation}
\begin{split}
&\partial_\rho\left\{\rho^5f{\tilde f}(1-\chi^2)\frac{\partial_\rho\chi}{\sqrt{1-\chi^2+\rho^2(\partial_\rho\chi)^2}}
  \left(\sqrt{1+\frac{8{\tilde {d_1}}^2}{\rho^6{\tilde f}^3(1-\chi^2)}} + \sqrt{1+\frac{8{\tilde{d_2}}^2}{\rho^6{\tilde f}^3(1-\chi^2)}}\right)\right\}\\
  =&-\frac{\rho^3f{\tilde f}\chi}{\sqrt{1-\chi^2+\rho^2(\partial_\rho\chi)^2}}\Bigg\{[3(1-\chi^2)+2\rho(\partial_\rho\chi)^2]
  \left(\sqrt{1+\frac{8{\tilde {d_1}}^2}{\rho^6{\tilde f}^3(1-\chi^2)}} + \sqrt{1+\frac{8 {\tilde{ d_2}}^2}{\rho^6{\tilde f}^3(1-\chi^2)}}\right)\\
  &-\frac{24}{\rho^6{\tilde f}^3(1-\chi^2)^3}(1-\chi^2+\rho^2(\partial_\rho\chi)^2
  \left(\frac{{\tilde d_1}^2}{\sqrt{1+\frac{8{\tilde{d_1}^2}}{\rho^6{\tilde f}^3(1-\chi^2)}}} 
  + \frac{{\tilde d_2}^2}{\sqrt{1+\frac{8{\tilde{d_2}^2}}{\rho^6{\tilde f}^3(1-\chi^2)}}}\right)\Bigg\}\,.
  \label{eq:chiEomBI}
\end{split}
\end{equation}
This is the equation of motion we need to solve numerically for the embedding function~$\chi(\rho,\tilde d_1,\tilde d_2)$.
The boundary conditions on~$\chi$ are unchanged to those in the purely baryonic case~$\chi(\rho_H)=\chi_0$ 
and~$\chi'(\rho_H)=0$.

\subsubsection{Thermodynamic quantities}
Let us collect the numerical results for thermodynamic quantities graphically here. We will use a few 
meaningful parameter combinations to produce plots versus the mass to energy ratio in order to 
understand how the finite baryon and isospin densities influence the
quark condensate, the themodynamic quantities entropy, internal energy, free energy,
and the hydrodynamic quantity speed of sound.

Let us start out by identifying the string theory objects which produce the spike which is always present
if any of the two (baryon or isospin) densities is non-zero. In the spirit of the 'strings from branes' approach
reviewed in section~\ref{sec:thermoBaryon} we 
Legendre-transformed action as
\begin{equation}
\tilde{S}=-\frac{T_{D7}}{\sqrt{2}}\int\,\dd^8\xi\frac{f}{\sqrt{\ft}}\sqrt{1+\frac{\vrho^2(\del_\vrho\chi)^2}{1-\chi^2}}\left(\sum_{i=1}^{N_f}\sqrt{\frac{d_i^2}{(2\pi\alpha')^2T_{D7}}+\frac{\vrho^6\ft^3(1-\chi^2)^3}{8}}\right)\,.
\end{equation}
Note that $\chi=\cos\theta$, which becomes $\chi\simeq 1$ if the embedding is very near to the axis. Therefore, the second factors in the square roots can be neglected and we get
\begin{align}
\tilde{S}&=-\frac{V_x\text{Vol}(S^3)}{2\pi\alpha'}\sum_{i=1}^{N_f}d_i\int\,\dd t\dd\rho \frac{f}{\sqrt{2\ft}}\sqrt{1+\frac{\vrho^2(\del_\vrho\chi)^2}{1-\chi^2}}\nonumber\\
&=-\frac{V_x\text{Vol}(S^3)}{2\pi\alpha'}\sum_{i=1}^{N_f}d_i\int\,\dd t\dd\vrho\sqrt{-g_{tt}(g_{\rho\rho}+g_{\theta\theta}(\del_\vrho\theta)^2)}\,.
\end{align}
Recognize the fact that the result above can be written as the Nambu-Goto action for a bundle of strings stretching in $\rho$ direction but free bending in the $\theta$ direction
\begin{equation}
\tilde{S}=-V_3\text{Vol}(S^3)\left(\sum_{i=1}^{N_f}d_i\right)S_{NG}\, ,
\end{equation}
where~$V_3$ is the Minkowski space volume while~$\text{Vol} (S^3)$ gives the volume of the~$S^3$.

As we have learned in section~\ref{sec:finiteTAdsCft} we need to compute the counter-terms 
\begin{equation}
S_{\text{ct}}=-\frac{\mathcal{N} N_f}{4}((\vrho_{\text{max}}^2-m^2)^2-4mc)\,,
\end{equation}
which holographically renormalize the supergravity action. This renormalized Euclideanized action is then 
identified with the free energy~\eqref{eq:actionFreeEnergy}. Here~$\varrho_{\text{max}}$ is the~$UV$-cutoff and the 
factor~$\mathcal N$ is given by
\begin{equation}
\mathcal{N}=\frac{T_{D7}V_3\text{Vol}(S^3)\vrho_H^4}{4T}=\frac{\lambda N_c V_3T^3}{32}\,,
\end{equation}
where~$V_3$ again is the Minkowski space volume.

\begin{figure}[t]
\centering
\psfrag{m}{$m$}
\psfrag{c}{$c$}
\psfrag{F}{$\mathcal{F}$}
\psfrag{S}{$S$}
\psfrag{E}{$E$}
\psfrag{dvs2}{$\delta\mathcal{v}_s^2$}
\parbox{0.45\textwidth}{
\includegraphics[width=0.45\textwidth]{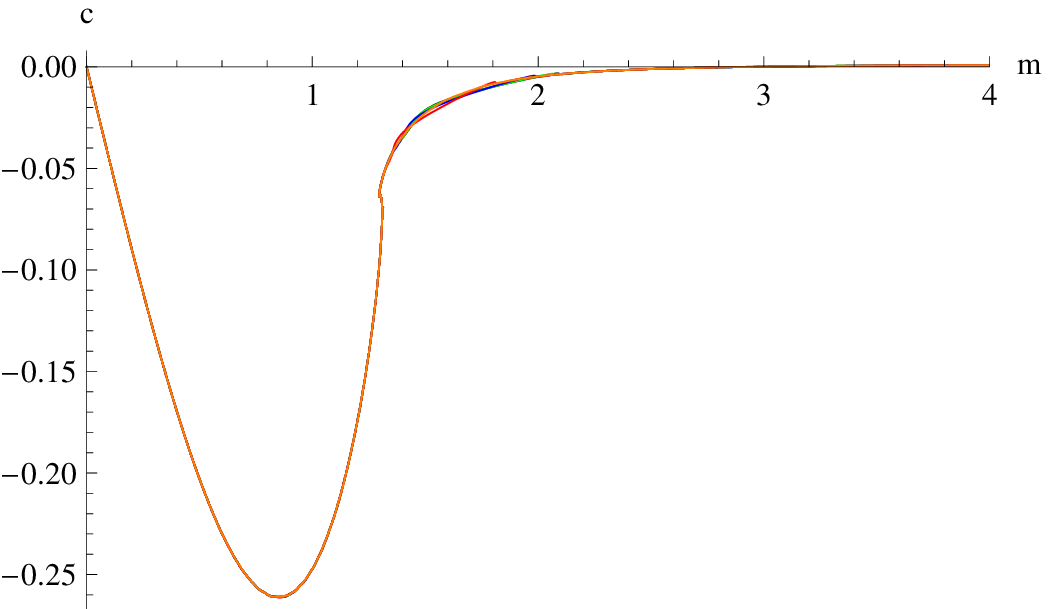}
}
\hfill
\parbox{0.45\textwidth}{
\includegraphics[width=0.45\textwidth]{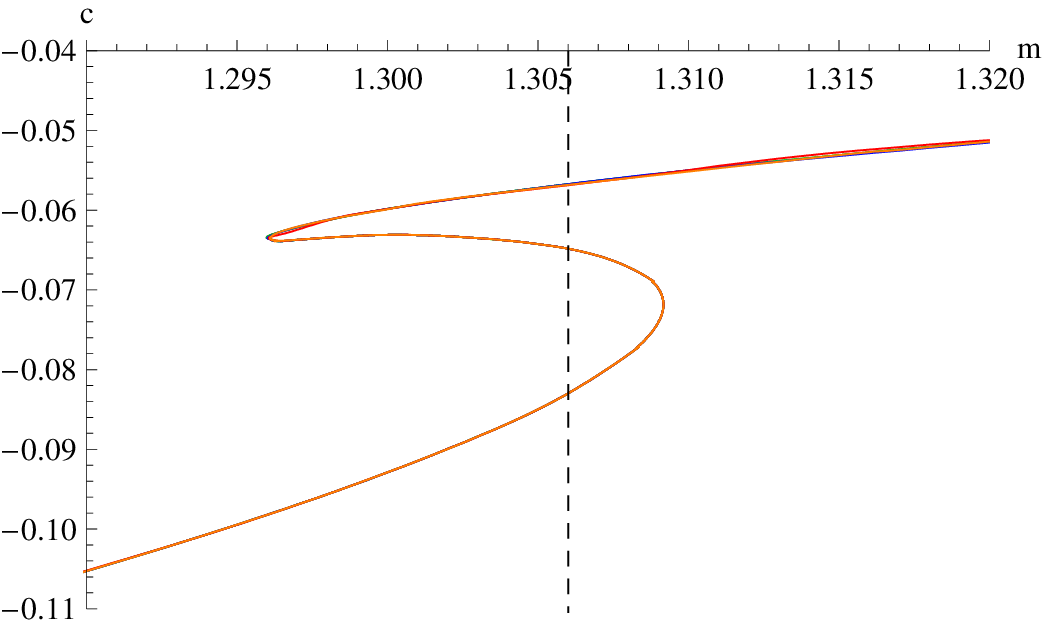}
}\\[5mm]
\parbox{0.45\textwidth}{
\includegraphics[width=0.45\textwidth]{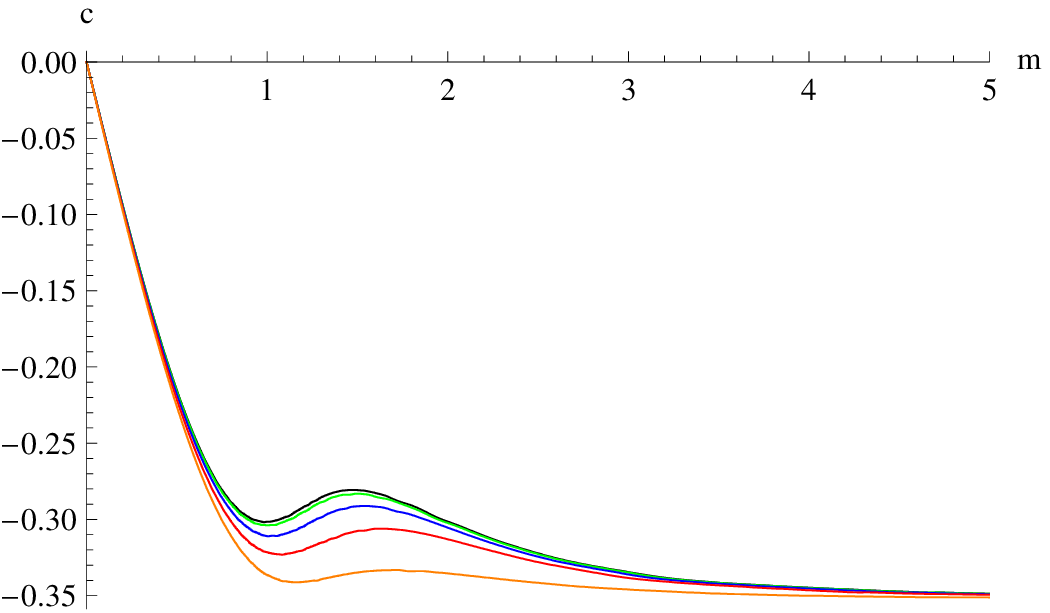}
}
\hfill
\parbox{0.45\textwidth}{
\includegraphics[width=0.45\textwidth]{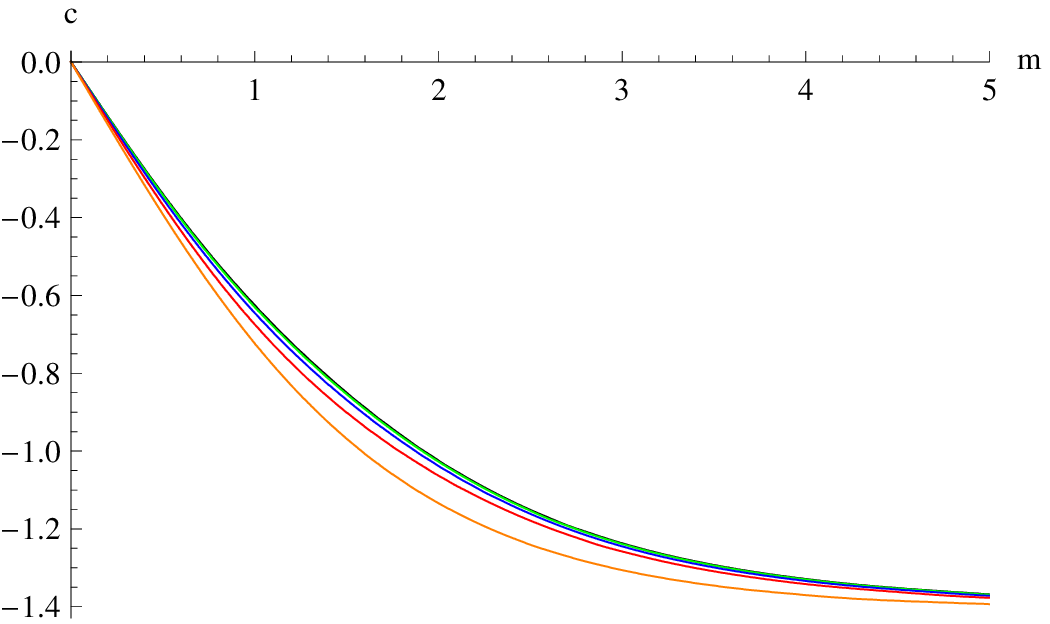}
}
\caption{
Numerical results in the canonical ensemble: The dependence of the quark
condensate on the scaled quark mass $m=\frac{2M_q}{\sqrt{\lambda}T}$ at baryon 
densities~$\tilde d^B = 0.00005$~(top left), the same value but zoomed into
the region near the black hole to black hole transition~(top right),
$\tilde d^B = 0.5$~(bottom left) and~$\tilde d^B = 2$~(bottom right). Differently colored curves in one plot show
distinct values of the isospin density in relation to the baryon density present:~$\tilde d^I = \tilde d^B$
in orange,~$\tilde d^I = 3/4 \tilde d^B$ in red,~$\tilde d^I = 1/2 \tilde d^B$ in blue,~$\tilde d^I = 1/4 \tilde d^B$
in green and~$\tilde d^I = 0$ in black. These plots were generated by Patrick Kerner~\cite{Kerner:2008diploma}. 
}
\label{fig:canCondensateBI}
\end{figure}

\begin{figure}[t]
\centering
\psfrag{dB}{$\tilde d^B$}
\psfrag{m}{$m$}
\psfrag{c}{$c$}
\psfrag{F}{$\mathcal{F}$}
\psfrag{S}{$S$}
\psfrag{E}{$E$}
\psfrag{dvs2}{$\delta\mathcal{v}_s^2$}
\parbox{0.45\textwidth}{
\includegraphics[width=0.45\textwidth]{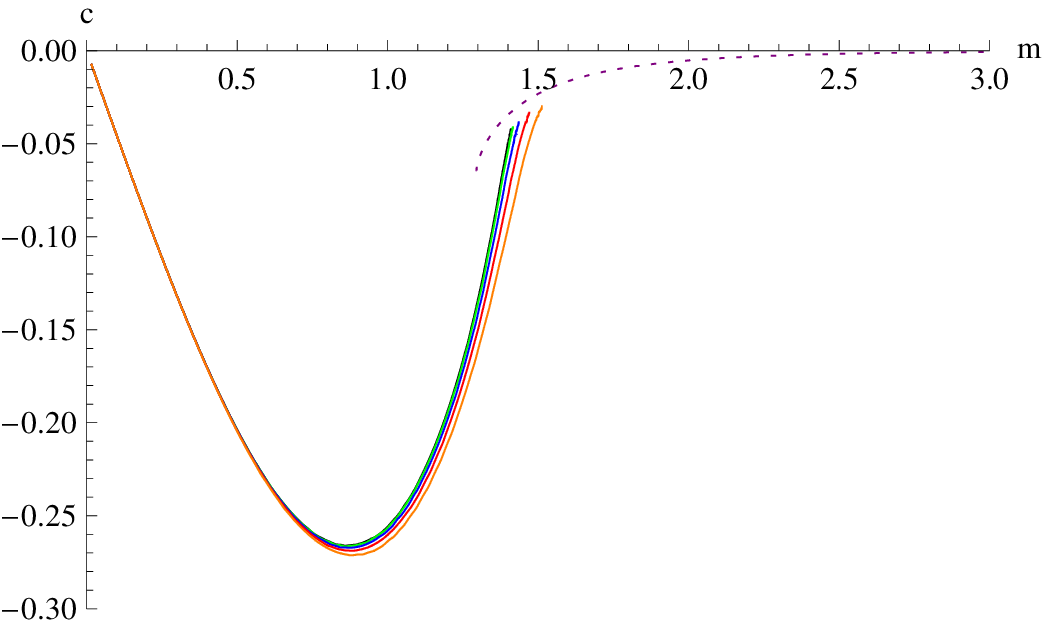}
}
\hfill
\parbox{0.45\textwidth}{
\includegraphics[width=0.45\textwidth]{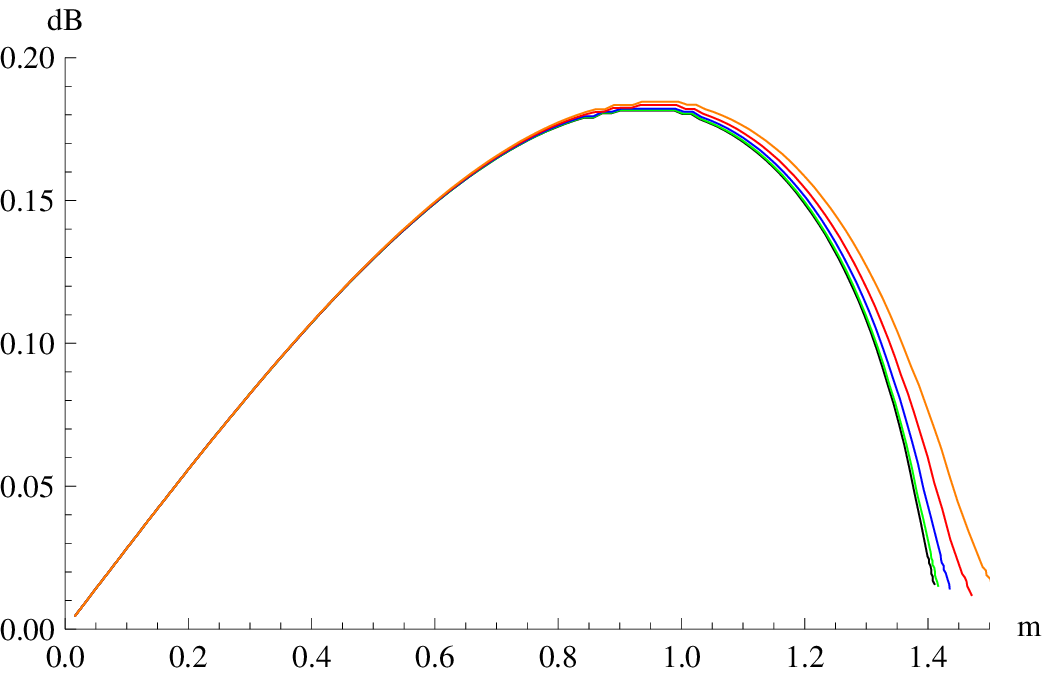}
}\\[5mm]
\parbox{0.45\textwidth}{
\includegraphics[width=0.45\textwidth]{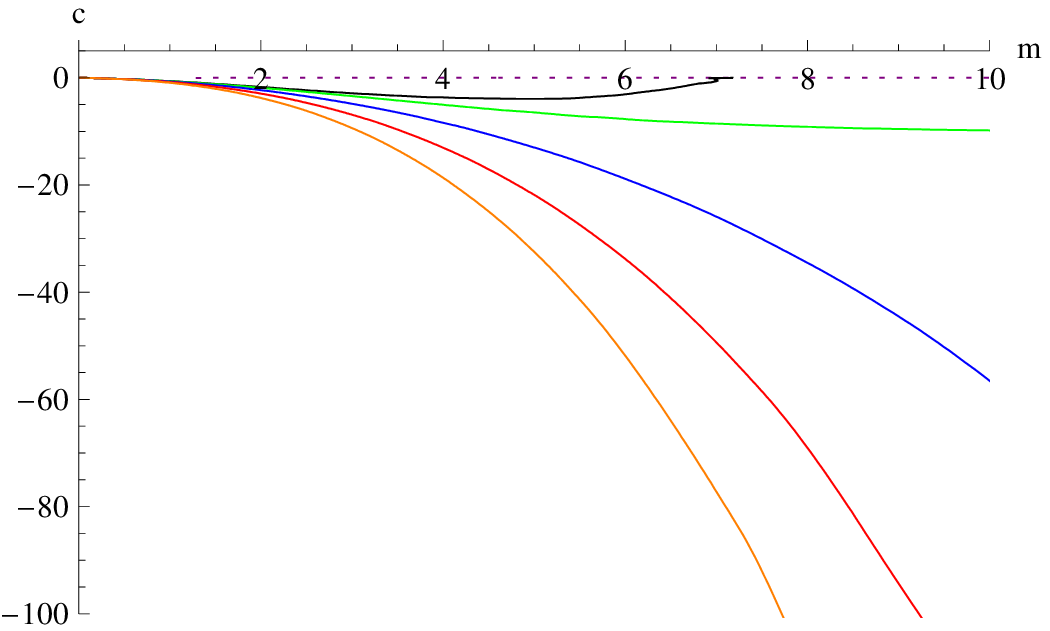}
}
\hfill
\parbox{0.45\textwidth}{
\includegraphics[width=0.45\textwidth]{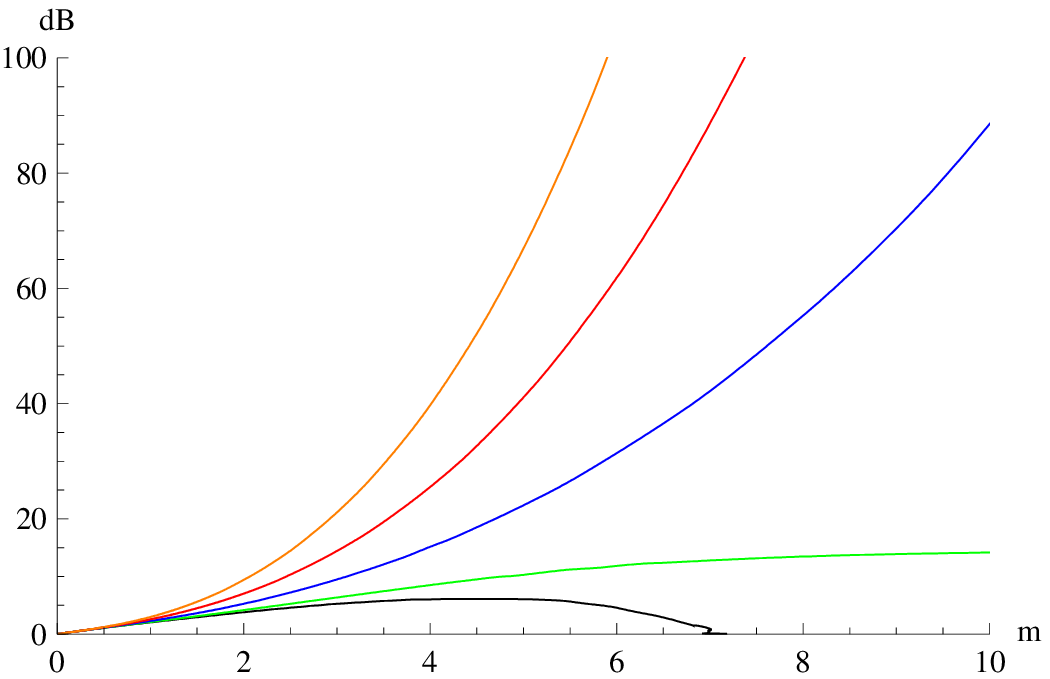}
}
\caption{
Numerical results in the grandcanonical ensemble: The dependence of the quark condensate on the scaled quark 
mass $m=\frac{2M_q}{\sqrt{\lambda}T}$ at baryon potentials~$\mu^B = 0.1 M_q$~(top) and~$\mu^B = 0.8 M_q$~(bottom) . 
Differently colored curves in one plot show distinct values of the isospin potential in relation to the 
baryon potential present:~$\mu^I/M_q = 0 \mu^B/M_q$~(black),~$\mu^I/M_q = \frac{1}{4} \mu^B/M_q$~(green),  
$\mu^I/M_q = \frac{1}{2} \mu^B/M_q$~(blue),~$\mu^I/M_q = \frac{3}{4} \mu^B/M_q$~(red), 
$\mu^I/M_q =  \mu^B/M_q$~(orange). The dotted purple curves correspond to Minkowski embeddings.
These plots were generated by Patrick Kerner~\cite{Kerner:2008diploma}.
}
\label{fig:grandcanCandDb}
\end{figure}

\begin{figure}[t]
\centering
\psfrag{m}{$m$}
\psfrag{c}{$c$}
\psfrag{F}{$\mathcal{F}$}
\psfrag{S}{$S$}
\psfrag{E}{$E$}
\psfrag{dvs2}{$\delta\mathcal{v}_s^2$}
\parbox{0.45\textwidth}{
\includegraphics[width=0.45\textwidth]{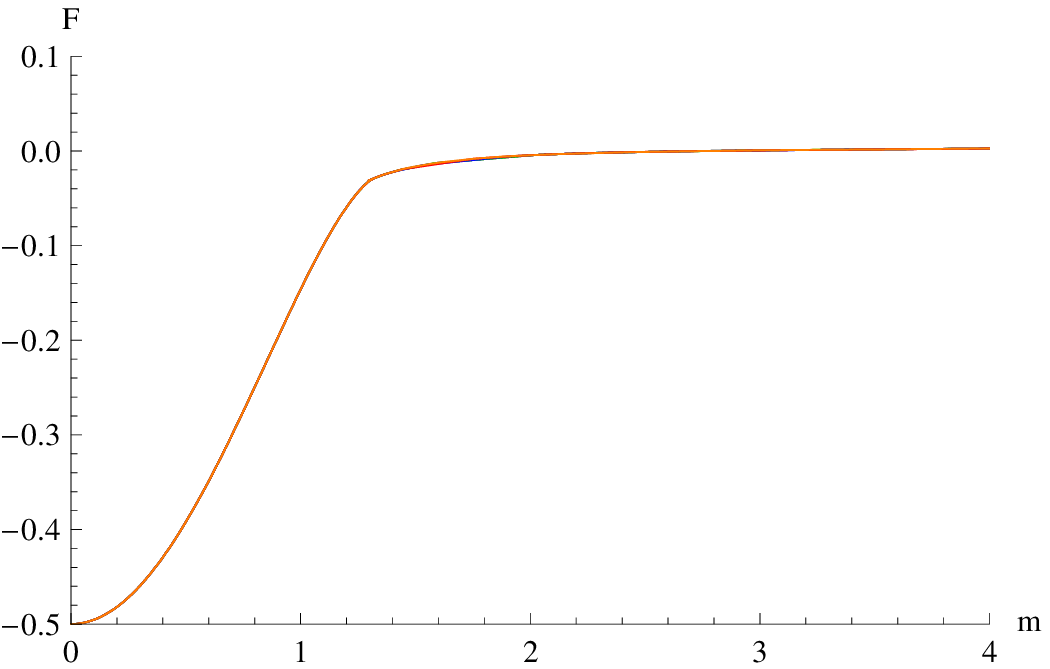}
}
\hfill
\parbox{0.45\textwidth}{
\includegraphics[width=0.45\textwidth]{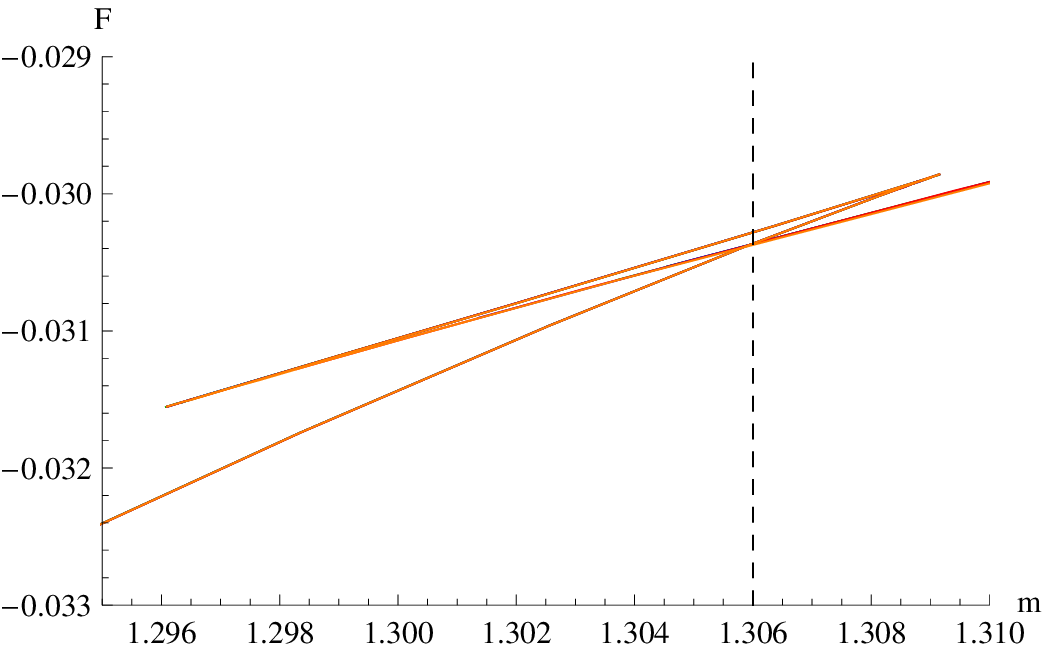}
}\\[5mm]
\parbox{0.45\textwidth}{
\includegraphics[width=0.45\textwidth]{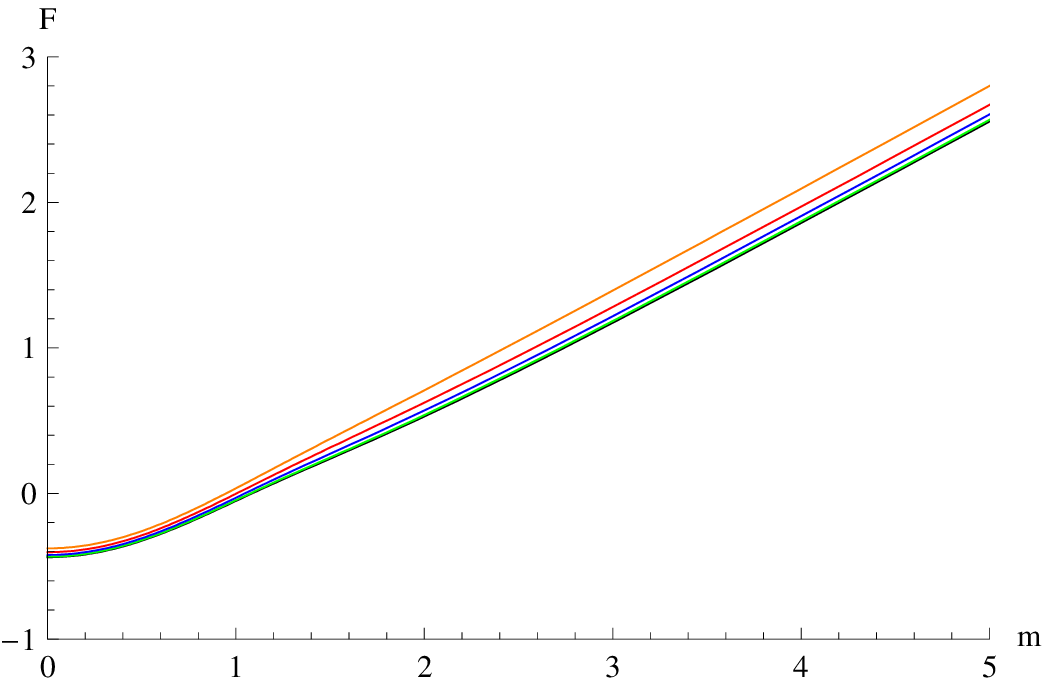}
}
\hfill
\parbox{0.45\textwidth}{
\includegraphics[width=0.45\textwidth]{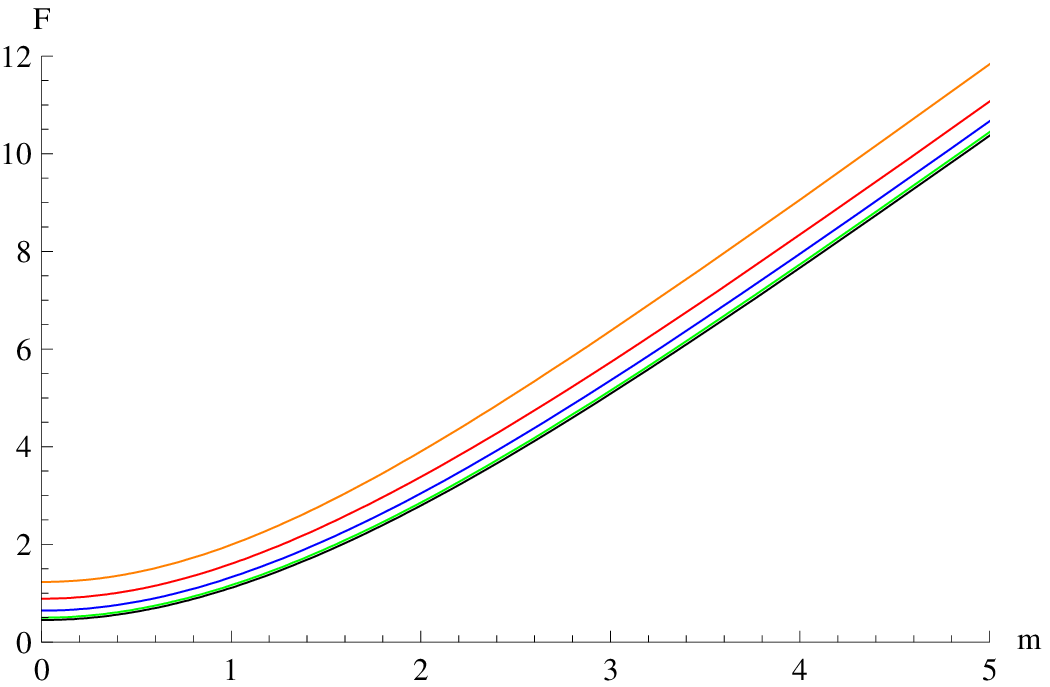}
}
\caption{
Numerical results in the canonical ensemble: The dependence of the free energy on the scaled quark 
mass $m=\frac{2M_q}{\sqrt{\lambda}T}$ at baryon densities~$\tilde d^B = 0.00005$~(top left), the same value but zoomed into
the region near the black hole to black hole transition~(top right),
$\tilde d^B = 0.5$~(bottom left) and~$\tilde d^B = 2$~(bottom right). Differently colored curves in one plot show
distinct values of the isospin density in relation to the baryon density present:~$\tilde d^I = \tilde d^B$
in orange,~$\tilde d^I = 3/4 \tilde d^B$ in red,~$\tilde d^I = 1/2 \tilde d^B$ in blue,~$\tilde d^I = 1/4 \tilde d^B$
in green and~$\tilde d^I = 0$ in black.These plots were generated by Patrick Kerner~\cite{Kerner:2008diploma}.
}
\label{fig:canFBI}
\end{figure}

\begin{figure}[t]
\centering
\psfrag{m}{$m$}
\psfrag{c}{$c$}
\psfrag{F}{$\mathcal{F}$}
\psfrag{S}{$\S$}
\psfrag{E}{$E$}
\psfrag{dvs2}{$\delta\mathcal{v}_s^2$}
\parbox{0.45\textwidth}{
\includegraphics[width=0.45\textwidth]{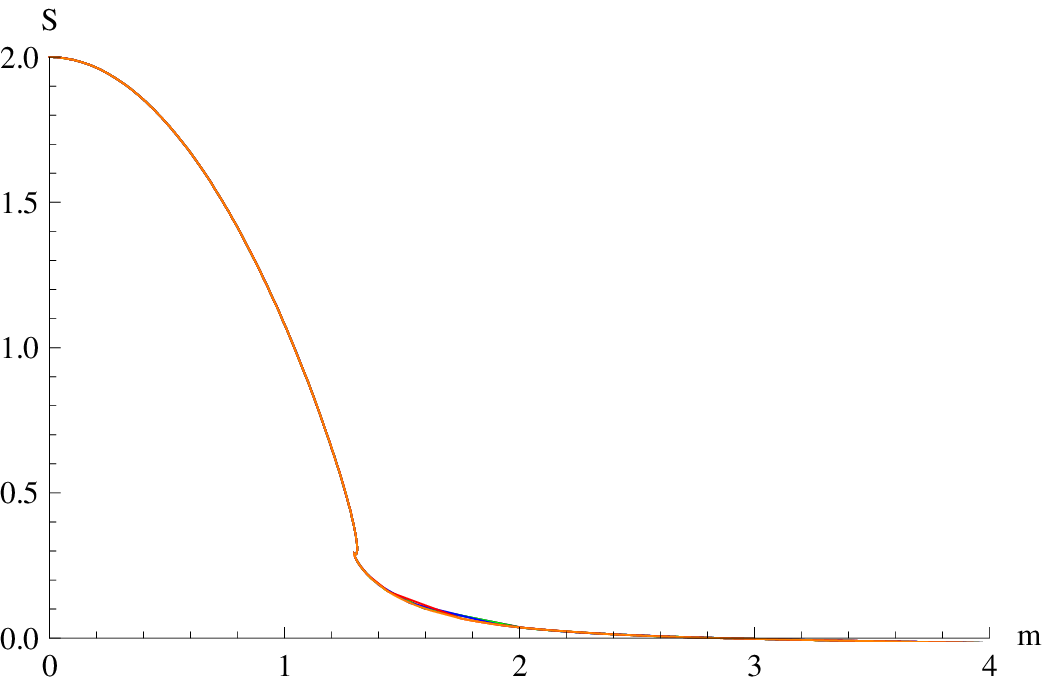}
}
\hfill
\parbox{0.45\textwidth}{
\includegraphics[width=0.45\textwidth]{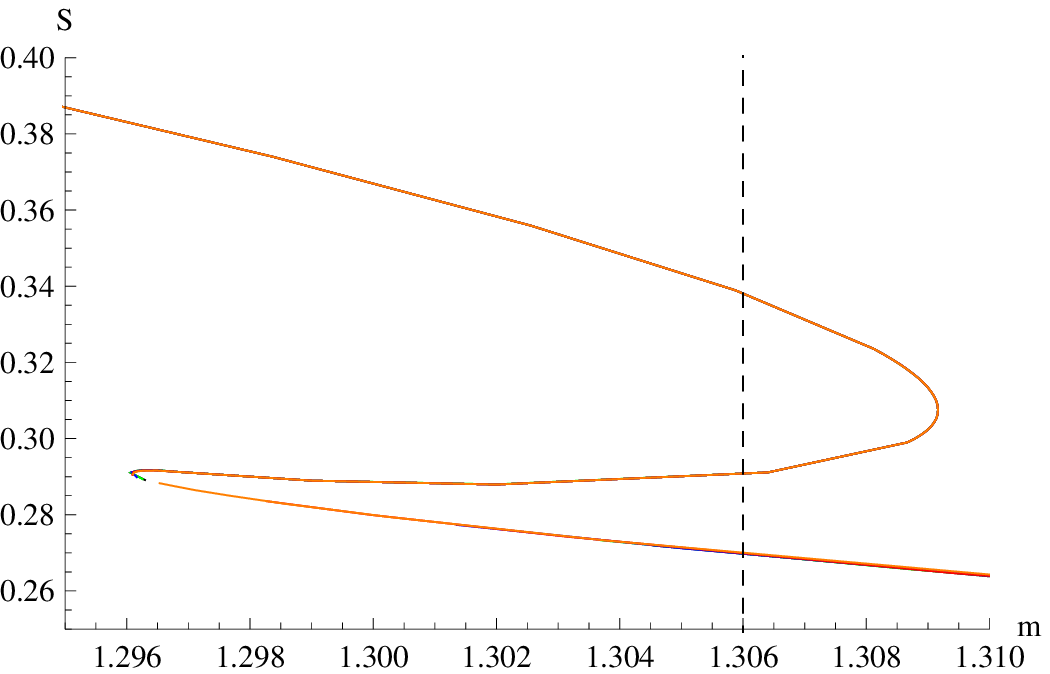}
}\\[5mm]
\parbox{0.45\textwidth}{
\includegraphics[width=0.45\textwidth]{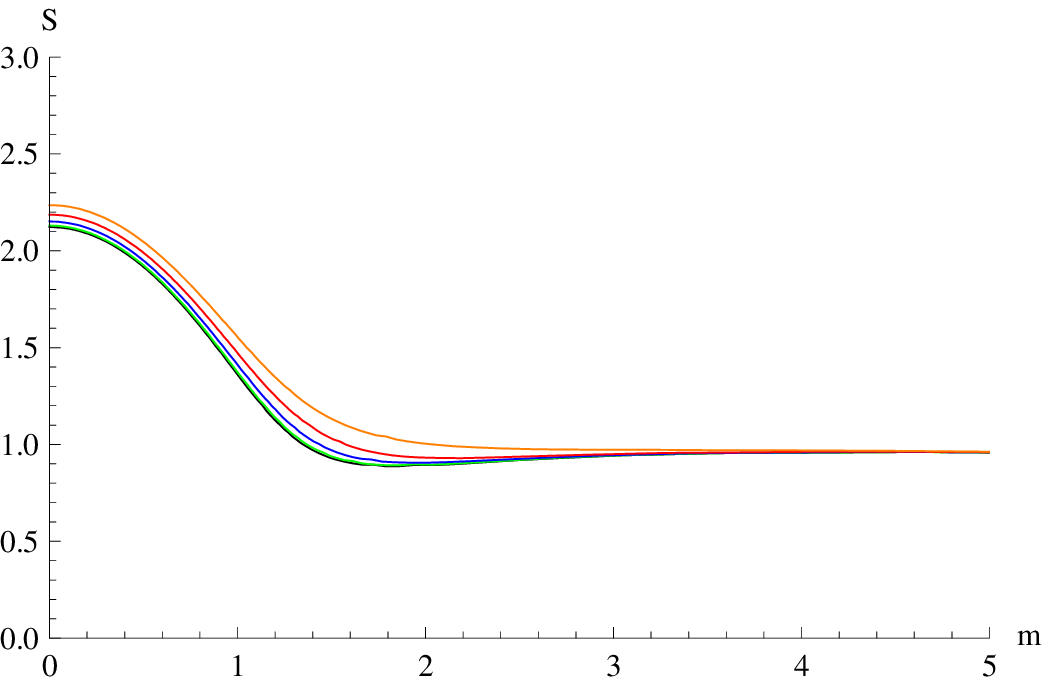}
}
\hfill
\parbox{0.45\textwidth}{
\includegraphics[width=0.45\textwidth]{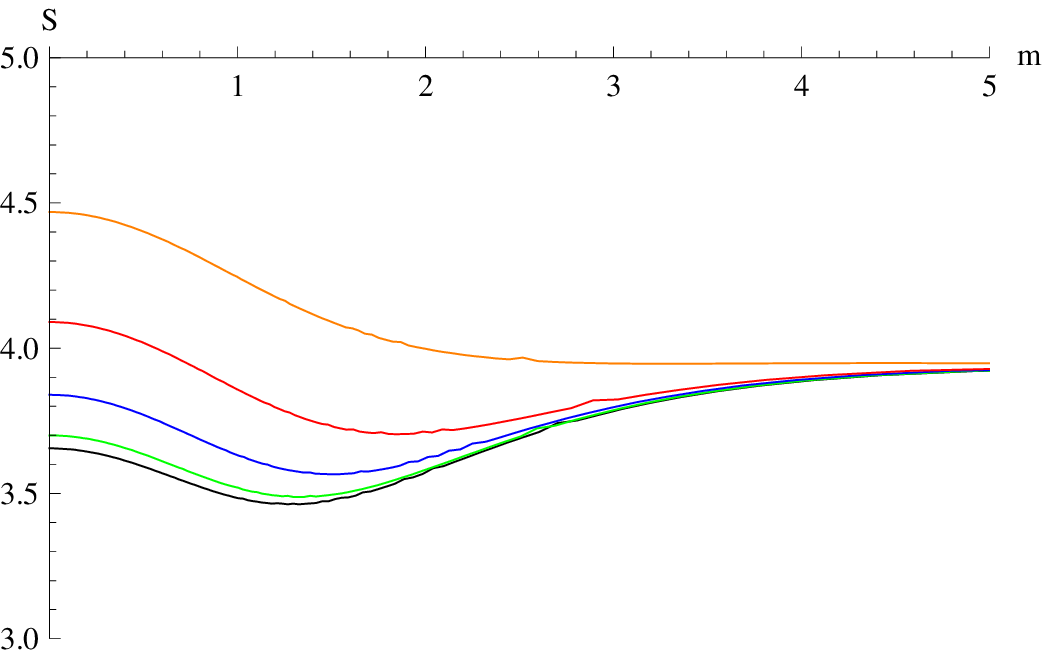}
}
\caption{
Numerical results in the canonical ensemble: The dependence of the entropy on the scaled quark 
mass $m=\frac{2M_q}{\sqrt{\lambda}T}$ at baryon densities~$\tilde d^B = 0.00005$~(top left), the same value but zoomed into
the region near the black hole to black hole transition~(top right),
$\tilde d^B = 0.5$~(bottom left) and~$\tilde d^B = 2$~(bottom right). Differently colored curves in one plot show
distinct values of the isospin density in relation to the baryon density present:~$\tilde d^I = \tilde d^B$
in orange,~$\tilde d^I = 3/4 \tilde d^B$ in red,~$\tilde d^I = 1/2 \tilde d^B$ in blue,~$\tilde d^I = 1/4 \tilde d^B$
in green and~$\tilde d^I = 0$ in black. These plots were generated by Patrick Kerner~\cite{Kerner:2008diploma}.
}
\label{fig:canEntropyBI}
\end{figure}

\begin{figure}
\centering
\psfrag{m}{$m$}
\psfrag{Omega}{$\Omega$}
\psfrag{c}{$c$}
\psfrag{F}{$\mathcal{F}$}
\psfrag{S}{$\S_7$}
\psfrag{E}{$E$}
\psfrag{dvs2}{$\delta\mathcal{v}_s^2$}
\parbox{0.45\textwidth}{
\includegraphics[width=0.45\textwidth]{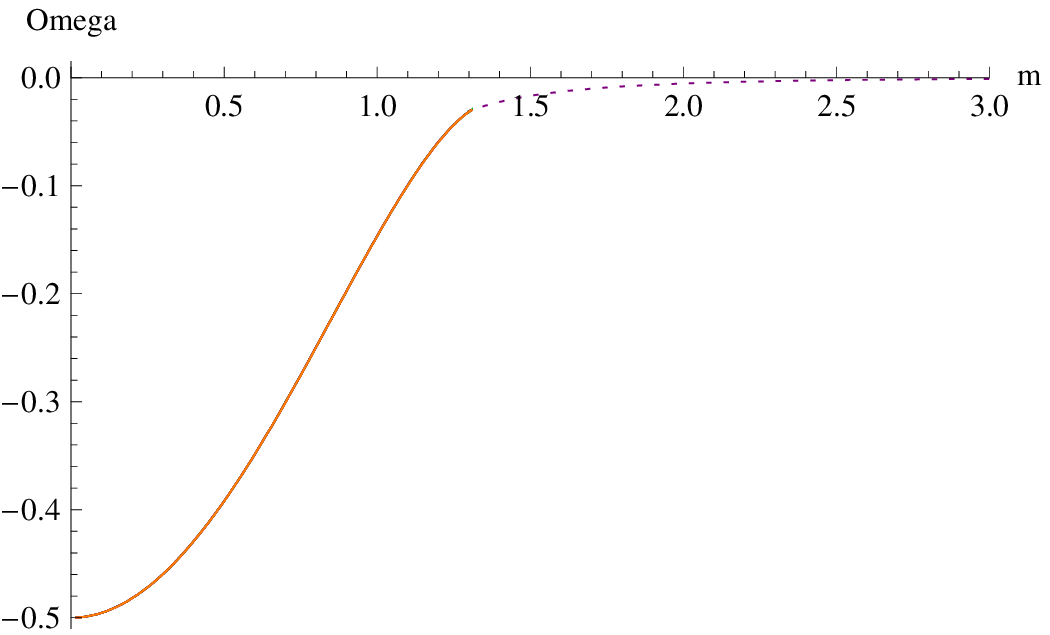}
}
\hfill
\parbox{0.45\textwidth}{
\includegraphics[width=0.45\textwidth]{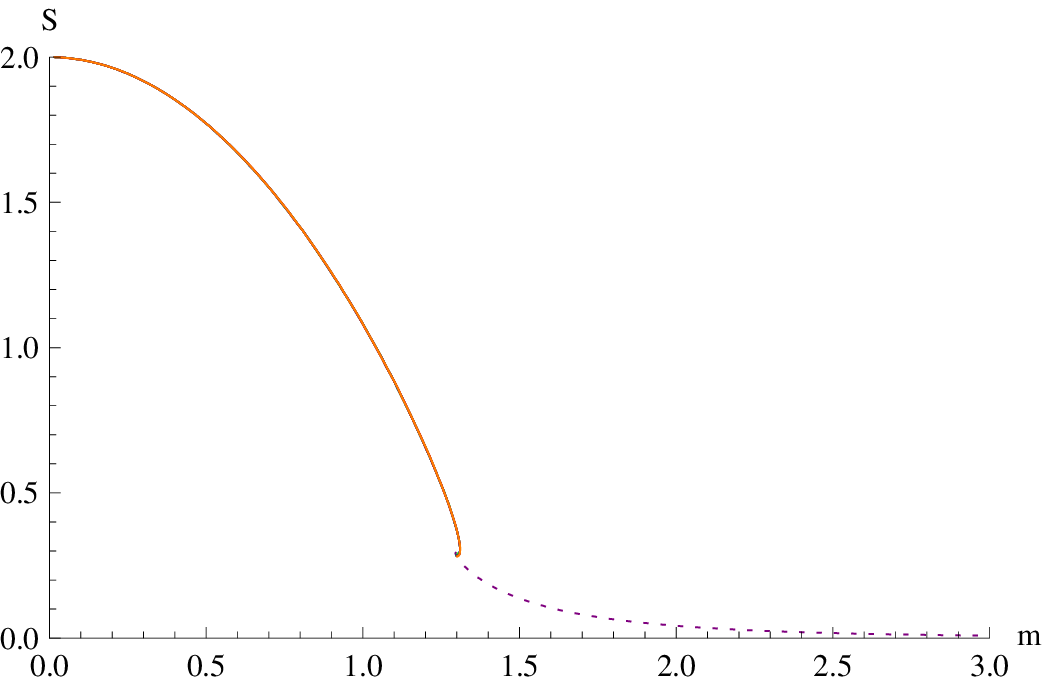}
}\\[5mm]
\parbox{0.45\textwidth}{
\includegraphics[width=0.45\textwidth]{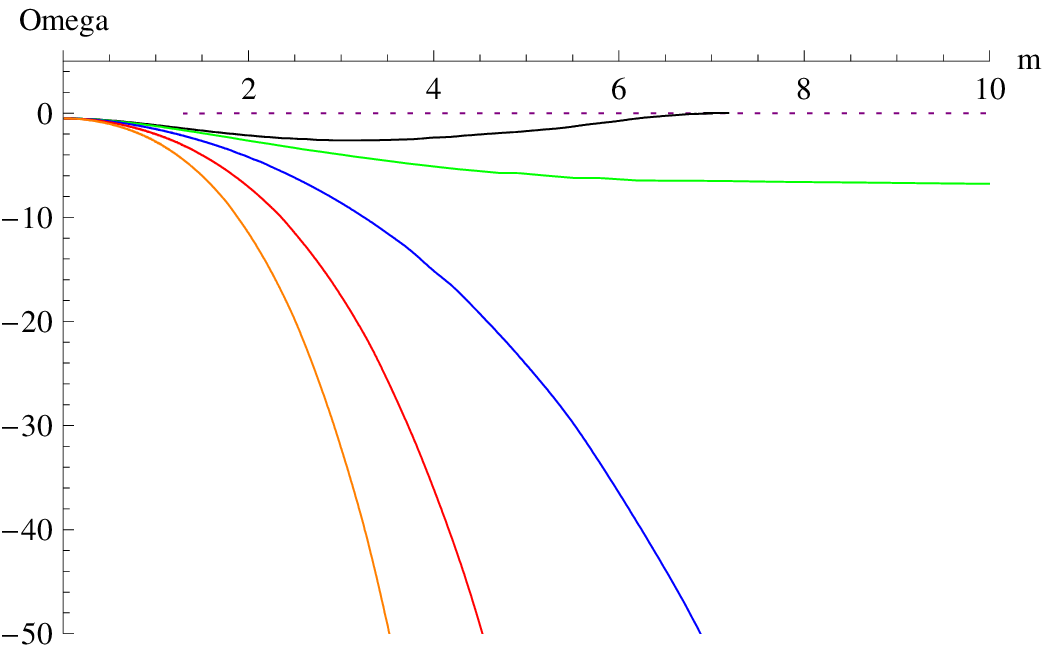}
}
\hfill
\parbox{0.45\textwidth}{
\includegraphics[width=0.45\textwidth]{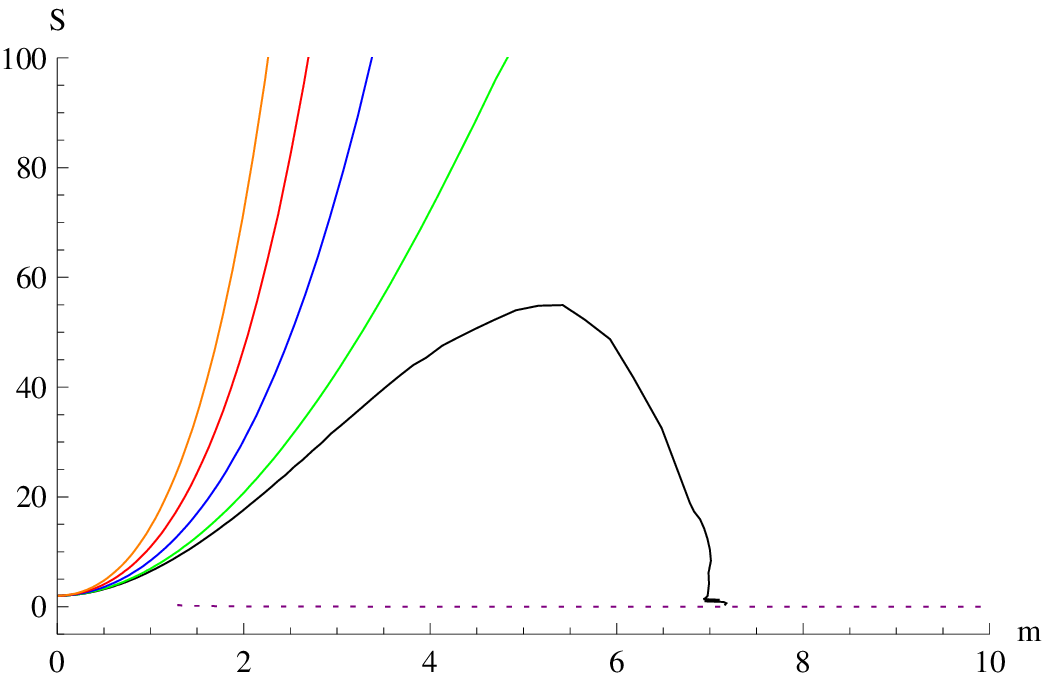}
}\\[5mm]
\caption{
Numerical results in the grandcanonical ensemble: The dependence of the grandcanonical potential~$\Omega$
and the entropy~$\mathcal{S}_7$ on the scaled quark mass $m=\frac{2M_q}{\sqrt{\lambda}T}$.
We have chosen $\tilde \mu_B/M_q = 0.01$ in the two plots on top and~$\tilde \mu_B/M_q = 0.8$ in the lower ones. 
Differently colored curves in one plot show distinct values of the isospin potential in relation to the 
baryon potential present:~$\mu^I = 0 $~(black),~$\mu^I = \frac{1}{4} \mu^B$~(green),  
$\mu^I= \frac{1}{2} \mu^B$~(blue),~$\mu^I = \frac{3}{4} \mu^B$~(red), 
$\mu^I=  \mu^B$~(orange). The dotted purple curves correspond to Minkowski embeddings. 
These plots were generated by Patrick Kerner~\cite{Kerner:2008diploma}. 
\label{fig:grandcanSandEBI}
}
\end{figure}

We have computed all thermodynamic quantities~(free energy, internal energy, entropy, speed of sound) in 
analogy to the case at vanishing densities~\cite{Mateos:2007vn}. 
In order to accomplish this we have made use of the thermodynamic relations given in 
equation~\eqref{eq:thermoRelFSE} and the equations following it. Nevertheless, here we only show selected quantities in order
to keep the overview. For details confer with~\cite{Erdmenger:2008yj} and~\cite{Kerner:2008diploma}.
Results in the canonical ensemble for the quark condensate are compared in figure~\ref{fig:canCondensateBI}, 
those for the entropy can be found in~\ref{fig:canEntropyBI},  
free energy in~\ref{fig:canFBI} 
. Results from the grandcanonical ensemble are displayed in figures~\ref{fig:grandcanSandEBI} 
and~\ref{fig:grandcanCandDb}.

\subsubsection{Discussion of numerical results}
As an analytical result we find an accidental symmetry in the numerical results which makes it possible to 
interchange baryon and isospin density. One result of
this is that the numerical embeddings are always black hole embeddings if either~$\tilde d^B\not =0$, or~$\tilde d^I\not =0$,
or both. Again these black hole embeddings mimic the behavior of Minkowski embeddings with a spike from the
brane to the horizon at small temperatures or large quark masses just like in the case with baryon density only.
The black hole to black hole phase transition found in the baryonic case continues to exist at finite isospin.
Nevertheless, there are some significant differences to the baryonic case showing in the quark condensate
and thermodynamical quantities upon introduction of isospin density or potential. In particular we find 
signatures of a new phase transition across the line of equal potential or density for isospin and baryon charge
resembling the phase diagram found in the case of 2-color QCD~\cite{Splittorff:2000mm}.

{\bf Condensates, chemical potentials and densities}
Figure~\ref{fig:canCondensateBI} shows the quark condensate~$c$ at different baryon densities. 
Different curves in the plots correspond
to different values for the isospin density in relation to the baryon density. The black curve is from now on always the
case with only baryon density. So in order to find out what the effect of isospin density is, we look for deviations from 
the black curves in all diagrams. We change the isospin density in quarter steps from~$\tilde d^I=0/4 \tilde d^B$ 
to~$ \tilde d^I=4/4 \tilde d^B$. Due to the accidental symmetry we can simply interchange~$\tilde d^B$ and~$\tilde d^I$
for all~$ \tilde d^I > \tilde d^B$ and we get the same pictures as for the case~$ \tilde d^I \le \tilde d^B$. 

At small~$\tilde d^B$ and~$\tilde d^I$ we still observe a phase transition between distinct
black hole embeddings~(see spiraling behavior in the top right plot in figure~\ref{fig:canCondensateBI}).
A look on the free energy diagram given in figure~\ref{fig:canFBI} confirms the existence of this 
transition near~$m=1.306$ where the branches of the free energy curve cross each other. Recall that this
is the phase transition discussed in the baryonic case which was found to be replaced by a transition from
the black hole phase to a mixed phase rather. We will study the dependence of the location of this transition on
isospin and baryon density below.
In the~$T\to 0$ limit any finite density breaks the supersymmetry and the chiral condensate asymptotes to a finite non-zero value.
We find that a larger baryon density produces a larger condensate in the limit~$T\to 0$. Furthermore we 
observe that the maximum appearing in the baryonic~(black) condensate curve in the bottom left plot from 
figure~\ref{fig:canCondensateBI}) vanishes with increasing isospin density. Adding larger and larger isospin density to 
the baryon density asymptotes to the case shown in the bottom right plot at large baryon density. Here 
the maximum has disappeared. In the limits~$T\to 0$ and~$T\to\infty$ introduction of isospin density does not
seem to have any effect on the condensate since all curves unify in these limits.

Calculating the baryon and isospin chemical potentials we find a discontinuity at the values~$\tilde d^B = \tilde d^I$.
We take this discontinuity as an indicator for the existence of a phase transition along the line~$\tilde d^B = \tilde d^I$.
In particular for~$\tilde d^B > \tilde d^I$ we find
\begin{equation}
\lim\limits_{m\to \infty} \mu^B = M_q\, ,\quad \lim\limits_{m\to \infty} \mu^I = 0\, .
\end{equation}
For the case~$\tilde d^B < \tilde d^I$ the accidental symmetry between baryon and isospin density allows to interchange
these two and we are back in the case we discussed before. Finally, in the crucial case~$\tilde d^B = \tilde d^I$ we 
can not distinguish between the two densities and both chemical potentials approach the same value
\begin{equation}
\lim\limits_{m\to \infty} \mu^B = \frac{M_q}{2}\, ,\quad \lim\limits_{m\to \infty} \mu^I = \frac{M_q}{2}\, .
\end{equation}
This means that the chemical potential has to change discontinuously when the case of equal densities is crossed
increasing or decreasing one of both densities.
We will discuss this phase transition further in~\cite{Erdmenger:2008yj} but we have indications that this transition is 
completely analogous to the one found in the condensates in the context of 2-color QCD~\cite{Splittorff:2000mm}.
Here we only collect more evidence for the transition from calculations in the grandcanonical ensemble.

In order to learn more about the structure of the isospin and baryon phase diagram, we investigate the setup in the
grandcanonical ensemble. Figure~\ref{fig:grandcanCandDb} shows the chiral condensate and the baryon density 
versus the mass parameter~$m$. The purple dotted curve in all grandcanonical plots shows the Minkowski embeddings
while the colored curves show results for different isospin chemical potentials and the black curve  always gives
the case of non-vanishing baryon chemical potential only. 

The condensate shows a discontinuity~(a gap) between the Minkowski and the black hole embeddings. Increasing
the baryon density the lower left plot in figure~\ref{fig:grandcanCandDb} shows that increasing the isospin density
there exist black hole embeddings for all values of~$m$, whereas the baryonic curve ends at a finite~$m$ where
the transition to Minkowski embeddings takes place. While the curves giving the baryon density 
(right column in figure~\ref{fig:grandcanCandDb}) for different
values of~$\mu^I$ have the same zero~$m$ limit, they split considerably increasing the mass parameter~$m$.
The isospin density shows a similar behavior except that the splitting between curves of different isospin potential is larger.
From the baryonic case we remember that we have no phase transition for~$\mu^B >M_q$~(compare the phase transition
line in figure~\ref{fig:canPhaseDiagB}). Looking at the case~$\mu^B=0.8 M_q$
with the orange~($\mu^I=\mu^B$), red~($\mu^I=3/4 \mu^B$) and blue~($\mu^I=1/2\mu^B$) curves in 
figure~\ref{fig:grandcanCandDb} we conclude from their monotonously ascending behavior that 
there is no phase transition for these combinations of potential values. In all these cases the sum of chemical potentials
satisfies~$(\mu^B+\mu^I) >M_q$ suggesting that compared to the baryonic case the same critical value for the 
phase transition to disappear exists, with the mere difference that the critical value~$M_q$ now has to be compared
to the sum of both chemical potentials. Since the black curve corresponds to~$(\mu^B +\mu^I)=(0.8 +0) M_q < M_q$
the black~(baryonic) curve shows a phase transition. Note that here the introduction and increase of isospin potential drives 
this system from a regime with a phase transition into a regime without a phase transition which is definitely a considerable
impact on the system. The condensate shows the same effect.

{\bf Thermodynamic quantities}
Coming to the thermodynamic quantities, we only mention a few exemplary points where the introduction of isospin has 
a significant impact on the quantity. The entropy in the canonical ensemble shows such an impact since the 
minimum present at vanishing isospin density in figure~\ref{fig:canEntropyBI} vanishes as the isospin density is increased. It 
is also worthwhile to note that in the large mass limit~$m\to\infty$ the baryonic entropy curve~(black) asymptotes to
zero while the finite densities generate entropy at any temperature or equivalently mass.

In the grandcanonical ensemble the entropy and internal energy have the same qualitative behavior shown in
figure~\ref{fig:grandcanSandEBI}. Similar to the condensate the purely baryonic curve in the black hole
phase~(black curve with~$\mu^I=0$) shows a maximum in entropy and energy near~$m=5$ before it 
ends near~$m=7$ and the system enters the Minkowski phase following the purple dotted line for larger 
mass parameter~$m$. Increasing the isospin chemical potential as in the condensate we 
see~(figure~\ref{fig:grandcanSandEBI}, bottom row) that the transition again vanishes since the system
remains in the black hole phase corresponding to the monotonously increasing entropy and energy curves.
This interpretation is confirmed by our studies~~\cite{Erdmenger:2008yj} of the grandcanonical potential shown 
in figure~\ref{fig:grandcanSandEBI}. 

\begin{figure}
\psfrag{m}{$m$}
\psfrag{dtB}{$\tilde d^B$}
\psfrag{dtI}{$\tilde d^I$}
\includegraphics[width=0.7 \textwidth]{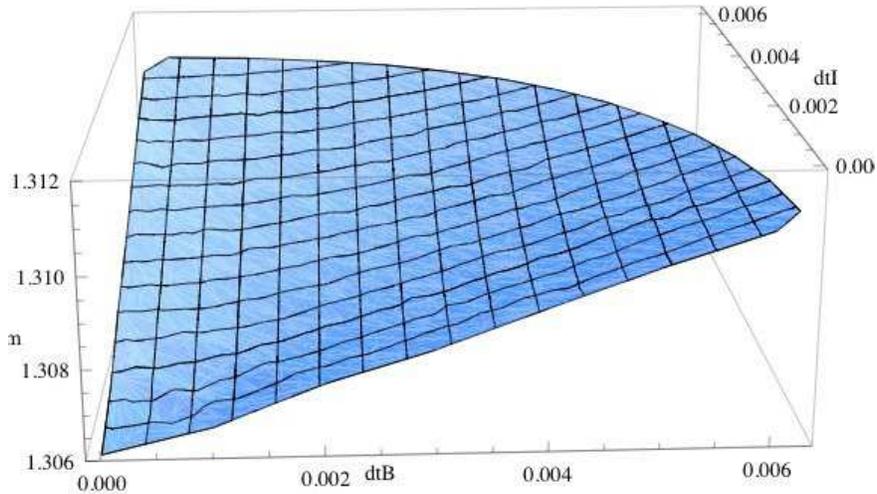}
\caption{\label{fig:canPhaseSurface} 
The location~$m$ of the black hole--black hole phase transition is shown as a surface over the baryon density -- isospin
density~($\tilde d^B$ --$\tilde d^I$)-plane. The approximately circular upper edge shows the line of
critical points where the transition vanishes. This plot was generated by Patrick Kerner~\cite{Kerner:2008diploma}.
}
\end{figure}

{\bf Black hole to black hole transition}
In figure~\ref{fig:canPhaseSurface} we trace the location of the black hole to black hole phase transition
in the volume spanned by baryon density~$\tilde d^B$,~isospin density~$\tilde d^I$ and the mass-temperature 
parameter~$m$. The result is a two-dimensional surface showing an apparent rotational~$SO(2)$-symmetry.
Note that we show only one quadrant since the accidental  symmetries between the charge densities 
mentioned earlier force the other three quadrants to be identical copies of this first one. The complete
phase transition surface would be nearly circular and finite since it terminates at the critical points on the upper edge.
A close study of the seemingly circular upper edge of this surface shows that the $SO(2)$-symmetry is actually broken.
This upper edge contains the critical points at which the phase transition disappears. An analysis of the 
inner region moving towards the origin we see that the surface asymptotes to being rotationally symmetric.

The phase transition line at finite baryon density only corresponds to the front edge~($\tilde d^I=0$) of the 
surface shown in figure~\ref{fig:canPhaseSurface}. Thus, together with the broken~$SO(2)$ symmetry
we conclude that the two differnent densities have actually a different effect than merely taking the 
baryon density to be larger. The broken symmetry shows a subtle interplay between isospin and baryon
density.

It would be interesting to study the stability of these phases~(or rather the stability of solutions in them). 
It is not impossible that the finite isospin also influences
the thermodynamics such that the baryonic black hole to mixed phase transition is qualitatively changed or it may not 
even be the favored transition anymore. We plan to study this in~\cite{Erdmenger:2008yj}.

The diffusive part of this system's hydrodynamics is examined in chapter~\ref{sec:transport}, section~\ref{sec:isospinDiffusion}.
We will extend the phase diagram from figure~\ref{fig:canPhaseSurface} there beyond the line of critical points tracing a 
minimum appearing in the diffusion
coefficient and claiming that this is a {\it hydrodynamic transition} being a softened version of the thermodynamical 
transition ending at the line of critical points. The rotational symmetry in that extended transition surface will be
obviously broken to a discrete~$\mathbf{Z}_4$-symmetry reflecting the accidental symmetries among the charges.

\subsection{Generalization to flavor number~$N_f>2$} \label{sec:hiNf}
In the previous section we restricted our study of the effects of a non-Abelian background gauge field on the thermodynamic 
quantities in a strongly coupled gauge theory on the case~$N_f=2$ for definiteness. In the present section we show how
this case can be systematically generalized to arbitrary flavor groups~$U(N_f > 2)$.

The first step to take is to find a generalization of the diagonal flavor representations which simplified taking the 
square root and the symmetrized trace over flavor representations in the example~$N_f=2$. Recall that there are
$(N_f-1)$~diagonal generators in a~$SU(N_f)$ which form the {\it Cartan subalgebra}. Inspired by the interpretation that
a diagonal generator of~$SU(N_f)$ should charge one brane differently with respect to all others, we write the diagonal
generators belonging to the Cartan algebra as
\begin{equation}
\lambda^i=\mathrm{diag}(1,\dots,\overbrace{-(N_f-1)}^{i\text{-th position}},\dots,1)\qquad i=1,\dots,N_f-1\,.
\end{equation}
For this choice of matrices the first flavor component is treated as the reference quantity to which all isospin charges are measured. 
We call the unity matrix for the baryonic part $\lambda^0$. Thus, we can generalize~$F^0_{40}\sigma^0 + F^3_{40} \sigma^3$ to
\begin{equation}
F^{N_f}_{\mu\nu}=F_{\mu\nu}=F^0_{\mu\nu}\lambda^0+\sum_{i=1}^{N_f-1}F^{i}_{\mu\nu}\lambda^i \, .
\end{equation}
Thus the general effective action for a geometry in which the background flavor field points along the diagonal directions only
then reads
\begin{equation}
\begin{split}
\label{eq:genSBI}
S_{\text{DBI}}&=-T_{D7}\int\,\dd\xi^8\text{Str}\left(\sqrt{|\det(g\lambda^0+2\pi\alpha'F)|}\right)\\
&=-T_{D7}\int\,\dd^8\xi\sqrt{-g}\text{Str}\left(\sqrt{\lambda^0+(2\pi\alpha')^2 g^{00}g^{44}(F_{40})^2}\right)\,,
\end{split}
\end{equation}
where in the second line the determinant is calculated. Since the action in~\eqref{eq:genSBI} is diagonal in the flavor space, we are able to evaluate the trace (for more details see~\cite{Kerner:2008diploma}). After a redefinition of the fields
\begin{equation}
X_0=A^0_0+\sum_{i=0}^{N_f-1}A^{i}_0\,,\qquad X_i=\sum_{j\not=i}A^{j}_0-(N_f-1)A^{i}_0\,,\quad i=1,\dots,(N_f-1)\,,
\end{equation}
where $X_i$ is the $i$-th component of the non-Abelian gauge field $A^{N_f}$, the non-Abelian DBI action becomes a sum of $N_f$ Abelian DBI actions
\begin{equation}
\label{eq:SBIX}
S=-T_{D7}\int\,\dd^8\xi\frac{\sqrt{h_3}}{4}\rho^3f\ft(1-\chi^2)\Bigg(\sum_{i=0}^{N_f-1}\sqrt{1-\chi^2+\vrho^2(\del_\vrho\chi)^2-2(2\pi\alpha')^2\frac{\ft}{f^2}(1-\chi^2)(\del_\vrho X_i)^2}\Bigg)\,,
\end{equation}
The constants of motion are given by
\begin{equation}
d_i=\frac{\delta S}{\delta(\del_\rho X_i)}=(2\pi\alpha')^2T_{D7}\frac{\sqrt{h_3}}{2}\rho^3\frac{\ft^2}{f}\frac{(1-\chi^2)^2\del_\vrho X_i}{\sqrt{1-\chi^2+\rho^2(\del_\rho\chi)^2-2(2\pi\alpha')^2\frac{\ft}{f^2}(1-\chi^2)(\del_\vrho X_i)^2}} \, .
\end{equation}
From the relations of the gauge fields we can read off the relations between the conjugate charge densities
\begin{equation}
d^B=d^{I_0}=\sum_{i=0}^{N_f-1}d_i\,,\qquad d^{I_i}=\sum_{j\not=i}d_j-(N_f-1)d_i\quad i=1,\dots,N_f-1\,.
\end{equation}
We now construct the Legendre transformation of the action~\eqref{eq:SBIX} to eliminate the fields $X_i$ in favor of the constants $d_i$
\begin{equation}
\begin{split}
\tilde{S}&=S-\int\,\dd^8\xi\sum_{i=1}^{N_f}X_i\frac{\delta S}{\delta X_i}\\
&=-T_{D7}\int\,\dd^8\xi\frac{\sqrt{h_3}}{4}\vrho^3f\ft(1-\chi^2)\sqrt{1-\chi^2+\vrho^2(\del_\vrho\chi)^2}\left(\sum_{i=0}^{N_f-1}\sqrt{1+\frac{8d_i^2}{(2\pi\alpha')^2T_{D7}^2\vrho^6\ft^3(1-\chi^2)^3}}\right)\,.
\end{split}
\end{equation}
Finally we obtain the equation of motion for the embedding~$\chi$ as
\begin{equation}
\begin{split}
&\del_\rho\left\{\rho^5f\ft(1-\chi^2)\frac{\del_\rho\chi}{\sqrt{1-\chi^2+\rho^2(\del_\rho\chi)^2}}\left(\sum_{i=0}^{N_f-1}\sqrt{1+\frac{8\dt_i^2}{\rho^6\ft^3(1-\chi^2)}}\right)\right\}\\
=&-\frac{\rho^3f\ft\chi}{\sqrt{1-\chi^2+\rho^2(\del_\rho\chi)^2}}\Bigg\{[3(1-\chi^2)+2\rho(\del_\rho\chi)^2]\left(\sum_{i=0}^{N_f-1}\sqrt{1+\frac{8\dt_i^2}{\rho^6\ft^3(1-\chi^2)}}\right)\\
&-\frac{24}{\rho^6\ft^3(1-\chi^2)^3}(1-\chi^2+\rho^2(\del_\rho\chi)^2\left(\sum_{i=0}^{N_f-1}\frac{\dt_i^2}{\sqrt{1+\frac{8\dt_i^2}{\rho^6\ft^3(1-\chi^2)}}}\right)\Bigg\}\,.
\end{split}
\end{equation}
This equation of motion completes the formulae describing the introduction of the non-Abelian part of the flavor 
group~$SU(N_f)$ in the gravity background for an arbitrary number~$N_f$ of flavors. This may be taken as 
the technical starting point for future investigations of the effects of non-Abelian chemical potentials with 
any desired flavor number~$N_f$ as long as we stay in the probe-brane~(or quenched) limit~$N_f\ll N_c$.

\subsection{Molecular dynamics} \label{sec:molecular}
Guided by the intuition obtained from dispersion effects in examples such as propagation of light through a prism,
we assume that perturbations inside the thermal medium, the plasma, with different frequencies and momenta will 
not all interact with the plasma in the same way and will not propagate in the same manner. Therefore it is reasonable 
that the constant transport coefficients we have considered so far should actually be modified to incorporate a
frequency and momentum dependence. On the thermal gauge theory side this idea is developed in the context of
{\it molecular dynamics}~\cite{Yip:1980}. For example the frequency-dependent generalization of the Kubo type formulae introduced
in section~\ref{sec:transportCoeffs} for the general transport coefficient~$\eta$ is given by
\begin{equation}
\label{eq:genKubo}
\eta ( \omega) = C \int\limits_0^\infty \dd t\, e^{i\omega t} \langle J_\eta(0), J_\eta (t) \rangle \, ,
\end{equation}
where~$C$ is a thermodynamic constant and~$J_\eta$ is the zero spatial momentum limit of the current relevant 
for this transport process. For example if~$\eta$ was the heat conductivity then~$J_\eta$ would be 
identified with the heat current. 

As described in chapter~\ref{sec:adsCft}, the gauge/gravity correspondence states that the full gauge theory is 
encoded in the gravity theory. Thus we can also assume that the momentum dependent transport coefficients 
are encoded in the gravity theory. In contrast to our hydrodynamic~(small frequency, long wave length) approach 
of section~\ref{sec:anaAdsG}, we can use the more general setup which will be described and applied in 
chapter~\ref{sec:thermalSpecFunc} for the computation of flavor current correlation functions. These are valid 
for perturbations with arbitrary four-momentum. So one way to find the momentum-dependent transport coefficients 
on the field theory side is to compute the correlators using a numerical gravity calculation. These then have to 
be substituted into expressions such as the generalized Kubo formula~\eqref{eq:genKubo}. 

It would also be interesting to fit these results to the analytic expressions from molecular dynamics. We may 
discover relations between the gravity and thermal gauge theory similar to the identification of correlator poles
with quasinormal frequencies. 

\subsection{Summary} \label{sec:sumHoloThermoHydro}
In this chapter I have presented some of the main results of this thesis including the analytic form of correlators being
connected to hydrodynamics. We have also seen the numerically found thermodynamics at finite non-Abelian flavor 
charge densities. 

The main result for the hydrodynamic case are the correlators which all are similar to
\begin{equation}
G^{XY}_{00} =
 \frac{N_c T_R T q^2}{8 \pi [i(\omega-\mu) - D q^2]} \text{for} \, \wn\ge\mn \, 
\, .
\end{equation}
The longitudinal and time component correlators all have the diffusion pole~$\omega=\pm\mu - i D q^2$ while 
transversal modes do not show this diffusive behavior. The correlators have different dependence on the 
frequency and spatial momentum~(cf.~\eqref{eq:GXY11} and the equations following it for details). The presence 
of an isospin potential mainly manifests itself in the pole structure of longitudinal~($0$ and $3$-component) correlators
through shifting the location of the pole in the complex $\omega$-plane by the amount of the chemical 
potential~$\pm\mu$ along the real axis. Thus the main effect of the isospin potential is that it splits 
the hydrodynamic diffusion pole located on the imaginary frequency axis into a triplet. 
This behavior is a direct consequence of the changed indicial 
structure with indices~$\beta=\pm i(\wn\pm\mn)/2$. Two directions in flavor space~($a=1,2$) are affected in this way
while the third flavor direction parallel to the chemical potential does not feel the potential. We have developed a physical
interpretation of this situation by analogy to the symmetry breaking which occurs in the case of {\it Larmor precession} 
of a spin inside a real-space magnetic field.

Since the poles of the correlator correspond to quasinormal frequencies in the gravity context, we have also analyzed
the structure of these poles using the imaginary part of the correlator in the complex frequency plane. 
We found an antisymmetry  around the pole which translates into an antisymmetry in the spectral function. 
The spectral function displays a low-energy cut-off at the value~$\wn=\mn$ which we interpret as a minimum
energy that perturbations in the plasma need to have in order to be produced. 
The spectral function also shows the structure of triplet splitting that we found in the poles. We will see exactly
this behavior again in chapter~\ref{sec:thermalSpecFunc} when we consider spectral functions at finite quark 
mass at arbitrary momentum.
In section~\ref{sec:discAnaHydro} we have discussed these results and compared to our earlier approach neglecting
terms of order~$\O(\mu^2)$ in~\cite{Erdmenger:2007ap}.

Furthermore, we have introduced the new concept of a full non-Abelian chemical potential, and we 
have developed the necessary techniques to analyze its dynamics and the thermodynamics produced by this setup. 
These methods include a flavor transformation to fields~$\propto (A^1\pm A^2)$ decoupling the flavor structure
in the corresponding background equations of motion. 
For definiteness we have applied our techniques to the example $N_f=2$ but section~\ref{sec:hiNf} generalizes these 
concepts and calculational methods to arbitrary flavor number~$N_f$. In particular we study the quark condensate,
the internal energy~$E$, the entropy~$\S$, free energy~$F$ and the speed of sound~$\mathcal{V}_s$.
In the two-flavor setup we find two different phase transitions. One is the black hole to black hole transition 
known from the baryonic case. However, the second transition is located at the line in the phase diagrams
where isospin and baryon density or potential are equal. We have strong indications that this transition is 
analogous to that one found for 2-flavor QCD in~\cite{Splittorff:2000mm}. It might also be worthwhile
to reduce our study of the phase structure to vanishing baryon but non-vanishing isospin density
in order to be able to more directly compare our results to lattice QCD or effective approaches
such as the Gross-Neveu model~\cite{Ebert:2008us}.

Finally, we have considered transport coefficients which depend on frequency and spatial momentum
of the disturbance in the context of {\it molecular dynamics} in~\ref{sec:molecular}. The gravity calculation
should contain all the information about this four-momentum dependence. Therefore, we suggest to obtain
correlators from gravity numerically for fixed frequency and momentum, and to substitute these correlators 
into Kubo formulae to obtain the transport coefficients. Repeating this procedure scanning through different
frequency and momentum values we should obtain the four-momentum dependence of the transport 
coefficient numerically.

\section{Thermal spectral functions at finite $U(N_f)$-charge density} \label{sec:thermalSpecFunc}
In this chapter we apply numerical techniques to compute the spectral function of vector currents at finite 
charge densities. We analyze the spectrum for the cases of vanishing densities, finite baryon 
density~(section~\ref{sec:mesonSpectraB}), finite isospin density~$N_f=2$~(section~\ref{sec:mesonSpectraI}), 
as well as finite baryon and isospin density at the same time~(section~\ref{sec:mesonSpectraB&I}). Especially the latter
case is motivated by the possible comparison to the phenomenology of effective two flavor models of QCD 
and lattice results. The spectra resulting from our gauge/gravity calculations show quasi-particle resonances 
which at low temperatures can be identified with vector mesons having survived the deconfinement transition. 
These mesons can be seen as analogs of the QCD rho-meson. A central point to this thesis is also the discovery
of a turning point in the frequency where the resonances appear when the mass-temperature parameter~$m\propto M_q/T$ 
is changed~(where~$M_q$ is the quark mass and~$T$ the temperature). At high temperatures the quasi-particle interpretation
of peaks in the spectral functions has to be modified as we speculate in section~\ref{sec:peakTurning} utilizing quasinormal 
modes. 

\subsection{Meson spectra at finite baryon density} \label{sec:mesonSpectraB}

{\bf Application of calculation method} 
We now compute the spectral functions of flavor currents at finite baryon
density~$n_B$, chemical potential~$\mu$ and temperature~$T$ in the `black hole
phase' which was discussed in section~\ref{sec:finiteTAdsCft}. Compared to the limit of vanishing
chemical potential treated in~\cite{Myers:2007we}, we discover a qualitatively
different behavior of the finite temperature oscillations corresponding to
vector meson resonances.

To obtain the spectral functions, we compute the correlations of flavor gauge
field fluctuations $A_\mu$ about the background given
by~\eqref{eq:actionEmbeddingsAt}, denoting the full gauge field by
\begin{equation}
\label{eq:hatA}  
        \hat A_\mu(\rho, \vec{x}) = \delta^0_\mu \tilde A_0(\rho) + A_\mu(\vec{x},\rho)\,.
\end{equation}
According to section~\ref{sec:thermoBaryon}, the background field has a
non-vanishing time component, which depends solely on $\rho$. The fluctuations
in turn are gauged to have non-vanishing components along the Minkowski
coordinates $\vec x$ only and only depend on these coordinates and on $\rho$.
Additionally they are assumed to be small, so that it suffices to consider their
linearized equations of motion. Note, that in these conventions the field strength 
fluctuations~$F_{\mu\nu} =2 \partial_{[\mu} A_{\nu]}$ only exist in directions~$\mu,\nu = 0,1,2,3,4$. Meanwhile
the anti-symmetric background field strength has only two non-vanishing components~$\tilde F_{40}=-\tilde F_{04}$.

The fluctuation equations of motion are obtained from the effective D7-brane action \eqref{eq:dbiSD7},
where we introduce small fluctuations $A$ by setting $\tilde F_{\mu\nu} \to \hat
F_{\mu\nu} = 2\, \partial_{[\mu}\hat A_{\nu]}$ with $\hat A = \tilde A + A$. The
background gauge field~$\tilde A$ is given by~\eqref{eq:eomD}. Note that from now on we denote field fluctuations
with the simple symbol~(e.g.$A$) and we provide the normalized background fields with a tilde~$\tilde A$. 
The main difference to the fluctuations considered in section~\ref{sec:anaHydroIso} is the fact that the 
present fluctuations now propagate on a non-symmetric background $G$ given by the symmetric and
diagonal metric part~$g$ summed with the anti-symmetric gauge field background~$\tilde F$
\begin{equation}
	G = g + \tilde F,
\end{equation}
and the fluctuation's dynamics is determined by the Lagrangian
\begin{equation}
	\mathcal{L} = \sqrt{\left| \det ( G + F )\right|},
\end{equation}
with the fluctuation field strength $F_{\mu\nu} = 2\partial_{[\mu}A_{\nu]}$.
Since the fluctuations and their derivatives are chosen to be small, we consider
their equations of motion only up to linear order, as derived from the part of
the Lagrangian $\mathcal{L}$ which is quadratic in the fields and their
derivatives. Denoting this part by $\mathcal{L}_2$, we get
\begin{equation}
\label{eq:L2B}
	\mathcal{L}_2 = -\frac{1}{4} \sqrt{\left|\det G\right|}\; \left (G^{\mu\alpha}G^{\beta\gamma} F_{\alpha\beta}F_{\gamma\mu}
	-\frac{1}{2} G^{\mu\nu}G^{\sigma\gamma} F_{\mu\nu}F_{\sigma\gamma} \right ) \, .
\end{equation}
Here and below we use upper indices on $G$ to denote elements of $G^{-1}$. The
equations of motion for the components of $A$ are
\begin{equation}
	\label{eq:eomFluct}
	0=\partial_\nu\left[ \sqrt{\left|\det G\right|} \left( G^{\mu\nu}G^{\sigma\gamma}-G^{\mu\sigma}G^{\nu\gamma} \right) \partial_{[\gamma}A_{\mu ]} +\frac{1}{2} G^{[\nu\sigma]} G^{\mu\gamma} F_{\mu\gamma} \right].
\end{equation}
Note, that the last term each in the quadratic Lagrangian~\eqref{eq:L2B} and 
in the equation of motion~\eqref{eq:eomFluct} comes from the term~$[\text{tr}(G^{-1} F)]^2$ in the determinant
expansion~\eqref{eq:detExpansion}. We recall that~$G^{-1}$ here including the background gauge field~$\tilde F$ is
not symmetric anymore and so the trace over the contraction with our anti-symmetric field strength~$F$ does not 
vanish in general. Nevertheless, in the geometry we have choosen here these extra terms are all proportional to 
the gauge fluctuation in time direction~$A_0$ which will drop out of our considerations by the time we set the 
spatial momentum of perturbations to zero. Let us keep these terms anyhow in order to be 
precise~\footnote{The author appreciates the comment on this notation issue given in~\cite{Myers:2008cj}.}.

The terms of the corresponding on-shell action at the $\rho$-boundaries are
(with $\rho$ as an index for the coordinate $\rho$, not summed)
\begin{equation}
\begin{split}
	S^{\text{on-shell}}_{\text{D7}} = \;& \varrho_H \pi^2 R^3 N_f T_{\text{D7}} \int\!\! \dd^4 x \sqrt{\left|\det G \right|}\\
& \times \left( \left( G^{04}\right)^2 A_0 \partial_\rho A_0 - G^{44} G^{ik} A_i \partial_\rho A_k \right)\Bigg|^{\rho_B}_{\varrho_H}.
\end{split}
\end{equation}
Note that on the boundary $\rho_B$ at $\rho\to\infty$, the background field strength~$\tilde F_{40}(\rho_B) = 0$ and the 
background matrix $G$ reduces to the induced D7-brane metric $g$. Therefore, the analytic expression
for the on-shell action is identical to the on-shell action found in
\cite{Myers:2007we}. There, the action was expressed in terms of the gauge
invariant field component combinations
\begin{equation}
E_x=\omega A_x+ q A_0,\qquad E_{y,z}=\omega A_{y,z}\, .
\end{equation}
In the case of vanishing spatial momentum $q\to 0$, the Green functions for the
different components coincide and were computed as \cite{Myers:2007we}
\begin{equation}
\label{eq:q0GreenFunction}
G^R = G^R_{xx} =G^R_{yy} = G^R_{zz} = 
 \frac{N_f N_c T^2}{8}\; \lim_{\rho\to\infty}\left(\rho^3 \frac{\partial_\rho E(\rho)}{E(\rho)} \right)\, ,  
\end{equation}
where the $E(\rho)$ in the denominator divides out the boundary value of the
field in the limit of large $\rho$ according to the recipe we developed and discussed in
section~\ref{sec:anaAdsG} and~\ref{sec:numAdsG}. 
The thermal correlators obtained in this way display hydrodynamic properties,
such as poles located at complex frequencies~(in particular when~$E(\rho)=0$ which is the boundary
condition on the equation of motion for~$E$ obeyed by quasinormal modes, cf.~\ref{sec:qnm}). 
They are used to compute the
spectral function \eqref{eq:spectralFunction}. We are going to compute the functions
$E(\rho,k) = E^{\text{bdy}}(k)\mathfrak{F}(\rho, k)$ numerically in the limit of vanishing spatial momentum $\bm q\to 0$.
The functions $\mathfrak{F}(\rho,\vec k)$ from the recipe in equation~\eqref{eq:anaGFormula} 
are then obtained by dividing out the boundary
value $E^{\text{bdy}}(\vec k) = \lim_{\rho\to\infty} E(\rho,\vec k)$.
Numerically we obtain the boundary value by computing the solution at a fixed
large $\rho$. 
Finally, the indices on the Green function denote the
components of the operators in the correlation function, in our case all
off-diagonal correlations (as~$G_{yz}$, for example) vanish.

In our case of finite baryon density, new features arise through the modified
embedding and gauge field background, which enter the equations of motion
\eqref{eq:eomFluct} for the field fluctuations. To apply the prescription to
calculate the Green function, we Fourier transform the fields as
\begin{equation}
\label{eq:fourierA}
A_\mu(\rho,\vec{x}) = \int\!\! \frac{\dd^4 k}{(2\pi)^4}\, e^{i\vec{k}\vec{x}} A_\mu(\rho,\vec{k}) \,.
\end{equation}

We choose our coordinate system to give us a momentum vector of the fluctuation
with nonvanishing spatial momentum only in a single direction, which we choose
to be the $x^1$ component, $\vec{k}=(\omega,q,0,0)$.

For simplicity we restrict ourselves to vanishing spatial momentum~$q=0$. In
this case the equations of motion for transversal fluctuations~$E_{y,z}$ match
those for longitudinal fluctuations~$E_x$. For a more detailed discussion see
\cite{Myers:2007we}. As an example consider the equation of motion obtained from
\eqref{eq:eomFluct} with $\sigma = 2$, determining $E=E_y=\omega A_2$,
\begin{equation}
\begin{split}
	\label{eq:eomEq0}
	0 =\,& E''+ \frac{\partial_\rho[\sqrt{|\det G|}G^{22}G^{44}]}{|\det G| G^{22}G^{44}}\,
          E'- \frac{G^{00}}{G^{44}}\, \varrho_H^2 \omega^2 E \\  
	 =\,& E'' + \partial_\rho \ln \Bigg( \frac{1}{8}\tilde f^2 f \rho^3 (1-\chi^2 + \rho^2{\chi'}^2)^{3/2} \\
	&\hphantom{E'' + \partial_\rho \ln \Bigg(\;} \times \sqrt{1-
	\frac{2\tilde f (1-\chi^2) (\partial_\rho\tilde A_0 )^2}{f^2(1-\chi^2 + \rho^2{\chi'}^2)}} \Bigg) E' \\
	&\hphantom{E''} + 8 \wn^2 \frac{\tilde f}{f^2} \frac{1-\chi^2 + \rho^2 {\chi'}^2 }{\rho^4 (1-\chi^2)}\,E.
\end{split}
\end{equation}
The symbol $\wn$ denotes the dimensionless frequency $\wn=\omega/(2\pi T)$,
and we made use of the dimensionless radial coordinate
$\rho=\varrho/\varrho_H$.

In order to numerically integrate this equation, we determine local solutions of
that equation near the horizon~$\rho=1$. These can be used to compute initial
values in order to integrate~(\ref{eq:eomEq0}) forward towards the boundary. The
equation of motion~(\ref{eq:eomEq0}) has coefficients which are singular at the
horizon. According to standard methods~\cite{Bender}, the local solution of this
equation behaves as~$(\rho-\rho_{\text{H}})^\beta$, where~$\beta$ is a so-called
`index' of the differential equation. We compute the possible indices to be
\begin{equation}
\label{eq:indices}
\beta=\pm i\,\wn .
\end{equation}
Only the negative one will be retained in the following, since it casts the
solutions into the physically relevant incoming waves at the horizon and
therefore satisfies the incoming wave boundary condition. The solution $E$ can
be split into two factors, which are $(\rho-1)^{-i\wn}$ and some function
$F(\rho)$, which is regular at the horizon. Note, that this~$F$ is different from the function~$\mathfrak{F}$
introduced earlier. While~$F$ results from splitting the full solution~$E$ into a regular and a 
regulating part~(see section~\ref{sec:anaAdsG}), the function~$\mathfrak{F}$ results from splitting the 
full solution~$E$ into a boundary and a bulk part. The first coefficients of a series
expansion of $F(\rho)$ can be found recursively as described
in~\cite{Teaney:2006nc,Kovtun:2006pf}. At the horizon the local solution then
reads
\begin{equation}
\begin{aligned}
\label{eq:localSolutions}
E(\rho)  =\; & (\rho-1)^{-i\wn}\, F(\rho)\\
          =\; & (\rho-1)^{-i\wn} \left [1+\frac{i\wn}{2}(\rho-1)+
\cdots \right ].
\end{aligned}
\end{equation}
So, $F(\rho)$ asymptotically assumes values
\begin{equation}
\label{eq:startingValues}
F(\rho=1)=1,\qquad \partial_\rho F(\rho)\Big|_{\rho=1}=\frac{i\wn}{2}\, .
\end{equation}

For the calculation of numbers, we have to specify the baryon density~$\tilde d$
and the mass parameter $\chi_0\sim M_q/T$ to obtain the embeddings $\chi$ used
in \eqref{eq:eomEq0}. Then we obtain a solution for a given frequency $\wn$
using initial values (\ref{eq:localSolutions}) and (\ref{eq:startingValues}) in
the equation of motion~(\ref{eq:eomEq0}). This eventually gives us the numerical
solutions for $E(\rho)$.

Spectral functions are then obtained by combining (\ref{eq:q0GreenFunction}) and 
\eqref{eq:spectralFunction},
\begin{equation}
\label{eq:specFuncFormula}
        \R(\omega,0) = - \frac{N_f N_c T^2}{4}\; \mathrm{Im}\lim_{\rho\to\infty}\left(\rho^3 \frac{\partial_\rho E(\rho)}{E(\rho)} \right).
\end{equation}

\begin{figure}
        \includegraphics[width=0.8\linewidth]{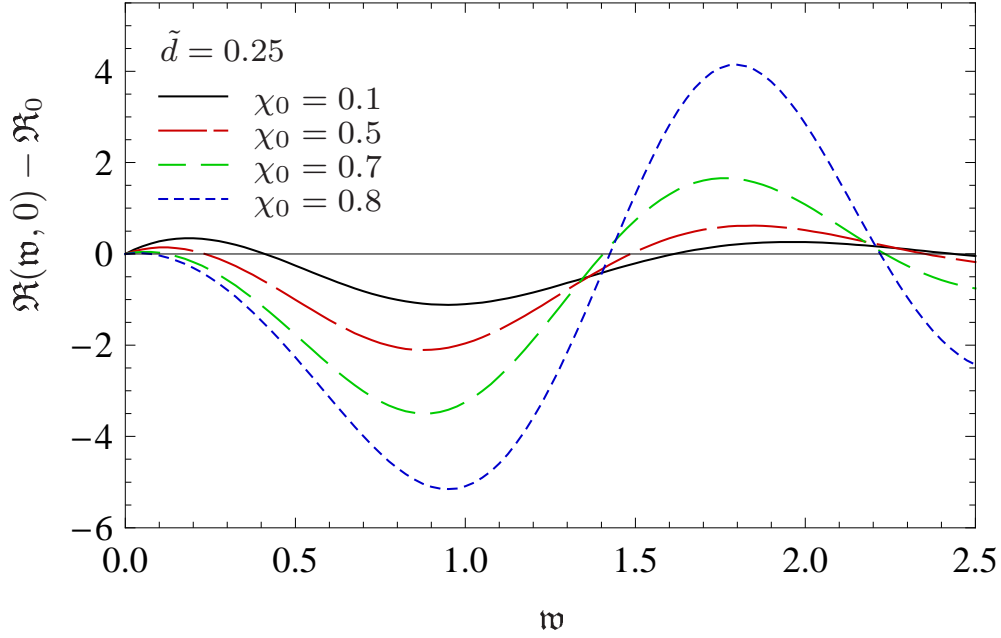}
        \caption{The finite temperature part of the spectral function~$\R-\R_0$
			(in units of~$N_f N_c T^2/4$) at finite baryon density~$\tilde d$.
			The maximum grows and shifts to smaller frequencies
			as~$\chi_0$ is increased towards~$\chi_0=0.7$, but then turns around
			to approach larger frequency values.}
        \label{fig:vectorPeaksDt025light}
\end{figure}

\begin{figure}
        \includegraphics[width=0.8\linewidth]{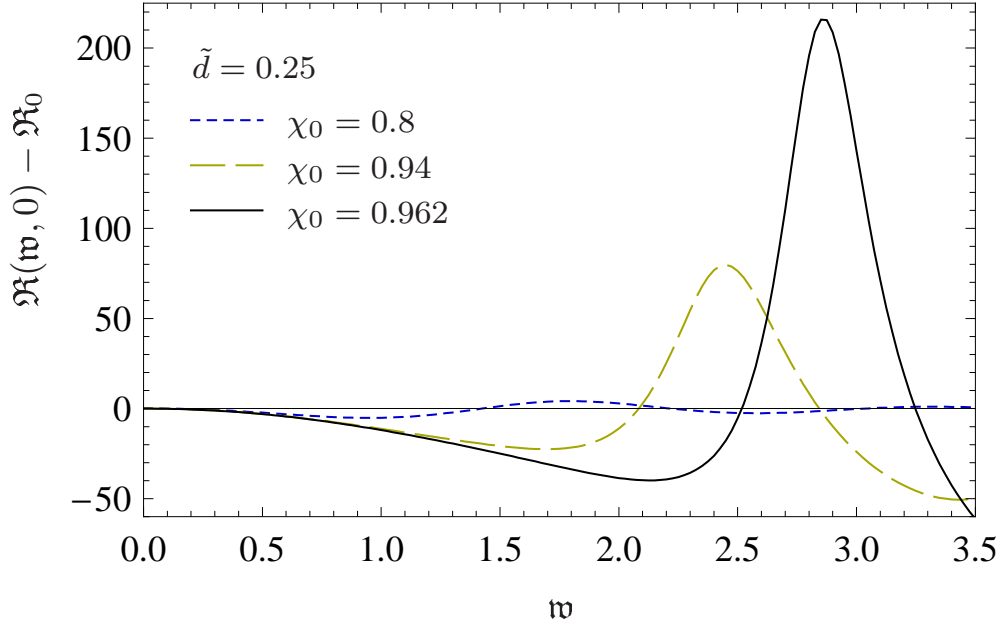}
        \caption{The finite temperature part of the spectral function~$\R-\R_0$
			(in units of~$N_f N_c T^2/4$) at finite baryon density~$\tilde d$.
			In the regime of~$\chi_0$ shown here, the peak shifts to larger
			frequency values with increasing $\chi_0$.}
        \label{fig:vectorPeaksDt025heavy}
\end{figure}

\begin{figure}
		\includegraphics[width=0.8\linewidth]{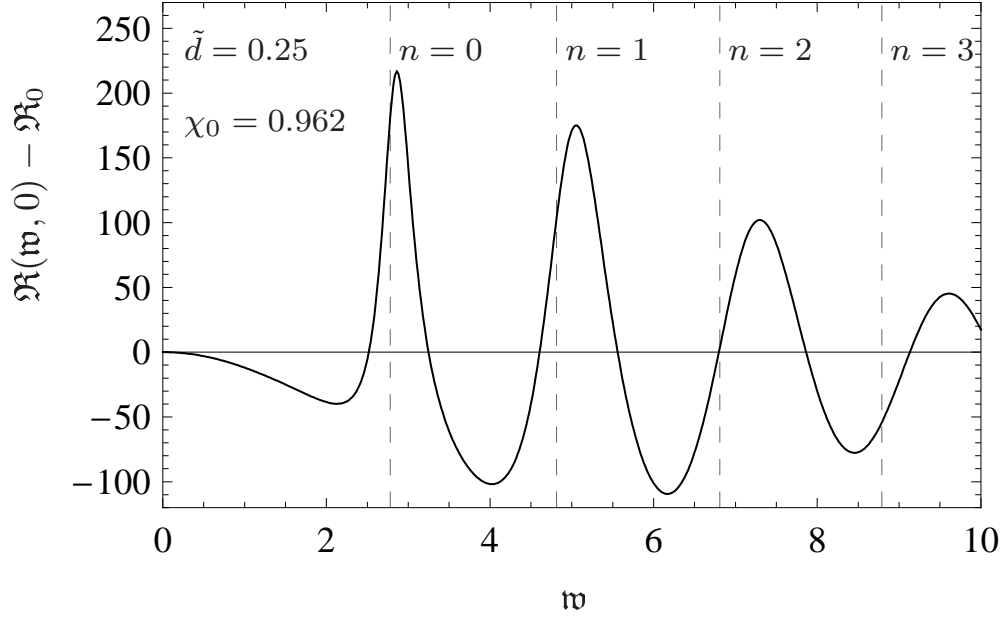}
		\caption{The finite temperature part~$\R-\R_0$ of the spectral function
			(in units of~$N_f N_c T^2/4$) at finite baryon density~$\tilde d$.
			The oscillation peaks narrow and get more pronounced compared to
			smaller~$\chi_0$. Dashed vertical lines show the meson mass spectrum
			given by equation~\eqref{eq:massFormula2}.}
        \label{fig:lineSpectrumDt025light}
\end{figure}

\begin{figure}
		\includegraphics[width=0.8\linewidth]{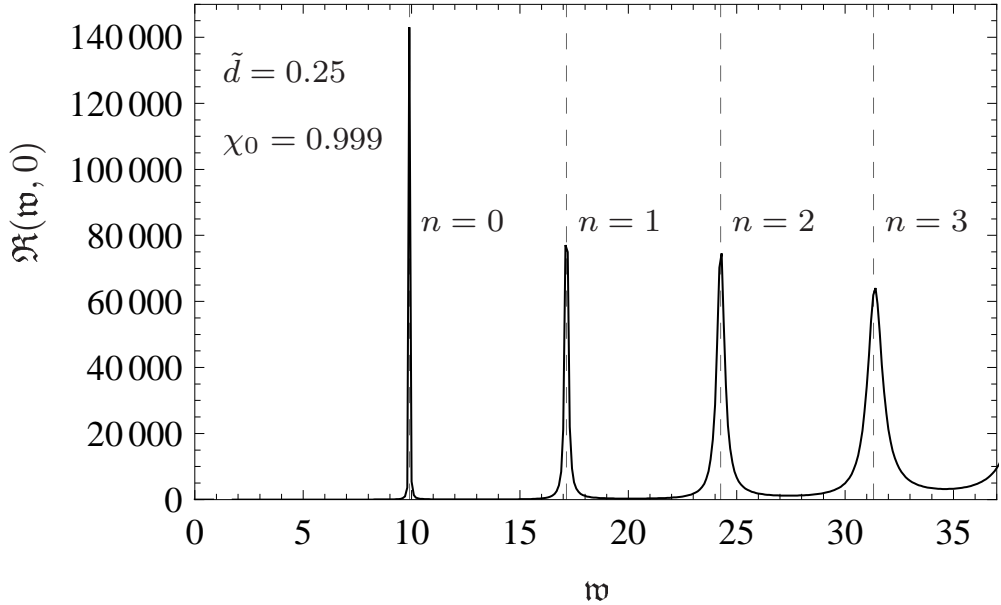}
        \caption{The spectral function $\mathfrak{R}$ (in units of~$N_f N_c
			T^2/4$) at finite baryon density~$\tilde d$. At large $\chi_0$, as
			here, the peaks approach the dashed drawn line spectrum given
			by~\eqref{eq:massFormula2}.}
        \label{fig:lineSpectrumDt025heavy}
\end{figure}

\begin{figure}
		\includegraphics[width=0.8\linewidth]{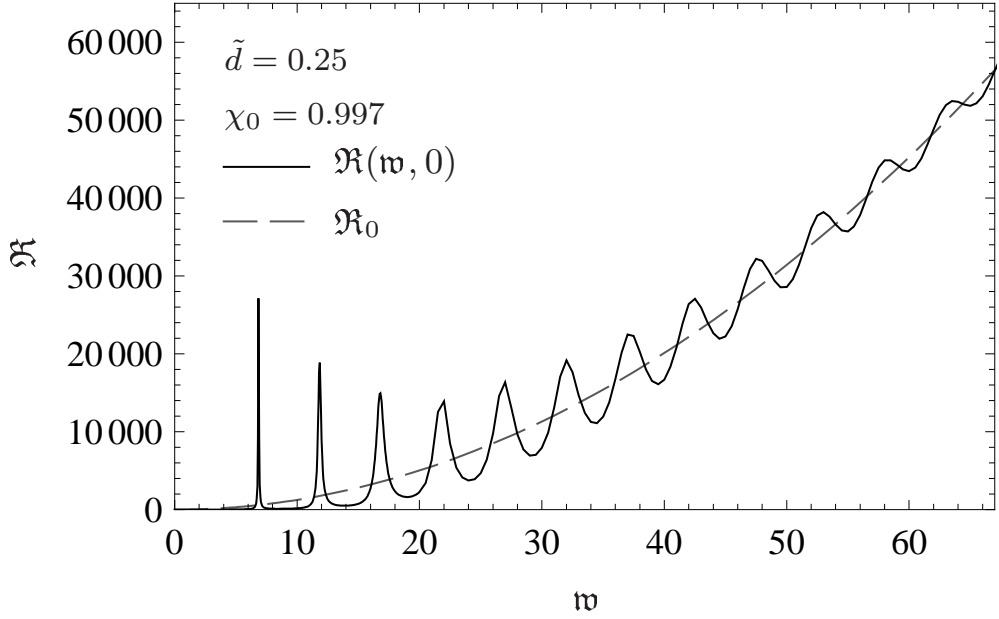}
        \caption{The thermal spectral function $\mathfrak{R}$ (in units
			of~$N_f N_c T^2/4$) compared to the zero temperature result
			$\mathfrak{R}_0$.}
		\label{fig:specTempZeroTemp}
\end{figure}

{\bf Results for spectral functions}
\label{sec:vectorResults}
We now discuss the resulting spectral functions at finite baryon density, and
observe crucial qualitative differences compared to the case of vanishing baryon
density. In figures~\ref{fig:vectorPeaksDt025light}
to~\ref{fig:lineSpectrumDt025heavy}, some examples for the spectral function at
fixed baryon density $n_B\propto \tilde d$ are shown. To emphasize the resonance
peaks, in some plots we subtract the quantity
\begin{equation}
	\mathfrak{R}_0 = N_f N_c T^2\, \pi\wn^2,
\end{equation}
around which the spectral functions oscillate,
cf.~figure~\ref{fig:specTempZeroTemp}.

The graphs are obtained for a value of $\tilde d$ above $\tilde d^*$ given by
\begin{equation}
\tilde d^* = 0.00315,\quad \tilde d = 2^{5/2} n_B / (N_f \sqrt{\lambda} T^3)\, , 
\end{equation}
where the fundamental phase transition does not occur. The
different curves in these plots show the spectral functions for different quark
masses, corresponding to different positions on the solid blue line in the phase
diagram shown in figure~\ref{fig:grandcanPhaseDiagB}. Regardless whether we chose
$\tilde d$ to be below or above the critical value $\tilde d^*$, we observe the
following behavior of the spectral functions with respect to changes in the
quark mass to temperature ratio.

Increasing the quark mass from zero to small finite values results in more and
more pronounced peaks of the spectral functions. This eventually leads to the
formation of resonance peaks in the spectrum. At small masses, though, there are
no narrow peaks. Only some broad maxima in the spectral functions are visible. At
the same time as these maxima evolve into resonances with increasing quark mass,
their position changes and moves to lower freqencies $\wn$, see
figure~\ref{fig:vectorPeaksDt025light}. This behavior was also observed for the
case of vanishing baryon density in ~\cite{Myers:2007we}.

However, further increasing the quark mass leads to a crucial difference to the
case of vanishing baryon density. Above a value $m^{\text{turn}}$ of the quark
mass, parametrized by $\chi_0^{\text{turn}}$, the peaks change their direction
of motion and move to larger values of $\wn$, see
figure~\ref{fig:vectorPeaksDt025heavy}. Still the maxima evolve into more and
more distinct peaks.

Eventually at very large quark masses, given by $\chi$ closer and closer to 1,
the positions of the peaks 
asymptotically reach exactly those frequencies which 
correspond to the masses of the vector mesons at zero temperature
\cite{Kruczenski:2003be}. In our coordinates, these masses are given by

\begin{equation}
\label{eq:massFormula2}
M= \frac{L_\infty}{R^2}\,\sqrt{2 (n+1)(n+2)}\, ,
\end{equation}
where $n$ labels the Kaluza-Klein modes arising from the D7-brane wrapping
$S^3$, and~$L_\infty$ is the radial distance in the~$(8\mathord{,}9)$-direction
between the stack of D3-branes and the D7, evaluated at the $AdS$-boundary,
\begin{equation}
\label{eq:L}
L_\infty=\lim_{\varrho\to\infty} \varrho\chi(\varrho)\propto \frac{M_q}{T}  \, .
\end{equation}

The formation of a line-like spectrum can be interpreted as the evolution of
highly unstable quasi-particle excitations in the plasma into quark bound
states, finally turning into nearly stable vector mesons,
cf.~figures~\ref{fig:lineSpectrumDt025light}
and~\ref{fig:lineSpectrumDt025heavy}.

\medskip 

We now consider the turning behavior of the resonance peaks shown in
figures~\ref{fig:vectorPeaksDt025light} and~\ref{fig:vectorPeaksDt025heavy}.
There are two different scenarios, depending on whether the quark mass is small
or large.

First, when the quark mass is very small~$M_q\ll T$, we are in the regime of the
phase diagram corresponding to the right half  of
figure~\ref{fig:grandcanPhaseDiagB}. In this regime the influence of the Minkowski
phase is negligible, as we are deeply inside the black hole phase. We therefore
observe only broad structures in the spectral functions, instead of peaks.

Second, when the quark mass is very large, $M_q\gg T$, or equivalently the
temperature is very small, the quarks behave just as they would at zero
temperature, forming a line-like spectrum. This regime corresponds to the left
side of the phase diagram in figure~\ref{fig:grandcanPhaseDiagB}, where all curves of
constant $\tilde d$ asymptote to the Minkowski phase.

The turning of the resonance peaks is associated to the existence of the 
two regimes. At~$\chi_0^{\text{turn}}$ the two regimes are connected to each other 
and none of them is dominant. 

The turning behavior is best understood by following a line of constant density
$\tilde d$ in the phase diagram of figure~\ref{fig:grandcanPhaseDiagB}. Consider for
instance the solid blue line in figure~\ref{fig:grandcanPhaseDiagB}, starting at large
temperatures/small masses on the right of the plot. First, we are deep in the
unshaded region ($n_B\not =0$), far inside the black hole phase. Moving along to
lower $T/\bar M$, the solid blue line in figure~\ref{fig:grandcanPhaseDiagB} rapidly
bends upwards, and asymptotes to both the line corresponding to the onset of the
fundamental phase transition, as well as to the separation line between black
hole and Minkowski phase (gray region).
This may be interpreted as the quarks joining in bound states. Increasing the
mass further, quarks form almost stable mesons, which give rise to resonance
peaks at larger frequency if the quark mass is increased. 

We also observe a dependence of $\chi_0^{\text{turn}}$ on the baryon density. As
the baryon density is increased from zero, the value of $\chi_0^{\text{turn}}$
decreases.

Figures~\ref{fig:lineSpectrumDt025heavy} and~\ref{fig:specTempZeroTemp} show
that higher~$n$ excitations from the Kaluza-Klein tower are less stable. While
the first resonance peaks in this plot are very narrow, the following peaks show
a broadening with decreasing amplitude.

\begin{figure}
	\includegraphics[width=.4\linewidth]{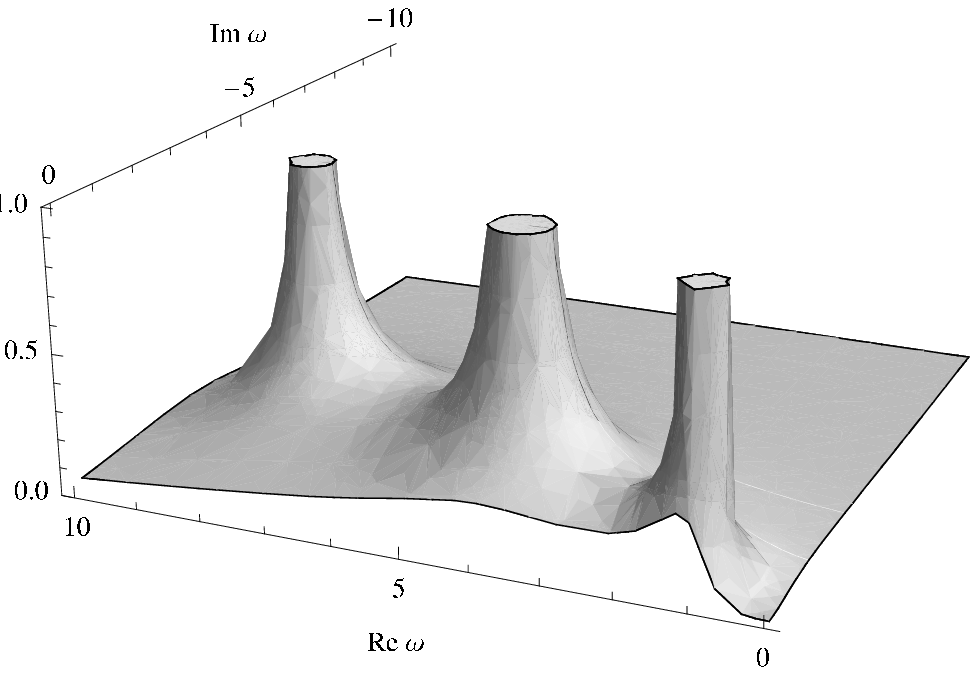}
	\hfill
	\includegraphics[width=.4\linewidth]{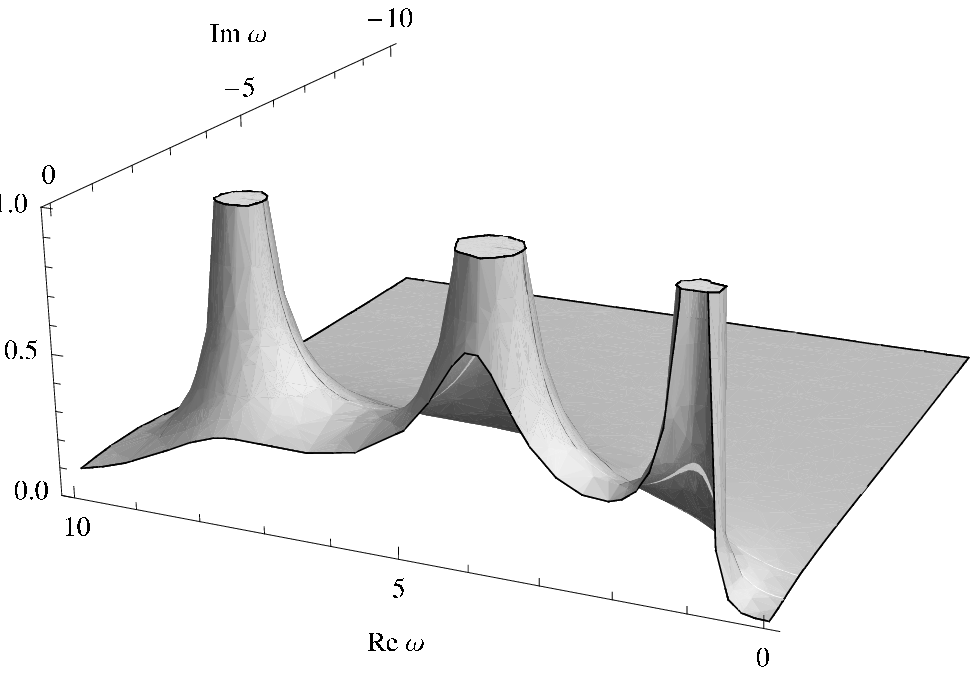}
	\caption{Qualitative relation between the location of the poles in the
		complex frequency plane and the shape of the spectral functions on the
		real $\omega$ axis. The function plotted here is an example for the
		imaginary part of a correlator. Its value on the real $\omega$ axis
		represents the spectral function. The poles in the right plot are closer
		to the real axis and therefore there is more structure in the spectral
		function. This figure was generated by Felix Rust~\cite{Rust:phd}.}
	\label{fig:polesExample}
\end{figure}

This broadening of the resonances is due to the behaviour of the quasinormal
modes of the fluctuations, which correspond to the poles of the correlators in
the complex~$\omega$ plane, as described in the example
\eqref{eq:quasiNormalModes} and sketched in figure~\ref{fig:polesExample}. The
location of the resonance peaks on the real frequency axis corresponds to the
real part of the quasinormal modes. It is a known fact that the the quasinormal
modes develop a larger real and \emph{imaginary} part at higher $n$. So the
sharp resonances at low~$\wn$, which correspond to quasi-particles of long
lifetime, originate from poles whith small imaginary part. For higher
excitations in $n$ at larger~$\wn$, the resonances broaden and get damped due to
larger imaginary parts of the corresponding quasinormal modes.

For increasing mass we described above that the peaks of the spectral functions
first move to smaller frequencies until they reach the turning point
$m^{\text{turn}}$. Further increasing the mass leads to the peaks moving to
larger frequencies, asymptotically approaching the line spectrum. This behavior
can be translated into a movement of the quasinormal modes in the complex plane.
It would be interesting to compare our results to a direct calculation of the
quasinormal modes of vector fluctuations in analogy to~\cite{Hoyos:2006gb}.
 
In~\cite{Hoyos:2006gb} the quasinormal modes 
are considered for scalar fluctuations exclusively, at vanishing
baryon density. The authors observe that starting 
from the massless case, the real part of the quasinormal frequencies
increases with the quark mass first, and then turns around to decrease.   
This behavior agrees with 
the peak movement for scalar spectral functions observed in~\cite[figure 9]{Myers:2007we}
(above the fundamental phase transition,~$\chi_0\le 0.94$) 
where the scalar meson resonances move to 
higher frequency first, turn around and
move to smaller frequency increasing the mass further. These results
do not contradict the present work since we consider vector modes exclusively.
The vector meson spectra considered in~\cite{Myers:2007we} at vanishing baryon density   
only show peaks moving to smaller frequency as the quark mass is increased.  
Note that the authors there continue to consider black hole embeddings below the 
fundamental phase transition which are only metastable, the Minkowski 
embeddings being thermodynamically favored. At small baryon density
and small quark mass our spectra are virtually coincident with those of~\cite{Myers:2007we}.  
In our case, at finite baryon density, black hole embeddings are favored
for all values of the mass over temperature ratio. At small values
of~$T/\bar M$ in the phase diagram of figure~\ref{fig:grandcanPhaseDiagB}, 
we are very close to the Minkowski regime, temperature effects
are small, and the meson mass is proportional to the quark
mass as in the supersymmetric case. Therefore, the peaks 
in the spectral function move to the right~(higher frequencies) as function
of increasing quark mass.   

The turning point in the location of the peaks is a consequence 
of the transition between two regimes, i.e.\ the temperature-dominated 
one also observed in~\cite{Myers:2007we}, and the potential-dominated  
one which asymptotes to the supersymmetric spectrum.  

We expect the physical interpretation of the left-moving of the 
peaks in the temperature-dominated regime to be related to the
strong dissipative effects present in this case. This is consistent 
with the large baryon diffusion coefficient present in this regime 
as discussed in section~\ref{sec:baryonDiffusion} and shown  
in figure~\ref{fig:diffusionConstants}. A detailed 
understanding of the physical picture in this regime requires a 
quantitative study of the quasipaticle behavior which we 
leave to future work.     

Let us emphasize that it is likely that the turning point 
behavior is not a consequence of the finite baryon density.   
In our approach it is just straightforward to investigate 
the $T\to 0$ limit since black hole embeddings are 
thermodynamically favored even near~$T=0$ at finite baryon density. 
We expect that a right-moving of the peaks consistent with
the SUSY spectrum should also be observable for Minkowski embeddings 
at vanishing baryon density for~$T\to 0$. However this has not been 
investigated for vector modes neither in~\cite{Hoyos:2006gb} 
nor in~\cite{Myers:2007we}. An extension of the analysis presented here to
perturbations with non-vanishing spatial momentum~$\qn\not = 0$ has appeared in~\cite{Myers:2008cj}.   

\subsection{Meson spectra at finite isospin density} \label{sec:mesonSpectraI}
{\bf Radially varying $SU(2)$-background gauge field} \label{sec:varyChemPot}
In order to examine the case $\mathcal{N}_f=2$ in the strongly coupled plasma,
we extend our previous analysis of vector meson spectral functions to a chemical
potential with $SU(2)$-flavor (isospin) structure. Starting from the general
action 
\begin{equation}
\label{eq:isoAction}
S_{\text{iso}}=-T_r T_{D7} \int\!\!\dd^8\xi\;\sqrt{|\det(g+\hat F)|} \, ,   
\end{equation} 
we now consider field strength tensors
\begin{equation}
\label{eq:isoF}
\hat F_{\mu\nu}=\sigma^a\,\left (2\partial_{[\mu}\hat A^a_{\nu]}
  +\frac{\varrho_H^2}{2\pi\alpha'}f^{abc}\hat A_\mu^b \hat A_\nu^c\right)\, ,
\end{equation}
with the Pauli matrices~$\sigma^a$ and $\hat A$ given by 
equation~\eqref{eq:hatA}. The factor~$\varrho_H^2/(2\pi\alpha')$ is due
to the introduction of dimensionless fields as described below~\eqref{eq:chemPotLimit}. 
In order to obtain a finite isospin-charge
density~$n_I$ and its conjugate chemical potential~$\mu_I$, we introduce an
$SU(2)$-background gauge field~$\tilde A$~\cite{Erdmenger:2007ap}  
\begin{equation}
\label{eq:isoBackgrd}
\tilde A^3_{0}\sigma^3= \tilde A_0(\rho) 
\left (\begin{array}{c c}
1 & 0\\ 
0& -1
\end{array}\right ) \, .
\end{equation}
This specific choice of the 3-direction in flavor space as well as  space-time
dependence simplifies the isospin background field strength, such that we get
two copies of the baryonic background~$\tilde F_{\rho 0}$ on the diagonal of the
flavor matrix,
\begin{equation}
\tilde F_{\rho 0}\,\sigma^3=
\left (\begin{array}{c c}
\partial_ \rho \tilde A_0& 0\\ 
0& -\partial_ \rho \tilde A_0
\end{array}\right )\,  .  
\end{equation}
The action for the isospin  
background differs from the action~(\ref{eq:actionEmbeddingsAt} for the baryonic
background only by a group theoretical factor:  
The factor~$T_r=1/2$~(compare  
\eqref{eq:isoAction}) replaces the
baryonic factor~$N_f$ in equation~(\ref{eq:dbiActionD7}), 
which arises by summation over the
$U(1)$ representations. We can thus use the
embeddings~$\chi(\rho)$ and background field solutions~$\tilde A_0(\rho)$ of  
the baryonic case of~\cite{Kobayashi:2006sb}, listed here in
section~\ref{sec:thermoBaryon}. As before, we collect the induced metric~$g$
and the background field strength~$\tilde F$ in the background tensor
$G=g+\tilde F$.

We apply the background field method in analogy to the baryonic case examined in
section~\ref{sec:mesonSpectraB}. As before, we obtain the quadratic action by
expanding the determinant and square root in fluctuations $A^a_\mu$.  The term
linear in fluctuations again vanishes by the equation of motion for
our background field. This leaves the quadratic action
\begin{multline}
\label{eq:quadIsoAction}
S^{(2)}_{\text{iso}} = {\varrho_H}  (2\pi^2 R^3) T_{D7} T_r
\int\limits_1^\infty\!\! \dd\rho\,\int\dd^4x\; \sqrt{\left|\det G\right|}\\   
 \times\Big[G^{\mu\mu'} G^{\nu\nu'} \Big (
       \partial_{[\mu}A^a_{\nu]}\partial_{[\mu'}A^a_{\nu']}
		\hphantom{G^{\mu\mu'}G^{\mu\mu'} G^{\mu\mu'} }\\ 
       \hphantom{G^{\mu\mu'} }+\frac{{\varrho_H}^4}{(2\pi\alpha')^2}(\tilde A_0^3)^2 f^{ab3} f^{ab'3} A_{[\mu}^b\delta_{\nu]0}
        A_{[\mu'}^{b'}\delta_{\nu']0} \Big)\\  
       +(G^{\mu\mu'} G^{\nu\nu'}\!\!\!\! -G^{\mu'\mu} G^{\nu'\nu})
       \frac{{\varrho_H}^2}{2\pi\alpha'} \tilde A_0^3 f^{ab3} \partial_{[\mu'}A^a_{\nu']} A_{[\mu}^{b}\delta_{\nu]0} \Big] .
\end{multline}
Note that besides the familiar Maxwell term, two other terms appear, which are 
due to the non-Abelian structure. One of the new terms  
depends linearly, the other quadratically on the background 
gauge field~$\tilde A$ and both contribute
nontrivially to the dynamics.
The equation of motion for gauge field fluctuations on the D7-brane is
\begin{align}
\label{eq:eomIsoFluct}  
0=\;&\partial_\kappa \left[ \sqrt{\left|\det G\right|}
 \left( G^{\nu\kappa} G^{\sigma\mu} - G^{\nu\sigma} G^{\kappa\mu} \right)
 \check F_{\mu\nu}^a \right] \\ \nonumber   
 & - \sqrt{\left|\det G\right|}
  \frac{{\varrho_H}^2}{2\pi\alpha'} \tilde A_0^3 f^{ab3} \left( G^{\nu 0} G^{\sigma\mu} 
  - G^{\nu\sigma} G^{0\mu} \right)\check F_{\mu\nu}^b
   \, ,  
\end{align}
with the modified field strength linear in fluctuations 
$\check{F}^a_{\mu\nu}=2\partial_{[\mu}A^a_{\nu]}+f^{ab3}
   \tilde A_0^3( \delta_{0\mu} A_\nu^b+ \delta_{0\nu} A_\mu^b){{\varrho_H}^2}/{(2\pi\alpha')}$.

Integration by parts of~\eqref{eq:quadIsoAction} and application of~\eqref{eq:eomIsoFluct} 
yields the on-shell action  
\begin{align}
\label{eq:onShellAction}
S^{\text{on-shell}}_{\text{iso}}=\; 
 &\varrho_H T_r T_{D7}  \pi^2 R^3 \int\!\! \dd^4x\, \sqrt{\left|\det G\right|}\nonumber \\
 & \times\left. \left (  
 G^{\nu 4} G^{\nu'\mu}- G^{\nu\nu'} G^{4\mu}
  \right ) A_{\nu'}^a \check F^a_{\mu\nu}\right |_{\rho_H}^{\rho_B}\, . 
\end{align}
The three flavor field equations of motion~(flavor index $a=1,2,3$) for fluctuations in 
transversal Lorentz-directions $\alpha=2,3$ can again be written in terms of the 
combination~$E^a_T=q A^a_0+\omega A^a_\alpha$.
At vanishing spatial momentum~$q=0$ we get
\begin{align}
\label{eq:eomAalphaSplitFlav1}
0 =\;& {E_T^1}''+\frac{\partial_\rho(\sqrt{\left|\det G\right|} G^{44} G^{22})}{\sqrt{\left|\det G\right|} G^{44} G^{22}} {E_T^1}' \\
     & -\frac{G^{00}}{G^{44}}\big[(\varrho_H \omega)^2+(\tilde A^3_0)^2\big]E^1_T   
        +\frac{2 i \varrho_H \omega G^{00}}{G^{44}} \tilde A^3_0 E^2_T\, ,\nonumber \\
\label{eq:eomAalphaSplitFlav2}
0 =\;& {E_T^2}'' +\frac{\partial_\rho(\sqrt{\left|\det G\right|} G^{44} G^{22})}{\sqrt{\left|\det G\right|} G^{44} G^{22}} {E_T^2}' \\
      & -\frac{G^{00}}{G^{44}}\big[(\varrho_H \omega)^2+(\tilde A^3_0)^2\big]E^2_T 
         - \frac{2 i \varrho_H \omega G^{00}}{G^{44}} \tilde A^3_0 E^1_T \, ,\nonumber \\
\label{eq:eomAalphaSplitFlav3}
\hphantom{.} 0 =\;&  {E_T^3}'' +\frac{\partial_\rho(\sqrt{\left|\det G\right|} G^{44} G^{22})}{\sqrt{\left|\det G\right|} G^{44} G^{22}} {E_T^3}' - \frac{G^{00} (\varrho_H\omega)^2}{G^{44}}E^3_T  \, .
\end{align}
Note that we use the dimensionless background gauge 
field~$\tilde A_0^3=\bar A_0^3 (2\pi \alpha')/\varrho_H$ and $\varrho_H=\pi T R^2$.
Despite the presence of the new non-Abelian terms, 
at vanishing spatial momentum the equations of motion for 
longitudinal fluctuations are the same as the transversal equations~\eqref{eq:eomAalphaSplitFlav1},
\eqref{eq:eomAalphaSplitFlav2} and \eqref{eq:eomAalphaSplitFlav3},  
such that~$E=E_T=E_L$.

Note at this point that there are two essential differences which distinguish this setup from the approach with a constant
potential~$\bar A_0^3$ at vanishing mass followed in~\cite{Erdmenger:2007ap}. First, the inverse metric coefficients $g^{\mu\nu}$ contain the 
embedding function $\chi(\rho)$ computed with varying background gauge field. Second,  
the background gauge field~$\bar A_0^3$, which gives rise to the chemical potential, now depends on $\rho$. 

Two of the ordinary second order differential equations~(\ref{eq:eomAalphaSplitFlav1}), 
(\ref{eq:eomAalphaSplitFlav2}), (\ref{eq:eomAalphaSplitFlav3}) are coupled through their flavor structure. Decoupling can
be achieved by transformation to the flavor combinations~\cite{Erdmenger:2007ap}
\begin{equation}
\label{eq:flavorTrafoNum}
X =E^1+ i E ^2,\;\; \; \;     
Y =E^1- i E ^2\, .
\end{equation}
The equations of motion for these fields are given by
\begin{align}
\label{eq:eomX}
0=&\;{X}'' +\frac{\partial_\rho(\sqrt{\left| \det G \right|} G^{44} G^{22})}{\sqrt{\left|\det G\right|} G^{44} G^{22}} {X}' -
4 \frac{G^{00} \left (\wn- \mathfrak{m}\right )^2}{G^{44}} X\, , \\    
\label{eq:eomY}
0=&\;{Y}'' +\frac{\partial_\rho(\sqrt{\left|\det G\right|} G^{44} G^{22})}{\sqrt{\left|\det G\right|} G^{44} G^{22}} {Y}' -
4 \frac{G^{00}\left (\wn+ \mathfrak m\right )^2}{G^{44}} Y\, , \\   
\label{eq:eomE3}
0=&\;{E^3}'' +\frac{\partial_\rho(\sqrt{\left|\det G\right|} G^{44} G^{22})}{\sqrt{\left|\det G\right|} G^{44} G^{22}} {E^3}' -
4 \frac{G^{00} \wn^2}{G^{44}}E^3\, , 
\end{align}
with dimensionless~$\mathfrak m= \bar A^3_0 /(2\pi T)$ and $\wn=\omega/(2\pi T)$.
Proceeding as described in section~\ref{sec:mesonSpectraB}, we determine the local
solution of~(\ref{eq:eomX}),~(\ref{eq:eomY}) and~(\ref{eq:eomE3}) at the
horizon. The indices turn out to be
\begin{equation}
\label{eq:isospinIndices}
\beta = \pm i \left[\wn \mp\frac{\bar A^3_0(\rho=1)}{(2\pi T)}\right ]\, .
\end{equation}
Since $\bar A^3_0(\rho=1)=0$ in the setup considered here, we are left with the
same index as in \eqref{eq:indices} for the baryon case. Therefore, here the
chemical potential does not influence the singular behavior of the fluctuations
at the horizon. The local solution coincides to linear order with the baryonic
solution given in~(\ref{eq:localSolutions}).

\begin{figure}
\begin{center}
        \includegraphics[width=.7\linewidth]{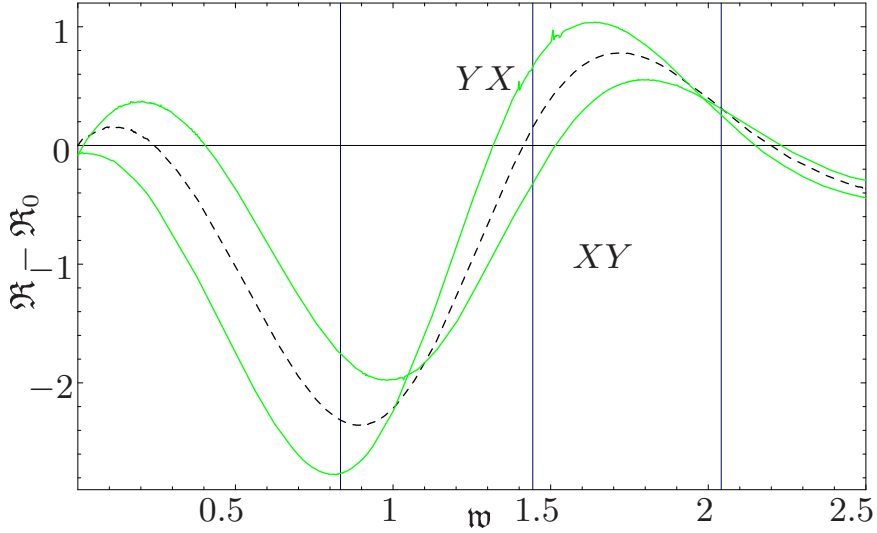}
        \caption{
        The finite temperature part of spectral
        functions~$\R_{\mathrm{iso}}-\mathfrak{R}_{0,\mathrm{iso}}$ (in units
        of~$N_c T^2 T_r/4$) of currents dual to fields~$X,\,Y$ 
        are shown versus~$\wn$. The dashed line shows the baryonic
        chemical potential case, the solid curves show the spectral functions in
        presence of an isospin chemical potential.
        Plots are generated for~$\chi_0=0.5$ and~$\tilde d=0.25$. 
        The combinations $X Y$ and $YX$ split in opposite directions 
        from the baryonic spectral function.
                       }
        \label{fig:isoLinesDt025}
        \end{center}
\end{figure}

\begin{figure}
		\includegraphics[width=.9\linewidth]{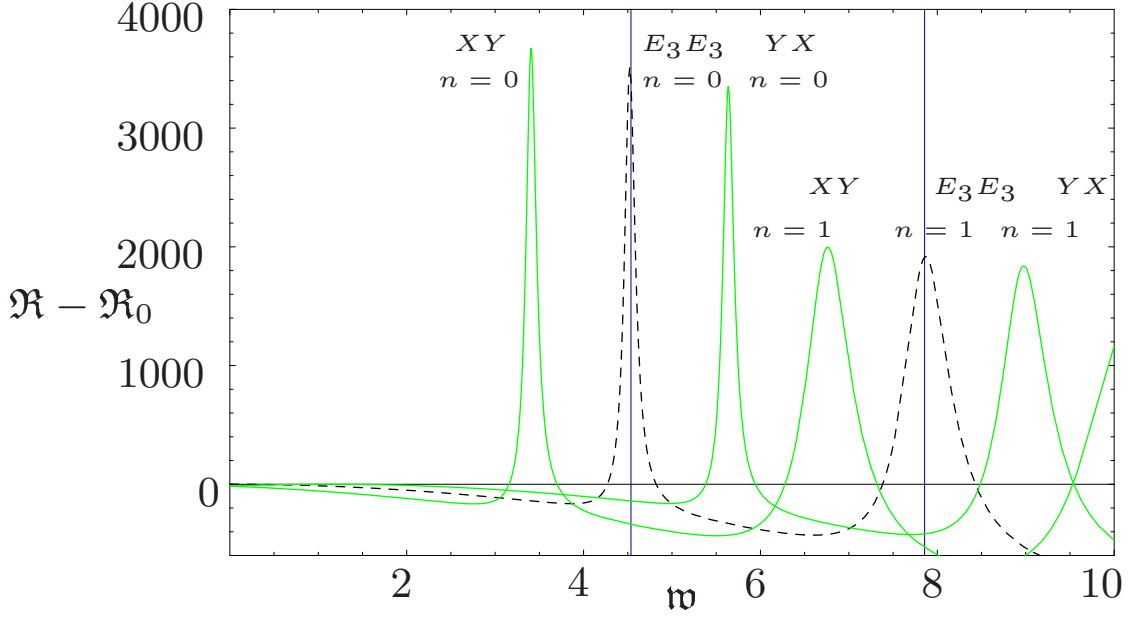}
        \caption{
        A comparison between the finite temperature part of the 
        spectral functions~$\mathfrak{R}_{XY}$ and $\mathfrak{R}_{YX}$~(solid lines)
         in the two flavor directions~$X$ 
        and~$Y$ transversal to the chemical potential
        is shown in units of~$N_c T^2 T_r/4$ for large quark mass to temperature
        ratio~$\chi_0=0.99$ and~$\tilde d=0.25$. The spectral 
        function~$\mathfrak{R}_{E^3E^3}$ along the~$a=3$-flavor direction is shown as 
        a dashed line. We observe a splitting of the line expected at the lowest 
        meson mass at~$\wn=4.5360$~($n=0$). The resonance is shifted to lower
		frequencies for~$\mathfrak{R}_{XY}$ 
        and to higher ones for~$\mathfrak{R}_{YX}$, while it remains in place
        for ~$\mathfrak{R}_{E^3E^3}$. The second meson resonance peak~($n=1$)
        shows a similar behavior. So the different flavor combinations
        propagate differently and have distinct quasi-particle resonances.
          }
        \label{fig:isoXYsplitDt025}
\end{figure}

Application of the recipe described in section~\ref{sec:anaAdsG},~\ref{sec:numAdsG}
and~\eqref{eq:spectralFunction} yields the
spectral functions of flavor current correlators shown in
figures~\ref{fig:isoLinesDt025} and~\ref{fig:isoXYsplitDt025}. Note that after
transforming to flavor combinations~$X$ and~$Y$, given in
\eqref{eq:flavorTrafoNum}, the diagonal elements of the propagation submatrix in
flavor-transverse~$X,\, Y$ directions vanish, $G_{XX}=G_{YY}=0$, while the
off-diagonal elements give non-vanishing contributions. The longitudinal
component~$E^3$ however is not influenced by the isospin chemical potential,
such that~$G_{E^3E^3}$ is nonzero, while other combinations with $E^3$
vanish~(see~\cite{Erdmenger:2007ap} for details).

Introducing the chemical potential as described above for a zero-temperature
$AdS_5\times S^5$ background, we obtain the gauge field correlators in analogy
to~\cite{Freedman:1998tz}. The resulting spectral function for the field theory
at zero temperature but finite chemical potential and density
$\mathfrak{R}_{0,\mathrm{iso}}$ is given by
\begin{equation}
\label{eq:zeroTspecFuncWithIsospin}
\mathfrak{R}_{0,\mathrm{iso}}=\frac{N_c T^2 T_r}{4}4\pi(\wn\pm \mn_\infty)^2   \, ,  
\end{equation}
with the dimensionless chemical 
potential~$\mn_\infty=\lim_{\rho\to\infty}\bar A_0^3/(2\pi T)=\mu/(2\pi T)$.
Note that~\eqref{eq:zeroTspecFuncWithIsospin} is independent of the temperature.
This part is always subtracted when we consider spectral functions at finite
temperature, in order to determine the effect of finite temperature separately,
as we did in the baryonic case.

{\bf Results at finite isospin density}
In figure~\ref{fig:isoLinesDt025} we compare typical spectral functions found
for the isospin case~(solid lines) with that found in the baryonic
case~(dashed line). While the qualitative behavior of the isospin spectral
functions agrees with the one of the baryonic spectral functions, there
nevertheless is a quantitative difference for the components $X,\,Y$, which are
transversal to the background in flavor space. We find that the propagator for
flavor combinations~$G_{YX}$ exhibits a spectral function for which the zeroes
as well as the peaks are shifted to higher frequencies, compared to the Abelian
case curve. For the spectral function computed from~$G_{XY}$, the opposite is
true. Its zeroes and peaks appear at lower frequencies. As seen from
figure~\ref{fig:isoXYsplitDt025}, also the quasi-particle resonances of these two
different flavor correlations show distinct behavior. The quasi-particle
resonance peak in the spectral function~$\R_{YX}$ appears at higher frequencies
than expected from the vector meson mass formula~$\eqref{eq:massFormula2}$~(shown
as dashed grey vertical lines in figure~\ref{fig:isoXYsplitDt025}). The other
flavor-transversal spectral function~$\R_{XY}$ displays a resonance at lower
frequency than observed in the baryonic curve. The spectral function for the
third flavor direction~$\R_{E^3E^3}$ behaves as~$\R$ in the baryonic case.

This may be viewed as a splitting of the resonance peak into three distinct
peaks with equal amplitudes. This is due to the fact that we explicitly break
the symmetry in flavor space by our choice of the background field~$\tilde
A^3_0$. Decreasing the chemical potential reduces the distance of the two outer
resonance peaks from the one in the middle and therefore the splitting is
reduced.

The described behavior resembles the mass splitting of mesons in presence of a
isospin chemical potential expected to occur in QCD according to calculations
in the Nambu-Jona-Lasinio model~\cite{He:2005nk,Chang:2007sr}. A linear dependence of the separation
of the peaks on the chemical potential is expected.
Our observations confirm this behavior. Since our vector mesons are
isospin triplets and we break the isospin symmetry explicitly, we 
see that in this respect our model is in qualitative
agreement with effective QCD models. Note also the complementary discussion of
this point in~\cite{Aharony:2007uu}.

To conclude this section, we comment on the relation of the present results to
those of our previous paper \cite{Erdmenger:2007ap} where we considered a
constant non-Abelian gauge field background for zero quark mass. From equation
\eqref{eq:isospinIndices}, the difference between a constant non-vanishing
background gauge field and the varying one becomes clear. In
\cite{Erdmenger:2007ap} the field is chosen to be constant in~$\rho$ and terms
quadratic in the background gauge field~$\tilde A_0^3\ll 1$ are neglected. This 
implies that the square~$(\wn\mp\mn)^2$ in~\eqref{eq:eomX} and~\eqref{eq:eomY} is replaced by~$\wn^2\mp 2\wn\mn$, 
such that we obtain the indices~$\beta=\pm \wn
\sqrt{1\mp\frac{\bar A^3_0(\rho=1)}{(2\pi T) \wn}}$ instead of~\eqref{eq:isospinIndices}. If we
additionally assume $\wn\ll \tilde A^3_0$, then the $1$ under the square root
can be neglected.
In this case the spectral function develops a non-analytic structure coming from
the $\sqrt \omega$ factor in the index.

However in the case considered here, the background gauge field is a
non-constant function of $\rho$ which vanishes at the horizon. Therefore the
indices have the usual form $\beta = \pm i \omega$ from
(\ref{eq:isospinIndices}), and there is no non-analytic behavior of the spectral
functions, at least none originating from the indices.

It will also be interesting to consider isospin diffusion in the setup of the
present paper. However, in order to see non-Abelian effects in the diffusion
coefficient, we need to give the background gauge field a more general direction
in flavor space or a dependence on further space-time coordinates
besides~$\rho$. In that case, we will have a non-Abelian term in the background
field strength~$\tilde F_{\mu\nu}=\partial_\mu\tilde A^a_\nu-\partial_\nu \tilde
A^a_\mu +f^{abc} \tilde A^b_\mu \tilde A^c_\nu {\varrho_H}^2/(2\pi\alpha')$ 
in contrast to~$\partial_\rho
\tilde A_0^a$ considered here.

\subsection{Peak turning behavior: quasinormal modes and meson masses} \label{sec:peakTurning}
This section serves to discuss the interpretation of resonance peaks appearing in the spectral functions we
computed previously. That interpretation is closely related to understanding the movement of the peaks as
the mass-temperature parameter~$m$ is changed, i.e. the turning of the resonance peaks observed in
section~\ref{sec:mesonSpectraB}. Also the quasinormal modes play an important role here since their 
location in the complex frequency plane is related to the resonance peaks appearing in the spectral function.
This is due to the fact that the quasinormal modes of a gravity field fluctuation are identical to the 
poles in the retarded correlator of the dual gauge theory operator, as was first noticed for metric
fluctuations in~\cite{Son:2002sd}.
Furthermore knowing the quasinormal modes precisely, we can quantify qualitative observations in the
spectral function's behavior. Note, that one important feature to remember about our setup is that 
the quark mass~$M_q$ and the temperature~$T$ do not appear independently but always together in 
the form of the mass parameter~$m = 2 M_q/(\sqrt{\lambda} T)$.

It should be kept in mind that in this present section we collect the intermediate outcomes of our investigation 
and we suggest a few
possible interpretations. Nevertheless, due to the intermediate state of our studies this section is very
speculative and we are working on testing the alternatives and making our line of argument concise.

\begin{figure}
	\psfrag{0}{$0$}
	\psfrag{dt000001}{$\tilde d=10^{-4}$}
	\psfrag{dt02}{$\tilde d=0.2$}
	\psfrag{dt1}{$\tilde d=1$}	
	\psfrag{chi0}{$\chi_0$}
	\psfrag{Omega}{$\wn_{n=0,l=0}$}
        \includegraphics[width=0.9\linewidth]{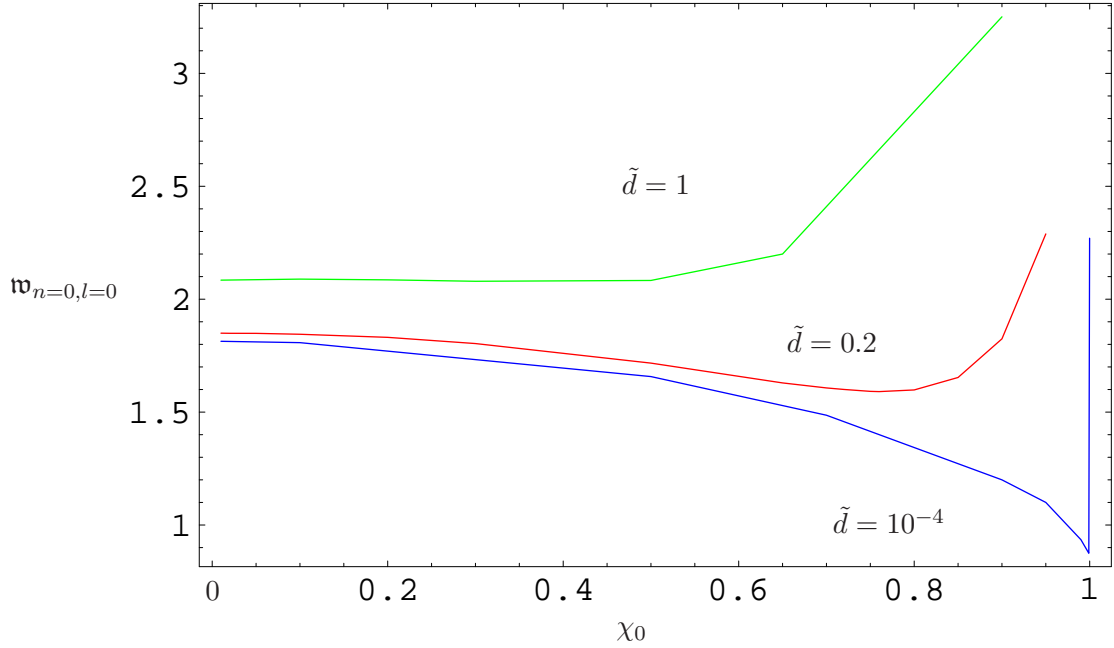}  
                \caption{
                The frequency of the first resonance peak~(mass of lightest vector meson) in the vector spectral function
                is shown depending on the mass of the quarks parametrized by $\chi_0$ for 
                different baryon densities~$\tilde d$. For the lower curves at small density we identify a clear turning 
                point~(minimum).
                }
                \label{fig:turnChi0}
\end{figure}

\begin{figure}
\begin{center}
	\psfrag{0}{$0$}
	\psfrag{dt000001}{$\tilde d=10^{-4}$}
	\psfrag{dt02}{$\tilde d=0.2$}
	\psfrag{dt1}{$\tilde d=1$}	
	\psfrag{chi0}{$\chi_0$}
	\psfrag{wPeak}{$\wn_{\text{peak}}$}	
	\psfrag{Omega}{$\wn_{n=0,l=0}$}
        \includegraphics[width=0.7\linewidth]{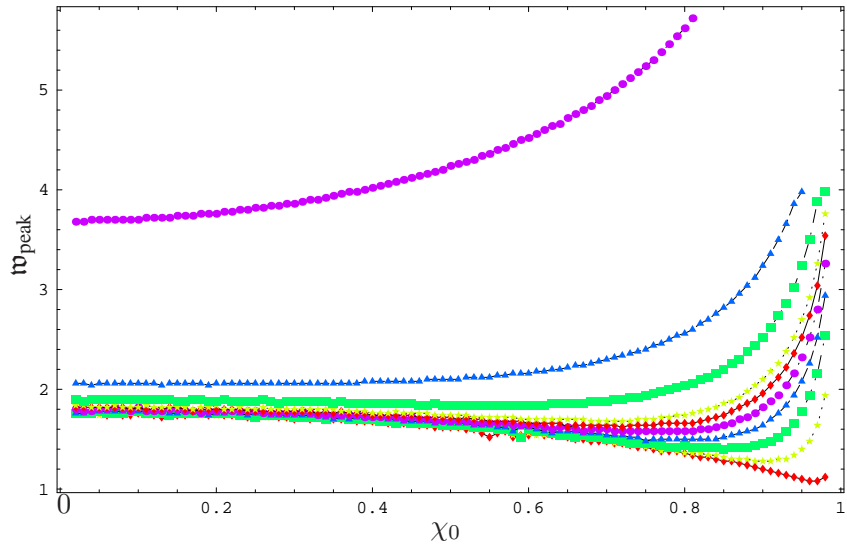}  
                \caption{
                The frequency of the first resonance peak~(mass of lightest vector meson) in the vector spectral function
                is shown depending on the mass of the quarks parametrized by $\chi_0$ for 
                different baryon densities~$\tilde d = 0.01,\, 0.05,\, 0.1,\, 0.15,\, 0.2,\, 0.25,\, 0.3,\, 0.5,\, 1,\,10$. 
                For the lower curves at small density we identify a clear turning point~(minimum) while the peaks
                at large density~$\tilde d >1$ move to higher frequency with increasing parameter~$\chi_0$.
                }
                \label{fig:turnDataChi0}
\end{center}
\end{figure}

\begin{figure}
	\psfrag{deTe}{$\tilde d$}
	\psfrag{chi0turn}{$\chi_0^\text{turn}$}
	\psfrag{0}{$0$}
	\psfrag{0.1}{$0.1$}	
	\psfrag{0.2}{$0.2$}
	\psfrag{0.25}{$0.25$}	
	\psfrag{0.4}{$0.4$}
	\psfrag{0.5}{$0.5$}	
	\psfrag{0.6}{$0.6$}
	\psfrag{0.8}{$0.8$}
	\psfrag{0.8}{$0.8$}		
	\psfrag{-0.2}{$-0.2$}
	\psfrag{-0.25}{$-0.25$}	
	\psfrag{-0.4}{$-0.4$}
	\psfrag{-0.5}{$-0.5$}	
	\psfrag{1}{$1$}
        \includegraphics[width=0.8\linewidth]{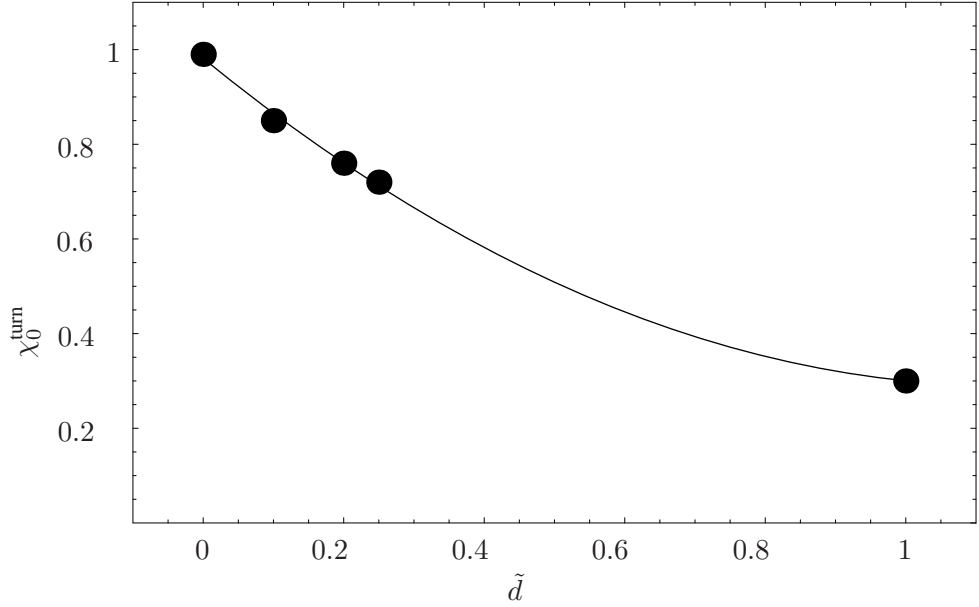}  
                \caption{
                The location of the turning point for the lowest vector meson mass is shown depending on
                the baryon density. Data points read from curves such as given in figure~\ref{fig:turnDataChi0}
                are displayed as dots, the line shows a quadratic fit~$0.98-1.21 \tilde d+0.53{\tilde d}^2$.
                The fit should not be taken too seriously since it is more reasonable to consider the equivalent plot
                for the turning point in terms of the physical parameter~$m$ shown for pseudoscalar excitations
                in figure~\ref{fig:compareDiffPeak}.
                }
                \label{fig:turnDt}
\end{figure}

{\bf Observations}
The three pictures~\ref{fig:turnChi0},~\ref{fig:turnDataChi0} and~\ref{fig:turnDt} summarize an analysis of the 
turning point appearing in 
vector meson spectral functions at finite baryon density at a distinct quark mass to temperature ratio~$m$~(roughly 
$m \propto\chi_0$ up to $\chi_0=0.6$ or $m=0.8$). 
In order to obtain the resonance frequency and decay width of the (quasi) mesons the spectral function peaks 
were locally (all the values of the peak which are above the horizontal axis) fitted to the Lorentz shape
\begin{equation}
\mathfrak{R}-\mathfrak{R}_0|_\text{near peak}=\frac{A \Gamma}{(\wn-\wn_{n=0,l=0})^2+\Gamma^2}\quad ,
\end{equation}
with the free parameter~$A$, the decay width~$\Gamma$ and the lowest vector meson resonance~$\wn_{n=0,l=0}$.
Although this is a crude approach~(the resonances do not have the Lorenz shape but are deformed, cf.~\cite{Amado:2007yr}) 
the location and width of the peaks are captured quite well~(optical check). 
The height of the peaks might be a subject to discussion since the unknown 
parameter~$A$ varies roughly between $0.1$ and $10$ over the scanned parameter range. Nevertheless, this
analysis is merely designed to find qualitative features and for quantitative results we plan to use a different approach
utilizing quasinormal modes.
\begin{figure}
	\psfrag{chi0}{$\chi_0$}
	\psfrag{Omega}{$\Omega$}
	\psfrag{Gamma}[lb]{\hspace{1cm}$\Gamma$}
	\psfrag{0.1}{$0.1$}	
	\psfrag{0.2}{$0.2$}
	\psfrag{0.25}{$0.25$}	
	\psfrag{0.4}{$0.4$}
	\psfrag{0.5}{$0.5$}	
	\psfrag{0.6}{$0.6$}
	\psfrag{0.8}{$0.8$}
	\psfrag{0.8}{$0.8$}		
	\psfrag{0.02}{$0.02$}
	\psfrag{0.06}{$0.06$}	
	\psfrag{0.04}{$0.04$}
	\psfrag{0.08}{$0.08$}	
	\psfrag{0.12}{$0.12$}	
	\psfrag{1}{$1$}	
        \includegraphics[width=0.8\linewidth]{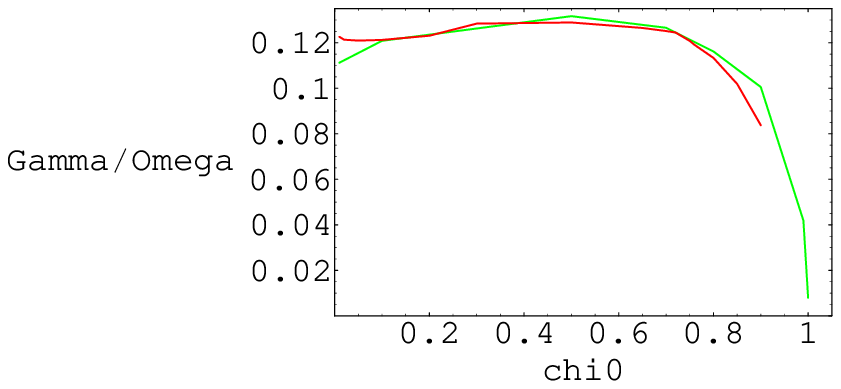}  
                \caption{
                The {\bf preliminary} Thirring coupling versus embedding parameter~$\chi_0$ in the case of $\tilde d=0$ (green) and
                $\tilde d=0.25$ (red).
                }
                \label{fig:thirring}
\end{figure}  

The movement of the resonance frequency visible in figure~\ref{fig:turnChi0} suggests two distinct limits.
First, there seems to be no turning point in the case of zero density. With increasing mass parameter~$m$
the resonance peak moves further and further to lower frequencies. Since the turning point should not be negative, 
we expect either that the curve goes back up or asymptotes to some finite value. The latter conclusion agrees with the spectra
shown in~\cite{Myers:2007we}, where for the case of vanishing baryon density the peaks were found to
approach a distinct small frequency as the mass parameter is increased towards its critical value~$\chi_0\to 1$.
The decrease of the turning point value with increasing baryon density as shown in figure~\ref{fig:turnDt} 
suggests that at vanishing density the turning point would lie at the critical embedding~$\chi_0=1$, corresponding
to a quark mass of~$m(\tilde d=0,\chi_0=1)\approx 1.3$.
 
Second, in the limit of large densities~$\tilde d\gg1$ we again find that the turning point disappears but
now there are only right-moving peaks approaching larger and larger frequencies as the mass parameter
is increased. Note, that this behavior agrees with what we expect if we are to identify the resonance peaks with
meson masses as discussed in section~\ref{sec:mesonSpectraB}. However, the peak movement towards smaller
frequencies in the limit of vanishing density as well as at intermediate densities is a rather unexpected
feature in the context of the meson mass interpretation. We may also say that the peak movement to
smaller frequencies causes the appearance of the peak turning point. For this reason later we will focus on explaining
the movement  of peaks to smaller frequencies and we will start with the vanishing density case for simplicity below in
the paragraph `Heuristic gravity interpretation'. 

\begin{figure}
	\psfrag{ReF}{$\mathrm{Re} F$}
	\psfrag{ImF}{$\mathrm{Im} F$}
	\psfrag{rho}{$\rho$}
	\psfrag{0}{\tiny$0$}
	\psfrag{0.1}{\tiny$0.1$}	
	\psfrag{0.2}{\tiny$0.2$}
	\psfrag{0.25}{\tiny$0.25$}	
	\psfrag{0.4}{\tiny$0.4$}
	\psfrag{0.5}{\tiny$0.5$}	
	\psfrag{0.6}{\tiny$0.6$}
	\psfrag{0.75}{\tiny$0.75$}
	\psfrag{0.8}{\tiny$0.8$}		
	\psfrag{-0.2}{\tiny$-0.2$}
	\psfrag{-0.25}{\tiny$-0.25$}	
	\psfrag{-0.4}{\tiny$-0.4$}
	\psfrag{-0.5}{\tiny$-0.5$}	
	\psfrag{-0.6}{\tiny$-0.6$}
	\psfrag{20000}{\tiny$20000$}
	\psfrag{40000}{\tiny$40000$}
	\psfrag{60000}{\tiny$60000$}
	\psfrag{80000}{\tiny$80000$}
	\psfrag{100000}{\tiny$100000$}										
        \includegraphics[width=0.45\linewidth]{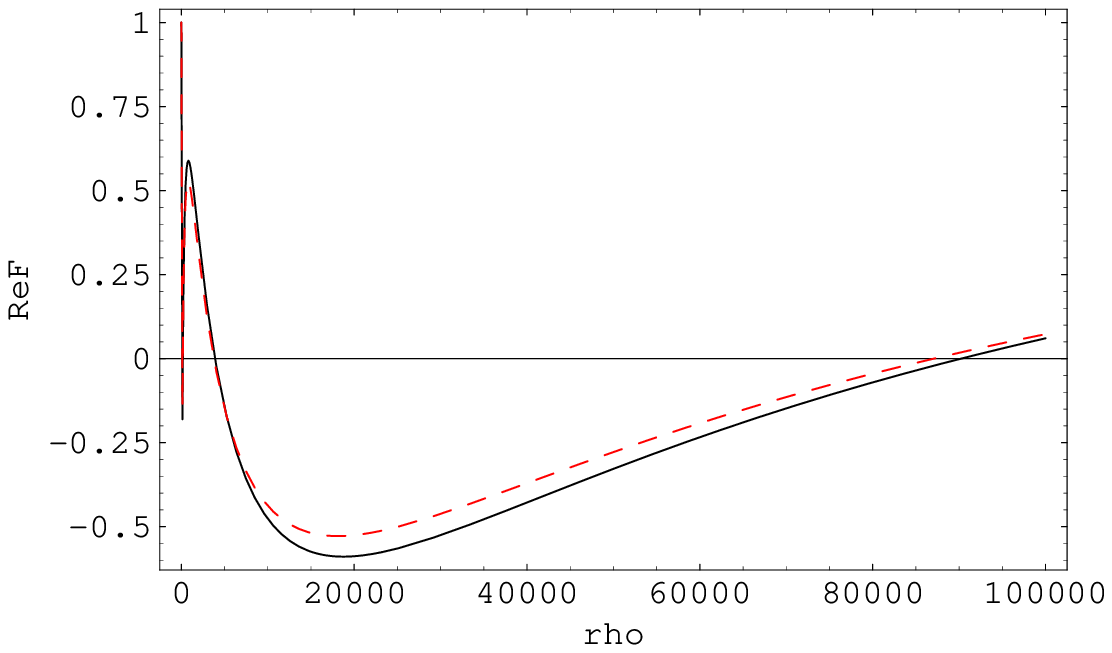}  \hfill
        \includegraphics[width=0.45\linewidth]{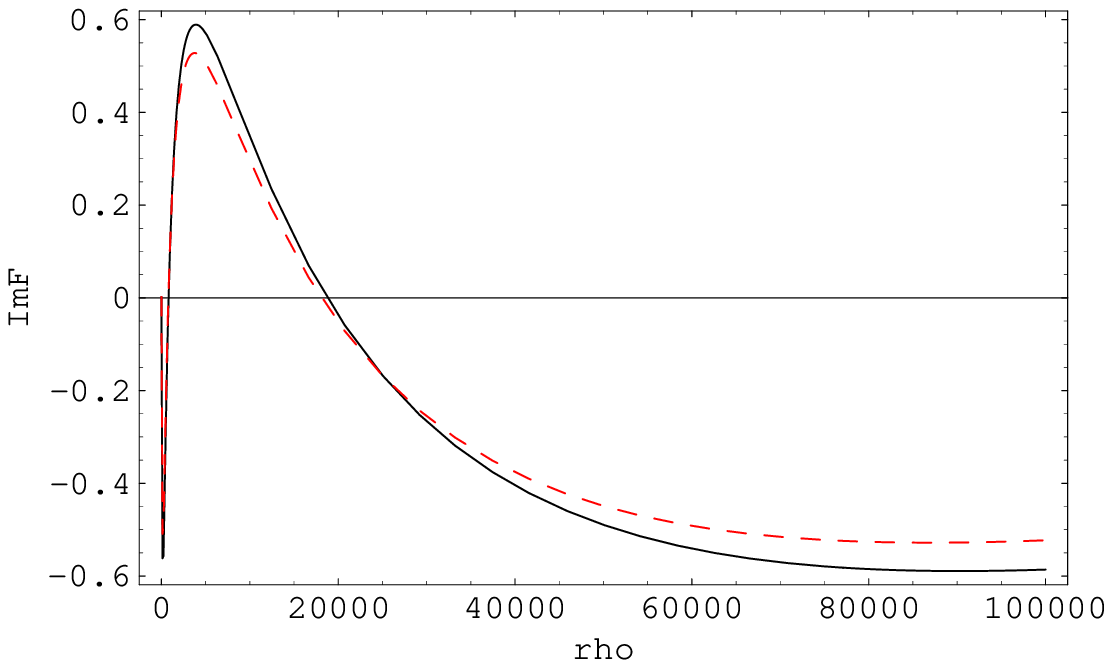}  \vfill
        \includegraphics[width=0.45\linewidth]{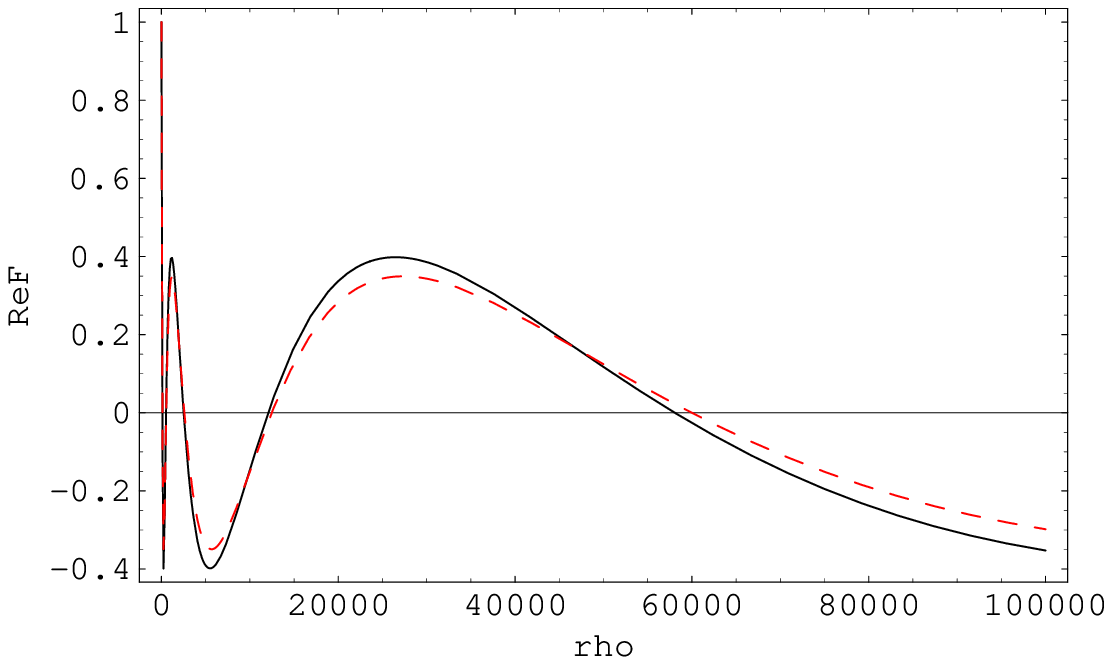}  \hfill
        \includegraphics[width=0.45\linewidth]{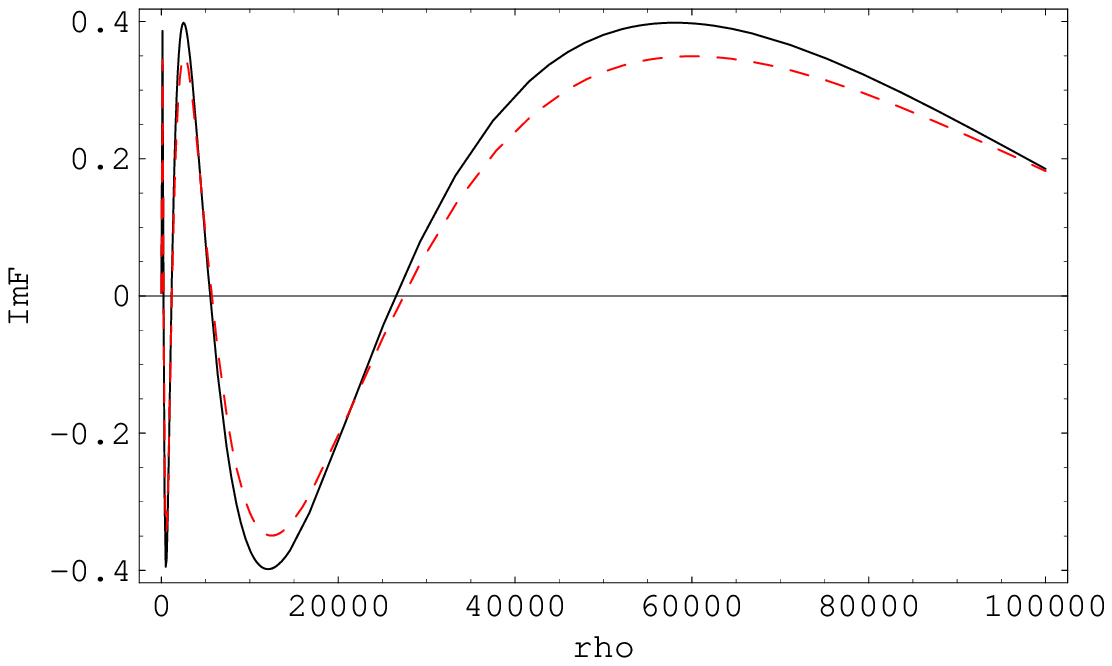}                          
                \caption{
                The real and imaginary part of solutions~$F(\rho)$ are shown versus the radial AdS-coordinate~$\rho$.
                Each plot shows two curves one of which is evaluated at vanishing mass~$\chi_0=0$~(solid black) 
                while the other is generated at a finite mass~$\chi_0=0.4$~(dashed red). The two plots on top are
                generated by a vector perturbation with energy~$\wn=1$ while the two lower plots show the equivalent
                results at the doubled energy~$\wn=2$. A quasinormal mode would satisfy the boundary condition
                $\lim\limits_{\rho\to\rho_{\text{bdy}}}|F| = 0$ at the boundary.
                }
                \label{fig:solutionsFChi0W}
\end{figure}

In order to understand what causes the resonance peaks in the spectral function to move towards smaller frequencies
with increasing mass parameter~$m$, we now examine the solutions to the regular function~$F$~(cf.~\eqref{eq:localSolutions}) 
which we found numerically and from which the spectral function is essentially computed using~\eqref{eq:specFuncFormula}. 
In figure~\ref{fig:solutionsFChi0W} the real and imaginary part of the regular function~$F(\rho)$ are shown versus the
radial coordinate~$\rho$. The two upper plots show the solution for a vector perturbation with energy~$\wn=1$, the two 
lower plots for~$\wn=2$. In all four plots the solid black line shows results for the flat~(massless) embedding~$\chi_0=0$,
the red dashed curve is evaluated at a finite mass~$\chi_0=0.4$. The real and imaginary part of~$F(\rho)$ show 
a similar oscillation behavior with decreasing frequency for larger~$\rho$. The lower curves at~$\wn=2$ display more
oscillations over the entire range of~$\rho$ than the upper ones at~$\wn =1$. Note that figure~\ref{fig:solutionsFChi0W}
shows the whole radial variable range since for the numerical solution we used the cutoffs:~$\rho^{\text{num}}_h =1.00001,\,
\rho^{\text{num}}_{\text{bdy}}= 10^5$. This means that figure~\ref{fig:solutionsFChi0W} shows all the oscillations 
which are present in the solutions over the whole AdS. This is a key observation for our interpretation since it
means that there is only a finite amount of oscillations in each solution and the number of oscillations increases
with increasing energy~$\wn$. We will come back to this observation in the paragraph~`Heuristic gravity interpretation'.
From figure~\ref{fig:solutionsFChi0W} it is also evident that the red dashed curve at larger~$\chi_0=0.4$ does
not reach the amplitude of the solid blac flat embedding curve at~$\chi_0=0$. One is tempted to interpret
that with growing mass parameter~$\chi_0$ or equivalently~$m$ the solution~$F(\rho)$ gets damped more and more.

Considering especially the real parts of the solutions displayed in the 
left column of figure~\ref{fig:solutionsFChi0W} we observe that the amplitude of this 'streched oscillation' 
near the horizon~$\rho=1$ first drops 
rapidly to remain almost constant in the rest of the variable range~$10\sim \rho \le\rho_{\text{bdy}}$. Note in particular, 
that all these features appear already in the massless embedding~(solid black line in figure~\ref{fig:solutionsFChi0W}).
Therefore we are lead to conclude that these features of the solutions are caused by the finite temperature background
(the pure AdS solution in terms of Bessel functions would show amplitude damping but no change of frequency). 

\begin{figure}
\begin{center}
	\psfrag{s}{\Huge $s$}
	\psfrag{Im F}{\Huge $\mathrm{Im} F$}
	\psfrag{Re F}{$\mathrm{Re} F$}	
	\psfrag{chi0.01}{ $\chi_0=0.01$}
	\psfrag{chi0.5}{ $\chi_0=0.5$}
	\psfrag{chi0.8}{$\chi_0=0.8$}
	\psfrag{chi0.9}{$\chi_0=0.9$}			
        \includegraphics[angle=-90,width=0.75\linewidth]{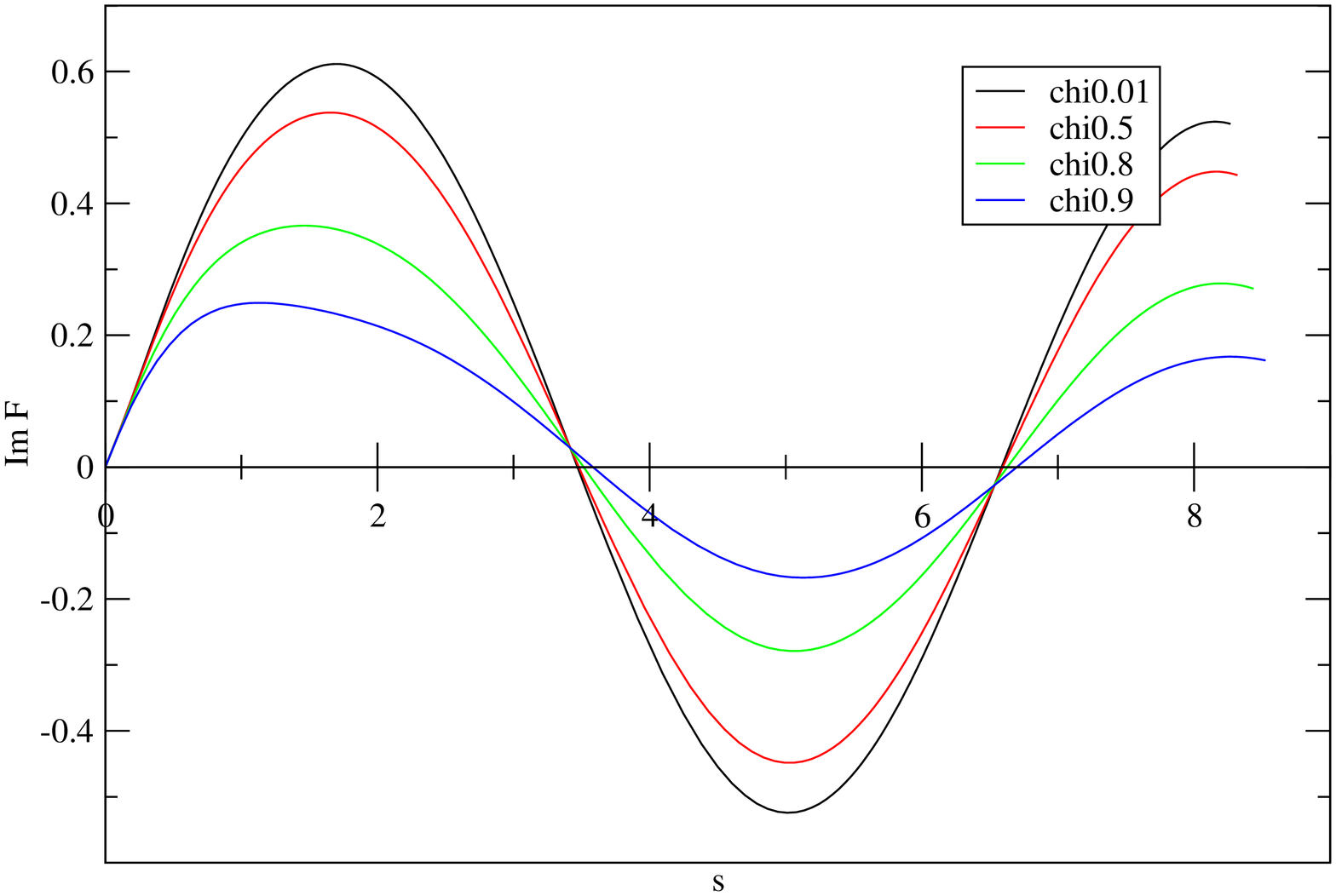}  
        \hfill
        \includegraphics[angle=-90,width=0.75\linewidth]{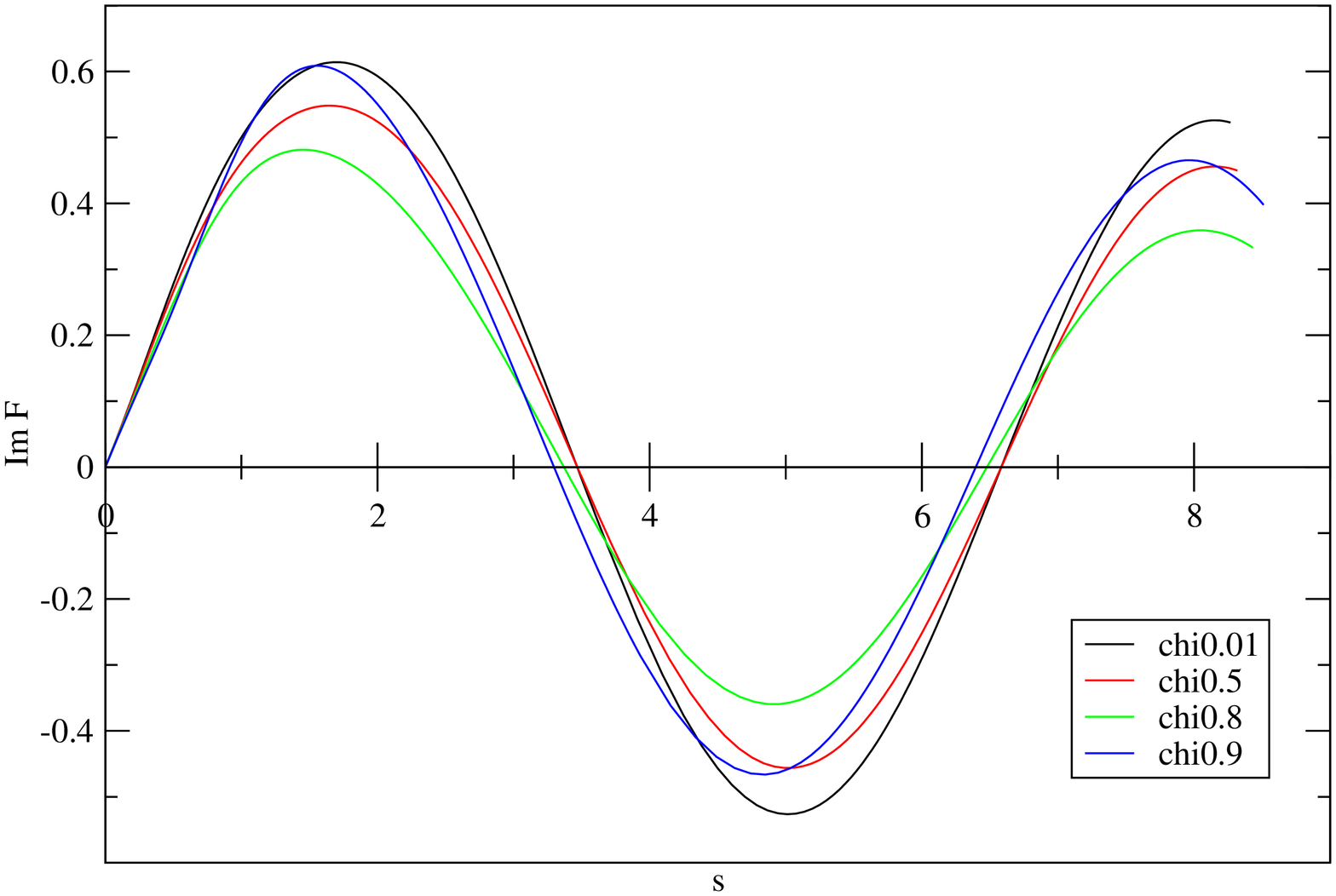}
                \caption{
                Imaginary part of the solution to the regular function~$F$ versus the proper radial 
                coordinate~$s$. 
                The upper plot at vanishing baryon density~$\tilde d=0$ shows that the initially sinusodial
                solution is deformed as the mass parameter~$\chi_0$ is increased. Furthermore, its amplitude  
                decreases while the wave length increases. The upper plot shows that introducing a finite
                baryon density~$\tilde d=0.2$ causes the solutions to change their behavior with increasing~$\chi_0$:
                While the first three curves for~$\chi_0=0.01,\,0.5,\, 0.8$ show the same qualitative behavior as 
                those in the upper plot, the blue curve for~$\chi_0=0.9$ clearly signals a qualitative change with its
                increased amplitude. Looking at the wave lengths in the lower plot we realize that already the 
                green curve~($\chi_0=0.8$) shows a decreased wave length as well as the blue curve~($\chi_0=0.9$).
                These plots have been generated by Patrick Kerner~\cite{Kerner:2008diploma}.
                }
                \label{fig:FOfS}
\end{center}                
\end{figure}

Nevertheless, we should not forget that the coordinate~$\rho$ displayed in figure~\ref{fig:solutionsFChi0W}
is not the radial distance which the mode experiences but the distance which is measured by an observer at 
infinity. Therefore the picture might be distorted. In order to get the physical distance which a comoving observer 
measures we have to transform to the proper radial coordinate
\begin{equation}
\label{eq:properRadial}
s = \int \dd \rho \sqrt{G^{44}(\rho,\chi)} \, ,
\end{equation}
where~$G$ is the metric induced on the D7-brane being a function of the variable~$\rho$ and the 
embedding~$\chi$ in general. Since we only have a numerical expression for~$\chi$ we can not find an
analytic expression for the coordinate~$s$. Either we get~$s$ numerically from the integration~\eqref{eq:properRadial} 
or we restrict ourselves to a near horizon approximation where we know that~$\chi(\rho) = \chi_0 + \chi_2 (\rho -1)^2 +\dots$.
We choose the numerical approach.
The solution~$F$ is plotted versus the proper coordinate in figure~\ref{fig:FOfS} near the horizon. Note that
the range of~$0\le s\le 9$ shown in these plots corresponds to a much larger range in the original coordinate
$1\le \rho \le 4000$. We observe
that the solution oscillates with apparently constant frequency and an evident decrease of the amplitude. Note that 
the decrease of the amplitude is very smooth here~(compare the first and second maximum for each curve). Increasing
the mass parameter~$\chi_0$ or equivalently~$m$ we find from the upper plot in figure~\ref{fig:FOfS} that the amplitude
is decreasing from curve to curve while the proper wave length grows. We argue that this wavelenght growth
is responsible for the shift of resonance peaks to smaller frequencies. A qualitative change of this situation which 
confirms our suspicion happens if we switch on a finite baryon density~(cf.~right plot in~\ref{fig:FOfS}). In this case
the decrease of the amplitude is diminished and the growth of the wavelenght is stopped and we observe a turning
behavior with growing amplitude and decreasing wave length for~$\chi_0=0.9$~(blue curve). 

The proper distance~(on the brane) between the horizon and a distinct point~$\rho$ in the bulk~$\Delta s=s-s_H$
depends on the embedding function~$\chi_0$ as seen from equation~\eqref{eq:properRadial}. In fact with increasing
mass parameter~$\chi_0$~(or~$m$) we find that the distance~$\Delta s$ also increases. This is already obvious
from the embeddings for increasing~$\chi_0$ shown in figure~\ref{fig:backgAt}. There the spike reaching from the 
brane to the horizon becomes larger and larger with increasing~$\chi_0$ and thus when traveling the same
distance in the coordinate~$\rho$, on the brane or rather on the spike one travels a longer and longer distance.

Before we consider possible explanations let us record one last observation comparing the movement
of the resonance peaks and the minimum appearing in the diffusion coefficient~(cf.~section~\ref{sec:baryonDiffusion}).
\begin{figure}
\begin{center}
	\psfrag{dt}{\Huge $\tilde d$}
	\psfrag{chiturn}{\large $\chi^{\text{turn}}$}
	\psfrag{mturn}{\large $m^{\text{turn}}$}	
	\psfrag{chimin}{\large $\chi^{\text{min}}$}
	\psfrag{mmin}{\large $m^{\text{min}}$}
        \includegraphics[angle=-90,width=0.9\linewidth]{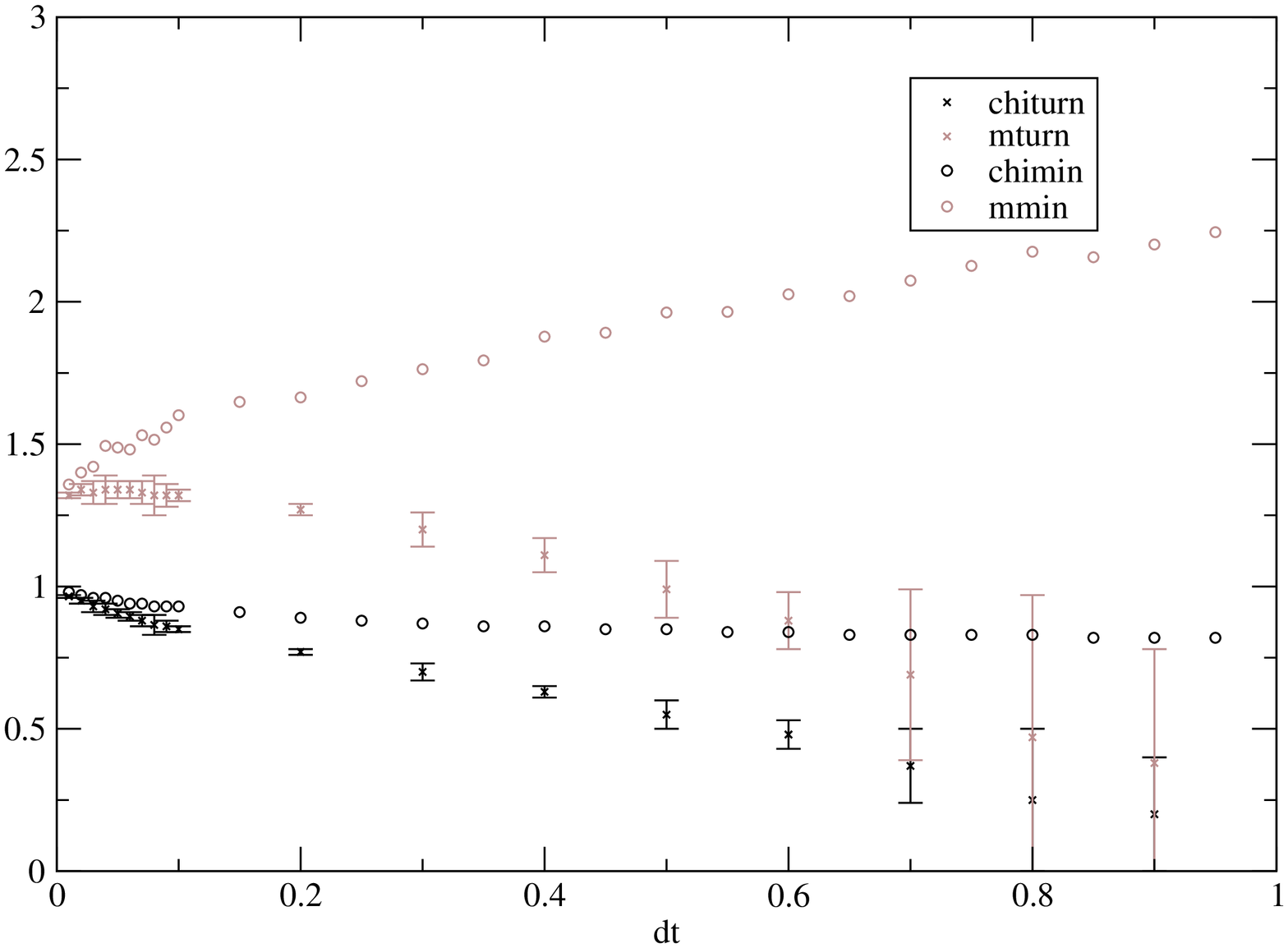}  
                \caption{
                Comparison of the resonance peak movement to the motion of the diffusion minimum
                versus the baryon density in the case of scalar fluctuations. This plot was generated
                by Patrick Kerner~\cite{Kerner:2008diploma}.
                }
                \label{fig:compareDiffPeak}
\end{center}                
\end{figure}  
In figure~\ref{fig:compareDiffPeak} the lower curve shows the location of the first resonance peak in the 
spectral function plotted against the density~$\tilde d$. With increasing~$\tilde d$
the peak moves to lower mass values~$m$. However, the upper curve shows that the location of 
the diffusion minimum with increasing density~$\tilde d$ moves to larger~$m$. This observation suggests
that these two quantities are driven apart from each other by an effect generated through the finite baryon density. 

\begin{figure}
\begin{center}
	\psfrag{m}{$m$}
	\psfrag{Omega}{$\frac{\Gamma}{\Omega}$}
	\psfrag{Gamma}{}
	\psfrag{Imww}{$\mathrm{Im}\, \wn$}
	\psfrag{Reww}{$\mathrm{Re}\, \wn$}	
        \includegraphics[width=0.65\linewidth]{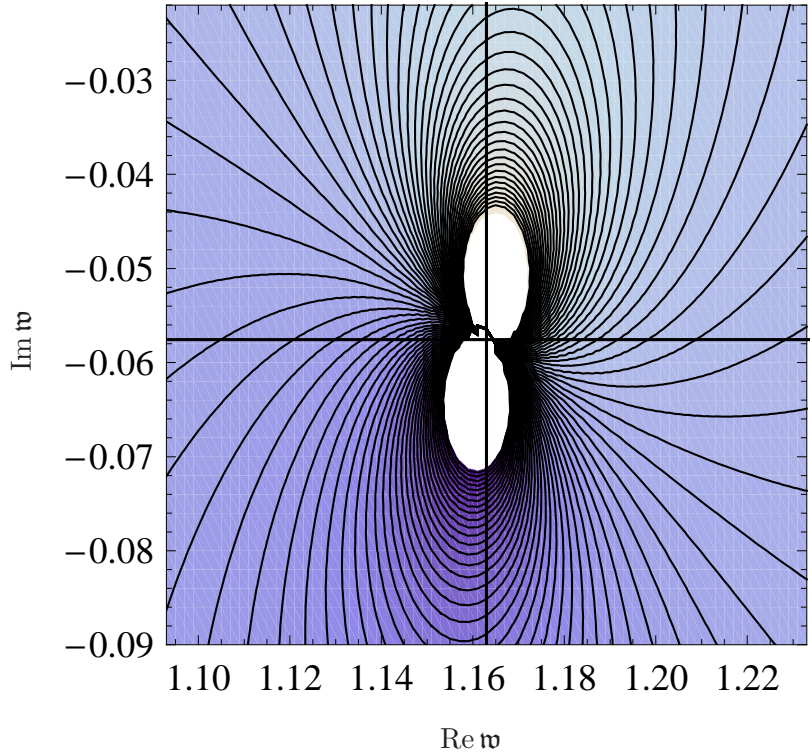}  
                \caption{
                Contour plot of the flavor current spectral function surface near the lowest quasinormal mode in
                the limit~$\qn=0$. Note, that this is not the diffusion pole. This plot has been generated
                by Felix Rust~\cite{Rust:phd}.
                }
                \label{fig:poleContour}
\end{center}                
\end{figure}

{\bf Heuristic gravity interpretation}
We now approach the interpretation of the left moving resonance peaks from the gravity side finding out 
how the solutions~$F$ to the equation of motion change with increasing mass parameter~$m$ and how
in turn this influences the spectral function peaks. So our task is to follow a distinct peak~(e.g. the first
resonance peak) appearing in the spectral function at a certain~$\mathrm{Re}\wn$ while we are
changing the mass parameter~$m$. The first problem that arises is how to identify those
solutions~$F$ which produce a particular peak in the spectral function. We would have to scan all possible~$\wn$
for each choice of~$m$. Therefore we take a more elegant detour via the quasinormal modes. 
As we have argued before in figure~\ref{fig:polesExample} 
the spectral function can be seen as the real frequency edge of a spectral function landscape over the 
complex frequency plane. The resonance peaks we observe in the spectral function over real~$\wn$ 
are caused by poles in the complex frequency plane appearing exactly at the quasinormal mode frequencies
of the equation of motion~\eqref{eq:eomEq0}. Although at the moment we do not have a concise quantitative 
relation between the quasinormal frequencies and the exact location of the resonance peaks in the spectral 
function at real~$\wn$, we assume that the qualitative motion of the resonance peaks is directly caused 
by the corresponding motion of the quasinormal frequencies as~$m$ is changed. In other words we assume
here that if we can show that the quasinormal frequencies are shifted to smaller~$\mathrm{Re}\wn$ 
as~$m$ is increased, then we have also shown that the resonance peaks move to smaller~$\mathrm{Re}\wn$.
This is confirmed by observations from contour plots of the spectral function near quasinormal mode locations
such as figure~\ref{fig:poleContour}.
At the moment we will just take this as an assumption motivated by our observations but we are 
momentarily working on a concise relation.

Quasinormal modes have a determined behavior at the  boundary since by 
definition~(cf.~section~\ref{sec:qnm}) they have to vanish there~$F_{\text{QNM}}(\rho=\rho_{\text{bdy}}) =0$. 
This means that if we keep this boundary condition satisfied by adjusting~$\wn$ as we dial through values of~$m$, we 
always pick that particular solution~$F$ which generates the pole in the spectral function at the 
quasinormal frequency. Thus we have solved the problem how to identify those solutions 
responsible for generating a peak in the spectral function.

Connecting the observation of finitely many oscillations in each solution in figure~\ref{fig:solutionsFChi0W} with
the distinct boundary condition at the AdS-boundary, we know that each quasinormal mode solution~$F_{QNM}$ at
the complex value~$\wn_{QNM}$ is fixed on both ends of the variable range~($\rho_H,\,\rho_{\text{bdy}}$) 
and shows a finite number of oscillations in between. This behavior is very similar to that of solutions
we would expect from quantum mechanics in a box. For this reason we start our line of argument with
the assumption that in the case at hand AdS-space in radial direction can be seen as a `box'. As we have
seen in figure~\ref{fig:solutionsFChi0W}, changing the mass parameter~$m$ or equivalently~$\chi_0$, 
causes the solution~$F$ to change. In our `box' picture we now attribute this change to the change of
the size of the AdS-`box'. Increasing~$m$ is equivalent to decreasing the temperature~$T$ which 
results in shifting the location of the horizon in the dimensionful coordinate~$\varrho=\varrho_H$ to a
smaller value since~$\varrho_H= \pi T R^2$. This means that we increase the distance between the
horizon and the boundary which makes the `box' larger. In order for the same number of 
oscillations~\footnote{Different numbers of oscillations correspond to the different quasinormal modes 
and according to our reasoning also to the different peaks appearing in the spectral function. Here 
we only want to follow one single peak in the spectral function and therefore we keep the number of oscillations constant.}
of~$F$ to fit into the larger box, the effective wave length has to grow and equivalently the effective 
frequency of the mode has to shrink. It is this shrinking effective frequency which we suspect to cause
a movement of the quasinormal frequency to smaller real parts and eventually to cause the left-motion of the 
resonance peaks versus real~$\wn$.

Note, that the heuristic description of AdS as a box with its size depending on the mass parameter
is supported by our discussion of the proper length~$s$~(cf.~equation~\eqref{eq:properRadial}) 
which the mode experiences on the brane. 

Looking at the problem even more generally, we notice that the peak motion to smaller frequencies 
appears exclusively at small values of the mass parameter~$m$ or equivalently at high temperatures. 
As we have seen in the analysis of (quasi)meson spectra in section~\ref{sec:mesonSpectraB} in this
parameter range it is no longer possible to identify the resonances as quasi-particles. Due to their
large decay width we should rather consider them to be short-lived mesonic excitations in the plasma. 
In this regime the finite temperature effects overcome the vacuum effects governed by supersymmetry.
Therefore it is natural to look for a thermal interpretation of the left-motion of these resonance peaks as 
an effect of the plasma interacting with the probe quarks. If this interaction on the gravity side could be
found to damp the function~$F$ and to become stronger as~$m$ is increased, this could give an 
explanation for the decreasing frequency in analogy to a damped harmonic oscillator. Exactly this
is the approach we take in the next paragraph to find an analytic solution.

{\bf Analytical results}
Motivated by the exact numerical solution to the fluctuation equations of motion shown in figure~\ref{fig:FOfS}, we suspect
that this damped oscillating curve near the horizon can be approximated by a damped {\it quasi-harmonic} 
oscillator, i.e. we should be able to find an approxiate equation of motion which is a generalization of the 
damped harmonic oscillator equation. By {\it quasi-harmonic} we mean that the oscillator is damped with the damping depending on the 
location of the mode in radial direction. From the observations in figure~\ref{fig:solutionsFChi0W} we 
have already concluded that the amplitude changes rapidly near the horizon and ceases to change very quickly 
in order to stay virtually constant until the boundary is reached. Thus it is reasonable to assume that the damping of
the mode~$F$ mainly takes place near the horizon and a near-horizon approximation can capture this effect.
In this spirit we take the near-horizon limit~$\varrho \sim 1$ and at the same time the high-frequency limit~$\wn \gg 1$. 

\begin{figure}
	\psfrag{rho}{$\rho$}
	\psfrag{Omega}{$\frac{\Gamma}{\Omega}$}
	\psfrag{0.1}{$0.1$}
	\psfrag{0.2}{$0.2$}	
	\psfrag{0.5}{$0.5$}
	\psfrag{0.175}{$0.175$}
	\psfrag{0.15}{$0.15$}
	\psfrag{0.125}{$0.125$}
	\psfrag{0.075}{$0.075$}
	\psfrag{0.05}{$0.05$}
	\psfrag{0.025}{$0.025$}	
	\psfrag{0}{$0$}
	\psfrag{1}{$1$}	
	\psfrag{-0.5}{$-0.5$}
	\psfrag{-1}{$-1$}	
	\psfrag{ImFAna}{$\mathrm{Im}\, F$}
        \includegraphics[width=0.9\linewidth]{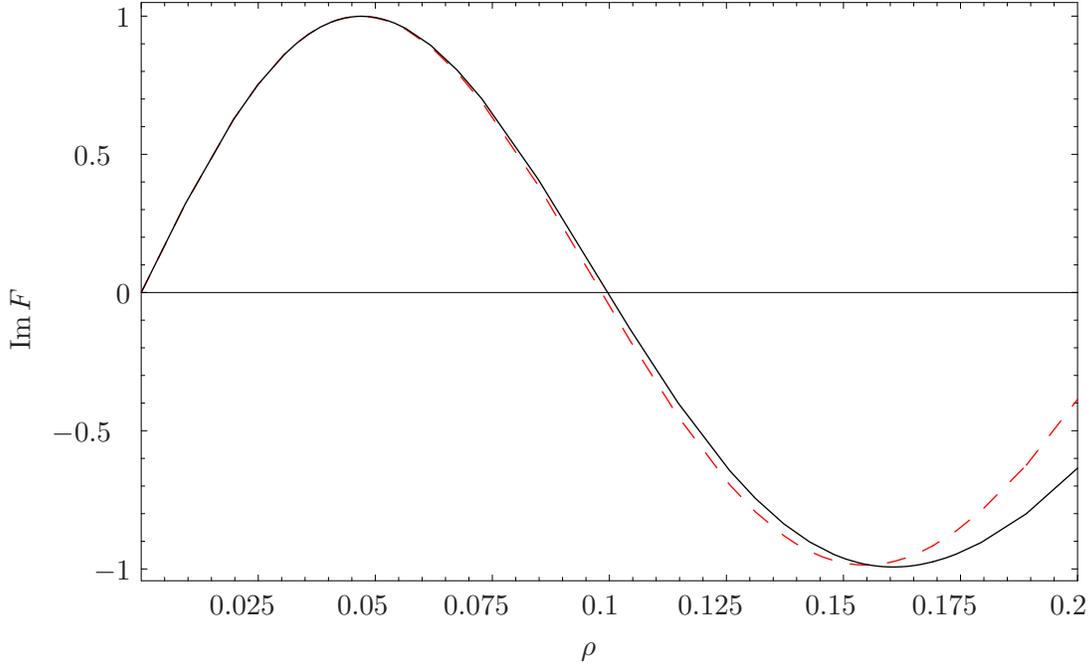}  
                \caption{
 		  Approximate analytic solution compared to the exact solution at~$\wn = 70 \, ,\tilde d = 0\, ,\chi_0= 0.4$.
                }
                \label{fig:FApprox}
\end{figure}

\begin{figure}
	\psfrag{RNearHorizon}{$\frac{\partial_\rho E}{E}$}
	\psfrag{w}{$\wn$}
        \includegraphics[width=0.5\linewidth]{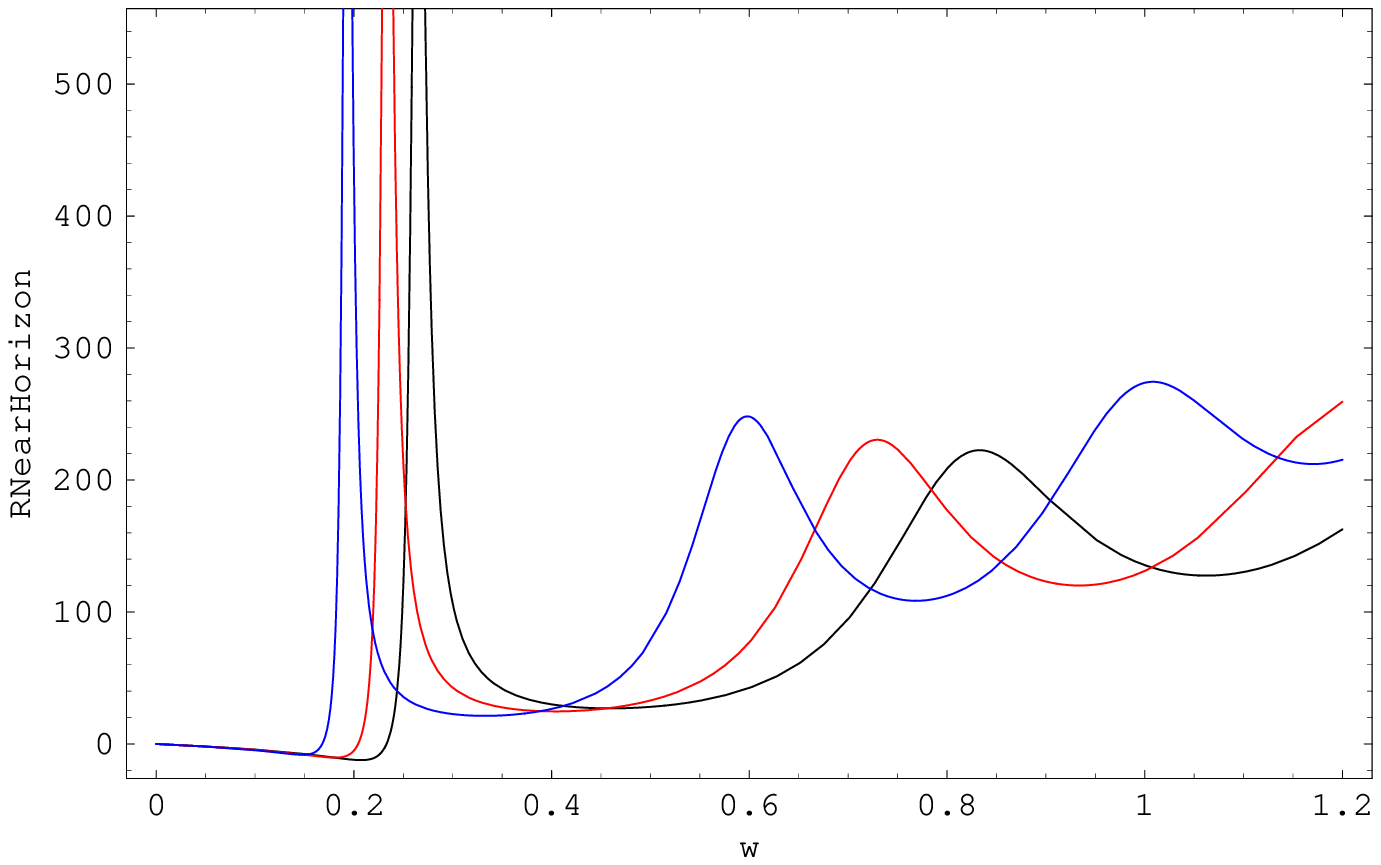}  
        \hfill
        \includegraphics[width=0.5\linewidth]{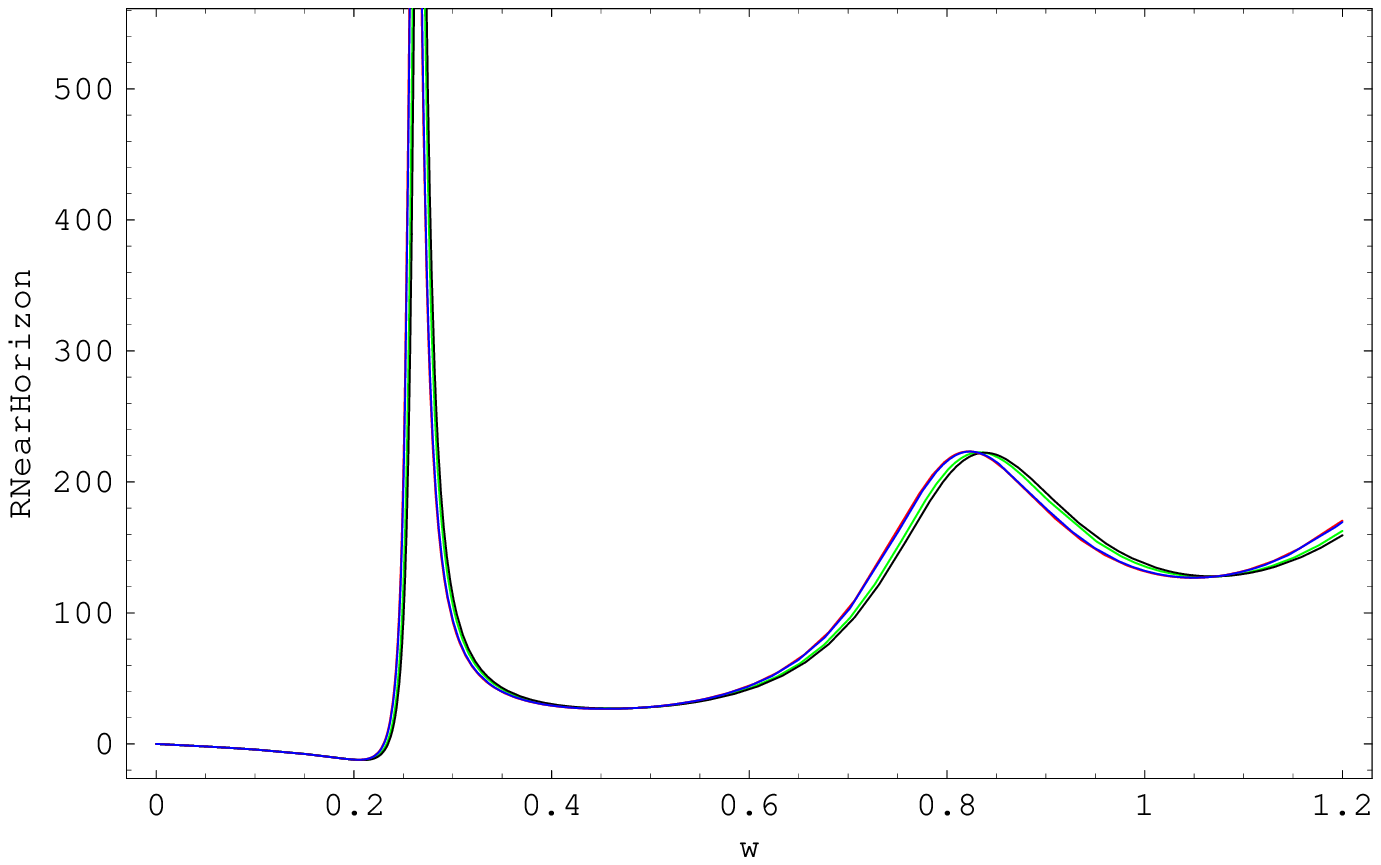}   
                \caption{
 		  Approximate spectral function fraction near the horizon computed with
		  the function~$E=(\rho-1)^\beta F(\rho)$ and~$F$ being the analytic approximation given
		  in equation~\eqref{eq:anaF}.
                }
                \label{fig:nearHorR}
\end{figure}

Applying these limits for the flat embedding~$\chi_0=0$ in the equation of motion~\eqref{eq:eomEq0} 
we obtain the simplified equation of motion
\begin{equation}
y H'' +\left (-2 i \wn -y \right ) H' + i\wn\left( \frac{1}{\sqrt{7}} + 1 \right ) H = 0 \, ,
\end{equation}
where the variable is~$y=2 i\wn\frac{\sqrt{7}}{2}(\varrho -1)$ and the regular 
function~$H(y)$ comes from the Ansatz~$E = (\varrho -1)^\beta F$ with the 
redefinition~$F=e^{-\sqrt{7}/2 \wn (\varrho-1)} H$. 
This equation of motion has the form of {\it Kummers equation}, which
is solved by the confluent hypergeometric function of first~$H={}_1 F_1[-i\wn(1/\sqrt{7}+1),-2i\wn,x]$ and second kind~$U$. 
Boundary conditions rule out the second kind solution which is non-regular at the horizon and therefore contradicts the 
assumptions put into the Ansatz~$E = (\varrho -1)^\beta F$. Since we are interested in how the solution changes
with decreasing~$m$, we need to choose~$\chi_0$ non-vanishing. Also with this complication we still get Kummers 
equation with changed parameters and the analytic solution for~$F$ is given by
\begin{equation}
\label{eq:anaF}
F= e^{-i \wn x \sqrt{\frac{7}{4} + \frac{4 \chi_2[\chi_0, \tilde d]^2}{1 - \chi_0^2}}} 
   {}_1F_1\left [-i \wn \left (\frac{1}{2 \sqrt{\frac{7}{4} + \frac{4 \chi_2[\chi_0, \tilde d]^2}{1 - \chi_0^2}}} + 
                1\right ),\, -2 i \wn,\, 2 i \wn x \sqrt{\frac{7}{4} + \frac{4 \chi_2[\chi_0, \tilde d]^2}{1 - \chi_0^2}}\right] \, ,
\end{equation}
with the near horizon expansion of the embedding function~$\chi=\chi_0+\chi_2[\chi_0,\tilde d] x^2 +\dots$ where
we determine recursively 
\begin{equation}
\chi_2[\chi_0, \tilde d] = 3 \chi_0 
 \frac{\chi_0^6 - 3 \chi_0^4 + 3 \chi_0^2 - 1}{4 (1 - 3 \chi_0^2 + 3 \chi_0^4 - \chi_0^6 + \tilde d^2)} \, .
\end{equation}
The approximate solution for~$F$ is shown in figure~\ref{fig:FApprox}. Furthermore we can calculate the 
fraction~$\partial_4 E/ E$ appearing in the spectral function near the horizon using this analytic solution. 
The result is displayed in figure~\ref{fig:nearHorR}. This near horizon limit is not the spectral function since 
we would have to evaluate it at the boundary which lies far beyond the validity of the near horizon 
approximation. Nevertheless, according to our initial assumptions that the effect of damping mainly
takes place near the horizon we further assume that the limit shown in figure~\ref{fig:nearHorR} already
contains the essential features of the spectral function. Indeed the fraction shows distinct resonance peaks
which move to lower frequencies if we increase the mass parameter~$m$. The right picture shows 
the same situation at a finite baryon density~$\tilde d=1$ and we see that the peaks do not move
to lower frequencies as much as before. Thus also the vanishing of the turning point at large densities
as observed before is captured by this approximate solution.

The fact that we find Kummer's equation to describe the high-frequency near-horizon dynamics of
our gravity problem is especially interesting in view of a recent thermodynamics work on the `Propagation of boundary of 
inhomogeneous heat conduction equation'~\cite{778835}. In this work exact analytical solutions of the 
heat conduction equation in an inhomogeneous medium are found. That diffusion equation which is
the analog of our gravity equation of motion reads
\begin{equation}
\partial_t J(\rho,t) = \rho^{1-s}\partial_\rho \left [ D(\rho) \rho^{s-1} \partial_\rho J(\rho,t) \right ] \, ,
\end{equation}
with the position dependent diffusivity~$D(\rho)$. The authors of~\cite{778835} show that this can be 
transformed into Kummer's differential equation. In our gravity equation of motion the metric factors
depend on the radial AdS postition~$\rho$ and therefore some combination of them can be seen 
as analog to the position-dependent diffusivity~$D(\rho)$. It might be no coincidence that our gravity
setup leads us directly to a diffusion equation where the diffusion  coefficient can be expressed in terms 
of metric factors since exactly this is what the membrane paradigm in the context of AdS/CFT 
predicts as we will discuss in section~\ref{sec:membranePara}.

{\bf Gauge theory speculations}
On the gravity side we have found some hints that the gravity solution can be viewed as a damped oscillation
with the damping depending dynamically on the radial AdS position and on the choice  of the mass parameter~$m$.
Increasing the mass parameter~$m$ we found that the solutions~$F$ are more damped. We attributed this damping 
to the metric background and found an analytic near-horizon solution for~$F$ which is damped by coefficients
in the near-horizon equation of motion which depend on the radial position, on the embedding function~$\chi_0$ and
on the finite baryon density~$\tilde d$. Now an open task is to translate this geometric gravity picture into a thermal
gauge theory phenomenology. Our basic assumption here will be that the damped gravity modes dissipating 
energy into the black hole horizon correspond to a dual current dissipating energy into the thermal plasma.

In order to see the peaks and their movement at all, we need to consider the background and the fluctuations at once.
We therefore suggest that the peaks and their motion are generated by the interaction of the metric components and
the gravity field fluctuations which translates to an interaction between the thermal plasma and the probe quarks we 
introduce. Our observations suggest that this interaction dominates the setup at small values of~$m$ and~$\chi_0$~(high
temperatures). The peak motion to lower frequencies while increasing~$m\propto M_q/T$ means that at fixed temperature~$T$ 
as we put more mass energy~$M_q$ into the excitations, the resulting plasma excitation~(at low temperatures identified
with a meson) is less and less energetic. Minding energy conservation we have to ask where the energy goes
which we put in. A possible explanation for this behavior is that the energy we put into the excitation is directly dissipated
into the plasma. This would happen if the coupling between the plasma and the quarks would become stronger and stronger
as the mass parameter is increased. 

One could try to put these speculations into a more rigorous form by assuming that we have a {\it Thirring model}-like
gauge theory here, which describes the self-interaction of our quarks~\footnote{The author is grateful to Karl Landsteiner
for suggesting this approach.}. So the idea here is that the quarks couple less and less
to each other and more and more to the plasma which could be seen as a decrease in the {\it Thirring coupling} 
\begin{equation}
g_{\text{Thirring}} \propto \frac{\Gamma}{\Omega}\, ,
\end{equation}
with the (quasi)meson decay width~$\Gamma$ and the (quasi)mesonic excitation energy~$\Omega$.
The Thirring coupling computed with our setup at~$\tilde d =0,\, 0.25$ is shown in figure~\ref{fig:thirring}
versus the quark mass parameter~$\chi_0$. At both finite and vanishing baryon density we observe that
the Thirring coupling decreases rapidly as the critical value~$\chi_0=1$ is approached. 

A more concise relation between the gravity and gauge setups shall be given soon~\cite{EKKR:2008tp}.

\subsection{Meson spectra at finite isospin and baryon density} \label{sec:mesonSpectraB&I}
We have successfully introduced finite baryon and isospin charges simultaneously into the dual thermal gauge theory
by the gravity background described in section~\ref{sec:thermoB&I}. In the two previous sections we have studied
fluctuations around the two limits~$\tilde d^B \not = 0,\, \tilde d^I =0$~(section~\ref{sec:mesonSpectraB}) 
or~$\tilde d^B = 0,\, \tilde d^I\not =0$~(section~\ref{sec:mesonSpectraI}). These setups allow for 
a comparison to lattice QCD results~\cite{Kogut:2002tm,Kogut:2004zg} and effective QCD models
with only one non-zero density. In order to compare our results to real experiments or to
computations in finite isospin {\it and} baryon QCD-models~\cite{Toublan:2003tt}~(two-flavor QCD), 
we need to compute fluctuations about the general 
case~$\tilde d^B \not = 0,\, \tilde d^I\not =0$.

We start from the Dirac-Born-Infeld action~\eqref{eq:dbiAction} at vanishing~$B$-field just like in the cases 
where only one density is non-zero
\begin{equation}
\label{eq:actionFlucBI}
S_{\text{DBI}} = -T_{D7} \int \dd^8 \xi \mathrm{Str} \left \{ \sqrt{-\det G}\left [ 
  1 + \frac{1}{2}\mathrm{tr} (G^{-1} \tilde F) -\frac{1}{4} (G^{-1} \tilde F)^2+ \frac{1}{8}\mathrm{tr} (G^{-1} \tilde F) +\dots
 \right ] \right\} \, ,
\end{equation}
where the field 
strength
\begin{equation}
\label{eq:flucBI} 
\tilde F_{\mu\nu}=\tilde F_{\mu\nu}^a T^a = 
 2\partial_{[\mu}A^a_{\nu]}T^a+f^{abc}A_\mu^b A_\nu^c T^a+2 f^{abc}\tilde A_{[\mu}^b A_{\nu]}^c T^a  \, ,
 \end{equation}
contains the non-Abelian gauge field fluctuations~$A$ as well as the background 
fields~$\tilde A^0,\, \tilde A^3$~(given by analogs of~\eqref{eq:backgroundAt} which can be derived from the
action~\eqref{eq:SBI3} using the transformation~\eqref{eq:flavorTrafoBackg}~\footnote{Note that in 
chapter~\ref{sec:thermalSpecFunc} the notation for background and fluctuations is reverse compared to 
chapter~\ref{sec:holoThermoHydro}. In this present chapter~$\tilde A$ denotes the background while $A$~denotes
fluctuations about the background. In chapter~\ref{sec:holoThermoHydro} we used~$A$ to denote the background
for simplicity since in that chapter there are no fluctuations.}). In the case of introduced isospin, i.e.~$N_f=2$ with
the Pauli sigma representations~$T^a=\sigma^a$ completed by unity~$\sigma^0=\mathbbm{1}$ the full background is 
collected in~$G$ given by
\begin{equation}
G =\left ( 
\begin{array}{c c c c c c c c}
g^{00}\sigma^0 & 0 & 0 & 0 & 2\partial_{[0} \tilde A^\alpha_{4]}\sigma^\alpha  & 0 & 0 & 0\\
0 & g^{11}\sigma^0 & 0 & 0 & 0 & 0 & 0 & 0\\
0 & 0 & g^{22} \sigma^0& 0 & 0 & 0 & 0 & 0\\
0 & 0 & 0 & g^{33}\sigma^0 & 0 & 0 & 0 & 0\\
2\partial_{[4} \tilde A^\alpha_{0]}\sigma^\alpha & 0 & 0 & 0 & g^{44}\sigma^0 & 0 & 0 & 0\\
0 & 0 & 0 & 0 & 0 & g^{55}\sigma^0 & 0 & 0\\
0 & 0 & 0 & 0 & 0 & 0 & g^{66}\sigma^0 & 0\\
0 & 0 & 0 & 0 & 0 & 0 & 0 & g^{77}\sigma^0\\
\end{array}
 \right ) \, ,
\end{equation}
where~$g$ is the metric~\eqref{eq:inducedMetric} induced on the D7-brane and~$\alpha=0,\,3$.
Note that we now have the complication of two different (diagonal) flavor representations in the determinant~$\sqrt{-\det G}$
and furthermore the operators~$F$ and~$G$ do not commute since there are flavor representations attached to each
of them. By our choice the background gauge fields come only with the diagonal representations, i.e.~$\alpha=0,3$ such that 
only~$\tilde A^0_0\not = 0$ and~$\tilde A^3_0\not = 0$, while the fluctuations are admitted in any flavor 
direction~$A_\mu^a\not = 0,\, \forall a=0,1,2,3$. In order to be able to compute the square roots and the
symmetrized flavor trace in the action~\eqref{eq:actionFlucBI} we need to simplify their arguments. 

In order to simplify the expressions appearing in the action, we need to commute the background~$G$ with 
fluctuations~$\tilde F$. It is reasonable to split the background into parts which live in distinct representations in 
order to have definite commutation rules. Taking into account that we only have background fields in flavor 
directions~$\alpha=0,3$ the background containing metric and background gauge fields reads
\begin{equation}
G_{\mu\nu} = \underbrace{\left ( g_{\mu\nu} +2\partial_{[\mu}\tilde A^0_{\nu]} \right )}_{= A_{\mu\nu}}\sigma^0 
  + \underbrace{2\partial_{[\mu}\tilde A^3_{\nu]}}_{=B_{\mu\nu}}\sigma^3\, .
\end{equation} 
Note that~$A_{\mu\nu}$ and~$B_{\mu\nu}$ both come with representations diagonal in flavor space 
but only~$A_{\mu\nu}$ commutes
with all flavor representations. $A_{\mu\nu}$ is further composed of the metric term being diagonal in Minkowski space 
and the antisymmetric~$2\partial_{[\mu}A^0_{\nu]}$ which has only two non-vanishing 
entries~$\pm \partial_{4}A^0_{0}$. The non-commuting term with the coefficient~$B_{\mu\nu}$
is anti-symmetric in the Minkowski indices and has only the two entries~$\pm \partial_{4}A^3_{0}$. 
We can make use of these properties later in order to simplify the action. For a simplified notation we 
abbreviate~${A^a_{0}}'=\partial_{4}A^a_{0}$.

Looking at the action~\eqref{eq:actionFlucBI} we learn that we need the inverse metric~$G^{-1}$ which we compute
by solving the defining equation~$G^{\mu\nu} G_{\nu\lambda} = \delta^\mu_\lambda \sigma^0$. The result is
\begin{equation}
G^{\mu\nu} = A^{\mu\nu} \sigma^0 + B^{\mu\nu} \sigma^3 \, ,
\end{equation}
with the inverse coefficients for the first~$5\times 5$ entries
\begin{equation}
A^{\mu\nu} =
\left ( \begin{array}{c c c c c}
\frac{g_{44}[({A_0^0}')^2+({A_0^3}')^2+g_{00} g_{44}]}{[({A_0^0}'-{A_0^3}')^2 +g_{00} g_{44}]
[ ({A_0^0}'+{A_0^3}')^2 +g_{00} g_{44}]} 
& 0 & 0 & 0 & 
\frac{{A_0^0}'[({A_0^0}')^2-({A_0^3}')^2+g_{00} g_{44}]}{[({A_0^0}'-{A_0^3}')^2 +g_{00} g_{44}]
[ ({A_0^0}'+{A_0^3}')^2 +g_{00} g_{44}]} \\
0 & g^{11} & 0 & 0 & 0  \\
0 & 0 & g^{22} & 0 & 0 \\
0 & 0 & 0 & g^{33} & 0 \\
-\frac{{A_0^0}'[({A_0^0}')^2-({A_0^3}')^2+g_{00} g_{44}]}{[({A_0^0}'-{A_0^3}')^2 +g_{00} g_{44}]
[ ({A_0^0}'+{A_0^3}')^2 +g_{00} g_{44}]} 
& 0 & 0 & 0 & 
\frac{g_{00}[({A_0^0}')^2+({A_0^3}')^2+g_{00} g_{44}]}{[({A_0^0}'-{A_0^3}')^2 +g_{00} g_{44}]
[ ({A_0^0}'+{A_0^3}')^2 +g_{00} g_{44}]}
\end{array} \right ) \, ,
\end{equation}
where the only other non-zero entries in the remaining directions are~$\text{diag}(g^{55},\,g^{66},\, g^{77})$ 
and the other coefficient is given by
\begin{equation}
B^{\mu\nu} =
\left ( \begin{array}{c c c c c}
-\frac{2 g_{44}{A_0^0}' {A_0^3}'}{[({A_0^0}'-{A_0^3}')^2 +g_{00} g_{44}]
[ ({A_0^0}'+{A_0^3}')^2 +g_{00} g_{44}]} & 0 & 0 & 0 &
\frac{{A_0^3}' [-({A_0^0}')^2+({A_0^3}')^2+g_{00} g_{44}]}{[({A_0^0}'-{A_0^3}')^2 +g_{00} g_{44}]
[ ({A_0^0}'+{A_0^3}')^2 +g_{00} g_{44}]}  \\
0 & 0 & 0 & 0 & 0 \\
0 & 0 & 0 & 0 & 0 \\
0 & 0 & 0 & 0 & 0 \\
-\frac{{A_0^3}' [-({A_0^0}')^2+({A_0^3}')^2+g_{00} g_{44}]}{[({A_0^0}'-{A_0^3}')^2 +g_{00} g_{44}]
[ ({A_0^0}'+{A_0^3}')^2 +g_{00} g_{44}]}
 & 0 & 0 & 0 & 
-\frac{2 g_{00}{A_0^0}' {A_0^3}'}{[({A_0^0}'-{A_0^3}')^2 +g_{00} g_{44}]
[ ({A_0^0}'+{A_0^3}')^2 +g_{00} g_{44}]}
\end{array} \right ) \, ,
\end{equation}
with all other entries vanishing.

In the isospin case the action can be simplified considerably since the representation matrices being spin
representations satisfy the Clifford algebra in addition to the commutation relations
\begin{equation}
\label{eq:sigmaAntiCommute}
\{ \sigma^a,\sigma^b \} = 2\delta^{ab}\,\,\,\text{and}\,\,\, [\sigma^a,\sigma^b] = i\epsilon^{abc}\sigma^c\,\forall a,b=1,2,3 \, ; 
\qquad [\sigma^0,\sigma^a] = 0 \; \forall a,b,c = 0,1,2,3 \, .
\end{equation}

The action~\eqref{eq:actionFlucBI} can now be written in terms of these inverse~$A^{\mu\nu},\,B^{\mu\nu}$ 
and the fluctuations~\eqref{eq:flucBI}. Using their properties along with the group structure 
simplifications~\eqref{eq:sigmaAntiCommute} we have to work out the commutation relation for~$G$ and~$F$
and apply these to simplify the action terms. For example the term proportional 
to~$G^{\mu\mu'}\tilde F_{\mu\nu} G^{\nu\nu'}\tilde F_{\mu'\nu'}$ can be brought to the standard 
form~$G^{\mu\mu'}G^{\nu\nu'}\tilde F_{\mu\nu}\tilde F_{\mu'\nu'} + \text{commutators}$.
These formulae may then be taken as the starting point for the calculation 
of fluctuations about the baryon and isospin background. I have performed 
the calculations necessary for this section in close collaboration with Patrick Kerner~\cite{Kerner:2008diploma}.

\subsection{Summary} \label{sec:sumSpecFunc}
In this chapter we have computed spectral functions to explore the thermal gauge theory dual to the D3/D7-brane setup 
with finite baryon and isospin densities. 

Upon the introduction of a finite baryon density we found resonance peaks in these spectral functions appearing at distinct
frequencies in section~\ref{sec:mesonSpectraB}. At small temperatures the energy~(frequency) of the resonances 
follows the vector meson mass formula
while their width becomes smaller and their resonance frequency larger when we decrease the temperature further. 
These facts suggest the interpretation that the resonance peaks correspond to mesonic quasi-particles formed inside 
the plasma. Having survived the deconfinement transition of the theory these vector mesons are analogous 
to the $\rho$-meson of QCD. 

However, at high temperatures the resonances become very broad and their frequency location does not relate 
to the mass formula. These peaks also move to lower frequencies if we decrease the temperature or equivalently 
increase the mass parameter~$m$. There exists a turning point at which the resonance peaks change their direction
along the frequency axis when the temperature-mass parameter~$m$ is changed. 
We speculate that in the same way in which the low-temperature~(large mass) regime is 
ruled by mass effects, the {\it thermal regime} is governed by temperature effects. In order to collect evidence for
this interpretation we examined the solutions of the gravity field dual to the flavor current relevant for our spectral
functions in section~\ref{sec:peakTurning}. We give an analytic solution for the gravity field equation of 
motion and the spectral function fraction in the high-frequency near-horizon limit. This solution is the confluent 
hypergeometric function~${}_1 F_1$ showing oscillatory and damped behavior. 

Introducing a finite isospin density in section~\ref{sec:mesonSpectraI} we discovered a triplet splitting of the peaks. 
This behavior agrees with the analytical results showing a triplet splitting of the correlator poles in the complex 
frequency plane at finite isospin chemical potential for massless quarks studied in section~\ref{sec:anaHydroIso}. 
The splitting depends on the size of the chemical potential. Note again that this behavior is reminiscent of the 
QCD $\rho$-meson which is a triplet under the QCD isospin symmetry.

Finally, in the last section~\ref{sec:mesonSpectraB&I} we introduced the concepts needed to compute gravity 
fluctuations and  to obtain from these the correlators at finite baryon and isospin densities.

\section{Transport processes at strong coupling} \label{sec:transport}
Experimental results obtained at the RHIC collider suggest that the plasma state generated there in collisions of 
gold ions behaves as a fluid~(rather than a gas as originally assumed) is microscopically governed by QCD 
at strong coupling and finite temperature. We thus use the AdS/CFT duality in the present chapter 
in order to compute transport properties of the strongly coupled plasma. In particular we focus on the diffusion
of conserved charges such as the baryon charge and isospin charge. Section~\ref{sec:membranePara} reviews
the general membrane paradigm approach to compute diffusion coefficients from the metric components only. 
We apply the formulae obtained there in sections~\ref{sec:baryonDiffusion} and~\ref{sec:isospinDiffusion} to 
find the baryon and the isospin diffusion coefficients, respectively. Since in the previous chapter we have found evidence 
for mesonic quasi-particle states to survive the deconfinement transition inside the plasma, we go on studying the
diffusion of such quarkonium states in section~\ref{sec:charmDiffusion}. Finally we consider the case of a background
gauge field in arbitrary flavor direction which induces three different isospin charges on the gauge theory side. In section~\ref{sec:diffusionMatrix}
we study gradients in these three charge densities which drive thermal currents.

\subsection{Membrane paradigm} \label{sec:membranePara}
Let us begin our study of the diffusion properties at finite baryon and isospin densities by motivating the so called
{\it membrane paradigm} which relates transport coefficients to components of the background metric tensor. In our
case this metric tensor will include contributions from the background gauge fields on the D7-brane. We also restate
the neccessary assumptions and put down the most important formulae.
A detailed derivation can be found in~\cite{Kovtun:2003wp} while a review of the complete subject may be 
found in~\cite{Son:2007vk}.

The basic idea behind the membrane paradigm approach is to relate the hydrodynamic normal modes on the
gauge theory side of the correspondence to a gravitational counterpart. This gravitational counterpart then has to 
fulfill the same dispersion relations as the hydrodynamic modes. For example the gravity mode dual to the 
diffusion mode should have the dispersion relation~$\omega = - i D q^2$. The approach used in~\cite{Kovtun:2003wp}
is to construct the gravity fluctuation with exactly this dispersion property. Having done this the authors 
identify the diffusion coefficient with an expression in the result depending on the metric components. 
This yields the diffusion formula for a charge coupled to a conserved vector current
\begin{equation}
\label{eq:diffusionConstant} 
D = \left .\frac{
        \sqrt{-g}}{
        g_{11}\sqrt{-g_{00} g_{44}}}\right |_{\rho=1}
\int\!\! \mathrm{d}\rho\;\frac{-g_{00} g_{44}}{\sqrt{-g}}\, .   
\end{equation}
Similar formulae can be found for the gravitational tensor fluctuations dual to the shear mode~\cite{Kovtun:2003wp}.
There are a few assumptions to be made in order for the derivation to work. First, the metric components all 
have to be independent from all coordinates but the radial AdS-coordinate~$x_4$. Second, the time component
of the gravity vector field~$A_0(\rho)$ can be expanded in a series over~$q^2/T^2\ll 1$~(at least if~$\rho$ is not
exponentially close to the horizon). Third, spatial gauge field components change slower with time than 
the time-component varies over space~$|\partial_0 A_1|\ll |\partial_1 A_0|$.

\subsection{Baryon diffusion} \label{sec:baryonDiffusion}
In this section we calculate the baryon diffusion coefficient and its dependence
on the baryon density. As discussed in \cite{Mateos:2007vc}, the baryon density
affects the location and the presence of the fundamental phase transition
between two black hole embeddings observed in~\cite{Kobayashi:2006sb}. This
first order transition is present only very close to the separation line between
the regions of zero and non-zero baryon density shown in
figure~\ref{fig:grandcanPhaseDiagB} as discussed before in section~\ref{sec:thermoBaryon}. 
We show that the fundamental phase transition may also be seen in the diffusion
coefficient for quark diffusion. It disappears at a critical baryon density.
Nevertheless, the diffusion coefficient shows a smoothened transition beyond this critical density, which we 
will call~{\it hydrodynamic transition} and which appears as a minimum in the diffusion coefficient
versus quark mass diagram.

In order to compute the diffusion using holography, we use the membrane paradigm approach reviewed in 
section~\ref{sec:membranePara} 
developed in~\cite{Kovtun:2003wp} and extended in~\cite{Myers:2007we}. 
This method allows to compute various transport coefficients in Dp/Dq-brane setups from the metric
coefficients. The resulting formula for our background is equation~\ref{eq:diffusionConstant} which is 
the same as in~\cite{Myers:2007we}.

\begin{figure}
        \includegraphics[width=.9\linewidth]{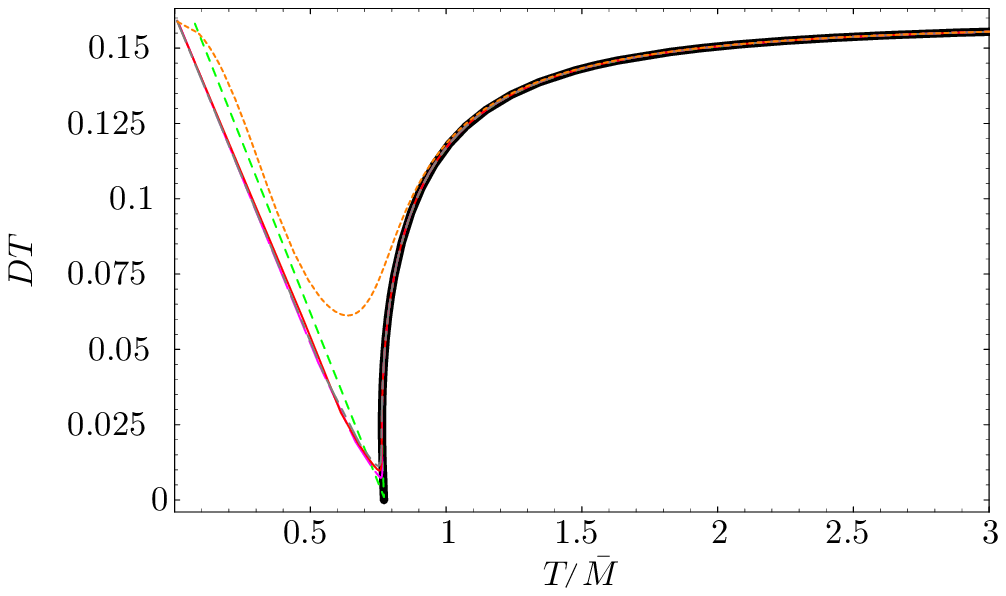}
\vfill
        \includegraphics[width=.9\linewidth]{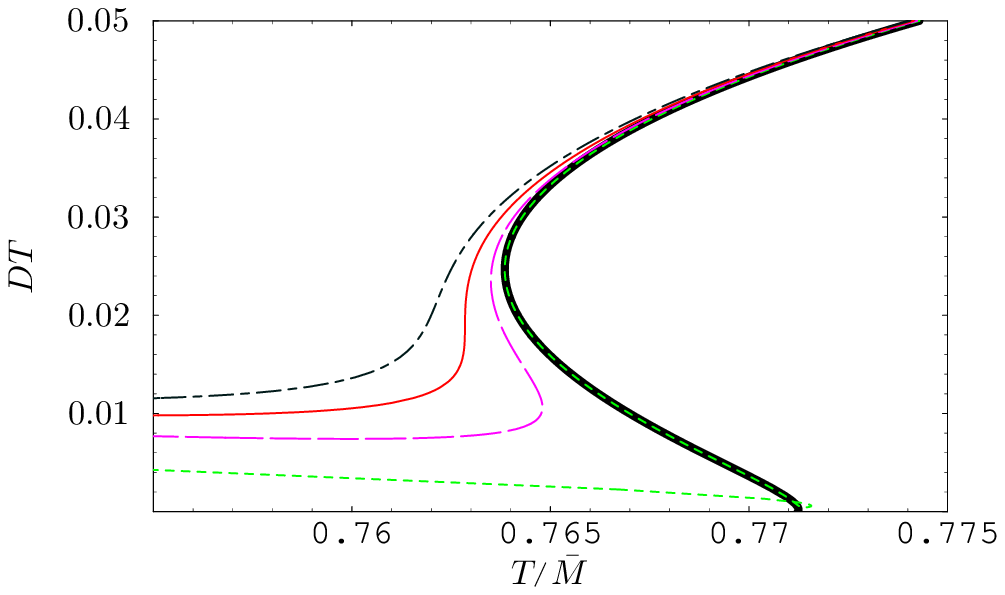}
        \caption{
                The diffusion coefficient times temperature is plotted against
                the mass-scaled temperature for diverse baryon densities
                parametrized by~$\tilde d=0.1$~(uppermost line in upper plot, not
		visible in lower plot),
                $0.004,$~(long-short-dashed), $0.00315$~(thin solid),
                $0.002$~(long-dashed), $0.000025$~(short-dashed) and
                $0$~(thick solid). The finite baryon density lifts the curves at
                small temperatures. Therefore the diffusion constant never
                vanishes but is only minimized near the phase transition. The
                lower plot zooms into the region of the transition. 
                The phase transition vanishes above a critical
                value~$\tilde d^*=0.00315$. The position of the
                transition shifts to smaller~$T/\bar M$, as~$\tilde d$ is
                increased towards its critical value.
		}
        \label{fig:diffusionConstants}
\end{figure}

The dependence of $D$ on the baryon density and on the quark mass originates
from the dependence of the embedding $\chi$ on these variables. The results
for~$D$ are shown in figure~\ref{fig:diffusionConstants}. 

{\bf Discussion}
The thick solid line shows the diffusion constant at vanishing baryon density found
in~\cite{Myers:2007we}, which reaches $D=0$ at the fundamental phase transition.
Increasing the baryon density, the diffusion coefficient curve is lifted up for
small temperatures, still showing a phase transition up to the critical
density $\tilde d^*=0.00315$. This is the same value as found
in~\cite{Kobayashi:2006sb} in the context of the phase transition of the quark
condensate.

The diffusion coefficient never vanishes for finite density. Both in the limit
of $T/\bar M\to 0$ and $T/\bar M\to \infty$, $D\cdot T$ converges to $1/2\pi$
for all densities, i.e.\ to the same value as for vanishing baryon density, as
given for instance in \cite{Kovtun:2003wp} for R-charge diffusion. Near the phase
transition, the diffusion constant develops a nonzero minimum at finite baryon
density. Furthermore, the location of the first order phase transition moves to
lower values of $T/\bar M$ while we increase~$\tilde d$ towards its critical
value.

In order to give a physical explanation for this behavior, we focus on the case
without baryon density first. We see that the diffusion coefficient vanishes at
the temperature of the fundamental deconfinement transition. This is simply due
to the fact that at and below this temperature, all charge carriers are bound
into mesons not carrying any baryon number.

For non-zero baryon density however, there is a fixed number of charge carriers
(free quarks) present at any finite temperature. This implies that the diffusion
coefficient never vanishes. Switching on a very small baryon density, even below
the phase transition, where most of the quarks are bound into mesons, by
definition there will still be a finite amount of free quarks. By increasing the
baryon density, we increase the amount of free quarks, which at some point
outnumber the quarks bound in mesons. Therefore in the large density limit the
diffusion coefficient approaches $D = 1/(2\pi T)$ for all values of $T/\bar
M$, because only a negligible fraction of the quarks is still bound in this
limit.

Note that as discussed in  \cite{Karch:2007br,Kobayashi:2006sb,Mateos:2007vc}
there exists a region in the $(n_B,T)$ phase diagram at small $n_B$ and $T$
where the embeddings are unstable. In figure~\ref{fig:diffusionConstants}, this
corresponds to the region just below the phase transition at small baryon
density. This instability disappears for large $n_B$~(compare also figure~\ref{fig:grandcanDVsTZoom}).

\subsection{Diffusion with isospin} \label{sec:isospinDiffusion}
\begin{figure}
  \subfigure[]{
        \psfrag{DT}{$\frac{DT}{2\pi}$}
        \psfrag{m}{$m$}
        \includegraphics[width=.45\linewidth]{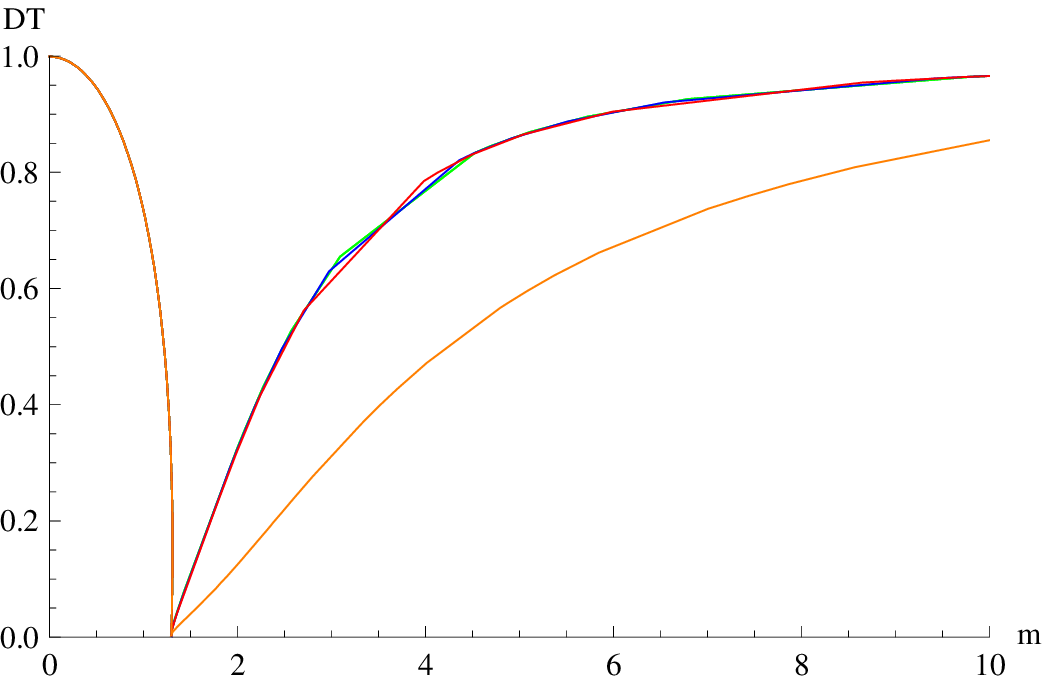}}
\hfill
  \subfigure[]{
        \psfrag{DT}{$\frac{DT}{2\pi}$}
        \psfrag{m}{$m$}
        \includegraphics[width=.45\linewidth]{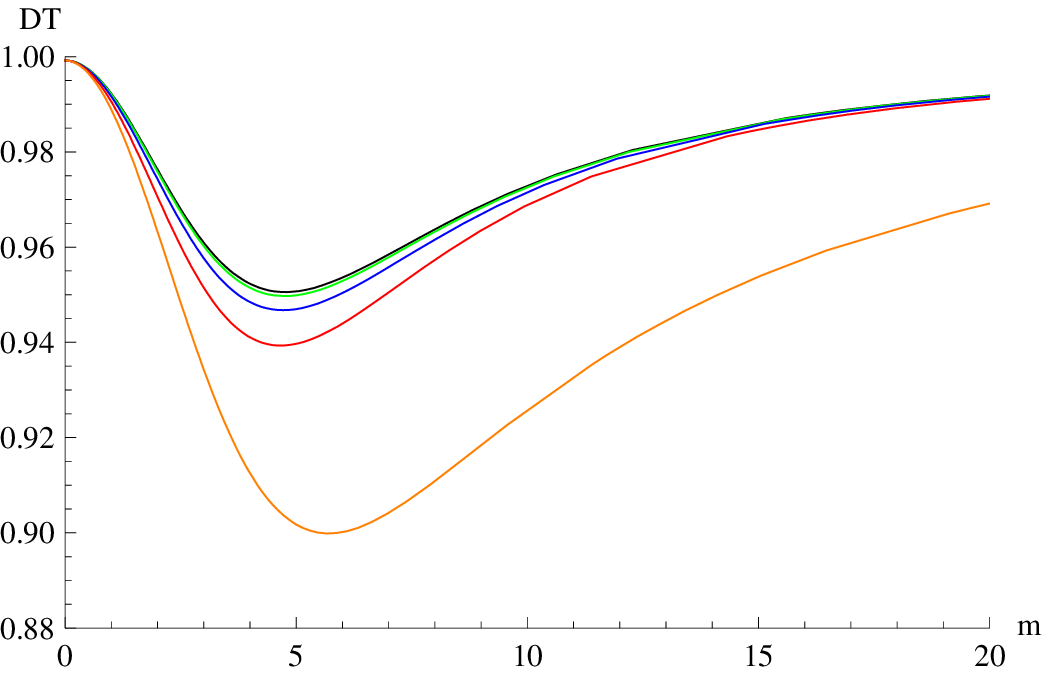}
        }
        \caption{
        The diffusion coefficient times temperature~$DT$ is plotted versus the mass-temperature parameter~$m$ at
        baryon density (a)~$\tilde d^B = 5\cdot 10^{-5}$ and (b)~$\tilde d^B = 20$. Different curves in one plot 
        show results for distinct values of the isospin density~(top down): $\tilde d^I = 0$~(black), 
        $\tilde d^I = 1/4\tilde d^B$~(green), $\tilde d^I = 1/2 \tilde d^B$~(blue), $\tilde d^I = 3/4 \tilde d^B$~(red),
        $\tilde d^I = \tilde d^B$~(orange). These plots were generated by Patrick Kerner~\cite{Kerner:2008diploma}.
                		}
        \label{fig:diffusionConstantsBI}
\end{figure}

In this section we consider the diffusion coefficient computed from the membrane paradigm 
formula~\eqref{eq:diffusionConstant} adding a finite isospin density to the finite baryon density exclusively
considered in the previous section. The gravity dual to such a theory has already been discussed in section~\ref{sec:thermoB&I}.
The finite isospin density enters the diffusion coefficient through the D7 embedding function~$\chi (\rho, \tilde d^B,\tilde d^I)$ which
appears in the metric components~$g_{\mu\nu}(\rho, \tilde d^B, \tilde d^I)$. We obtain the explicit embedding function by 
solving its equation of motion~\eqref{eq:chiEomBI} and then simply plug in the metric coefficients~\eqref{eq:inducedMetric} into the
diffusion formula~\eqref{eq:diffusionConstant}. This procedure yields the plots given in figure~\ref{fig:diffusionConstantsBI}.
The physical significance of this diffusion coefficient will be discussed at the end of this section and for now we refer
to it as the {\it effective baryon diffusion coefficient}.  

{\bf Discussion}
The diffusion coefficient in this background with finite baryon density and with finite isospin density~(thermodynamical
conjugate of the chemical potential in the third flavor direction) behaves very similar to the case with finite baryon
density only. In the limit of vanishing densities~$\tilde d^B = 0 = \tilde d^I$ the diffusion coefficient reduces to the 
thick black line shown in figure~\ref{fig:diffusionConstants} showing a sharp transition from the diffusive black hole
phase to the non-diffusive Minkowski phase at the critical mass-temperature value~$m_{\text{crit}}$. Again the
explanation is that neither baryon nor isospin charges are available below the critical~$m$. We now switch on a small
baryon density and increase the isospin density in quarter steps from zero~(black curve) 
to~$\tilde d^I = \tilde d^B$~(orange curve) in figure~\ref{fig:diffusionConstantsBI},~(a). At these small densities 
only the case in which both densities are equal differs significantly from the only baryon density case. The diffusion
curve for this case drops up to 50 percent below the baryonic value above the transition and follows the baryonic
case closely below the transition. Zooming in on figure~\ref{fig:diffusionConstantsBI},~(a) would show a spiraling
behavior for all the curves near the location of the former phase transition. The new location of the transition
shifts to smaller~$m$ as the isospin density is increased. This qualitative behavior has also been observed 
when we increased the baryon density at vanishing isospin density in the previous section. Thus we can 
summarize that the introduction of any of the two densities shifts the location of the phase transition to 
lower values of the mass-temperature parameter~$m$~(as may be seen in the phase diagram~\ref{fig:canPhaseSurface}). 
At a critical combination of densities the transition again vanishes.

\begin{figure}[h!]
\includegraphics[width=.7\linewidth]{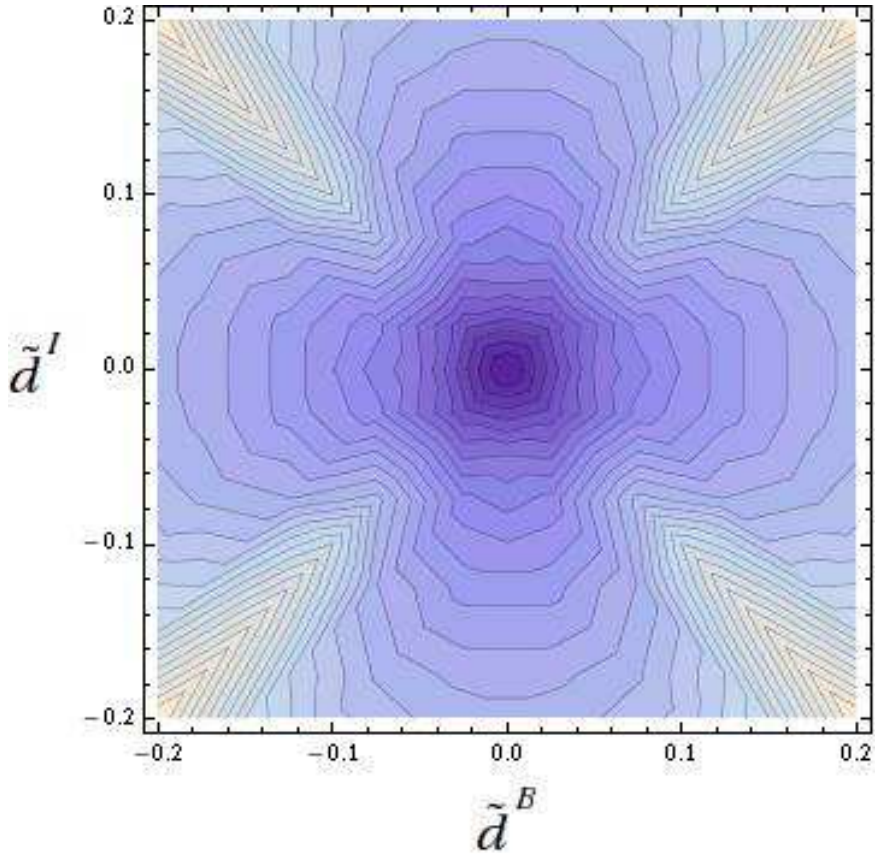}
\caption{\label{fig:contourPhaseDiagBI}
Contour plot of the location of the phase transition/crossover
mass parameter over the baryon density-isospin density plane.
This plot was generated by Patrick Kerner~\cite{Kerner:2008diploma}.
}
\end{figure}

Increasing the baryon density to~$\tilde d^B = 20$ in figure~\ref{fig:diffusionConstantsBI},~(b) we observe a more
pronounced splitting between the different isospin density curves. Again the case~$\tilde d^I = \tilde d^B$ drops 
significantly below the other isospin value curves. All the curves show a clear minimum near the location of the 
former phase transition. We interpret this minimum structure as a smoothed version of the previously sharp 
phase transition and call it~{\it hydrodynamic transition}. We further identify this transition as a 
crossover. Following the location~$m_{\text{min}}$ 
of the minimum when varying the two densities we observe that the rotational symmetry~$O(2)$ formerly
present at small densities in the phase diagram~\ref{fig:canPhaseSurface}, 
now at large densities is broken to a discrete~$\mathbf{Z}_4$.
All the diffusion curves approach the value~$1/(2\pi)$ in both the large and small mass limit. This 
evolution is shown in the contour plot~\ref{fig:contourPhaseDiagBI}. Contours correspond to equal values
of the mass parameter~$m$ at the phase transition.

We now come back to the question which diffusion coefficient we have actually calculated applying the
membrane paradigm formula~\eqref{eq:diffusionConstant}. Since we have not changed the formula at all 
and taking a closer look at its derivation, we are lead to the conclusion that we have again calculated the baryon
diffusion coefficient. Since the metric background now includes isospin density in addition to the baryon density,
this formula also incorporates the effect of the finite isospin density on the diffusion of our baryons. So we 
conclude that the coefficient we have computed is the baryon diffusion coefficient taking in account the 
isospin-driven baryon diffusion, thus we call it the {\it effective baryon diffusion coefficient}. In order to study
the effects of baryon and isospin diffusion seperately, we have to modify the membrane paradigm and 
also our setup to include a non-Abelian structure in the metric background as briefly discussed in the
outlook section~\ref{sec:diffusionMatrix}.

\subsection{Charmonium diffusion} \label{sec:charmDiffusion}
In accordance with recent QCD lattice results~\cite{Asakawa:2003xj} and investigations of~\cite{Shuryak:2004tx} we observed 
in the previous chapter that in our model D3/D7-theory stable quasi-particle states 
of quarks and antiquarks survive the deconfinement transition of the thermal field theory~($\N=4$ SYM in our case)
which governs the thermal plasma under investigation. 
After having studied the diffusion of individual quarks considering their isospin and baryon charge in the previous 
sections~(see also~\cite{Herzog:2006gh,CasalderreySolana:2006rq,Gubser:2006bz}),
we now turn to the diffusion of those quark-antiquark bound states having survived in the thermal plasma
at vanishing densities.
We will find that similar to the viscosity bound~$\eta/s$, the quarkonium diffusion at strong coupling is 
also significantly smaller than at weak coupling.
The energy loss of heavy quarks and their bound states is experimentally of high 
interest~\cite{Adare:2006nq,Bielcik:2005wu,Adare:2008sh,Adare:2006ns,Adler:2005ph,Arnaldi:2006ee}. The most 
prominent example of such bound states in QCD is charmonium~($c\bar c$), or rather its first excited state 
called~$J/\psi$. In our holographic setup we examine an analogous 
configuration of fundamental fields in SYM theory at strong coupling. 
We start by illustrating the general idea of our calculation with a review of the analogous QCD calculation. 
Afterwards we translate the problem to SYM theory and solve it by the calculation of correlation functions in the
dual gravity theory. The result will be the quark-antiquark bound state diffusion coefficient at strong coupling.

The content of this section collects intermediate results of the work~\cite{Dusling:2008} which is based on 
an idea by Derek Teaney. Note that during the publication process of this thesis all the open checks
mentioned in this section have been completed and the final result can now be found in~\cite{Dusling:2008}.

{\bf Summary of QCD results }
Our task is to describe the interactions of a heavy meson with the QCD medium. 
We accomplish this by a dipole approximation which has yielded a good estimate of the~$J/\psi$ coupling to
nuclei~\cite{Luke:1992tm}.
Following the effective field theory calculation first carried out in \cite{Luke:1992tm}, we consider the sum of the 
pure QCD Lagrangian  and an interaction Lagrangian describing color-electric~(index~$E$) and color-magnetic~(
index~$B$) interactions
\begin{equation}
\L = \phi_v^\dagger i v\cdot \partial \phi_v \, +  \frac{c_E}{N^2} \phi_v^\dagger \O_E \phi_v+ 
  \frac{c_B}{N^2} \phi_v^\dagger\O_B \phi_v \, ,
\end{equation}
with~$v$ being the fixed velocity of the heavy (scalar) meson described by the 
field~$\phi_v$ 
\begin{gather}
\O_E = - G^{\mu \alpha A}G_{\alpha}{}^{\nu A}  v_\mu v_\nu \, , \quad 
\O_B = {\textstyle \frac{1}{2}}  G^{\alpha \beta A} G_{\alpha
  \beta}{}^A  - G^{\mu \alpha A}G_{\alpha}{}^{\nu A}  v_\mu v_\nu \, . 
\end{gather}
$G$ is the non-Abelian field strength of QCD, and $c_E$ and $c_B$ are
coefficients to be determined from the QCD dynamics.
This Lagrangian may be used for describing
bound states of heavy quarks with four-velocity $v_\mu$. 
In the rest frame of a heavy quark bound state, $v=(1,0,0,0)$, the
operators $\O_E$ and $\O_B$ are
\begin{gather}
\O_E = {\bf E}^A \cdot {\bf E}^A \, , \qquad \O_B = {\bf B}^A \cdot {\bf B}^A \, ,
\end{gather}
where ${\bf E}^A$ and ${\bf B}^A$ are the color electric and magnetic
fields.  
If the constituents of the charmonium dipole are non-relativistic it is expected that
the magnetic polarizability~$c_B$ is of second order in the four-velocity~$\O(v^2)$
relative to the electric polarizability~$c_E$. For heavy quarks we assume
that~$c_B$ can be neglected and set~$c_B=0$. For heavy quarks and
large~$N$ Peskin found~\cite{Peskin:1979va,Bhanot:1979vb} 
\begin{equation}
c_E = \frac{28\pi}{3\Lambda_B^3} \, ,
\end{equation}
with the inverse Bohr radius~$\Lambda_B \equiv 1/a_0$.
Below we will generalize these results to $\N=4$ Super-Yang-Mills theory.

We expect the kinetics of the heavy meson dipole in the medium to be described by
Langevin equations for long time scales compared to the medium correlations
\begin{eqnarray}
\label{eq:langevin}
\frac{d p_i}{d t} &= \xi_i(t) - \eta_D p_i \, , \\
\langle \xi_i (t) \xi_j (t') \rangle &= \kappa \delta_{ij} \delta(t-t') \, .
\end{eqnarray}
Here the~$\xi_i$ are components of an arbitrary force acting on the heavy dipole.
The coefficient~$\kappa$ is the second moment of the force applied to the dipole. 
The drag coefficient~$\eta_D$ and the fluctuation coefficient~$\kappa$ are related by the Einstein equation
\begin{equation}
\eta_D = \frac{\kappa}{2 M T}\, ,
\end{equation}
with the mass~$M$ and temperature~$T$.

In the regime of times long compared to medium correlations but short compared to the time the
system needs to equilibrate, we can neglect the drag coefficient in equation~\eqref{eq:langevin}. Then
the fluctuation coefficient $\kappa$ is obtained from the correlation of the microscopic 
forces~$\F^i$ on the dipole,
\begin{equation}  
\kappa = \frac{1}{3}\int\dd t \langle \F^i (t) \F^i (0) \rangle \, .    
\end{equation} 
The thermodynamical force~$\F$ acting on the heavy dipole is determined by the 
gradient~$\F=-\nabla U$ of the potential~$U$ identified as the interaction part of the Lagrangian   
\begin{equation}
U = \L_{\text{int}}=\int \dd^3 x \phi_v^\dagger c_E  \frac{E^2}{2} (\mathbf{x}, t) \phi_v (\mathbf{x},t) \, .  
\end{equation}
  
So the fluctuation coefficient is given by
\begin{gather}
\label{eq:qcdDiffusion}  
\kappa = \, - \, \lim_{\omega \rightarrow 0} \frac{2T}{3 \omega}\frac{c_E^2}{N^4}
  \int \frac{\dd^3 q}{(2\pi)^3}  
  \bm{q}^2     
  \mathrm{Im}  G_{E^2 E^2}^R (\omega,\bm{q}) \, ,  
\end{gather}
where 
\begin{gather} 
\label{eq:qcdCorrelator}
G_{E^2 E^2}^R (\omega,\bm{k}) = 
-i\int\dd^4 x e^{-i \vec k\cdot \vec x}\Theta (x^0) 
 \langle \frac{\O_E}{2} (\mathbf{x}, t) \frac{\O_E}{2}(0) \rangle     
\, . 
\end{gather}
The three-momentum factor~$\bm{q}^2$ in~\eqref{eq:qcdDiffusion}
comes from the derivative in the potential gradient~$\nabla U$ and
the term proportional to~$\omega^2$ vanishes in the zero-frequency limit.  

In the case of QCD the integral in~\eqref{eq:qcdDiffusion} evaluates to 
\begin{equation}
\kappa = \frac{c_E^2}{N^2} \frac{64 \pi^5}{135} T^9 \, .
\end{equation}
The fluctuation coefficient~$\kappa$ which we identified with the second moment of the 
force acting on the dipole gives the rate of momentum broadening. 
We also identify the coefficients~$c_E,\, c_B$ as the 
electric and magnetic polarizabilities. These and analogous 
coefficients in the following are called~$\alpha$ with an
appropriate index~(e.g. $\alpha_F,\, \alpha_T$).  

{\bf Linear perturbations of $\N=4$ Super-Yang-Mills theory}
Our aim is to calculate the heavy meson diffusion coefficient $\kappa$
from gauge/gravity duality. This requires the calculation of the
two-point correlators as well as of the polarizabilities in $\N=4$ 
Super-Yang-Mills theory.

To set the scene we transfer the results of the preceding section to
$\N=4$ $SU(N)$ Super-Yang-Mills theory.
We consider the effective Lagrangian
\begin{equation}
\label{eq:effSYML}
\L = -\phi_v^\dagger i v\cdot \partial \phi_v + \, \frac{\alpha_T}{N^2}\phi_v^\dagger T^{\mu\nu}\phi_v v_\mu v_\nu \, 
  + \, \frac{\alpha_F}{N^2} \phi_v^\dagger \tr F^2 \phi_v \, ,
\end{equation}
which is a linear perturbation of $\N=4$ Super-Yang-Mills theory by
two composite operators.  
The polarization coefficients $\alpha_T$, $\alpha_F$ will be determined below
from meson mass shifts in gauge/gravity duality. 

The force on the dipole now becomes 
\begin{gather} \label{eq:adsCftForce}
\F(t) = \int \dd^3 \mathbf{x} \phi_v^\dagger\nabla \left [ \frac{\alpha_T}{N^2} T^{\mu\nu} v_\mu v_\nu
  +\frac{\alpha_F}{N^2} \mathrm{tr}\,{F^2} \right ] \phi_v \, .  
\end{gather}
Again there will be no cross-terms.
In the gauge/gravity duality this is reflected in the fact that at tree
level in supergravity, there is no contribution to
$\langle T_{00} (x) \tr F^2 (y) \rangle = \frac{\delta^2}{\delta
  g^{00} (x) \delta \Phi (y)} W=0$, with $g^{00}$ the metric component
  and $\Phi$ the dilaton. 

We proceed by calculating the stress tensor and $\tr F^2$ correlators
from graviton and dilaton propagation through the AdS-Schwarzschild
black hole background. Moreover we determine the polarizability $\alpha_T$
by considering the linear response of the meson mass to switching on
the black hole. The polarizability $\alpha_F$ is obtained by determining
the linear response of the meson mass to a perturbation of the
dilaton. As an example we choose the dilaton deformation of Liu and
Tseytlin \cite{Liu:1999fc}. 

{\bf AdS/CFT setup} 
We consider two different gravity backgrounds, the thermal and the dilaton one. Starting with the
gravity dual of $\N=4$ theory at finite temperature given by the AdS-Schwarzschild black hole with Minkowski signature
(see e.g.~\cite{Policastro:2002se}). Asymptotically near the horizon the
corresponding metric returns to $AdS_5 \times S^5$. The black hole
background is needed in the subsequent both for calculating the
necessary two-point correlators $\langle T_{00} T_{00} \rangle$ and
$\langle \tr F^2 \tr F^2 \rangle$, as well as for obtaining the
polarizability contribution for the linear response of the meson
masses to the temperature. 

We make use of the coordinates of \cite{Babington:2003vm} to write the
AdS-Schwarzschild background
in Minkowski signature as
\begin{equation}
\label{eq:adsBHMetric}
\dd s^2 =   
  \left(\frac{w}{R}\right)^2
  \left( -\frac{f^2}{\tilde f}\dd t^2 + \tilde{f} \dd \bm{x}^2 \right) 
  \left(\frac{R}{w}\right)^2
  \left( \dd \varrho^2 + \varrho^2 \dd \Omega_3^2 +\dd w_5^2 + \dd w_6^2 \right)\, , 
\end{equation}
with the metric $\mathrm d \Omega_3^2$ of the unit $3$-sphere, where
\begin{equation}
\begin{gathered}
f(r)=1-\frac{r_H^4}{4 w^{4}}, \qquad \tilde f(r)=1+\frac{r_H^4}{4w^{4}},
  \qquad w^2=\varrho^2+w_5^2+w_6^2,\qquad  r_H = T \pi R^2,\\
R^4=4\pi g_s N_c  {\alpha'}^2, \qquad \lambda=4\pi N g_s,\qquad g_{YM}^2=4\pi g_s, \qquad w_H=\frac{r_H}{\sqrt{2}}\, .
\end{gathered}
\end{equation}
In the next section we will work in a coordinate system with inverted 
radial $AdS$-coordinate~$u=R^2/r^2$  
used in e.g.~\cite{Policastro:2002se}. In these coordinates, the deformed 
$AdS_5$ part of the metric~\eqref{eq:adsBHMetric} reads          
\begin{equation} 
\dd s_5^2 = \frac{(\pi T R)^2}{u} 
  \left (-f(u) \dd {x_0}^2 +\dd \bm{x}^2 \right )+ 
  \frac{R^2}{4 u^2 f(u)} \dd u^2 \, ,     
\end{equation}
with~$f(u)=1-u^2$ and the determinant square 
root~$\sqrt{-g_5}=\frac{R^{10}(\pi T)^8}{4 u^3}$.  

A further necessary ingredient is the polarizability contribution
obtained from the linear response of the meson mass to $\tr F^2$. 
The gravity dual of the operator $\tr F^2$~(and its $\N=2$ supersymmetric completion)
is the dilaton field. Therefore, we consider a dual gravity background
with a non-trivial dilaton flow. We choose the dilaton flow
of Liu and Tseytlin \cite{Liu:1999fc} which corresponds to a
configuration of D3 and D(-1) branes.
In order to fix notation, we write down the string frame metric of \cite{Liu:1999fc}
in the form
\begin{equation}  
\label{eq:liuTseytlinBackground}
{\dd s}^2_{\text{string}} = 
\text{e}^{\Phi/2} {\dd s}^2_{\text{Einstein}}   
    = e^{\Phi}   
     \left [ 
      \left (\frac{r}{R}\right )^2 \dd \vec{x}^2 + 
      \left ( \frac{R}{r}
      \right )^2 
      \left (\dd r^2 + r^2 \dd\Omega_5^2\right ) 
     \right ]  \, .
\end{equation}       
The type IIB action in the Einstein frame for the dilaton~$\Phi$,
the axion~$C$ and the self-dual gauge field strength~$F_5=\star F_5$
reads
\begin{equation}
\label{eq:IIBAction}
S=\frac{1}{2\kappa_{10}^2}\int \dd^{10}x\sqrt{-g} \left[
  \mathcal{R}-\frac{1}{2}(\partial \Phi)^2 -\frac{1}{2}e^{2\Phi}  
   (\partial C)^2 -  \frac{1}{4\cdot 5!} (F_5)^2 +\dots
\right] \quad ,
\end{equation}
with the curvature scalar~$\mathcal{R}$ and the ten-dimensional gravity constant
\begin{equation}  
\label{eq:gravityConst10} 
\frac{1}{2\kappa_{10}^2}=\frac{1}{(2\pi)^7 (\alpha')^4 g_s^2} \quad .
\end{equation}
Solving the equations of motion derived from~\eqref{eq:IIBAction}, 
we obtain the dilaton solution  
\begin{equation}
\label{eq:dilatonSolution}  
e^\Phi=g_s( 1+ \frac{q_{}}{r^4} ) \, .   
\end{equation} 
Note that the parameter~$q$ we are using here differs from that given in~\cite{Liu:1999fc} in the following 
way~$q=\frac{R^8}{\lambda}q_{\text{Liu\&Tseytlin}}$.  

The dilaton is dual to the field theory operator~$\text{tr} F^2$ 
appearing in the gauge theory 
action~$S_{\text{gauge}}=\int\dd^4 x\tr F^2+\dots$. 
So the expectation value or one-point function of this operator 
is given by    
\begin{equation} 
\langle \text{tr} F^2 \rangle = \lim\limits_{r\to\infty}\frac{\delta S}{\delta \Phi}=  
  \frac{N^2}{2\pi^2 R^8} q_{\text{}} \, .   
\end{equation}  
Note that we use the Minkowski version of the originally Euclidean Liu-Tseytlin background.
 
{\bf Correlators} \label{sec:adsCftCorrelators}
According to~\eqref{eq:qcdDiffusion} and~\eqref{eq:adsCftForce}
the heavy meson momentum broadening is given by    
\begin{equation}
\label{eq:adsCftDiffusion}  
\kappa = -\lim_{\omega\to 0}\left (\frac{2T}{3\omega}\right) 
  \int\frac{\dd^3q}{(2\pi)^3} {\bm{q}^2}    
  \left [  
     \left (\frac{\alpha_F}{N^2}\right )^2 \mathrm{Im} G^R_{F^2 F^2}(\omega,\bm{q}) + 
     \left (\frac{\alpha_T}{N^2}\right )^2 \mathrm{Im} G^R_{TT}(\omega,\bm{q})  
  \right ] \,  ,       
\end{equation}  
where the bracket is the imaginary part of the force~\eqref{eq:adsCftForce} correlator   
$G_{\F\F}^R$. 
We need to 
calculate the retarded momentum space correlator~$G^R_{TT}$   
of the energy momentum tensor component~$T^{00}$ which is 
dual to the metric perturbation~$h_{00}$, and the 
2-point correlator~$G^R_{F^2 F^2}$ of the operator~$\text{tr} F^2$  
dual to the dilaton~$\Phi$. On the gravity side both field  
correlators are computed in the black hole background~\eqref{eq:adsBHMetric}  
placing the dual gauge theory operator correlation functions 
at finite temperature. 

For simplicity in this section we work in the conventions and
coordinates of~\cite{Policastro:2002se}. Especially the radial coordinate is changed 
from~$r$ to~$u$ with the horizon at~$u=1$. These are the same coordinates we have used in section~\ref{sec:anaAdsG}. 
We apply the method of~\cite{Son:2002sd} to find the 
two-point Minkowski space correlators from the classical supergravity action as described in section~\ref{sec:anaAdsG} .   

The classical gravity action for the graviton and 
dilaton is obtained from~\eqref{eq:IIBAction} as   
\begin{equation}
\label{eq:dilatonGravitonAction}
S=\frac{1}{2\kappa_{5}^2}\int \dd u\dd^{4}x\sqrt{-g_5} \left[
  (\mathcal{R}- 2\Lambda)- \frac{1}{2}(\partial \Phi)^2 
    +\dots
\right] \quad , 
\end{equation}
where
\begin{equation}  
\frac{1}{\kappa_5^2} = \frac{\Omega_5}{\kappa_{10}^2} 
  = \frac{N^2}{4\pi^2 R^3} \, .  
\end{equation}
So comparing to~\eqref{eq:recipeAction} we get  
\begin{equation}  
\B_\Phi= -\frac{1}{4\kappa_5^2}\sqrt{-g_5} g^{uu}  \, .    
\end{equation}   
The equation of motion derived from~\eqref{eq:dilatonGravitonAction} 
in momentum space reads  
\begin{equation} 
\label{eq:dilatonEom}  
\Phi ''-\frac{1+u^2}{u f(u)}\Phi' + 
 \frac{\wn^2-\qn^2 f(u)}{u f(u)^2}\Phi = 0\, ,   
\end{equation} 
with the dimensionless frequency~$\wn=\omega/2\pi T$ and 
spatial momentum component~$\qn=q/2\pi T$.  
The equation of motion~\eqref{eq:dilatonEom} has to be 
solved numerically with incoming wave boundary 
condition at the black hole horizon.    
Computing the indices and expansion coefficients near the boundary and horizon as done 
in~\cite{Teaney:2006nc,Kovtun:2006pf}, we obtain the asymptotic
behavior as linear combination of two solutions. We get the correlators by applying the {\it matching
method} described in section~\ref{sec:numAdsG}.      
Solving~\eqref{eq:dilatonEom} and matching the asymptotic solutions, we obtain 
\begin{equation}  
\label{eq:forceCorrelatorF} 
\lim_{\omega\to 0}\int\frac{\dd^3 q}{(2\pi)^3}
  \frac{\bm{q}^2}{3 \omega}[-2 T \mathrm{Im}  G^R_{F^2 F^2}(\omega,\bm{q})]         
  = N^2 T^9 C_1 \, .   
\end{equation}    
The corresponding result for the energy-momentum component correlator 
is obtained in an analogous way from the action and equations of 
motion already discussed in~\cite{Policastro:2002tn} . The final result is
\begin{equation}  
\label{eq:forceCorrelatorT} 
\lim_{\omega\to 0}\int\frac{\dd^3 q}{(2\pi)^3}
  \frac{\bm{q}^2}{3 \omega}[-2 T \mathrm{Im}  G^R_{TT}(\omega,\bm{q})]         
  = N^2 T^9 C_2 \, . 
\end{equation}  
The real numbers~$C_1,\, C_2$ here are numerical values which are currently being checked. 
\footnote{Note that during the publication process of this thesis our checks have been
completed and the final results are given in~\cite{Dusling:2008}.}  

{\bf Polarizabilities}
Looking at the meson diffusion formula~\eqref{eq:adsCftDiffusion} we realize that we have to determine the
polarizabilities~$\alpha_T,\, \alpha_F$. In analogy to the QCD calculation we consider the effective SYM 
Lagrangian~\eqref{eq:effSYML} leading to the meson mass shift 
\begin{equation}
	\delta M = -\frac{\alpha_T}{N^2} \left<T^{00}\right> - \frac{\alpha_F}{N^2}\left<\tr F^2\right>\, .
\end{equation}
On the other hand the mesons are dual to the gravity field fluctuations describing the embedding of 
our D7-brane~(cf.~section~\ref{sec:mesons}) and their masses are determined by the dynamics of the
gravity fluctuations. 
We have already reviewed how to compute meson masses from D7-brane embeddings in section~\ref{sec:mesons}. 
One of the major results there is the meson mass formula for scalar excitations~\eqref{eq:Ms} which depends on the
angular excitation number~$l$ as well as on the radial excitation~$n$. From here on we will consider the case of 
the lowest angular excitation~$l=0$ only. Picking up the QCD idea that the interaction with external color-fields shifts the
meson mass linearly~(cf.~equation~\eqref{eq:MShiftQCD}) we write down an analogous relation for the gauge condensate~$
\langle \textrm{tr} F^2 \rangle$
\begin{equation}
\label{eq:massShiftT}
\delta M = -\frac{\alpha_F}{N^2} \langle \mathrm{tr} F^2 \rangle \, .
\end{equation}
The constant of proportionality~$\alpha_F$ is identified with the polarization. It can be calculated by determining the 
meson mass shift~$\delta M$ at a given value of the gauge condensate~$\langle \mathrm{tr} F^2 \rangle \propto q$.
Let us now determine the mass shift analytically. This requires the further assumption that~$\bar q = q/L^4$ is small.
Next we derive the equation of motion for D7-brane fluctuations as shown in~\cite{Ghoroku:2004sp} 
and subsequently linearize that equation in~$q$, which then gives
 \begin{equation}
	-\partial_\rho \rho^3 \partial_\rho \phi(\rho) = \bar M^2\frac{\rho^3}{(\rho+1)^2}\,\phi(\rho) + \Delta(\rho) \phi(\rho)\, ,
\end{equation}
where the operator $\Delta(\rho)$ is given by
\begin{equation}
	\Delta(\rho) = - 4\bar q \frac{\rho^4}{(\rho^2 +1)^3}\,\partial_\rho \, .
\end{equation}
Setting the operator~$\Delta\equiv 0$ returns the case of vanishing gauge condensate~$\langle \mathrm{tr} F^2 \rangle 
\equiv 0$. So the term~$\Delta \phi$ describes the meson mass shift generated by the condensate on the level of the
equation of motion. We consider the lightest of the mesons by choosing the lowest radial excitation number~$n=0$ and
the solution at vanishing condensate is~$\phi_0$. Any deviation
$\delta\phi_0$ from the solution $\phi_0$ of the case $q=0$ may
be written as a linear combination of the functions $\phi_n$, which are a basis
of the function space of all solutions,
\begin{align}
\label{eq:phiPerturbed}
	\phi(\rho) &=  \phi_0(\rho) + \sum_{n=0}^\infty a_n \phi_n(\rho), & a_n &\ll 1,\\
	\bar M^2   &= \bar M_0^2 + \delta \bar M_0^2, &  \delta \bar M_0^2 &\ll 1\, ,
\end{align}
with the meson mass~$M_0$.
Plug this Ansatz into the equation of motion derived in~\cite{Ghoroku:2004sp}, make use of
the radial fluctuation equation of motion at vanishing~$q$~\eqref{eq:w56EOM} and keep terms up to linear order 
in the small parameters $a_n$, $\bar q_{\text{}}$ and $\delta M_0^2$ to get
\begin{equation}
	\frac{\rho^3}{(\rho^2+1)^2}\,\sum_{n=0}^\infty a_n \bar M^2_n \phi_n(\rho) =  
	\delta \bar M_0^2 \frac{\rho^3}{(\rho^2+1)^2}\,\phi_0(\rho) + \bar M_0^2 \frac{\rho^3}{(\rho^2+1)^2} \sum_{n=0}^\infty a_n\phi_n(\rho)
	 + \Delta(\rho) \phi_0(\rho).
\end{equation}
We now multiply this equation by $\phi_0(\rho)$, integrate over
$\rho \in [0,\infty[$ and make use of the fact that the~$\phi_n$ are orthonormal and of the non-interacting lowest 
mode solution~$\phi_0 = \sqrt{12}/(\rho^2 + 1)$ in order to rewrite
\begin{equation}
\begin{aligned}
	\delta \bar M_0^2 &= - \int\limits_0^\infty \!\! \dd\rho \; \phi_0(\rho)\, \Delta(\rho) \phi_0(\rho) \\
		 &= 4 \bar q_{\text{}} \int\limits_0^\infty \!\!\dd \rho\; \frac{\rho^4}{(\rho^2+1)^3}\,\phi_0(\rho)\partial_\rho\phi_0(\rho)\\
		 &= -\frac{8}{5}\,\bar q_{\text{}}.
\end{aligned}
\end{equation}
From $\delta\bar M^2_0 = 2 \bar M_0 \delta\bar M_0$ we therefore obtain
\begin{equation}
	\delta M_0 = \frac{L}{2R^2} \frac{\delta \bar M_0^2}{\bar M_0} 
	  =- \frac{8}{5\pi}\left (\frac{2\pi}{M_0}\right)^3 \frac{1}{N^2}\left<\tr F^2\right>,
\end{equation}
where we inserted the meson mass formula~\eqref{eq:Ms} and switched back to dimensionful
quantities. By comparison with \eqref{eq:massShiftT} we may now identify
$\alpha_F$
\begin{equation}
\label{eq:alphaF}  
	\alpha_F =  \frac{8}{5\pi}\left (\frac{2\pi}{M_0}\right)^3\,.
\end{equation}

The calculation of the polarizability $\alpha_T$ is completely analogous. We
are now looking for the proportionality constant of meson mass shifts with
respect to deviations from zero temperature,
\begin{equation}
	\label{eq:deltaMalphaT}
	\delta M = -\frac{\alpha_T}{N^2} \left<T^{00}\right>.
\end{equation}
The vacuum expectation value 
\begin{equation}
	\left< T^{00}\right> = \frac{1}{2}\pi^2 N^2 T^4
\end{equation}
is proportional to $(\text{temperature})^4$. We eventually obtain the polarizability $\alpha_T$ as
\begin{equation}
\label{eq:alphaT}  
	\alpha_T= \frac{12}{5\pi} \left(\frac{2\pi}{M_0}\right)^3\, .
\end{equation}
These results~\eqref{eq:alphaF} and~\eqref{eq:alphaT} for small values of~$\bar q$ agree very well with the numerical calculation 
we performed in parallel~(not shown here, see~\cite{Dusling:2008} for details) relaxing the assumption that~$\bar q$ needs 
to be small. 

{\bf Result}
Substituting our polarizations~\eqref{eq:alphaF} and~\eqref{eq:alphaT}, as well as the 
correlators~\eqref{eq:forceCorrelatorT} and~\eqref{eq:forceCorrelatorF}
into the Kubo equation for the heavy meson momentum broadening~\eqref{eq:adsCftDiffusion} yields
\begin{equation}  
\label{eq:strongK}
\kappa = 
\frac{T^3}{N^2}\left (\frac{2\pi T}{M_0}\right)^6 \left [\left(\frac{8}{5\pi}\right)^2 C_1 
+\left(\frac{12}{5\pi}\right)^2 C_2 \right ] = 
 C_3\;\frac{T^3}{N^2}\left (\frac{2\pi T}{M_0}\right)^6\, ,       
\end{equation}  
with numerical values~$C_1,\,C_2,\,C_3$ which are currently being checked. 
\footnote{Note that during the publication process of this thesis our checks have been
completed and the final results are given in~\cite{Dusling:2008}.} 

This strong coupling result resembles the weak coupling result obtained from a 
perturbative calculation at large~$N_c$ very closely
\begin{equation}  
\label{eq:weakK}
\kappa =  \tilde C_3 \frac{T^3}{N^2}\left (\frac{\pi T}{\Lambda_B}\right)^6 \, ,
\end{equation}  
where the inverse Bohr radius~$\Lambda_B$ replaces the meson mass~$M_0$ as the characteristic energy scale.
In order to compare the weak coupling result~\eqref{eq:weakK} to the strong coupling result~\eqref{eq:strongK},
we need to divide by the corresponding mass shifts~$(\delta M)^2$ such that the Bohr radius and the quark 
mass cancel from the results. The number~$\tilde C_3$ is still being checked. 
Nevertheless, our preliminary results indicate that the ratio~$\kappa/(\delta M)^2$ is about five times smaller at
strong coupling compared to its value at weak coupling. It is reassuring that the viscosity to entropy quotient 
shows an analogous behavior being much smaller at strong coupling~\cite{Policastro:2001yc}.
After the exact values~$C_3, \tilde C_3$ are confirmed we will draw a more precise conclusion.
\footnote{Note that during the publication process of this thesis all the open checks have been completed
and the final result can now be found in~\cite{Dusling:2008}.}

\subsection{Diffusion matrix} \label{sec:diffusionMatrix}
This section collects a few ideas and formulae which result from working towards the computation of 
diffusion matrices. The basic idea here is motivated by the fact that the diffusion of a certain charge can
be induced by different gradients. A setup in which such an effect might occur is a thermal plasma in which
the three flavor charge densities are fixed to three different values. 
For example in section~\ref{sec:isospinDiffusion} we found that the
simultaneous presence of finite baryon and isospin density changes the baryon diffusion coefficient 
in a different way than increasing a finite baryon density alone. This motivates the idea to arrange the diffusion coefficients
relating distinct gradients to distinct currents in a matrix~\footnote{The author is grateful to Christopher Herzog
and Laurence Yaffe for valuable discussions on this topic.}.
Analogous matrix structures appear in the context of Ohm's law at strong coupling for the heat and charge
conductivity~\cite{Hartnoll:2007ip}. In the context of QCD calculations the importance of
flavor diffusion matrices has been stressed for example in~\cite{Arnold:2000dr,Arnold:2003zc}. 
We will use the D3/D7-system with a non-vanishing isospin density in 
all three flavor directions as a sample setup to study.
We collect a few intermediate results and ideas in order to develop the basic plan of the calculation.

Up to now in this work we have chosen a chemical potential~$\mu^I$ along the third flavor direction~$\sigma^3$ in isospin space. 
The thermodynamically conjugate quantity is the charge density~$\tilde d^I\equiv \tilde d^{I_3}$ coupling to this particular 
flavor direction. Now in general there are two more charge densities~$\tilde d^{I_1},\,\tilde d^{I_2}$ to which the 
corresponding chemical potentials are conjugate. On the gravity side of the correspondence the flavor gauge field
components~$A^1$ and~$A^2$ couple to the isospin charge densities~$\tilde d^{I_1},\,\tilde d^{I_2}$, respectively.
The action relevant for this approach including all three isospin directions has been given in equation~\eqref{eq:SI} already.
In this section we are interested in how a gradient in one of the three isospin charge densities~$\tilde d^{I_1},\,\tilde d^{I_2},
\,\tilde d^{I_3}$ influences the current of a different one of these charge densities. In other words, our goal is to compute the 
components of the diffusion matrix
\begin{equation}
D = \left(
\begin{array}{c c c}
D_{11} & D_{12} & D_{13} \\
D_{21} & D_{22} & D_{23} \\
D_{31} & D_{32} & D_{33} 
\end{array}
\right) \, ,
\end{equation}
appearing in the diffusion equation for three isospin charges
\begin{equation}
\partial_0 \left (\begin{array}{c}
J^1_0\\
J^2_0\\
J^3_0
\end{array}\right ) = \left(
\begin{array}{c c c}
D_{11} & D_{12} & D_{13} \\
D_{21} & D_{22} & D_{23} \\
D_{31} & D_{32} & D_{33} 
\end{array}
\right)
\partial^i \partial_i \left (\begin{array}{c}
J^1_0\\
J^2_0\\
J^3_0
\end{array}\right )
 \, .
\end{equation}
In this general setup all three flavor directions are equal. We have not picked any one of them to be special as in our
previous approach. So it is reasonable to assume that the diffusion induced by a charge gradient e.g. in 1-direction
drives a current in 2-direction with the same strength as a charge gradient in 2-direction drives a 1-current, which
implies~$D_{12}=D_{21}$. Therefore, the diffusion matrix is assumed to be symmetric and can thus be diagonalized.
To check this heuristic argument we have at least two paths which could bring us to our goal: 
Either we extend the membrane paradigm by flavor indices,
such that we have an equation similar to~\eqref{eq:diffusionConstant} but with flavor 
indices~$D^{ab} = D^{ab}(g^a_{\mu\nu},g^b_{\rho\sigma})$,
or we compute the fluctuations and read the diffusion coefficient from the zero frequency limit of the spectral functions
as described in equation~\eqref{eq:spectralFunctionDiffusion}. No matter which of these two approaches we choose,
either one has to incorporate the flavor structure of fields.

{\bf A trivial result?}
To conclude this section we briefly discuss the probable outcomes for the calculation of different 
diffusion matrices.
We could bring up the argument that due to the rotational symmetry in flavor space we can always rotate to a 
flavor coordinate system in which the isospin points along the third flavor direction for example. This is the 
restriction from three degrees of freedom~$\tilde d^I_1,\,\tilde d^I_2,\,\tilde d^I_3$ to only one~$\tilde d^I_3$. 
If this is true then the isospin diffusion matrix would have to be proportional to unity. 
However, in the case of finite baryon and isospin density we have explicitly seen that the baryon diffusion coefficient 
changes differently if a finite isospin density is present. Thus we assume that at least in that case the diffusion matrix can not be proportional to unity. In this case the explanation is that we introduce the physical baryon and isospin densities and give them different values, so that we can distinguish between them. Then we transform to non-physical densities in which the problem 
simplifies. But here we keep all the degrees of freedom we have since we transform~$(\tilde d^B,\tilde d^I)\to(\tilde d^1,\tilde d^2)$. 
After solving the problem with these simpler flavor coordinates we transform back to the physical
baryon and isospin densities which we had introduced in the beginning. Since we had chosen the physical baryon and 
isospin density to be different, this difference reflects in the response of the system given by the baryon
diffusion coefficient being changed compared to the case of vanishing isospin density. Now we argue that 
in the case of vanishing baryon but non-zero isospin densities a similar effect might occur if we 
admit different, non-zero, physical charge densities for all three flavor directions.

\subsection{Summary} \label{sec:sumTransport}
In this chapter we have studied the diffusion of quarks and their bound states inside a thermal plasma at strong coupling.
We started this study by reviewing the membrane paradigm, a holographic method to find transport coefficients merely 
knowing the metric components on the gravity dual side in section~\ref{sec:membranePara}. 
With this calculational tool at hand we found in section~\ref{sec:baryonDiffusion} the coefficient of baryon charge 
diffusion in the thermal theory at finite
baryon charge density which is dual to our D3/D7-setup. That diffusion coefficient approaches a fixed value of~$D=1/(2\pi T)$
at low and at high temperatures. At intermediate temperatures the baryon diffusion coefficient shows a minimum
which shifts to lower temperatures as the density is increased. At vanishing baryon density the diffusion coefficient
still asymptotes to the value~$D=1/(2\pi T)$ at large temperatures while it vanishes at the phase transition temperature
and for all temperatures below it. We interpret this by the baryon charge carriers, the quarks to vanish below the 
transition because they get bound into quasi-meson states carrying no net-baryon charge. At finite baryon density 
by definition we always have a finite amount of baryon charge carriers so the diffusion coefficient can not vanish for this reason.

In section~\ref{sec:isospinDiffusion} we additionally introduced a finite isospin density to the baryon density and
studied their combined effect on baryon charge diffusion. The baryon diffusion coefficient qualitatively behaves
as in the pure baryon density case studied before and increasing the isospin density appears to have the same
qualitative effect as adding more baryon density. That this is not the case can be seen from the study of the 
extended baryon-isospin density phase diagram~\ref{fig:contourPhaseDiagBI}. In this diagram we have first
traced the location of the~(black hole to black hole) phase transition present at small densities. Then we extended
it by following the minimum in the diffusion coefficient mentioned above. Since the rotational symmetry in this 
phase diagram over the baryon-isospin density plane is obviously broken to~$\mathbf{Z}_4$, we clearly see that baryon
and isospin density have different effects on hydrodynamics of this theory, so there is a subtle interplay between them.

Section~\ref{sec:charmDiffusion} extends our considerations of quark diffusion to the diffusion of their bound states.
In particular motivated by experimental and lattice results hinting at charmonium bound states having survived the
deconfinement phase transition of QCD, we examine the mesonic bound states which we have found in 
chapter~\ref{sec:mesonSpectraB}. We find the charmonium diffusion to meson mass-shift quotient~$\kappa/(\delta M)^2$
to be significantly smaller at strong coupling compared to its value at weak coupling. 
The calculation is still being checked, but will be published soon~\cite{Dusling:2008}.

Collecting basic ideas and proposing some technical starting points in section~\ref{sec:diffusionMatrix} we 
suggest how to introduce the concept of a flavor diffusion matrix. The matrix structure is based on the idea
that a charge density in one flavor direction might drive a current in another. In analogy to similar effects 
present in classical systems with different charges studying this matrix may also elucidate the different 
(baryon and isospin density-induced)~contributions to the effective baryon diffusion coefficient found in
section~\ref{sec:isospinDiffusion}.

\section{Conclusion}
\label{sec:conclusion}
This final chapter summarizes what we have learned in the course of this thesis about the thermal 
gauge theory at strong coupling holographically dual to the D3/D7-setup described in sections~\ref{sec:mesons} 
and~\ref{sec:thermoBaryon}. In particular we have studied the background introducing finite baryon and 
isospin densities and chemical potentials, as well as the fluctuations around this background. The  
strongly coupled thermal Super-Yang-Mills theory with finite densities or potentials serves as our model theory for the 
quark-gluon plasma produced at present and future colliders~(RHIC at Brookhaven, LHC at CERN).
I list all of my results and discuss their interrelations. Finally, I give my conclusions and an outlook.
Recall for the discussion that our D3/D7-setup at finite temperature is controlled by the 
parameter~$m\propto M_q/T$, thus increasing the quark mass~$M_q$ is equivalent to decreasing the
temperature~$T$, and vice versa.
 
\begin{figure}
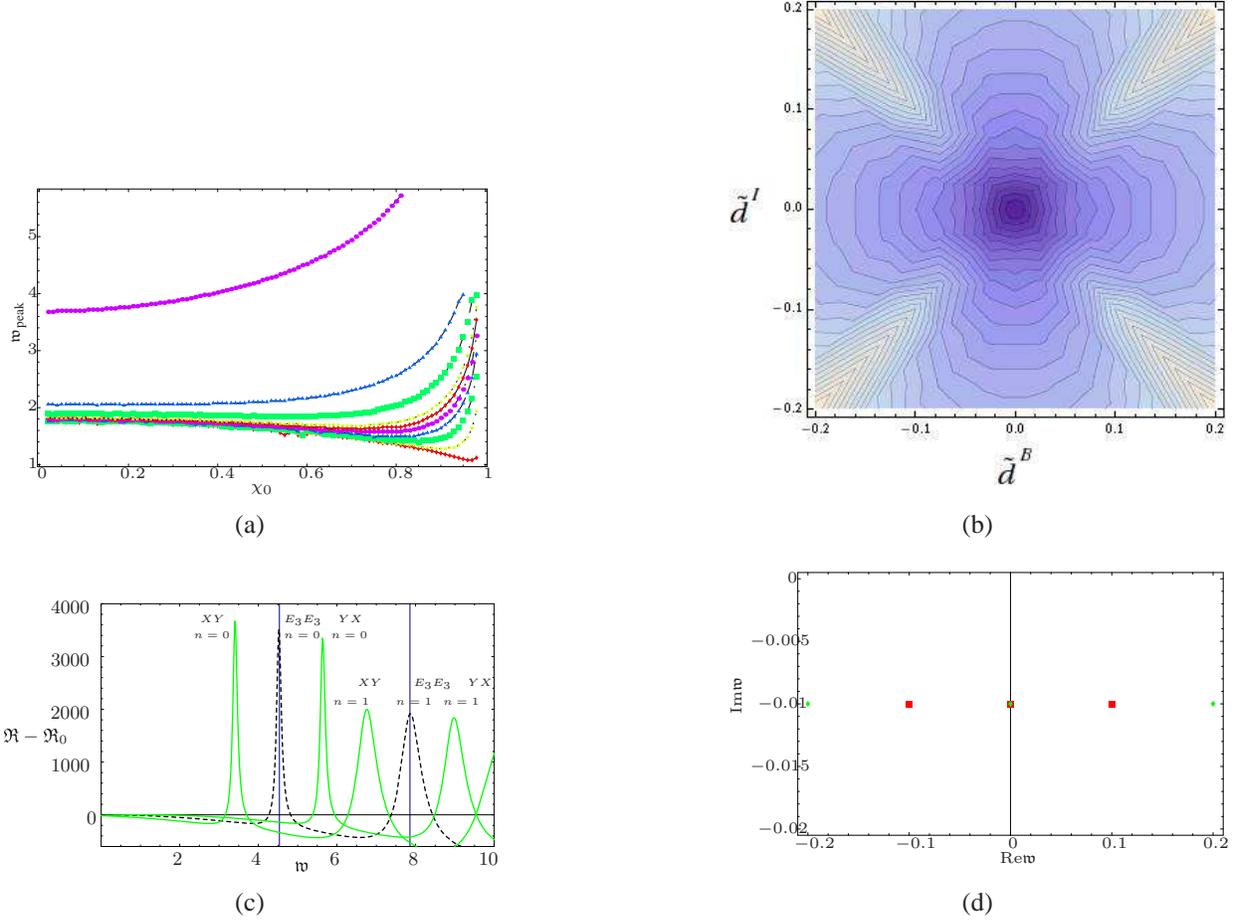

  \subfigure[]{
  	\psfrag{0}{\tiny $0$}
	\psfrag{0.1}{\tiny $0.1$}	
	\psfrag{0.2}{\tiny $ 0.2$}
	\psfrag{0.25}{\tiny $ 0.25$}	
	\psfrag{0.4}{\tiny $ 0.4$}
	\psfrag{0.5}{\tiny $ 0.5$}	
	\psfrag{0.6}{\tiny $ 0.6$}
	\psfrag{0.8}{\tiny $ 0.8$}
	\psfrag{0.8}{\tiny $ 0.8$}		
	\psfrag{-0.2}{\tiny $ -0.2$}
	\psfrag{-0.25}{\tiny $ -0.25$}	
	\psfrag{-0.4}{\tiny $ -0.4$}
	\psfrag{-0.5}{\tiny $ -0.5$}	
	\psfrag{1}{\tiny $ 1$}
  	\psfrag{2}{\tiny $ 2$}
  	\psfrag{3}{\tiny $ 3$}
  	\psfrag{4}{\tiny $ 4$}
  	\psfrag{5}{\tiny $ 5$}				
  	\psfrag{wPeak}{\tiny $ \wn_{\text{peak}}$}
	\psfrag{chi0}{\tiny $ \chi_0$}
        \includegraphics[width=.4\linewidth]{./figures/turningDataVsChi0.eps}
}
\hfill
  \subfigure[]{
        \includegraphics[width=.4\linewidth]{./figures/phasediagramdiffcontoursymbreak.eps}
        }
\vfill
  \subfigure[]{
        \includegraphics[width=.4\linewidth]{./figures/isospinSplitting.eps}
        }
\hfill
  \subfigure[]{
  	\psfrag{0}{\tiny $0$}
	\psfrag{-0.005}{\tiny $-0.005$}	
	\psfrag{-0.01}{\tiny $-0.01$}	
	\psfrag{-0.015}{\tiny $-0.015$}		
	\psfrag{-0.02}{\tiny $-0.02$}		
	\psfrag{-0.1}{\tiny $-0.1$}	
	\psfrag{0.1}{\tiny $0.1$}	
	\psfrag{0.2}{\tiny $0.2$}
	\psfrag{0.25}{\tiny $0.25$}	
	\psfrag{0.4}{\tiny $0.4$}
	\psfrag{0.5}{\tiny $0.5$}	
	\psfrag{0.6}{\tiny $0.6$}
	\psfrag{0.8}{\tiny $0.8$}
	\psfrag{0.8}{\tiny $0.8$}		
	\psfrag{-0.2}{\tiny $-0.2$}
	\psfrag{-0.25}{\tiny $-0.25$}	
	\psfrag{-0.4}{\tiny $-0.4$}
	\psfrag{-0.5}{\tiny $-0.5$}	
	\psfrag{1}{\tiny $1$}
	\psfrag{ImW}{\tiny $\mathrm{Im}\wn$}
	\psfrag{ReW}{\tiny $\mathrm{Re}\wn$}
        \includegraphics[width=.4\linewidth]{./figures/anaQNM.eps}
        }
\caption{
These four plots visualize some of the main results of this thesis. 
(a) Frequency location of the lowest resonance peak in the spectral function at finite baryon 
density~$\tilde d_B\not =0$. The baryon density is increased from~$\tilde d_B=0.01$~(the bottom curve) 
to~$\tilde d_B=10$~(the top curve). For details confer figure~\ref{fig:turnDataChi0} and its discussion in the text.
(b) Contour plot of the location of the phase transition/crossover
mass parameter over the (baryon density, isospin density)-plane.  
(c) Triplet splitting of resonance peaks at finite isospin density~$\tilde d_I\not =0$ for vanishing
spatial momentum~$\qn\equiv 0$. This splitting corresponds to a triplet splitting of the 
corresponding poles in the complex frequency plane.
(d) Location of the diffusion pole for the three different flavor combinations~$XY$, $YX$ and 
$33$~(cf.~section~\ref{sec:mesonSpectraI})
computed analytically in the hydrodynamic limit~$\wn,\,\qn^2,\,\mn \ll 1$ at finite spatial momentum~$\qn\not =0$.
The diffusion pole shows a triplet splitting as well.
\label{fig:mainResults}
}
\end{figure}

{\bf Results at a glance and discussion}
At finite baryon density we have discovered mesonic quasi-particle resonances in the thermal spectral 
functions of flavor currents in section~\ref{sec:mesonSpectraB}~(see figure~\ref{fig:lineSpectrumDt025heavy}). 
These resonance peaks follow the holographic meson mass formula~\cite{Kruczenski:2003be}
\begin{equation}
M= \frac{L_\infty}{R^2}\,\sqrt{2 (n+1)(n+2)}\, ,
\end{equation}
at large masses or equivalently at low temperature. This means that increasing the quark mass (which 
increases~$L_\infty$ as well) the resonance peaks move towards higher frequency. Since also their width~(inversely
proportional to the lifetime of that exciatation) compared to
their energy is narrow, we identify these resonances with stable vector mesons in the plasma having survived the
deconfinement phase transition of the theory. This is in qualitative agreement with the lattice 
calculation given in~\cite{Asakawa:2003xj} and also with~\cite{Shuryak:2004tx}.
On the other hand, in the small mass/high temperature regime the interpretation of spectral function maxima is 
still controversial~(see also~\cite{Paredes:2008nf,Myers:2008cj}). 
In this high temperature regime we find broad maxima as opposed to narrow low-temperature resonance peaks. 
Moreover, these maxima do not follow the meson mass formula at all~(see figure~\ref{fig:vectorPeaksDt025light}). 
Quite the contrary is true since we observe
the maxima to move towards lower frequencies as we increase the quark mass. A (stable) particle interpretation is no 
longer justified in this high-temperature/ small mass regime. Decreasing the temperature in order to approach
the low-temperature regime, we discover a turning point, where the maxima of the spectral functions change their
direction along the frequency axis as discussed in section~\ref{sec:peakTurning}. The location of the lowest lying 
resonance peak is shown in figure~\ref{fig:mainResults}~(a) versus the mass parameter~$\chi_0$~(cf.~figure~\ref{fig:mOfChi0}).
Different curves correspond to distinct baryon densities, with the bottom curve corresponding to the lowest 
density~(cf.~figure~\ref{fig:turnDataChi0} for details). 
Thus, we claim that we have to distinguish between the temperature-dominated and the mass-dominated regime.
In section~\ref{sec:peakTurning} we have worked towards an explanation for the high temperature behavior and for the 
peak turning we observe. In the limit of high frequencies we have found an analytical solution near the horizon in terms
of the confluent hypergeometric function. This analytical solution (as well as the numerical solution for arbitrary momenta
and radial coordinate values) shows oscillatory behavior and damping in agreement with
our hypothesis: In the high temperature regime there are no stable bound states of quarks, but merely
unstable excitations in the plasma which quickly dissipate their energy to the plasma. Our analytical solution
reproduces the effect of resonance peaks in the 'spectral function fraction'~(see~\ref{sec:peakTurning}) 
moving towards lower frequencies when the mass parameter is increased. We have also related these
thoughts to quasinormal modes. Further, we commented on that we could learn more about the inner workings of the 
gauge/gravity correspondence in this example by studying how to relate the bulk solutions generating the peaks 
in the spectral function to the spectral functions explicitely~(see discussion of the quasinormal mode solutions
in section~\ref{sec:peakTurning} contained in the paragraph 'Heuristic gravity interpretation').

We have studied the fluctuations around an $SU(2)$ isospin background as well in section~\ref{sec:mesonSpectraI}. 
The resulting spectral functions at finite isospin density are shown in figure~\ref{fig:mainResults}~(c). 
We clearly observe a triplet splitting of the resonance peaks. Introducing a chemical potential in a specific
flavor direction we have broken the $SU(2)$-symmetry and we clearly observe the splitting 
because our vector mesons are triplets under the isospin group (analogous to the $\rho$-meson in QCD).
As a methodical achievement we have generalized all formulae describing this 
setup to include $U(N_f)$-chemical potentials with arbitrary~$N_f$ in section~\ref{sec:hiNf}.
Note, that all the spectral functions we have computed numerically are evaluated for perturbations with
vanishing spatial momentum~$\qn=0$. In this limit the correlators for transversal and longitudinal directions
coincide. One effect of this is that we are not able to identify the lowest one of the poles, i.e. the hydrodynamic diffusion 
pole which should appear in the longitudinal correlators. However, in the analytical calculation in 
section~\ref{sec:anaHydroIso} we consider exclusively this pole. 

In the hydrodynamic approximation, i.e. at small frequencies and spatial momenta we are able to find 
correlators analytically at finite isospin chemical potential~(see section~\ref{sec:anaHydroIso}). 
The longitudinal correlators are particularly interesting since the diffusion pole appears in them. 
We have observed a triplet splitting~(see figure~\ref{fig:mainResults}~(d)) of this diffusion pole which can also be seen from the dispersion
relation which we read off the longitudinal correlation functions
\begin{eqnarray}
\omega = & -i D q^2 \pm \mu \quad &\text{for} \quad \wn\ge\mn \, , \\
\omega = &  i D q^2 + \mu \quad &\text{for} \quad \wn < \mn\quad \text{and only in} \quad G^{XY}\, ,
\end{eqnarray}
where the positive sign of~$\mu$ corresponds to the dispersion of the flavor combination~$G^{XY}$ and the negative
sign of~$\mu$ corresponds to~$G^{YX}$. For the third flavor direction correlators~$G^{33}$ there is no chemical potential 
contribution in the dispersion relation~$\omega= -i D q^2$. We have argued that by introducing a chemical potential 
along the third flavor direction and considering the fluctuations in any flavor direction the setup in flavor space 
resembles that of Larmor precession in real space. The fluctuations precede around the designated third flavor
direction with the Larmor frequency~$\omega_L = \mu$. This frequency we also interpret as the minimal energy any
excitation needs to have in order to be produced in the plasma. In this hydrodynamic limit we have also computed the
spectral functions corresponding to the diffusion poles, discussed the quasinormal modes and the residues.
We have also discussed the reconciliation of these present results with the approach taken in~\cite{Erdmenger:2007ap} in 
section~\ref{sec:anaHydroIso}.

From our discussion in section~\ref{sec:mesonSpectraB} we know that the poles of a correlation function in the complex frequency
plane generate the structure in thermal spectral functions~(cf.~figure~\ref{fig:polesExample}).
It is convincing that upon introduction of isospin we observe the same behavior of triplet splitting in 
both the analytical approximation for the diffusion pole shown in figure~\ref{fig:mainResults}~(d) and for the 
mesonic resonances in the numerically computed spectral functions shown in figure~\ref{fig:mainResults}~(c).
We are not able to see the effect of the diffusion pole itself in the numerical results because there we simplified to~$\qn=0$.
But the higher frequency poles obviously have the same triplet splitting as the diffusion pole, as we can
infer by looking at the spectral function peaks splitting more and more when we increase the isospin density and with
it the chemical isospin potential as well.

We have studied diffusion of quarks and their quarkonium bound states as specific examples for transport 
phenomena in chapter~\ref{sec:transport}.
Utilizing the membrane paradigm in section~\ref{sec:baryonDiffusion} we have found the coefficient of baryon 
or equivalently quark charge diffusion in the thermal theory at finite
baryon charge density which is dual to our D3/D7-setup~(see figure~\ref{fig:diffusionConstants}). 
That diffusion coefficient approaches a fixed value of~$D=1/(2\pi T)$
at low and at high temperatures. At intermediate temperatures the baryon diffusion coefficient shows a minimum
which shifts to lower temperatures as the density is increased. The minimum is also lifted if the density is increased. 
At vanishing baryon density the diffusion coefficient
still asymptotes to the value~$D=1/(2\pi T)$ at large temperatures while it vanishes at the phase transition temperature
and for all temperatures below it. This effect is caused by the baryon charge carriers, the quarks which vanish below the 
transition because they get bound into meson states carrying no net-baryon charge. At finite baryon density 
this effect is still present at sufficiently low temperature since there the quarks are also bound into mesonic
states as we have learned from our study of the spectral functions.
Nevertheless, by definition we always have a finite amount of baryon charge carriers so the diffusion coefficient can never vanish.

The black hole to black hole phase transition present at finite and increasing baryon density is shifted to a 
lower temperature as we see for example in the diffusion coefficient in figure~\ref{fig:diffusionConstants}. As mentioned above,
the transition is lifted in the sense that the minimum in the diffusion coefficient increases from zero at vanishing
baryon density towards~$1/(2\pi)$ at large densities. This black hole to black hole transition continues to 
exist also if a small isospin density is introduced additionally.

Simultaneously introducing baryon and isospin density in the background we have discovered a further phase transition
indicated by discontinuities in thermodynamic quantities. For example the quark condensate and the baryon and 
isospin densities are discontinuous on the line of points~$\mu^B =\mu^I$. This transition resembles that one found
in the case of 2-flavor QCD found in~\cite{Splittorff:2000mm}.
In addition we found significant changes in thermodynamic quantities through simultaneous isospin in 
section~\ref{sec:thermoB&I}. These changes are of qualitative nature, i.e. introducing isospin charge or potential
is {\it not} identical to merely introducing more baryon density. The distinct effects of baryon and isospin charge 
or potential become obvious in the hydrodynamic regime. In figure~\ref{fig:mainResults}(b) we see
a contour plot of the transition temperature parametrized by~$m$ over the (baryon density, isospin density)-plane.
This means that the contours are contours of equal transition temperature. Only the innermost part of this
diagram traces the black hole to black hole transition at small densities. This transition vanishes
for baryon densities above~$\tilde d^B_* =0.00315$~(see discussion in~\ref{sec:thermoBaryon}). 
For larger densities we have simply traced the location
of the minimum in the diffusion coefficient which we identify as the temperature at which a softened version
of the thermodynamic transition, i.e. a hydrodynamic 
transition occurs. From the contour plot in figure~\ref{fig:mainResults}(b) we clearly see that an initial
rotational symmetry at small densities suggests that baryon and isospin density have the same effect.
However at large densities the outermost contours clearly show that the rotational symmetry is broken 
to a~$\mathbf{Z}_4$ symmetry. This means that baryon and isospin density have different effects on the hydrodynamics
of this theory.

Extending our studies of transport phenomena to bound states of quarks, we have computed the diffusion of quarkonium
in section~\ref{sec:charmDiffusion}. Our results indicate that the diffusion to meson mass-shift quotient~$\kappa/(\delta M)^2$ 
is significantly smaller at strong coupling than at weak coupling. This resembles the case of the viscosity to entropy density
quotient which takes on significantly smaller values at strong coupling as well~\cite{Policastro:2001yc}.

{\bf Conclusions \& Outlook}
In conclusion we have reached the goal of this thesis outlined in the introduction on page~\pageref{page:theMission}. 
We have successfully incorporated the concepts of baryon/isospin
chemical potentials and densities in the D3/D7-gravity dual modeling quarks and mesons. We have studied the rich 
phenomenology of this model on a qualitative level and we have found many interesting signatures being consistent with lattice results
and effective QCD calculations. Nevertheless, we have also found novel structures, which had not been predicted previously.
Based on our experience with its qualitative behavior it would be interesting to study this 
model also on a quantitative level. In this analysis quotients of quantities could prove to be useful, 
which show universal behavior, such as the viscosity to entropy ratio. Our preliminary quantitative result on the 
charmonium diffusion to meson mass-shift ratio clearly confirms this belief.
 
Constructing the phase diagram we have shown that isospin density/potential has effects significantly different from baryon 
density/potential. Also in the analysis of spectral functions isospin effects such as the triplet splitting distinguish the isospin 
phenomenology clearly from the baryonic signatures. One important extension of the work presented here will
be the computation of meson spectra at finite baryon and isospin density as described in section~\ref{sec:mesonSpectraB&I}.
Having both the rich effects of the baryon and isospin background and the interaction with fluctuations about
it will produce a potentially rich phenomenology. The technical considerations in section~\ref{sec:mesonSpectraB&I}
show that this calculation is complex but feasible. Furthermore, we have restricted our analysis to vector mesons, 
but it is easy to extend it to scalars and pseudoscalars as well (see~\cite{Kerner:2008diploma} for parts of 
an equivalent analysis). Elaborated results on the baryon and isospin background will we published soon~\cite{Erdmenger:2008yj}.

Nevertheless, also our studies of the setup with baryon density only, brought up interesting relations. 
For example the analysis of the resonance peak turning point gives us a deeper insight in how the gauge/gravity
correspondence works in terms of a correspondence between the gravity bulk solutions and the gauge theory 
spectral functions. The further study of this topic will either confirm our speculations about the thermal origin of
the resonance turning point or prove it wrong. In any case the analytic gravity solutions which we seek to construct
and their direct relation to the gauge theory spectral functions encodes valuable information about the gauge/gravity 
correspondence. We will develop this analysis in~\cite{EKKR:2008tp}.

Analytical and numerical studies of the charge diffusion coefficients have consistently confirmed the interpretations we 
developed for our spectral functions. For example the decreasing baryon charge diffusion coefficient at small temperature
confirms the meson interpretation of the formation of narrow resonance peaks. A further way of testing this interpretation
would be the computation of the diffusion coefficient for the quasi-mesons corresponding to the peaks in the spectral
functions. If this quasi-meson diffusion coefficient vanishes above the hard phase transition at zero densities, this would confirm
that these mesons simply vanish there. At finite density we expect this quasi-meson diffusion only to decrease as 
the temperature is increased well above the transition temperature.

The baryon charge diffusion coefficient has been computed both at finite baryon density only and at finite baryon
and isospin densities. We found that the isosin density changes the baryon charge diffusion coefficient significantly.
Due to our computational method using the membrane paradigm we have not been able to separate the  
diffusive contributions generated by the finite isospin from those generated by the finite baryon density. 
Therefore we suggest to study these different contributions developing the framework of a diffusion matrix
as desribed in section~\ref{sec:diffusionMatrix}. This computation will also answer the question if the effect of
finite isospin density is simply additive, i.e. if we could get its contribution to the diffusion coefficient by subtracting the
diffusion coefficient in the purely baryonic background. Based on our observations of the minimum in the
baryon diffusion coefficient shown in figure~\ref{fig:mainResults}(b), we suspect a more subtle interplay
between baryon and isospin densities. Note, that in section~\ref{sec:diffusionMatrix} we develop 
the relevant formulae for three different isospin charges rather than for one isospin and one baryon charge. 
Nevertheless, the framework once developed should easily generalize to that case as well.

Now after considering the possibility that modes with different flavor might behave differently inside the
thermal plasma, we should also worry about the fact that modes with different frequencies or spatial momenta
propagate through the plasma in different ways. We have commented on the possible incorporation
of this idea into our setup in the context of molecular dynamics discussed in section~\ref{sec:molecular}.

Finally, we collect a few pronounced signals which the rich phenomenology explored here predicts to be seen
at colliders. A clear signature are the stable meson resonances having survived the deconfinement transition,
showing a turning behavior in their energy as the temperature is decreased. At sufficiently high
isospin density in the plasma a resonance peak triplet splitting depending on the amount of isospin density
should be visible. We further expect discontinuities in thermodynamical quantities to show up across
the line of equal baryon and isospin densities or potentials due to the phase transition we discovered across that line. 
Nevertheless, since our supersymmetric model is not QCD we should not be surprised to see 
different behavior in some cases in the collider experiments. However, the high-temperature regime of the baryon 
diffusion coefficient down to the thermodynamic or hydrodynamic phase transition should be taken seriously.
Also the small value of the charmonium diffusion coefficient is a very interesting effect to look for,
given that it resembles the viscosity to entropy ratio in its strong coupling behavior.

\acknowledgements
The author would like to thank Johanna Erdmenger, Dieter L\"ust, 
Felix Rust, Patrick Kerner, Derek Teaney, Martin Ammon, Stephan H\"ohne and Rene Meyer. 
I am grateful to Christopher Herzog, Steven Gubser, Clifford Johnson, Igor Klebanov, 
Andreas Karch, Karl Landsteiner, Robert C. Myers, 
David Mateos, Dam T. Son, Andrei O. Starinets and Laurence Yaffe for valuable correspondence and discussions.
My work has partly been supported by {\it The Cluster of Excellence for Fundamental Physics --Origin and Structure
of the Universe}.

\begin{appendix}
\section{Notation}
Here we give an overview of the notation which we use in this thesis if not specified otherwise. 
We denote three-vectors in spatial directions by minuscule letters in bold face such as~$\mathbf{x}$, four vectors
including the time component are given by~$\vec x$, higher-dimensional vectors are given by the plain minuscule 
letter, e.g.~$x$. 
If any of the momentum components~$\wn,\, \qn$ appears in an order relation such as~$\wn\ll 1$ we actually 
mean to denote the real part~$\mathrm{Re}\wn\ll 1$. The chemical potential is assumed to take real 
values~$\mu, \mn\in \mathbf{R}$ throughout the whole thesis. 
All mathematical sets of numbers are given in bold face font. For example the whole numbers are given by~$\mathbf{Z}$,
the real numbers are given by~$\mathbf{R}$ and the complex numbers by~$\mathbf{C}$. 
We work in natural units, i.e. we set the reduced Planck's constant~$\hbar =1$ and the speed of light~$c=1$. 
Additionally the Boltzmann constant is chosen~$k_B=1$ for convenience. 

{\bf Symbols}
If not specified otherwise in the text, the following symbols have been used to denote the quantities listed below
in arbitrary order
\begin{eqnarray}
p & \text{local pressure} \\
\epsilon & \text{internal energy} \\
T_{\mu\nu} & \text{energy momentum tensor} \\
P_\mu & \text{heat current} \\ 
T & \text{local temperature} \\
\beta\equiv\frac{1}{T} & \text{inverse temperature} \\  
\F & \text{free energy in canonical ensemble} \\  
\Omega & \text{grandcanonical potential} \\  
\S & \text{entropy} \\  
s & \text{entropy density} \\  
S & \text{action} \\  
\mu & \text{chemical potential} \\  
n & \text{charge density} \\  
d & \text{conserved charge density} \\  
u_\mu & \text{four-velocity of a fluid volume} \\  
c & \text{quark condensate}~\langle \bar q q\rangle \\  
Q & \text{charge} \\       
\R & \text{thermal spectral function} \\ 
D & \text{diffusion coefficient} \\  
\eta & \text{shear viscosity (coefficient)} \\  
\kappa & \text{quarkonium diffusion coefficient}  
\end{eqnarray}
\begin{eqnarray}
T^a & \text{Lie group generator} \\  
T_r\in\mathbf{R} & \text{representation factor} \\  
N_f & \text{number of flavors / D7-branes} \\  
N \equiv N_c & \text{number of colors / D3-branes} \\  
\rho_H& \text{horizon value of the dimensionless radial AdS coordinate} \\  
\rho_B\equiv\rho_{\text{bdy}} & \text{boundary value of the dimensionless radial AdS coordinate} \\  
\varrho_H& \text{horizon value of the dimensionful radial AdS coordinate} \\  
\varrho_B\equiv\varrho_{\text{bdy}} & \text{boundary value of the dimensionful radial AdS coordinate} 
\end{eqnarray}
\begin{eqnarray}
m_q\equiv M_q & \text{quark mass} \\  
\chi_0 & \text{horizon value of embedding function~$\chi$} \\  
u & \text{dimensionless radial AdS coordinate with~$0\le u\le 1$} \\  
\wn\equiv\frac{\omega}{2\pi T} & \text{dimensionless frequency} \\  
\qn\equiv\frac{q}{2\pi T} & \text{dimensionless spatial momentum} \\  
\mn\equiv\frac{\mu}{2\pi T} & \text{dimensionless chemical potential}
\end{eqnarray}
Furthermore, the indices~$B$ or~$I$ in~$d^B,\, d^I,\, \mu^B,\, \mu^I$ denote baryon or isospin charge densities 
and chemical potentials, respectively.
\end{appendix}  

\begin{thebibliography}{10%
0}

\bibitem{Erdmenger:2007ap}
J.~Erdmenger, M.~Kaminski, and F.~Rust, \emph{ Isospin diffusion in thermal
  AdS/CFT with flavor}, Phys. Rev. {\bf D76} (2007) 046001,
\href{http://www.slac.stanford.edu/spires/find/hep/www?texkey=Erdmenger:2007ap%
}{arXiv:0704.1290 [hep-th]}.

\bibitem{Erdmenger:2007ja}
J.~Erdmenger, M.~Kaminski, and F.~Rust, \emph{ {Holographic vector mesons from
  spectral functions at finite baryon or isospin density}}, Phys. Rev. {\bf
  D77} (2008) 046005,
\href{http://www.slac.stanford.edu/spires/find/hep/www?texkey=Erdmenger:2007ja%
}{arXiv:0710.0334 [hep-th]}.

\bibitem{Erdmenger:2008yj}
J.~Erdmenger, M.~Kaminski, P.~Kerner, and F.~Rust, \emph{ {Finite baryon and
  isospin chemical potential in AdS/CFT with flavor}},
\href{http://www.slac.stanford.edu/spires/find/hep/www?texkey=Erdmenger:2008yj%
}{arXiv:0807.2663 [hep-th]}.

\bibitem{Dusling:2008}
K.~Dusling, J.~Erdmenger, M.~Kaminski, F.~Rust, D.~Teaney, and C.~Young, \emph{
  {Quarkonium transport in thermal AdS/CFT}},
\href{http://www.slac.stanford.edu/spires/find/hep/www?texkey=Dusling:2008}{ar%
Xiv:0808.0957 [hep-th]}.

\bibitem{PDBook}
W.-M. {Yao}, C.~{Amsler}, D.~{Asner}, R.~{Barnett}, J.~{Beringer},
  P.~{Burchat}, C.~{Carone}, C.~{Caso}, O.~{Dahl}, G.~{D'Ambrosio},
  A.~{DeGouvea}, M.~{Doser}, S.~{Eidelman}, J.~{Feng}, T.~{Gherghetta},
  M.~{Goodman}, C.~{Grab}, D.~{Groom}, A.~{Gurtu}, K.~{Hagiwara}, K.~{Hayes},
  J.~{Hern\'andez-Rey}, K.~{Hikasa}, H.~{Jawahery}, C.~{Kolda}, K.~Y.,
  M.~{Mangano}, A.~{Manohar}, A.~{Masoni}, R.~{Miquel}, K.~{M\"onig},
  H.~{Murayama}, K.~{Nakamura}, S.~{Navas}, K.~{Olive}, L.~{Pape},
  C.~{Patrignani}, A.~{Piepke}, G.~{Punzi}, G.~{Raffelt}, J.~{Smith},
  M.~{Tanabashi}, J.~{Terning}, N.~{T\"ornqvist}, T.~{Trippe}, P.~{Vogel},
  T.~{Watari}, C.~{Wohl}, R.~{Workman}, P.~{Zyla}, B.~{Armstrong}, G.~{Harper},
  V.~{Lugovsky}, P.~{Schaffner}, M.~{Artuso}, K.~{Babu}, H.~{Band},
  E.~{Barberio}, M.~{Battaglia}, H.~{Bichsel}, O.~{Biebel}, P.~{Bloch},
  E.~{Blucher}, R.~{Cahn}, D.~{Casper}, A.~{Cattai}, A.~{Ceccucci},
  D.~{Chakraborty}, R.~{Chivukula}, G.~{Cowan}, T.~{Damour}, T.~{DeGrand},
  K.~{Desler}, M.~{Dobbs}, M.~{Drees}, A.~{Edwards}, D.~{Edwards}, V.~{Elvira},
  J.~{Erler}, V.~{Ezhela}, W.~{Fetscher}, B.~{Fields}, B.~{Foster},
  D.~{Froidevaux}, T.~{Gaisser}, L.~{Garren}, H.-J. {Gerber}, G.~{Gerbier},
  L.~{Gibbons}, F.~{Gilman}, G.~{Giudice}, A.~{Gritsan}, M.~{Gr\"unewald},
  H.~{Haber}, C.~{Hagmann}, I.~{Hinchliffe}, A.~{H\"ocker}, P.~{Igo-Kemenes},
  J.~{Jackson}, K.~{Johnson}, D.~{Karlen}, B.~{Kayser}, D.~{Kirkby},
  S.~{Klein}, K.~{Kleinknecht}, I.~{Knowles}, R.~{Kowalewski}, P.~{Kreitz},
  B.~{Krusche}, Y.~{Kuyanov}, O.~{Lahav}, P.~{Langacker}, A.~{Liddle},
  Z.~{Ligeti}, T.~{Liss}, L.~{Littenberg}, L.~{Liu}, K.~{Lugovsky},
  S.~{Lugovsky}, T.~{Mannel}, D.~{Manley}, W.~{Marciano}, A.~{Martin},
  D.~{Milstead}, M.~{Narain}, P.~{Nason}, Y.~{Nir}, J.~{Peacock}, S.~{Prell},
  A.~{Quadt}, S.~{Raby}, B.~{Ratcliff}, E.~{Razuvaev}, B.~{Renk},
  P.~{Richardson}, S.~{Roesler}, G.~{Rolandi}, M.~{Ronan}, L.~{Rosenberg},
  C.~{Sachrajda}, S.~{Sarkar}, M.~{Schmitt}, O.~{Schneider}, D.~{Scott},
  T.~{Sj\"ostrand}, G.~{Smoot}, P.~{Sokolsky}, S.~{Spanier}, H.~{Spieler},
  A.~{Stahl}, T.~{Stanev}, R.~{Streitmatter}, T.~{Sumiyoshi}, N.~{Tkachenko},
  G.~{Trilling}, G.~{Valencia}, K.~{van Bibber}, M.~{Vincter}, D.~{Ward},
  B.~{Webber}, J.~{Wells}, M.~{Whalley}, L.~{Wolfenstein}, J.~{Womersley},
  C.~{Woody}, A.~{Yamamoto}, O.~{Zenin}, J.~{Zhang}, and R.-Y. {Zhu}, \emph{
  {Review of Particle Physics}}, {Journal of Physics G} {\bf 33} (2006) 1+.

\bibitem{Polchinski:1998rq}
J.~Polchinski, \emph{ {String theory. Vol. 1: An introduction to the bosonic
  string}},. Cambridge, UK: Univ. Pr. (1998) 402 p.

\bibitem{Polchinski:1998rr}
J.~Polchinski, \emph{ {String theory. Vol. 2: Superstring theory and beyond}},.
  Cambridge, UK: Univ. Pr. (1998) 531 p.

\bibitem{Maldacena:1997re}
J.~M. Maldacena, \emph{ {The large N limit of superconformal field theories and
  supergravity}}, Adv. Theor. Math. Phys. {\bf 2} (1998) 231--252,
\href{http://www.slac.stanford.edu/spires/find/hep/www?texkey=Maldacena:1997re%
}{hep-th/9711200}.

\bibitem{Son:2007vk}
D.~T. Son and A.~O. Starinets, \emph{ {Viscosity, Black Holes, and Quantum
  Field Theory}}, Ann. Rev. Nucl. Part. Sci. {\bf 57} (2007) 95--118,
\href{http://www.slac.stanford.edu/spires/find/hep/www?texkey=Son:2007vk}{arXi%
v:0704.0240 [hep-th]}.

\bibitem{Gyulassy:2004zy}
M.~Gyulassy and L.~McLerran, \emph{ {New forms of QCD matter discovered at
  RHIC}}, Nucl. Phys. {\bf A750} (2005) 30--63,
\href{http://www.slac.stanford.edu/spires/find/hep/www?texkey=Gyulassy:2004zy}%
{nucl-th/0405013}.

\bibitem{Peskin:1995ev}
M.~E. Peskin and D.~V. Schroeder, \emph{ {An Introduction to quantum field
  theory}},. Reading, USA: Addison-Wesley (1995) 842 p.

\bibitem{Higgs:1964pj}
P.~W. Higgs, \emph{ {Broken symmetries and the masses of gauge bosons}}, Phys.
  Rev. Lett. {\bf 13} (1964)
508--509.

\bibitem{Djouadi:1998di}
{\bf MSSM Working Group} Collaboration, A.~Djouadi {\em et al.}, \emph{ {The
  minimal supersymmetric standard model: Group summary report}},
\href{http://www.slac.stanford.edu/spires/find/hep/www?texkey=Djouadi:1998di}{%
hep-ph/9901246}.

\bibitem{Hooper:2007qk}
D.~Hooper and S.~Profumo, \emph{ {Dark matter and collider phenomenology of
  universal extra dimensions}}, Phys. Rept. {\bf 453} (2007) 29--115,
\href{http://www.slac.stanford.edu/spires/find/hep/www?texkey=Hooper:2007qk}{h%
ep-ph/0701197}.

\bibitem{Calmet:2001na}
X.~Calmet, B.~Jurco, P.~Schupp, J.~Wess, and M.~Wohlgenannt, \emph{ {The
  standard model on non-commutative space-time}}, Eur. Phys. J. {\bf C23}
  (2002) 363--376,
\href{http://www.slac.stanford.edu/spires/find/hep/www?texkey=Calmet:2001na}{h%
ep-ph/0111115}.

\bibitem{Najafabadi:2008sa}
M.~M. Najafabadi, \emph{ {Noncommutative Standard Model in Top Quark Sector}},
\href{http://www.slac.stanford.edu/spires/find/hep/www?texkey=Najafabadi:2008s%
a}{0803.2340}.

\bibitem{Alboteanu:2007bp}
A.~Alboteanu, T.~Ohl, and R.~Ruckl, \emph{ {The Noncommutative Standard Model
  at $O(theta^2)$}}, Phys. Rev. {\bf D76} (2007) 105018,
\href{http://www.slac.stanford.edu/spires/find/hep/www?texkey=Alboteanu:2007bp%
}{0707.3595}.

\bibitem{Chamseddine:2007ia}
A.~H. Chamseddine and A.~Connes, \emph{ {Conceptual Explanation for the Algebra
  in the Noncommutative Approach to the Standard Model}}, Phys. Rev. Lett. {\bf
  99} (2007) 191601,
\href{http://www.slac.stanford.edu/spires/find/hep/www?texkey=Chamseddine:2007%
ia}{0706.3690}.

\bibitem{Buric:2007qx}
M.~Buric, D.~Latas, V.~Radovanovic, and J.~Trampetic, \emph{ {Nonzero Z $\to$
  gamma gamma decays in the renormalizable gauge sector of the noncommutative
  standard model}}, Phys. Rev. {\bf D75} (2007)
097701.

\bibitem{Georgi:2007ek}
H.~Georgi, \emph{ {Unparticle Physics}}, Phys. Rev. Lett. {\bf 98} (2007)
  221601,
\href{http://www.slac.stanford.edu/spires/find/hep/www?texkey=Georgi:2007ek}{h%
ep-ph/0703260}.

\bibitem{Ashtekar:2007tv}
A.~Ashtekar, \emph{ {An Introduction to Loop Quantum Gravity Through
  Cosmology}}, Nuovo Cim. {\bf 122B} (2007) 135--155,
\href{http://www.slac.stanford.edu/spires/find/hep/www?texkey=Ashtekar:2007tv}%
{gr-qc/0702030}.

\bibitem{Back:2004je}
B.~B. Back {\em et al.}, \emph{ {The PHOBOS perspective on discoveries at
  RHIC}}, Nucl. Phys. {\bf A757} (2005) 28--101,
\href{http://www.slac.stanford.edu/spires/find/hep/www?texkey=Back:2004je}{nuc%
l-ex/0410022}.

\bibitem{Fischer:2006xq}
W.~Fischer, \emph{ {RHIC operational status and upgrade plans}},. Prepared for
  European Particle Accelerator Conference (EPAC 06), Edinburgh, Scotland,
  26-30 Jun 2006.

\bibitem{Aharony:1999ti}
O.~Aharony, S.~S. Gubser, J.~M. Maldacena, H.~Ooguri, and Y.~Oz, \emph{ {Large
  N field theories, string theory and gravity}}, Phys. Rept. {\bf 323} (2000)
  183--386,
\href{http://www.slac.stanford.edu/spires/find/hep/www?texkey=Aharony:1999ti}{%
hep-th/9905111}.

\bibitem{Erdmenger:2007cm}
J.~Erdmenger, N.~Evans, I.~Kirsch, and E.~Threlfall, \emph{ {Mesons in
  Gauge/Gravity Duals - A Review}},
\href{http://www.slac.stanford.edu/spires/find/hep/www?texkey=Erdmenger:2007cm%
}{arXiv:0711.4467 [hep-th]}.

\bibitem{Policastro:2001yc}
G.~Policastro, D.~T. Son, and A.~O. Starinets, \emph{ The shear viscosity of
  strongly coupled N = 4 supersymmetric Yang-Mills plasma}, Phys. Rev. Lett.
  {\bf 87} (2001) 081601,
\href{http://www.slac.stanford.edu/spires/find/hep/www?texkey=Policastro:2001y%
c}{hep-th/0104066}.

\bibitem{Son:2002sd}
D.~T. Son and A.~O. Starinets, \emph{ Minkowski-space correlators in AdS/CFT
  correspondence: Recipe and applications}, JHEP {\bf 09} (2002) 042,
\href{http://www.slac.stanford.edu/spires/find/hep/www?texkey=Son:2002sd}{hep-%
th/0205051}.

\bibitem{Policastro:2002se}
G.~Policastro, D.~T. Son, and A.~O. Starinets, \emph{ From AdS/CFT
  correspondence to hydrodynamics}, JHEP {\bf 09} (2002) 043,
\href{http://www.slac.stanford.edu/spires/find/hep/www?texkey=Policastro:2002s%
e}{hep-th/0205052}.

\bibitem{Policastro:2002tn}
G.~Policastro, D.~T. Son, and A.~O. Starinets, \emph{ From AdS/CFT
  correspondence to hydrodynamics. II: Sound waves}, JHEP {\bf 12} (2002) 054,
\href{http://www.slac.stanford.edu/spires/find/hep/www?texkey=Policastro:2002t%
n}{hep-th/0210220}.

\bibitem{Herzog:2002pc}
C.~P. Herzog and D.~T. Son, \emph{ {Schwinger-Keldysh propagators from AdS/CFT
  correspondence}}, JHEP {\bf 03} (2003) 046,
\href{http://www.slac.stanford.edu/spires/find/hep/www?texkey=Herzog:2002pc}{h%
ep-th/0212072}.

\bibitem{Kovtun:2003wp}
P.~Kovtun, D.~T. Son, and A.~O. Starinets, \emph{ Holography and hydrodynamics:
  Diffusion on stretched horizons}, JHEP {\bf 10} (2003) 064,
\href{http://www.slac.stanford.edu/spires/find/hep/www?texkey=Kovtun:2003wp}{h%
ep-th/0309213}.

\bibitem{Kovtun:2004de}
P.~Kovtun, D.~T. Son, and A.~O. Starinets, \emph{ Viscosity in strongly
  interacting quantum field theories from black hole physics}, Phys. Rev. Lett.
  {\bf 94} (2005) 111601,
\href{http://www.slac.stanford.edu/spires/find/hep/www?texkey=Kovtun:2004de}{h%
ep-th/0405231}.

\bibitem{Teaney:2006nc}
D.~Teaney, \emph{ Finite temperature spectral densities of momentum and R-
  charge correlators in N = 4 Yang Mills theory}, Phys. Rev. {\bf D74} (2006)
  045025,
\href{http://www.slac.stanford.edu/spires/find/hep/www?texkey=Teaney:2006nc}{h%
ep-ph/0602044}.

\bibitem{Kovtun:2006pf}
P.~Kovtun and A.~Starinets, \emph{ Thermal spectral functions of strongly
  coupled N = 4 supersymmetric Yang-Mills theory}, Phys. Rev. Lett. {\bf 96}
  (2006) 131601,
\href{http://www.slac.stanford.edu/spires/find/hep/www?texkey=Kovtun:2006pf}{h%
ep-th/0602059}.

\bibitem{Son:2006em}
D.~T. Son and A.~O. Starinets, \emph{ Hydrodynamics of R-charged black holes},
  JHEP {\bf 03} (2006) 052,
\href{http://www.slac.stanford.edu/spires/find/hep/www?texkey=Son:2006em}{hep-%
th/0601157}.

\bibitem{Karch:2002sh}
A.~Karch and E.~Katz, \emph{ Adding flavor to AdS/CFT}, JHEP {\bf 06} (2002)
  043,
\href{http://www.slac.stanford.edu/spires/find/hep/www?texkey=Karch:2002sh}{he%
p-th/0205236}.

\bibitem{Babington:2003vm}
J.~Babington, J.~Erdmenger, N.~J. Evans, Z.~Guralnik, and I.~Kirsch, \emph{
  Chiral symmetry breaking and pions in non-supersymmetric gauge / gravity
  duals}, Phys. Rev. {\bf D69} (2004) 066007,
\href{http://www.slac.stanford.edu/spires/find/hep/www?texkey=Babington:2003vm%
}{hep-th/0306018}.

\bibitem{Kruczenski:2003be}
M.~Kruczenski, D.~Mateos, R.~C. Myers, and D.~J. Winters, \emph{ Meson
  spectroscopy in AdS/CFT with flavour}, JHEP {\bf 07} (2003) 049,
\href{http://www.slac.stanford.edu/spires/find/hep/www?texkey=Kruczenski:2003b%
e}{hep-th/0304032}.

\bibitem{Kruczenski:2003uq}
M.~Kruczenski, D.~Mateos, R.~C. Myers, and D.~J. Winters, \emph{ Towards a
  holographic dual of large-N(c) QCD}, JHEP {\bf 05} (2004) 041,
\href{http://www.slac.stanford.edu/spires/find/hep/www?texkey=Kruczenski:2003u%
q}{hep-th/0311270}.

\bibitem{Kirsch:2004km}
I.~Kirsch, \emph{ Generalizations of the AdS/CFT correspondence}, Fortsch.
  Phys. {\bf 52} (2004) 727--826,
\href{http://www.slac.stanford.edu/spires/find/hep/www?texkey=Kirsch:2004km}{h%
ep-th/0406274}.

\bibitem{Mateos:2006nu}
D.~Mateos, R.~C. Myers, and R.~M. Thomson, \emph{ Holographic phase transitions
  with fundamental matter}, Phys. Rev. Lett. {\bf 97} (2006) 091601,
\href{http://www.slac.stanford.edu/spires/find/hep/www?texkey=Mateos:2006nu}{h%
ep-th/0605046}.

\bibitem{Kobayashi:2006sb}
S.~Kobayashi, D.~Mateos, S.~Matsuura, R.~C. Myers, and R.~M. Thomson, \emph{
  Holographic phase transitions at finite baryon density}, JHEP {\bf 02} (2007)
  016,
\href{http://www.slac.stanford.edu/spires/find/hep/www?texkey=Kobayashi:2006sb%
}{hep-th/0611099}.

\bibitem{Ghoroku:2005kg}
K.~Ghoroku and M.~Yahiro, \emph{ Holographic model for mesons at finite
  temperature}, Phys. Rev. {\bf D73} (2006) 125010,
\href{http://www.slac.stanford.edu/spires/find/hep/www?texkey=Ghoroku:2005kg}{%
hep-ph/0512289}.

\bibitem{Maeda:2006by}
K.~Maeda, M.~Natsuume, and T.~Okamura, \emph{ Viscosity of gauge theory plasma
  with a chemical potential from AdS/CFT}, Phys. Rev. {\bf D73} (2006) 066013,
\href{http://www.slac.stanford.edu/spires/find/hep/www?texkey=Maeda:2006by}{he%
p-th/0602010}.

\bibitem{Peeters:2006iu}
K.~Peeters, J.~Sonnenschein, and M.~Zamaklar, \emph{ Holographic melting and
  related properties of mesons in a quark gluon plasma}, Phys. Rev. {\bf D74}
  (2006) 106008,
\href{http://www.slac.stanford.edu/spires/find/hep/www?texkey=Peeters:2006iu}{%
hep-th/0606195}.

\bibitem{Mateos:2006yd}
D.~Mateos, R.~C. Myers, and R.~M. Thomson, \emph{ Holographic viscosity of
  fundamental matter}, Phys. Rev. Lett. {\bf 98} (2007) 101601,
\href{http://www.slac.stanford.edu/spires/find/hep/www?texkey=Mateos:2006yd}{h%
ep-th/0610184}.

\bibitem{Nakamura:2006xk}
S.~Nakamura, Y.~Seo, S.-J. Sin, and K.~P. Yogendran, \emph{ {A new phase at
  finite quark density from AdS/CFT}},
\href{http://www.slac.stanford.edu/spires/find/hep/www?texkey=Nakamura:2006xk}%
{hep-th/0611021}.

\bibitem{Hoyos:2006gb}
C.~Hoyos-Badajoz, K.~Landsteiner, and S.~Montero, \emph{ Holographic Meson
  Melting}, JHEP {\bf 04} (2007) 031,
\href{http://www.slac.stanford.edu/spires/find/hep/www?texkey=Hoyos:2006gb}{he%
p-th/0612169}.

\bibitem{Amado:2007yr}
I.~Amado, C.~Hoyos-Badajoz, K.~Landsteiner, and S.~Montero, \emph{ {Residues of
  Correlators in the Strongly Coupled N=4 Plasma}}, Phys. Rev. {\bf D77} (2008)
  065004,
\href{http://www.slac.stanford.edu/spires/find/hep/www?texkey=Amado:2007yr}{07%
10.4458}.

\bibitem{Nakamura:2007nx}
S.~Nakamura, Y.~Seo, S.-J. Sin, and K.~P. Yogendran, \emph{ Baryon-charge
  Chemical Potential in AdS/CFT},
\href{http://www.slac.stanford.edu/spires/find/hep/www?texkey=Nakamura:2007nx}%
{arXiv:0708.2818 [hep-th]}.

\bibitem{Parnachev:2007bc}
A.~Parnachev, \emph{ Holographic QCD with Isospin Chemical Potential},
\href{http://www.slac.stanford.edu/spires/find/hep/www?texkey=Parnachev:2007bc%
}{arXiv:0708.3170 [hep-th]}.

\bibitem{Mateos:2007vc}
D.~Mateos, S.~Matsuura, R.~C. Myers, and R.~M. Thomson, \emph{ Holographic
  phase transitions at finite chemical potential}, JHEP {\bf 11} (2007) 085,
\href{http://www.slac.stanford.edu/spires/find/hep/www?texkey=Mateos:2007vc}{a%
rXiv:0709.1225 [hep-th]}.

\bibitem{Karch:2007br}
A.~Karch and A.~O'Bannon, \emph{ Holographic Thermodynamics at Finite Baryon
  Density: Some Exact Results}, JHEP {\bf 11} (2007) 074,
\href{http://www.slac.stanford.edu/spires/find/hep/www?texkey=Karch:2007br}{ar%
Xiv:0709.0570 [hep-th]}.

\bibitem{Ghoroku:2007re}
K.~Ghoroku, M.~Ishihara, and A.~Nakamura, \emph{ D3/D7 holographic Gauge theory
  and Chemical potential},
\href{http://www.slac.stanford.edu/spires/find/hep/www?texkey=Ghoroku:2007re}{%
arXiv:0708.3706 [hep-th]}.

\bibitem{Erdmenger:2007bn}
J.~Erdmenger, R.~Meyer, and J.~P. Shock, \emph{ AdS/CFT with Flavour in
  Electric and Magnetic Kalb-Ramond Fields},
\href{http://www.slac.stanford.edu/spires/find/hep/www?texkey=Erdmenger:2007bn%
}{arXiv:0709.1551 [hep-th]}.

\bibitem{Mateos:2007vn}
D.~Mateos, R.~C. Myers, and R.~M. Thomson, \emph{ Thermodynamics of the brane},
\href{http://www.slac.stanford.edu/spires/find/hep/www?texkey=Mateos:2007vn}{h%
ep-th/0701132}.

\bibitem{Mateos:2007yp}
D.~Mateos and L.~Patino, \emph{ Bright branes for strongly coupled plasmas},
  JHEP {\bf 11} (2007) 025,
\href{http://www.slac.stanford.edu/spires/find/hep/www?texkey=Mateos:2007yp}{a%
rXiv:0709.2168 [hep-th]}.

\bibitem{Aharony:2007uu}
O.~Aharony, K.~Peeters, J.~Sonnenschein, and M.~Zamaklar, \emph{ Rho meson
  condensation at finite isospin chemical potential in a holographic model for
  QCD},
\href{http://www.slac.stanford.edu/spires/find/hep/www?texkey=Aharony:2007uu}{%
arXiv:0709.3948 [hep-th]}.

\bibitem{Myers:2007we}
R.~C. Myers, A.~O. Starinets, and R.~M. Thomson, \emph{ Holographic spectral
  functions and diffusion constants for fundamental matter}, JHEP {\bf 11}
  (2007) 091,
\href{http://www.slac.stanford.edu/spires/find/hep/www?texkey=Myers:2007we}{ar%
Xiv:0706.0162 [hep-th]}.

\bibitem{Evans:2008tv}
N.~Evans and E.~Threlfall, \emph{ {Mesonic quasinormal modes of the
  Sakai-Sugimoto model at high temperature}},
\href{http://www.slac.stanford.edu/spires/find/hep/www?texkey=Evans:2008tv}{ar%
Xiv:0802.0775 [hep-th]}.

\bibitem{Myers:2008cj}
R.~C. Myers and A.~Sinha, \emph{ {The fast life of holographic mesons}},
\href{http://www.slac.stanford.edu/spires/find/hep/www?texkey=Myers:2008cj}{08%
04.2168}.

\bibitem{Splittorff:2000mm}
K.~Splittorff, D.~T. Son, and M.~A. Stephanov, \emph{ QCD-like theories at
  finite baryon and isospin density}, Phys. Rev. {\bf D64} (2001) 016003,
\href{http://www.slac.stanford.edu/spires/find/hep/www?texkey=Splittorff:2000m%
m}{hep-ph/0012274}.

\bibitem{Asakawa:2003xj}
M.~Asakawa and T.~Hatsuda, \emph{ {Charmonia above the deconfinement phase
  transition}}, Nucl. Phys. Proc. Suppl. {\bf 129} (2004) 584--586,
\href{http://www.slac.stanford.edu/spires/find/hep/www?texkey=Asakawa:2003xj}{%
hep-lat/0309001}.

\bibitem{Shuryak:2004tx}
E.~V. Shuryak and I.~Zahed, \emph{ {Towards a theory of binary bound states in
  the quark gluon plasma}}, Phys. Rev. {\bf D70} (2004) 054507,
\href{http://www.slac.stanford.edu/spires/find/hep/www?texkey=Shuryak:2004tx}{%
hep-ph/0403127}.

\bibitem{Mohaupt:2002py}
T.~Mohaupt, \emph{ {Introduction to string theory}}, Lect. Notes Phys. {\bf
  631} (2003) 173--251,
\href{http://www.slac.stanford.edu/spires/find/hep/www?texkey=Mohaupt:2002py}{%
hep-th/0207249}.

\bibitem{Witten:1995ex}
E.~Witten, \emph{ {String theory dynamics in various dimensions}}, Nucl. Phys.
  {\bf B443} (1995) 85--126,
\href{http://www.slac.stanford.edu/spires/find/hep/www?texkey=Witten:1995ex}{h%
ep-th/9503124}.

\bibitem{Horava:1995qa}
P.~Horava and E.~Witten, \emph{ {Heterotic and type I string dynamics from
  eleven dimensions}}, Nucl. Phys. {\bf B460} (1996) 506--524,
\href{http://www.slac.stanford.edu/spires/find/hep/www?texkey=Horava:1995qa}{h%
ep-th/9510209}.

\bibitem{Becker:2007zj}
K.~Becker, M.~Becker, and J.~H. Schwarz, \emph{ {String theory and M-theory: A
  modern introduction}},. Cambridge, UK: Cambridge Univ. Pr. (2007) 739 p.

\bibitem{Kirsch:2006he}
I.~Kirsch, \emph{ {Spectroscopy of fermionic operators in AdS/CFT}}, JHEP {\bf
  09} (2006) 052,
\href{http://www.slac.stanford.edu/spires/find/hep/www?texkey=Kirsch:2006he}{h%
ep-th/0607205}.

\bibitem{Polchinski:1995mt}
J.~Polchinski, \emph{ {Dirichlet-Branes and Ramond-Ramond Charges}}, Phys. Rev.
  Lett. {\bf 75} (1995) 4724--4727,
\href{http://www.slac.stanford.edu/spires/find/hep/www?texkey=Polchinski:1995m%
t}{hep-th/9510017}.

\bibitem{Nahm:1977tg}
W.~Nahm, \emph{ {Supersymmetries and their representations}}, Nucl. Phys. {\bf
  B135} (1978)
149.

\bibitem{Erdmenger:1996yc}
J.~Erdmenger and H.~Osborn, \emph{ {Conserved currents and the energy-momentum
  tensor in conformally invariant theories for general dimensions}}, Nucl.
  Phys. {\bf B483} (1997) 431--474,
\href{http://www.slac.stanford.edu/spires/find/hep/www?texkey=Erdmenger:1996yc%
}{hep-th/9605009}.

\bibitem{Carroll:2004st}
S.~M. Carroll, \emph{ {Spacetime and geometry: An introduction to general
  relativity}},. San Francisco, USA: Addison-Wesley (2004) 513 p.

\bibitem{tHooft:1993gx}
G.~'t~Hooft, \emph{ {Dimensional reduction in quantum gravity}},
\href{http://www.slac.stanford.edu/spires/find/hep/www?texkey=tHooft:1993gx}{g%
r-qc/9310026}.

\bibitem{Susskind:1994vu}
L.~Susskind, \emph{ {The World as a hologram}}, J. Math. Phys. {\bf 36} (1995)
  6377--6396,
\href{http://www.slac.stanford.edu/spires/find/hep/www?texkey=Susskind:1994vu}%
{hep-th/9409089}.

\bibitem{Pernici:1985ju}
M.~Pernici, K.~Pilch, and P.~van Nieuwenhuizen, \emph{ {Gauged N=8 D=5
  Supergravity}}, Nucl. Phys. {\bf B259} (1985)
460.

\bibitem{Gunaydin:1985cu}
M.~Gunaydin, L.~J. Romans, and N.~P. Warner, \emph{ {Compact and Noncompact
  Gauged Supergravity Theories in Five-Dimensions}}, Nucl. Phys. {\bf B272}
  (1986)
598.

\bibitem{Gibbons:1983aq}
G.~W. Gibbons, C.~M. Hull, and N.~P. Warner, \emph{ {The Stability of Gauged
  Supergravity}}, Nucl. Phys. {\bf B218} (1983)
173.

\bibitem{Maldacena:2001ss}
J.~M. Maldacena, G.~W. Moore, and N.~Seiberg, \emph{ {D-brane charges in
  five-brane backgrounds}}, JHEP {\bf 10} (2001) 005,
\href{http://www.slac.stanford.edu/spires/find/hep/www?texkey=Maldacena:2001ss%
}{hep-th/0108152}.

\bibitem{Klebanov:1997kc}
I.~R. Klebanov, \emph{ {World-volume approach to absorption by non-dilatonic
  branes}}, Nucl. Phys. {\bf B496} (1997) 231--242,
\href{http://www.slac.stanford.edu/spires/find/hep/www?texkey=Klebanov:1997kc}%
{hep-th/9702076}.

\bibitem{Gubser:1997yh}
S.~S. Gubser, I.~R. Klebanov, and A.~A. Tseytlin, \emph{ {String theory and
  classical absorption by three-branes}}, Nucl. Phys. {\bf B499} (1997)
  217--240,
\href{http://www.slac.stanford.edu/spires/find/hep/www?texkey=Gubser:1997yh}{h%
ep-th/9703040}.

\bibitem{Lee:1998bxa}
S.~Lee, S.~Minwalla, M.~Rangamani, and N.~Seiberg, \emph{ {Three-point
  functions of chiral operators in D = 4, N = 4 SYM at large N}}, Adv. Theor.
  Math. Phys. {\bf 2} (1998) 697--718,
\href{http://www.slac.stanford.edu/spires/find/hep/www?texkey=Lee:1998bxa}{hep%
-th/9806074}.

\bibitem{Constable:1999ch}
N.~R. Constable and R.~C. Myers, \emph{ {Exotic scalar states in the AdS/CFT
  correspondence}}, JHEP {\bf 11} (1999) 020,
\href{http://www.slac.stanford.edu/spires/find/hep/www?texkey=Constable:1999ch%
}{hep-th/9905081}.

\bibitem{Witten:1998zw}
E.~Witten, \emph{ {Anti-de Sitter space, thermal phase transition, and
  confinement in gauge theories}}, Adv. Theor. Math. Phys. {\bf 2} (1998)
  505--532,
\href{http://www.slac.stanford.edu/spires/find/hep/www?texkey=Witten:1998zw}{h%
ep-th/9803131}.

\bibitem{Das:1997gg}
A.~K. Das, \emph{ {Finite temperature field theory}},. Singapore, Singapore:
  World Scientific (1997) 404 p.

\bibitem{Skenderis:2002wp}
K.~Skenderis, \emph{ {Lecture notes on holographic renormalization}}, Class.
  Quant. Grav. {\bf 19} (2002) 5849--5876,
\href{http://www.slac.stanford.edu/spires/find/hep/www?texkey=Skenderis:2002wp%
}{hep-th/0209067}.

\bibitem{Brigante:2007nu}
M.~Brigante, H.~Liu, R.~C. Myers, S.~Shenker, and S.~Yaida, \emph{ {Viscosity
  Bound Violation in Higher Derivative Gravity}},
\href{http://www.slac.stanford.edu/spires/find/hep/www?texkey=Brigante:2007nu}%
{0712.0805}.

\bibitem{Brigante:2008gz}
M.~Brigante, H.~Liu, R.~C. Myers, S.~Shenker, and S.~Yaida, \emph{ {The
  Viscosity Bound and Causality Violation}},
\href{http://www.slac.stanford.edu/spires/find/hep/www?texkey=Brigante:2008gz}%
{0802.3318}.

\bibitem{Adare:2006nq}
{\bf PHENIX} Collaboration, A.~Adare {\em et al.}, \emph{ {Energy Loss and Flow
  of Heavy Quarks in Au+Au Collisions at $\sqrt{(s_{NN})} = 200$ GeV}}, Phys.
  Rev. Lett. {\bf 98} (2007) 172301,
\href{http://www.slac.stanford.edu/spires/find/hep/www?texkey=Adare:2006nq}{nu%
cl-ex/0611018}.

\bibitem{Romatschke:2007mq}
P.~Romatschke and U.~Romatschke, \emph{ {Viscosity Information from
  Relativistic Nuclear Collisions: How Perfect is the Fluid Observed at
  RHIC?}}, Phys. Rev. Lett. {\bf 99} (2007) 172301,
\href{http://www.slac.stanford.edu/spires/find/hep/www?texkey=Romatschke:2007m%
q}{0706.1522}.

\bibitem{Buchel:2004qq}
A.~Buchel, \emph{ {On universality of stress-energy tensor correlation
  functions in supergravity}}, Phys. Lett. {\bf B609} (2005) 392--401,
\href{http://www.slac.stanford.edu/spires/find/hep/www?texkey=Buchel:2004qq}{h%
ep-th/0408095}.

\bibitem{Starinets:2008fb}
A.~O. Starinets, \emph{ {Quasinormal spectrum and the black hole membrane
  paradigm}},
\href{http://www.slac.stanford.edu/spires/find/hep/www?texkey=Starinets:2008fb%
}{0806.3797}.

\bibitem{Baier:2007ix}
R.~Baier, P.~Romatschke, D.~T. Son, A.~O. Starinets, and M.~A. Stephanov,
  \emph{ {Relativistic viscous hydrodynamics, conformal invariance, and
  holography}}, JHEP {\bf 04} (2008) 100,
\href{http://www.slac.stanford.edu/spires/find/hep/www?texkey=Baier:2007ix}{07%
12.2451}.

\bibitem{Natsuume:2007ty}
M.~Natsuume and T.~Okamura, \emph{ {Causal hydrodynamics of gauge theory
  plasmas from AdS/CFT duality}}, Phys. Rev. {\bf D77} (2008) 066014,
\href{http://www.slac.stanford.edu/spires/find/hep/www?texkey=Natsuume:2007ty}%
{0712.2916}.

\bibitem{Natsuume:2007tz}
M.~Natsuume and T.~Okamura, \emph{ {Comment on ``Viscous hydrodynamics
  relaxation time from AdS/CFT correspondence'}},
\href{http://www.slac.stanford.edu/spires/find/hep/www?texkey=Natsuume:2007tz}%
{0712.2917}.

\bibitem{Natsuume:2008iy}
M.~Natsuume and T.~Okamura, \emph{ {A note on causal hydrodynamics for M-theory
  branes}},
\href{http://www.slac.stanford.edu/spires/find/hep/www?texkey=Natsuume:2008iy}%
{0801.1797}.

\bibitem{Erdmenger:2006bg}
J.~Erdmenger, N.~Evans, and J.~Grosse, \emph{ {Heavy-light mesons from the
  AdS/CFT correspondence}}, JHEP {\bf 01} (2007) 098,
\href{http://www.slac.stanford.edu/spires/find/hep/www?texkey=Erdmenger:2006bg%
}{hep-th/0605241}.

\bibitem{Erdmenger:2007vj}
J.~Erdmenger, K.~Ghoroku, and I.~Kirsch, \emph{ Holographic heavy-light mesons
  from non-Abelian DBI}, JHEP {\bf 09} (2007) 111,
\href{http://www.slac.stanford.edu/spires/find/hep/www?texkey=Erdmenger:2007vj%
}{arXiv:0706.3978 [hep-th]}.

\bibitem{Herzog:2008bp}
C.~P. Herzog, S.~A. Stricker, and A.~Vuorinen, \emph{ {Remarks on Heavy-Light
  Mesons from AdS/CFT}},
\href{http://www.slac.stanford.edu/spires/find/hep/www?texkey=Herzog:2008bp}{0%
802.2956}.

\bibitem{Evans:2007sf}
N.~Evans and A.~Tedder, \emph{ {A holographic model of hadronization}},
\href{http://www.slac.stanford.edu/spires/find/hep/www?texkey=Evans:2007sf}{07%
11.0300}.

\bibitem{Castorina:2008gf}
P.~Castorina, D.~Grumiller, and A.~Iorio, \emph{ {The Exact String Black-Hole
  behind the hadronic Rindler horizon?}},
\href{http://www.slac.stanford.edu/spires/find/hep/www?texkey=Castorina:2008gf%
}{0802.2286}.

\bibitem{Albash:2006ew}
T.~Albash, V.~Filev, C.~V. Johnson, and A.~Kundu, \emph{ A topology-changing
  phase transition and the dynamics of flavour},
\href{http://www.slac.stanford.edu/spires/find/hep/www?texkey=Albash:2006ew}{h%
ep-th/0605088}.

\bibitem{Albash:2006bs}
T.~Albash, V.~Filev, C.~V. Johnson, and A.~Kundu, \emph{ Global currents, phase
  transitions, and chiral symmetry breaking in large N(c) gauge theory},
\href{http://www.slac.stanford.edu/spires/find/hep/www?texkey=Albash:2006bs}{h%
ep-th/0605175}.

\bibitem{Mazu:2007tp}
V.~Mazu and J.~Sonnenschein, \emph{ {Non critical holographic models of the
  thermal phases of QCD}},
\href{http://www.slac.stanford.edu/spires/find/hep/www?texkey=Mazu:2007tp}{071%
1.4273}.

\bibitem{Dhar:2008um}
A.~Dhar and P.~Nag, \emph{ {Tachyon condensation and quark mass in modified
  Sakai- Sugimoto model}},
\href{http://www.slac.stanford.edu/spires/find/hep/www?texkey=Dhar:2008um}{080%
4.4807}.

\bibitem{Kovtun:2005ev}
P.~K. Kovtun and A.~O. Starinets, \emph{ Quasinormal modes and holography},
  Phys. Rev. {\bf D72} (2005) 086009,
\href{http://www.slac.stanford.edu/spires/find/hep/www?texkey=Kovtun:2005ev}{h%
ep-th/0506184}.

\bibitem{Bender}
C.~M. Bender and S.~Orszag, \emph{ Advanced mathematical methods for scientists
  and engineers},.

\bibitem{Landau}
L.~D. Landau and E.~M. Lifshitz, \emph{ {Fluid Mechanics}},. Pergamon Press
  (1959) 536 p.

\bibitem{Forster}
D.~Forster, \emph{ {Hydrodynamic Fluctuations, Broken Symmetry, and Correlation
  Functions}},. Frontiers in physics; 47 (1983) 326 p.

\bibitem{Kapusta:1989tk}
J.~I. Kapusta, \emph{ Finite temperature field theory},.

\bibitem{Hosoya:1983id}
A.~Hosoya, M.-a. Sakagami, and M.~Takao, \emph{ Nonequilibrium thermodynamics
  in field theory: Transport coefficients}, Ann. Phys. {\bf 154} (1984)
229.

\bibitem{Zubarev:1974}
D.~N. Zubarev, \emph{ {Noneuqilibrium statistical thermodynamics}},. New York,
  Consultants Bureau; Studies in Soviet science (1974) 489 p.

\bibitem{Horowitz:1999jd}
G.~T. Horowitz and V.~E. Hubeny, \emph{ {Quasinormal modes of AdS black holes
  and the approach to thermal equilibrium}}, Phys. Rev. {\bf D62} (2000)
  024027,
\href{http://www.slac.stanford.edu/spires/find/hep/www?texkey=Horowitz:1999jd}%
{hep-th/9909056}.

\bibitem{Freedman:1998tz}
D.~Z. Freedman, S.~D. Mathur, A.~Matusis, and L.~Rastelli, \emph{ Correlation
  functions in the CFT($d$)/AdS($d+1$) correspondence}, Nucl. Phys. {\bf B546}
  (1999) 96--118,
\href{http://www.slac.stanford.edu/spires/find/hep/www?texkey=Freedman:1998tz}%
{hep-th/9804058}.

\bibitem{Georgi:1982jb}
H.~Georgi, \emph{ {Lie algebras in particle physics. From isospin to unified
  theories}}, {Front. Phys.} {\bf 54} (1982)
1--255.

\bibitem{Bohm:2001yx}
M.~Bohm, A.~Denner, and H.~Joos, \emph{ {Gauge theories of the strong and
  electroweak interaction}},. Stuttgart, Germany: Teubner (2001) 784 p.

\bibitem{Kerner:2008diploma}
P.~Kerner, \emph{ {Diploma thesis}},. work in progress.

\bibitem{Yip:1980}
J.~P. Boon and S.~Yip, \emph{ {Molecular hydrodynamics}},. Dover Publications
  Inc., New York (1991) 417p.

\bibitem{Ebert:2008us}
D.~Ebert, K.~G. Klimenko, A.~V. Tyukov, and V.~C. Zhukovsky, \emph{ {Finite
  size effects in the Gross-Neveu model with isospin chemical potential}},
\href{http://www.slac.stanford.edu/spires/find/hep/www?texkey=Ebert:2008us}{08%
04.4826}.

\bibitem{Rust:phd}
F.~Rust, \emph{ {PhD thesis}},. work in progress.

\bibitem{He:2005nk}
L.-y. He, M.~Jin, and P.-f. Zhuang, \emph{ Pion superfluidity and meson
  properties at finite isospin density}, Phys. Rev. {\bf D71} (2005) 116001,
\href{http://www.slac.stanford.edu/spires/find/hep/www?texkey=He:2005nk}{hep-p%
h/0503272}.

\bibitem{Chang:2007sr}
S.~Chang, J.~Liu, and P.~Zhuang, \emph{ Nucleon mass splitting at finite
  isospin chemical potential},
\href{http://www.slac.stanford.edu/spires/find/hep/www?texkey=Chang:2007sr}{nu%
cl-th/0702032}.

\bibitem{778835}
R.~G. Abdel-Rahman, \emph{ Propagation of boundary of inhomogeneous heat
  conduction equation}, Appl. Math. Comput. {\bf 141} (2003), no.~2-3,
  231--239.

\bibitem{EKKR:2008tp}
J.~Erdmenger, M.~Kaminski, P.~Kerner, and F.~Rust, \emph{ {Interpretation of
  peaks in holographic spectral functions and their movement}},. work in
  progress.

\bibitem{Kogut:2002tm}
J.~B. Kogut and D.~K. Sinclair, \emph{ Quenched lattice QCD at finite isospin
  density and related theories}, Phys. Rev. {\bf D66} (2002) 014508,
\href{http://www.slac.stanford.edu/spires/find/hep/www?texkey=Kogut:2002tm}{he%
p-lat/0201017}.

\bibitem{Kogut:2004zg}
J.~B. Kogut and D.~K. Sinclair, \emph{ The finite temperature transition for
  2-flavor lattice QCD at finite isospin density}, Phys. Rev. {\bf D70} (2004)
  094501,
\href{http://www.slac.stanford.edu/spires/find/hep/www?texkey=Kogut:2004zg}{he%
p-lat/0407027}.

\bibitem{Toublan:2003tt}
D.~Toublan and J.~B. Kogut, \emph{ Isospin chemical potential and the QCD phase
  diagram at nonzero temperature and baryon chemical potential}, Phys. Lett.
  {\bf B564} (2003) 212--216,
\href{http://www.slac.stanford.edu/spires/find/hep/www?texkey=Toublan:2003tt}{%
hep-ph/0301183}.

\bibitem{Herzog:2006gh}
C.~P. Herzog, A.~Karch, P.~Kovtun, C.~Kozcaz, and L.~G. Yaffe, \emph{ Energy
  loss of a heavy quark moving through N = 4 supersymmetric Yang-Mills plasma},
  JHEP {\bf 07} (2006) 013,
\href{http://www.slac.stanford.edu/spires/find/hep/www?texkey=Herzog:2006gh}{h%
ep-th/0605158}.

\bibitem{CasalderreySolana:2006rq}
J.~Casalderrey-Solana and D.~Teaney, \emph{ {Heavy quark diffusion in strongly
  coupled N = 4 Yang Mills}}, Phys. Rev. {\bf D74} (2006) 085012,
\href{http://www.slac.stanford.edu/spires/find/hep/www?texkey=CasalderreySolan%
a:2006rq}{hep-ph/0605199}.

\bibitem{Gubser:2006bz}
S.~S. Gubser, \emph{ {Drag force in AdS/CFT}}, Phys. Rev. {\bf D74} (2006)
  126005,
\href{http://www.slac.stanford.edu/spires/find/hep/www?texkey=Gubser:2006bz}{h%
ep-th/0605182}.

\bibitem{Bielcik:2005wu}
{\bf STAR} Collaboration, J.~Bielcik, \emph{ {Centrality dependence of heavy
  flavor production from single electron measurement in s(NN)**(1/2) = 200-GeV
  Au + Au collisions}}, Nucl. Phys. {\bf A774} (2006) 697--700,
\href{http://www.slac.stanford.edu/spires/find/hep/www?texkey=Bielcik:2005wu}{%
nucl-ex/0511005}.

\bibitem{Adare:2008sh}
{\bf PHENIX} Collaboration, A.~Adare {\em et al.}, \emph{ {J/psi Production in
  $\sqrt{(s_{NN})}= 200$ GeV Cu+Cu Collisions}},
\href{http://www.slac.stanford.edu/spires/find/hep/www?texkey=Adare:2008sh}{08%
01.0220}.

\bibitem{Adare:2006ns}
{\bf PHENIX} Collaboration, A.~Adare {\em et al.}, \emph{ {J/psi production vs
  centrality, transverse momentum, and rapidity in Au + Au collisions at
  s(NN)**(1/2) = 200- GeV}}, Phys. Rev. Lett. {\bf 98} (2007) 232301,
\href{http://www.slac.stanford.edu/spires/find/hep/www?texkey=Adare:2006ns}{nu%
cl-ex/0611020}.

\bibitem{Adler:2005ph}
{\bf PHENIX} Collaboration, S.~S. Adler {\em et al.}, \emph{ {J/psi production
  and nuclear effects for d + Au and p + p collisions at s(NN)**(1/2) =
  200-GeV}}, Phys. Rev. Lett. {\bf 96} (2006) 012304,
\href{http://www.slac.stanford.edu/spires/find/hep/www?texkey=Adler:2005ph}{nu%
cl-ex/0507032}.

\bibitem{Arnaldi:2006ee}
{\bf NA60} Collaboration, R.~Arnaldi {\em et al.}, \emph{ {Anomalous J/psi
  suppression in In-In collisions at 158- GeV/nucleon}}, Nucl. Phys. {\bf A774}
  (2006)
711--714.

\bibitem{Luke:1992tm}
M.~E. Luke, A.~V. Manohar, and M.~J. Savage, \emph{ A QCD Calculation of the
  interaction of quarkonium with nuclei}, Phys. Lett. {\bf B288} (1992)
  355--359,
\href{http://www.slac.stanford.edu/spires/find/hep/www?texkey=Luke:1992tm}{hep%
-ph/9204219}.

\bibitem{Peskin:1979va}
M.~E. Peskin, \emph{ {Short Distance Analysis for Heavy Quark Systems. 1.
  Diagrammatics}}, Nucl. Phys. {\bf B156} (1979)
365.

\bibitem{Bhanot:1979vb}
G.~Bhanot and M.~E. Peskin, \emph{ {Short Distance Analysis for Heavy Quark
  Systems. 2. Applications}}, Nucl. Phys. {\bf B156} (1979)
391.

\bibitem{Liu:1999fc}
H.~Liu and A.~A. Tseytlin, \emph{ D3-brane D-instanton configuration and N = 4
  super YM theory in constant self-dual background}, Nucl. Phys. {\bf B553}
  (1999) 231--249,
\href{http://www.slac.stanford.edu/spires/find/hep/www?texkey=Liu:1999fc}{hep-%
th/9903091}.

\bibitem{Ghoroku:2004sp}
K.~Ghoroku and M.~Yahiro, \emph{ {Chiral symmetry breaking driven by dilaton}},
  Phys. Lett. {\bf B604} (2004) 235--241,
\href{http://www.slac.stanford.edu/spires/find/hep/www?texkey=Ghoroku:2004sp}{%
hep-th/0408040}.

\bibitem{Hartnoll:2007ip}
S.~A. Hartnoll and C.~P. Herzog, \emph{ {Ohm's Law at strong coupling: S
  duality and the cyclotron resonance}}, Phys. Rev. {\bf D76} (2007) 106012,
\href{http://www.slac.stanford.edu/spires/find/hep/www?texkey=Hartnoll:2007ip}%
{0706.3228}.

\bibitem{Arnold:2000dr}
P.~Arnold, G.~D. Moore, and L.~G. Yaffe, \emph{ {Transport coefficients in high
  temperature gauge theories. I: Leading-log results}}, JHEP {\bf 11} (2000)
  001,
\href{http://www.slac.stanford.edu/spires/find/hep/www?texkey=Arnold:2000dr}{h%
ep-ph/0010177}.

\bibitem{Arnold:2003zc}
P.~Arnold, G.~D. Moore, and L.~G. Yaffe, \emph{ {Transport coefficients in high
  temperature gauge theories. II: Beyond leading log}}, JHEP {\bf 05} (2003)
  051,
\href{http://www.slac.stanford.edu/spires/find/hep/www?texkey=Arnold:2003zc}{h%
ep-ph/0302165}.

\bibitem{Paredes:2008nf}
A.~Paredes, K.~Peeters, and M.~Zamaklar, \emph{ {Mesons versus quasi-normal
  modes: undercooling and overheating}}, JHEP {\bf 05} (2008) 027,
\href{http://www.slac.stanford.edu/spires/find/hep/www?texkey=Paredes:2008nf}{%
0803.0759}.

\end{thebibliography}
\providecommand{\href}[2]{#2}\begingroup\raggedright\endgroup

\end{document}